\pdfoutput=1
\documentclass[12pt]{article}
\usepackage{tikz}

\usetikzlibrary{decorations.pathmorphing}

\usepackage{jheppub}

\usepackage{pgfplots}
\usepackage{pgfplotstable}
\usepackage{pgfkeys}
\allowdisplaybreaks
\usepackage{ytableau}
\usepackage{comment}
\usepackage[vcentermath]{youngtab}

\usepackage{bm}

\newcommand\be{\begin{equation}}
\newcommand\ee{\end{equation}}

\newcommand\bt{\mathfrak{t}}
\usepackage{braket}
\usepackage{subfigure}
\newcommand\tu{\widetilde{u}}

\title{ADHM Wilson line defect indices}
\abstract{
The Coulomb and Higgs indices of the 3d $\mathcal{N}=4$ $U(N)$ ADHM theories can be decorated by line defect operators as the line defect correlators. 
We obtain exact closed-form expressions and various non-trivial algebraic relations for the correlators of the Wilson lines in the fundamental and (anti)symmetric representations 
by means of the Hall-Littlewood expansion, the Fermi-gas method and the residue calculation. 
From the large $N$ limit of the correlators we obtain the single particle gravity indices which are expected to encode the spectra of fluctuation modes on the gravity duals of line operators in M2-brane SCFTs.
}
\author[a]{Hirotaka Hayashi,}
\emailAdd{h.hayashi@tokai.ac.jp}
\affiliation[a]{Department of Physics, School of Science, Tokai University,\\
4-1-1 Kitakaname, Hiratsuka-shi, Kanagawa 259-1292, Japan}
\author[b]{Tomoki Nosaka}
\emailAdd{nosaka@yukawa.kyoto-u.ac.jp}
\affiliation[b]{Kavli Institute for Theoretical Sciences, University of Chinese Academy of Sciences,\\
No.~3, Nanyitiao, Zhongguancun, Haidian District, Beijing 100190, China}
\author[c]{and Tadashi Okazaki}
\emailAdd{tokazaki@seu.edu.cn}
\affiliation[c]{
Shing-Tung Yau Center of Southeast University,\\
Yifu Architecture Building, No.2 Sipailou, Xuanwu district, Nanjing, Jiangsu, 210096, China}

\begin{document}
\maketitle

\ytableausetup{boxsize=1.5mm}

\section{Introduction and summary}
The supersymmetric indices of 3d  
supersymmetric field theories encode the protected spectra of BPS local operators \cite{Bhattacharya:2008zy,Bhattacharya:2008bja,Kim:2009wb,Imamura:2011su,Kapustin:2011jm,Dimofte:2011py}. 
They can be viewed as supersymmetric partition functions on $S^1\times S^2$ from the UV descriptions. 
One can decorate them as correlation functions of line operators 
by introducing BPS line operators wrapping the $S^1$ and localized on the $S^2$ \cite{Drukker:2012sr}, 
which we call the \textit{line defect indices}. 
The line defect indices can be made topological in that they do not depend on locations of the BPS line operators in certain delicate setups with enhanced supersymmetry. 
They can also encode the spectra of BPS local operators living at junctions of line operators. The spectra of the theories in the presence of line operators is particularly attractive 
in the study of dualities of field theories with line operators and the spectra of quantum fluctuations on the gravity dual geometries. 
The derivation of the exact closed-form expressions of line defect indices is highly desirable. 

In this paper we study the line defect indices of the 3d $\mathcal{N}=4$ $U(N)$ ADHM theory, 
i.e.~3d $\mathcal{N}=4$ gauge theory of gauge group $U(N)$ with an adjoint hypermultiplet and $l$ fundamental hypermultiplets \cite{deBoer:1996mp,deBoer:1996ck}.
The theory describes the low-energy dynamics of a stack of $N$ M2-branes probing $\mathbb{C}^2\times (\mathbb{C}^2/\mathbb{Z}_l)$ 
and it is holographically dual to the M-theory geometry $AdS_4\times S^7/\mathbb{Z}_l$ \cite{Benini:2009qs,Bashkirov:2010kz}. 
The BPS line operators can preserve one-dimensional superconformal symmetry group whose bosonic subgroup is $SL(2,\mathbb{R})$. 
The detailed information about the spectra of the theory is expected to reveal quantum fluctuations on the gravity dual M-theory geometry with the $AdS_2$ factor 
(see e.g.~\cite{Drukker:2008jm,Drukker:2008zx,Rey:2008bh,Farquet:2013cwa,Chen:2014gta,Aguilera-Damia:2014bqa,Muck:2016hda,Cookmeyer:2016dln,
Lietti:2017gtc,David:2019lhr,Correa:2019rdk,Giombi:2020mhz,Giombi:2023vzu} for the study of the gravity duals of Wilson lines in the M2-brane SCFTs). 
Extended $\mathcal{N}\ge4$ supersymmetry admits two half-BPS limits of supersymmetric indices, the Coulomb limit and the Higgs limit \cite{Razamat:2014pta}  
in which the indices become the Hilbert series enumerating the Coulomb branch operators and Higgs branch operators. 
Our overarching goal in this paper is to derive exact closed-form expressions 
and algebraic relations of the Wilson line defect indices of the 3d $\mathcal{N}=4$ $U(N)$ ADHM theory in the Coulomb and Higgs limits. 







We investigate the closed-form expressions for the line defect indices in the Coulomb/Higgs limit through various approaches.
The first method is the Hall-Littlewood expansion \cite{MR1354144,Crew:2020psc,Dorey:2016mxm,Barns-Graham:2017zpv,Okazaki:2017sbc} which is applicable for any representations in both limits.
In this method we reduce the integration over the gauge holonomies to discrete infinite summations by expanding the integrand with respect to the Hall-Littlewood functions.
In many examples we consider in this paper, the infinite summations can be performed explicitly and hence gives the closed-form expressions for the line defect indices.

It is also convenient to consider the generating function for the line defect indices with respect to the representations, 
which enables us to study the closed-form expressions more directly through the following methods.
In the Coulomb limit, we can reformulate the holonomy integrations into 
the canonical partition function of a one-dimensional quantum mechanics of ideal Fermi gas on a circle with a common one-particle density matrix for all topological charges \cite{Hayashi:2022ldo} (see also \cite{Kazakov:1998ji,Marino:2011eh,Bourdier:2015wda,Klemm:2012ii,Mezei:2013gqa,Chester:2020jay,Gaiotto:2020vqj,Hatsuda:2021oxa,Hatsuda:2022xdv,Hatsuda:2023iwi}).
In this way the generating function is expressed as a finite sum of the partition functions of Fermi-gas with different particle numbers, each of which can be calculated in a closed-form.

In the Higgs limit, the generating function can be evaluated by applying the Jeffery-Kirwan (JK) residue prescription \cite{MR1318878}, 
which again results into a finite sums over the poles. 
We also discover a purely combinatorial formula to classify the poles and evaluate the residues which is a slight generalization of the Nekrasov formula for the instanton partition function of 5d ${\cal N}=1$ Yang-Mills theories
\cite{Nekrasov:2002qd,Nekrasov:2003rj,Hwang:2014uwa}.
This enables us to obtain the closed-form expressions for higher ranks $N$ and flavors $l$ which are practically hard in the Hall-Littlewood expansion.
Moreover, when the number of flavors is $l=1$, by making use of the (refined) topological vertex formalism \cite{Iqbal:2002we,Aganagic:2003db,Awata:2005fa,Iqbal:2007ii,Taki:2007dh} for the instanton partition function, 
we can further obtain the generating function for line defect indices both with respect to the representations and the ranks $N$.

\subsection{Main results}
We present exact closed-form expressions of the flavored Wilson line defect indices for the ADHM theory in the main text 
while we list those for the unflavored indices in Appendix \ref{app_closedform}. 
On the other hand, here we summarize other remarkable features of the line defect indices we find in this work.
For the notations, see section \ref{sec_general}.

\subsubsection{Algebraic relations}
The Wilson line defect indices turn out to satisfy algebraic relations.
Here we list some outstanding relations.
In the Coulomb limit we find
\begin{align}
&\langle W_{(1^k)}W_{(\overline{1^k})}\rangle^{U(N)\text{ADHM-}[l](C)}(z;\mathfrak{t})=
{\cal I}^{U(k)\text{ ADHM-}[l](C)}(z;\mathfrak{t})
{\cal I}^{U(N-k)\text{ ADHM-}[l](C)}(z;\mathfrak{t}).
\label{factorizationC}
\end{align}
On the other hand, in the Higgs limit we find
\begin{align}
&\langle W_{(1^k)}\rangle^{U(N)\text{ ADHM-}[1](H)}(x;\mathfrak{t})
=\langle W_{(1^k)}\rangle^{U(k)\text{ ADHM-}[1](H)}(x;\mathfrak{t})
{\cal I}^{U(N-k)\text{ ADHM-}[1](H)}(x;\mathfrak{t}),\label{factorizationH} \\
&\langle W_{(1^N)}\rangle^{U(N)\text{ ADHM-}[1](H)}(x;\mathfrak{t})\nonumber \\
&=-(-\mathfrak{t})^N{\cal I}^{U(N)\text{ ADHM-}[1](H)}(x;\mathfrak{t})\nonumber \\
&\quad -\sum_{n=1}^{N-1} (-\mathfrak{t})^n{\cal I}^{U(n)\text{ ADHM-}[1](H)}(x;\mathfrak{t})
\langle W_{(1^{N-n})}\rangle^{U(N-n)\text{ ADHM-}[1](H)}(x;\mathfrak{t}),\label{Higgsl1algebraicreln1}
\end{align}
and
\begin{align}
&\sum_{\substack{N_1,N_2\\
0\le N_1\le \text{min}(k_1,N-k_2)\\
0\le N_2\le \text{min}(N-k_1,k_2)\\
N_1+N_2\le N
}}
(-\mathfrak{t})^{N_1+N_2}
\langle
W_{(1^{k_1-N_1})}
W_{\overline{(1^{k_2-N_2})}}
\rangle^{U(N-N_1-N_2)\text{ ADHM-}[1](H)}(x;\mathfrak{t})\nonumber \\
&\quad\quad\quad\quad\times {\cal I}^{U(N_1)\text{ ADHM-}[1](H)}(x;\mathfrak{t})
{\cal I}^{U(N_2)\text{ ADHM-}[1](H)}(x;\mathfrak{t})\nonumber \\
&=\delta_{k_1,k_2}
{\cal I}^{U(k_1)\text{ ADHM-}[1](H)}(x;\mathfrak{t})
{\cal I}^{U(N-k_1)\text{ ADHM-}[1](H)}(x;\mathfrak{t}).
\label{Higgsl1algebraicreln2}
\end{align}
Here we have set $y_1=1$
since the $y_1$-dependences of the correlators are trivial as
$
\langle W_\lambda W_{\overline{\rho}}\rangle^{U(N)\text{ ADHM-}[1](H)}(x,y_1;\mathfrak{t})
=
y_1^{-|\lambda|+|\rho|}
\langle W_\lambda W_{\overline{\rho}}\rangle^{U(N)\text{ ADHM-}[1](H)}(x;\mathfrak{t})
$.
Note that the first two relations \eqref{factorizationH} and \eqref{Higgsl1algebraicreln1} in the Higgs limit follows from the third relation \eqref{Higgsl1algebraicreln2}.

These relations \eqref{factorizationC}-\eqref{Higgsl1algebraicreln2} assert that even though the Wilson lines transform in the irreducible representations of gauge group, 
the associated line defect indices can be factorized into some irreducible ones. 

\subsubsection{Large $N$ limits}
The exact closed-form expressions of the large $N$ (connected) $2$-point functions of the Wilson lines in a pair of conjugate representations are particularly interesting 
as they are expected to capture the quantum fluctuation modes on the gravity dual geometries. 

For a fundamental representation one expects that the Wilson line is dual to an M2-brane wrapping 
the $AdS_2$ in the global $AdS_4$ and the M-theory circle. 
We obtain 
\begin{align}
&\langle {\cal W}_{\ydiagram{1}} {\cal W}_{\overline{\ydiagram{1}}}\rangle^{U(\infty)\text{ ADHM-}[l](C)}(z;\mathfrak{t})
=\frac{1-\mathfrak{t}^{2l}}{(1-\mathfrak{t}^2)(1-\mathfrak{t}^lz^l)(1-\mathfrak{t}^lz^{-l})},\\
&\langle {\cal W}_{\ydiagram{1}} {\cal W}_{\overline{\ydiagram{1}}}\rangle_{c}^{U(\infty)\text{ ADHM-}[l](H)}(x;\mathfrak{t})\nonumber \\
&
:=\langle {\cal W}_{\ydiagram{1}} {\cal W}_{\overline{\ydiagram{1}}}\rangle^{U(\infty)\text{ ADHM-}[l](H)}(x,y_\alpha;\mathfrak{t})\nonumber \\
&\quad
-\langle {\cal W}_{\ydiagram{1}}\rangle^{U(\infty)\text{ ADHM-}[l](H)}(x,y_\alpha;\mathfrak{t})
\langle
{\cal W}_{\overline{\ydiagram{1}}}\rangle^{U(\infty)\text{ ADHM-}[l](H)}(x,y_\alpha;\mathfrak{t})\nonumber \\
&=\frac{1}{(1-\mathfrak{t}x)(1-\mathfrak{t}x^{-1})}. 
\end{align}
They encode a finite number of quantum fluctuation fields on the gravity dual geometry. 
In particular, when $l=1$, the theory is self-mirror and it is dual to the ABJM theory \cite{Hosomichi:2008jd,Aharony:2008ug}. 
Both of the gravity indices resulting from the Coulomb and Higgs limits enumerate two bosonic excitations, 
which would correspond to the quantum fluctuations on the quantum M2-brane wrapping the $AdS_2$. 

As the rank of representation of Wilson line increases, the degeneracy of states grows. 
In particular, we find that 
the large $N$ normalized $2$-point function of the Wilson line in the rank-$k$ antisymmetric representation for the $U(N)$ ADHM theory agrees with 
the Coulomb index of the $U(k)$ ADHM theory 
\begin{align}
\langle {\cal W}_{(1^k)} {\cal W}_{\overline{(1^k)}}\rangle^{U(\infty)\text{ ADHM-}[l](C)}(z;\mathfrak{t})
&=\mathcal{I}^{U(k)\text{ ADHM-}[l](C)}(z;\mathfrak{t}). 
\end{align}
The large $N$ normalized $2$-point function of the Wilson lines in the rank-$k$ (anti)symmetric representation in $\mathcal{N}=8$ $U(N)$ SYM theory coincides with 
the Higgs index (or equivalently Coulomb index) of the $U(k)$ ADHM theory with one flavor
\begin{align}
\langle \mathcal{W}_{(k)}\mathcal{W}_{\overline{(k)}}\rangle^{\textrm{$\mathcal{N}=8$ $U(\infty) (H)$}}(x;\mathfrak{t})
&=\langle \mathcal{W}_{(1^k)}\mathcal{W}_{\overline{(1^k)}}\rangle^{\textrm{$\mathcal{N}=8$ $U(\infty) (H)$}}(x;\mathfrak{t})
\nonumber\\
&=\mathcal{I}^{\textrm{$U(k)$ ADHM-$[1] (H)$}}(x;\mathfrak{t}). 
\end{align}
Also we find that the large $N$ normalized $2$-point function of the Wilson line in the rank-$k$ antisymmetric representation for the $U(N)$ ADHM theory with one flavor is given by
\begin{align}
&
\langle {\cal W}_{(1^k)} {\cal W}_{\overline{(1^k)}}\rangle^{U(\infty)\text{ ADHM-}[1](H)}(x;\mathfrak{t})
\nonumber\\
&=
\mathcal{I}^{U(k)\text{ ADHM-}[1](H)}(x;\mathfrak{t})
+
\sum_{l=1}^{k}
\langle W_{(1^l)} \rangle^{U(k)\text{ ADHM-}[1](H)}(x;\mathfrak{t})
\langle W_{(1^l)} \rangle^{U(l)\text{ ADHM-}[1](H)}(x;\mathfrak{t}), 
\end{align}
where we have set $y_1=1$.
In the large representation limit, we find 
\begin{align}
&
\langle {\cal W}_{(1^{\infty})} {\cal W}_{\overline{(1^{\infty})}}\rangle^{U(\infty)\text{ ADHM-}[1](H)}(x;\mathfrak{t})
\nonumber\\
&=
\mathcal{I}^{U(\infty)\text{ ADHM-}[1](H)}(x;\mathfrak{t})
\left(
1+
\sum_{l=1}^{\infty}
{\langle \mathcal{W}_{(1^l)} \rangle^{U(\infty)\text{ ADHM-}[1](H)}(x;\mathfrak{t})}^2
\right). 
\end{align}
Besides, we find that the large representation limit of the large $N$ normalized $2$-point function of the symmetric Wilson lines agrees with 
\begin{align}
&
\langle {\cal W}_{(\infty)} {\cal W}_{\overline{(\infty)}}\rangle^{U(\infty)\text{ ADHM-}[1](H)}(x;\mathfrak{t})
\nonumber\\
&=
\mathcal{I}^{U(\infty)\text{ ADHM-}[1](H)}(x;\mathfrak{t})
\left(
1+
\sum_{l=1}^{\infty}
{\langle \mathcal{W}_{(l)} \rangle^{U(\infty)\text{ ADHM-}[1](H)}(x;\mathfrak{t})}^2
\right). 
\end{align}
Since the large representation limits of the large $N$ normalized $2$-point functions contain the large $N$ Higgs or Coulomb indices, 
the spectra of quantum fluctuations around the geometries 
dual to the Wilson lines in the large antisymmetric and symmetric representations contain an infinite tower of excitations. 

\subsection{Structure}
The structure of this paper is straightforward. 
In section \ref{sec_general} we introduce the Wilson line defect indices of 3d $\mathcal{N}=4$ $U(N)$ ADHM theory 
which decorate the supersymmetric indices. 
In section \ref{sec_Coulomb} we study the Coulomb limit of the line defect indices 
by using the Hall-Littlewood expansions and the Fermi-gas method. 
In section \ref{sec_Higgs} we examine the Higgs limit of the  line defect indices 
by means of the Hall-Littlewood expansions, JK-residue sums and refined topological vertex. 
The definitions and formulae of the $q$-factorial and $q$-analogs are summarized in appendix \ref{app_qpoch}. 
The notation of the Young diagram is presented in appendix \ref{app_Young}.  
In appendix \ref{app_chargedCoulombderivation} we give the details of derivation of the multi-point functions of the charged Wilson lines in the Coulomb limit. 
The formalism of the refined topological vertex is summarized in appendix \ref{app:top}.
The exact closed-form expressions in the unflavored limit are listed in appendix \ref{app_closedform}. 

\subsection{Future works}

\begin{itemize}

\item 
The Wilson lines in the ADHM theories can be further explored by taking different gauge groups and representations. 
One of the intriguing setups is the irreducible representation labeled by the rectangular Young diagram, 
for which 4d $\mathcal{N}=4$ $U(N)$ SYM theory has the gravity dual as the bubbling geometry in the large $N$ and large representation limit \cite{Yamaguchi:2006te,Lunin:2006xr,Okuda:2007kh,DHoker:2007mci,Okuda:2008px,Gomis:2008qa,Benichou:2011aa,Fiol:2013hna,Aguilera-Damia:2017znn,Hatsuda:2023iof}. 

\item 
It would be intriguing to figure out physical interpretations and explanations of the algebraic relations,
in particular the
factorizations 
\eqref{factorizationC} and \eqref{factorizationH}, by engineering the Wilson lines in the ADHM theory in the brane configurations. 

\item 
While we focus on the Wilson line defect indices in this work, 
it would also be interesting to examine the vortex line defect indices as they were examined for the M2-brane SCFTs in \cite{Drukker:2008jm,Lee:2010hk,Drukker:2023bip}. 
As mirror symmetry exchanges the Wilson lines and the vortex lines \cite{Assel:2015oxa,Dimofte:2019zzj,Nawata:2021nse,Nawata:2023rdx}, 
it is tempting to extend webs of dualities of the M2-brane SCFTs in our previous work \cite{Hayashi:2022ldo}. 

\item 
The giant graviton expansions of supersymmetric indices have been recently examined (see e.g.~\cite{Arai:2020uwd,Gaiotto:2021xce,Arai:2020qaj,Imamura:2021ytr,Murthy:2022ien,Lee:2022vig,Imamura:2022aua}). For the line defect indices of 4d $\mathcal{N}=4$ SYM theory they 
have been recently studied in \cite{Imamura:2024lkw,Beccaria:2024oif,Imamura:2024pgp,Beccaria:2024dxi,Hatsuda:2024uwt}. 
The exact closed-form expressions of line defect indices in this paper should be crucial for the detailed analysis of the giant graviton expansions of the line defect indices for the M2-brane SCFTs. 
We hope to report our results in future work. 


\item We propose a factor which gives a ratio between the generating function of the 2-point Wilson line correlators in the antisymmetric representations and the corresponding integral evaluated by the JK residue prescription for some cases of the ADHM theory with $l$ flavors. The factor contains a part of so-called extra factor \cite{Bergman:2013ala,Bao:2013pwa,Hayashi:2013qwa,Bergman:2013aca} which we remove to obtain the Nekrasov partition function of a UV complete 5d $SU(l)$ gauge theory. 
It would be interesting to find an interpretation of the factor we find in terms of the 5d gauge theory, as well as an  interpretation of the 5d extra factor in terms of the 3d theory.
\end{itemize}

\section{ADHM line defect indices}
\label{sec_general}

\subsection{Matrix integral}
The 3d $U(N)$ $\mathcal{N}=4$ ADHM theory with $l$ flavors 
consists of $\mathcal{N}=4$ $U(N)$ vector multiplet 
coupled to a single hypermultiplet in the adjoint representation and $l$ hypermultiplets in the fundamental representation. 
It is a low-energy effective description of $N$ coincident M2-branes propagating in 
the $\mathbb{C}^2\times (\mathbb{C}^2/\mathbb{Z}_l)$. 
When $l=1$ the theory has enhanced $\mathcal{N}=8$ supersymmetry and is equivalent to the $U(N)_{1}\times U(N)_{-1}$ ABJM theory \cite{Bashkirov:2010kz}.

We study the \textit{ADHM line defect indices}, 
correlation functions of Wilson line operators which decorate the supersymmetric partition function of 3d $\mathcal{N}=4$ supersymmetric $U(N)$ ADHM theory with $l$ flavors on $S^1\times S^2$ 
with the insertion of BPS Wilson line operators wrapped on the $S^1$ and localized on the $S^2$. 
According to the supersymmetric localization, they can be evaluated by an $N$-dimensional matrix integral \cite{Kim:2009wb,Imamura:2011su}
\begin{align}
&\langle
W_{{\cal R}_1}
W_{{\cal R}_2}
\cdots
W_{{\cal R}_k}
\rangle^{U(N)\text{ ADHM-}[l]}(t,x;y_{\alpha},z;q)
\nonumber \\
&=\frac{1}{N!}\sum_{{\bm m}\in\mathbb{Z}^N}
q^{\frac{l}{4}\sum_i|m_i|}t^{-l\sum_i|m_i|}z^{l\sum_im_i}
\oint
\prod_{i=1}^N
\frac{ds_i}{2\pi is_i}
\prod_{j=1}^k
\chi_{{\cal R}_j}(s)\nonumber \\
&\quad\times \prod_{i\neq j}^N\Bigl(1-q^{\frac{|m_i-m_j|}{2}}\frac{s_i}{s_j}\Bigr)
\prod_{i,j=1}^{N}\frac{
(q^{\frac{1}{2}+\frac{|m_i-m_j|}{2}}t^2\frac{s_j}{s_i};q)_\infty
}{
(q^{\frac{1}{2}+\frac{|m_i-m_j|}{2}}t^{-2}\frac{s_i}{s_j};q)_\infty
}
\prod_{i,j=1}^N
\frac{
(q^{\frac{3}{4}+\frac{|m_i-m_j|}{2}}t^{-1}\frac{s_j}{s_i}x^{\mp};q)_\infty
}{
(q^{\frac{1}{4}+\frac{|m_i-m_j|}{2}}t\frac{s_i}{s_j}x^{\pm};q)_\infty
}\nonumber \\
&\quad\times \prod_{i=1}^N\prod_{\alpha=1}^l
\frac{
(q^{\frac{3}{4}+\frac{|m_i|}{2}}t^{-1}s_i^{\mp}y_\alpha^{\mp};q)_\infty
}{
(q^{\frac{1}{4}+\frac{|m_i|}{2}}ts_i^{\pm}y_\alpha^{\pm};q)_\infty
},
\label{Wilsonfullindex}
\end{align}
where we have used the convention $(x^{\pm};q)_{\infty}$ $:=$ $(x;q)_{\infty}(x^{-1};q)_{\infty}$. 
Here $\chi_{\mathcal{R}_j}(s)$ is a character of the representation $\mathcal{R}_j$ in which the $j$-th Wilson line transforms. 
While they are topological in that they are independent of the locations of inserted points of the line operators on the $S^2$, 
they depend on the fugacities $t$, $x$, $y_{\alpha}$, $z$ and $q$, 
which are coupled to the R-charges, the flavor charge for the adjoint hypermultiplet, 
the flavor charges for the fundamental hypermultiplets, the topological charge and the superconformal generators respectively.\footnote{
See \cite{Okazaki:2019ony,Hayashi:2022ldo} for the convention.
}
The Wilson lines $W_{\mathcal{R}_j}$ in the representation $\mathcal{R}_j$ can consecutively collide at the north pole. 
When complex conjugates $W_{\overline{\mathcal{R}_j}}$ labeled by the characters $\chi_{\mathcal{R}_j}(s^{-1})$ collide, they can sit at the south pole. 

Upon the conformal map from $S^1\times S^2$ to $\mathbb{R}^3$, 
the lines can form a junction in such a way that each of them maps to a semi-infinite line originating from the origin in $\mathbb{R}^3$. 
Accordingly, the line defect indices can count the BPS local operators sitting at the junction of lines. 
The additional degrees of freedom due to the insertion of lines are obtained from the normalized correlation function
\begin{align}
\label{normalize_Wilson}
&
\langle
\mathcal{W}_{{\cal R}_1}
\mathcal{W}_{{\cal R}_2}
\cdots
\mathcal{W}_{{\cal R}_k}
\rangle^{U(N)\text{ ADHM-}[l]}(t,x;y_{\alpha},z;q)
\nonumber\\
&:=\frac{
\langle
W_{{\cal R}_1}
W_{{\cal R}_2}
\cdots
W_{{\cal R}_k}
\rangle^{U(N)\text{ ADHM-}[l]}(t,x;y_{\alpha},z;q)}
{\mathcal{I}^{U(N)\text{ ADHM-}[l]}(t,x;y_{\alpha},z;q)}, 
\end{align}
where 
\begin{align}
\mathcal{I}^{U(N)\text{ ADHM-}[l]}(t,x;y_{\alpha},z;q)
&:=
\langle
1
\rangle^{U(N)\text{ ADHM-}[l]}(t,x;y_{\alpha},z;q)
\end{align}
is the supersymmetric index. 

\subsection{Symmetric functions}
An irreducible representation $\mathcal{R}$ of $U(N)$ is labeled by a Young diagram $\lambda$ 
and its character $\chi_{\mathcal{R}}(s)$ is given by the Schur function $s_{\lambda}(s)$. 
For the rank-$k$ antisymmetric representation, the character is the elementary symmetric function of degree $k$
\begin{align}
e_{k}(s)&=\sum_{1<i_1<i_2<\cdots<i_k}
s_{i_1}s_{i_2}\cdots s_{i_k}. 
\end{align}
The generating function for the elementary symmetric function is 
\begin{align}
E(x;s)&=\sum_{k=0}^{\infty}e_k(s) x^k=\prod_{i=1}^N(1+xs_i). 
\end{align}
For the rank-$k$ symmetric representation, the character is the complete homogeneous symmetric function of degree $k$
\begin{align}
h_{k}(s)&=\sum_{1\le i_1\le i_2\le \cdots\le i_k}
s_{i_1}s_{i_2}\cdots s_{i_k}. 
\end{align}
The generating function for the complete homogeneous symmetric function is
\begin{align}
H(x;s)&=\sum_{k=0}^{\infty}h_k(s) x^k=\prod_{i=1}^N\frac{1}{1-xs_i}.
\end{align}

While the Schur functions form the basis of the algebra of symmetric functions, 
one can alternatively use the power sum symmetric functions
\begin{align}
p_n(s)&=\sum_{i=1}^{N}s_i^n
\label{powersum}
\end{align}
as the basis labeled by conjugacy classes of the symmetric group. 
We will call the Wilson line operator labeled by the power sum symmetric function $p_n(s)$ of degree $n$ the \textit{charged Wilson line} $W_n$. 
The correlators of the charged Wilson line operators can be simply obtained by setting the characters 
to the associated power sum symmetric functions in the integrand of (\ref{Wilsonfullindex}). 

\section{Coulomb line defect indices}
\label{sec_Coulomb}
In this section we consider the Wilson line defect correlator \eqref{Wilsonfullindex} in the Coulomb limit 
\begin{align}
q,t\rightarrow 0\text{ with }\mathfrak{t}=q^{\frac{1}{4}}t^{-1}\text{ fixed}.
\end{align}
In this limit the correlation function reduces to
\begin{align}
&\langle
W_{{\cal R}_1}
W_{{\cal R}_2}
\cdots
W_{{\cal R}_k}
\rangle^{U(N)\text{ ADHM-}[l](C)}(z;\mathfrak{t})
\nonumber \\
&=\frac{1}{N!}\sum_{{\bm m}\in\mathbb{Z}^N}
\mathfrak{t}^{l \sum_i|m_i|}z^{l\sum_im_i}
\oint
\prod_{i=1}^N
\frac{ds_i}{2\pi is_i}
\frac{
\displaystyle \prod_{\substack{i\neq j\\ (m_i=m_j)}}
\Bigl(1-\frac{s_i}{s_j}\Bigr)}
{\displaystyle \prod_{\substack{i,j\\ (m_i=m_j)}}
\Bigl(1-\mathfrak{t}^2\frac{s_i}{s_j}\Bigr)}
\prod_{j=1}^{k}
\chi_{{\cal R}_j}(s), 
\label{wilsonsCoulomb}
\end{align}
where the $N$-dimensional matrix integral factorizes into 
a product of the lower-dimensional ones when the corresponding magnetic fluxes are distinct. 

Note that in the Coulomb limit the 1-point function of the Wilson line vanishes trivially, since there is no contribution from the (anti)fundamental hypermultiplets in this limit.
In particular, there is no non-trivial correlators of Wilson lines when $N=1$.
In the following we mainly focus on the 2-point functions for $N\ge 2$. 

\subsection{Hall-Littlewood expansions}
\label{sec_C_HL}
To proceed with the exact calculation, 
we observe that the matrix integral in (\ref{wilsonsCoulomb}) can be evaluated 
by making use of the inner product of the Hall-Littlewood symmetric functions \cite{MR1354144}\footnote{
The similar expansions are obtained in \cite{Crew:2020psc} for the Higgs indices of the ADHM theory 
and in \cite{Dorey:2016mxm,Barns-Graham:2017zpv,Okazaki:2017sbc} for the partition functions of matrix Chern-Simons models.
}
\begin{align}
P_{\lambda}(s;\mathfrak{t})
&=\sum_{
w\in S_n/S_n^{\lambda}
}
w\left(
s^{\lambda}
\prod_{\lambda_i>\lambda_j}
\frac{s_i-\mathfrak{t}s_j}
{s_i-s_j}
\right), 
\end{align}
where $S_n^{\lambda}$ is the subgroup of the permutations $w$ $\in$ $S_n$ which consists of elements keeping $\lambda$ invariant in such a way that $\lambda_{w(i)}$ $=$ $\lambda_i$ for $1\le i\le n$. 
It follows that 
\begin{align}
\label{HL_inner}
\frac{1}{N!}\oint 
\prod_{i=1}^{N}\frac{ds_i}{2\pi is_i}
\frac{\prod_{i\neq j}(1-\frac{s_i}{s_j})}
{\prod_{i,j}(1-\mathfrak{t}\frac{s_i}{s_j})}
P_{\mu}(s;\mathfrak{t})P_{\nu}(s^{-1};\mathfrak{t})
&=\frac{\delta_{\mu\nu}}{(\mathfrak{t};\mathfrak{t})_{N-\ell(\mu)}\prod_{j\ge}(\mathfrak{t};\mathfrak{t})_{m_j(\mu)}}, 
\end{align}
where $\ell(\mu)$ is the length of $\mu$ and $m_j(\mu)$ is the multiplicity of $\mu$. 
From the orthogonality of the Hall-Littlewood functions, the $1$-point function is shown to vanish. 

\subsubsection{$U(2)$}
\label{sec_C_HL_u2}
For $N=2$ the Coulomb line defect indices are given by 
\begin{align}
\label{C_u2}
&\langle
W_{{\cal R}_1}
W_{{\cal R}_2}
\cdots
W_{{\cal R}_k}
\rangle^{U(2)\text{ ADHM-}[l](C)}(z;\mathfrak{t})
\nonumber\\
&=
\frac12 
\frac{1}{(1-\mathfrak{t}^2)^2}
\sum_{m_1\in \mathbb{Z}}
\mathfrak{t}^{2l|m_1|}
z^{2lm_1}
\oint \frac{ds_1}{2\pi is_1} \frac{ds_2}{2\pi is_2}
\frac{(1-\frac{s_1}{s_2}) (1-\frac{s_2}{s_1})}
{(1-\mathfrak{t}^2 \frac{s_1}{s_2}) (1-\mathfrak{t}^2 \frac{s_2}{s_1})}
\prod_{j=1}^{k}
\chi_{{\cal R}_j}(s)
\nonumber\\
&\quad +\frac12
\frac{1}{(1-\mathfrak{t}^2)^2}
\sum_{
\begin{smallmatrix}
m_1,m_2\in \mathbb{Z}\\
m_1\neq m_2\\
\end{smallmatrix}}
\mathfrak{t}^{l(|m_1|+|m_2|)}
z^{l(m_1+m_2)}
\oint \frac{ds_1}{2\pi is_1}
\oint \frac{ds_2}{2\pi is_2}
\prod_{j=1}^{k}
\chi_{{\cal R}_j}(s). 
\end{align}
We have
\begin{align}
\sum_{m_1\in \mathbb{Z}}\mathfrak{t}^{2l |m_1|}z^{2l m_1}&=\frac{1-\mathfrak{t}^{4l}}
{(1-\mathfrak{t}^{2l}z^{2l})(1-\mathfrak{t}^{2l}z^{-2l})}, \\
\sum_{
\begin{smallmatrix}
m_1,m_2\in \mathbb{Z}\\
m_1\neq m_2\\
\end{smallmatrix}}
\mathfrak{t}^{l(|m_1|+|m_2|)}
z^{l(m_1+m_2)}&=
2\frac{\mathfrak{t}^{l}(z^l+z^{-l})-\mathfrak{t}^{2l}-\mathfrak{t}^{3l}(z^l+z^{-l})+\mathfrak{t}^{6l}}
{(1-\mathfrak{t}^{l}z^{l})(1-\mathfrak{t}^{2l}z^{2l})(1-\mathfrak{t}^{l}z^{-l})(1-\mathfrak{t}^{2l}z^{-2l})}. 
\end{align}
Without any insertion of Wilson line, it reduces to the Coulomb index \cite{Hayashi:2022ldo}
\begin{align}
&
\mathcal{I}^{U(2)\text{ ADHM-}[l](C)}(z;\mathfrak{t})
\nonumber\\
&=\frac{1-\mathfrak{t}^{4l}}
{(1-\mathfrak{t}^2)(1-\mathfrak{t}^4)(1-\mathfrak{t}^{2l}z^{2l}) (1-\mathfrak{t}^{2l}z^{-2l})}
\nonumber\\
&\quad +\frac{(z^l+z^{-l})\mathfrak{t}^l-\mathfrak{t}^{2l}-(z^l+z^{-l})\mathfrak{t}^{3l}+\mathfrak{t}^{6l}}
{(1-\mathfrak{t}^2)^2 (1-\mathfrak{t}^{l}z^l)(1-\mathfrak{t}^{2l}z^{2l})  (1-\mathfrak{t}^{l}z^{-l})(1-\mathfrak{t}^{2l}z^{-2l})}
\nonumber\\
&=\frac{(1-\mathfrak{t}^{2l})(1+(z^l+z^{-l})\mathfrak{t}^{l+2}+\mathfrak{t}^{2l}-\mathfrak{t}^{2l+2}-(z^l+z^{-l})\mathfrak{t}^{3l}-\mathfrak{t}^{4l+2})}
{(1-\mathfrak{t}^2)(1-\mathfrak{t}^4)(1-\mathfrak{t}^lz^l)(1-\mathfrak{t}^lz^{-l})(1-\mathfrak{t}^{2l}z^{2l})(1-\mathfrak{t}^{2l}z^{-2l})}. 
\end{align}
In the presence of the Wilson line in the fundamental representation at the north pole and that in the antifundamental representation at the south pole, we have $\chi_{\mathcal{R}_1}(s)$ $=$ $P_{(1)}(s;\mathfrak{t}^2)$ and $\chi_{\mathcal{R}_2}(s)$ $=$ $P_{(1)}(s^{-1};\mathfrak{t}^2)$. 
The $2$-point function of the fundamental Wilson lines is evaluated as
\begin{align}
&
\langle W_{\ydiagram{1}}W_{\overline{\ydiagram{1}}} \rangle^{U(2)\text{ ADHM-}[l](C)}(z;\mathfrak{t})
\nonumber\\
&=\frac{1-\mathfrak{t}^{4l}}
{(1-\mathfrak{t}^2)^2(1-\mathfrak{t}^{2l}z^{2l}) (1-\mathfrak{t}^{2l}z^{-2l})}
\nonumber\\
&\quad +\frac{2(z^l+z^{-l})\mathfrak{t}^l-2\mathfrak{t}^{2l}-2(z^l+z^{-l})\mathfrak{t}^{3l}+2\mathfrak{t}^{6l}}
{(1-\mathfrak{t}^2)^2 (1-\mathfrak{t}^{l}z^l)(1-\mathfrak{t}^{2l}z^{2l})  (1-\mathfrak{t}^{l}z^{-l})(1-\mathfrak{t}^{2l}z^{-2l})}
\nonumber\\
&=\frac{(1-\mathfrak{t}^{2l})^2}
{(1-\mathfrak{t}^2)^2 (1-\mathfrak{t}^lz^l)^2 (1-\mathfrak{t}^lz^{-l})^2}. 
\end{align}
Note that it is factorized into the two $U(1)$ Coulomb indices
\begin{align}
\label{u2C_fund2pt}
\langle W_{\ydiagram{1}}W_{\overline{\ydiagram{1}}} \rangle^{U(2)\text{ ADHM-}[l](C)}(z;\mathfrak{t})
&=\langle 1 \rangle^{U(1)\text{ ADHM-}[l](C)}(z;\mathfrak{t})^2.
\end{align}
As we will see, such a relation is generalized to higher rank gauge groups. 

For the Wilson lines in more general representations at the north and south poles, 
the inserted characters or equivalently the Schur functions are not identical to the Hall-Littlewood functions. 
In general, they can be expanded as
\begin{align}
\label{schur_HL}
s_{\lambda}(s)&=\sum_{\mu}K_{\lambda\mu}(\mathfrak{t})P_{\mu}(s;\mathfrak{t}), 
\end{align}
where $K_{\lambda\mu}(\mathfrak{t})$ are the Kostka-Foulkes polynomials. 
The Kostka-Foulkes polynomial can be viewed as the generating function for semistandard Young tableau with the charge statistics. Let $\textrm{STY}(\lambda,\mu)$ be a set of all semistandard Young tableaux of shape $\lambda$ and weight $\mu$. For each of $T$ $\in$ $\textrm{STY}(\lambda,\mu)$ the non-negative integer $c(T)$ which is called the charge can be defined by summing over the indices (see \cite{MR1354144} for a precise definition). 
According to the theorem of Lascoux and Sch\"{u}tzenberger \cite{MR472993,MR646486}, one has a combinatorial description of the Kostka-Foulkes polynomial
\begin{align}
K_{\lambda\mu}(\mathfrak{t})
&=\sum_{T\in \textrm{STY}(\lambda,\mu)}
\mathfrak{t}^{c(T)}. 
\end{align}
For example, we have
\begin{align}
K_{(k)\mu}&=\mathfrak{t}^{n(\lambda)}
\end{align}
for any partition $\mu$ of $k$, where $n(\lambda)$ is defined by \eqref{nlambda}.
By means of the relation (\ref{schur_HL}), we can compute more general correlation functions. 
As an example, let us consider the $2$-point function of the Wilson lines in the rank-$2$ symmetric representation. 
It can be obtained from the expansion
\begin{align}
h_2(s)&=P_{\ydiagram{2}}(s;\mathfrak{t}^2)+\mathfrak{t}^2P_{\ydiagram{1,1}}(s;\mathfrak{t}^2). 
\end{align}
We find 
\begin{align}
&
\langle W_{\ydiagram{2}}W_{\overline{\ydiagram{2}}} \rangle^{U(2)\text{ ADHM-}[l](C)}(z;\mathfrak{t})
\nonumber\\
&=\frac{1-\mathfrak{t}^{4l}}
{(1-\mathfrak{t}^2)^2(1-\mathfrak{t}^{2l}z^{2l}) (1-\mathfrak{t}^{2l}z^{-2l})}
+
\frac{\mathfrak{t}^4(1-\mathfrak{t}^{4l})}
{(1-\mathfrak{t}^2)(1-\mathfrak{t}^4)(1-\mathfrak{t}^{2l}z^{2l}) (1-\mathfrak{t}^{2l}z^{-2l})}
\nonumber\\
&\quad +\frac{3(z^l+z^{-l})\mathfrak{t}^l-3\mathfrak{t}^{2l}-3(z^l+z^{-l})\mathfrak{t}^{3l}+3\mathfrak{t}^{6l}}
{(1-\mathfrak{t}^2)^2 (1-\mathfrak{t}^{l}z^l)(1-\mathfrak{t}^{2l}z^{2l})  (1-\mathfrak{t}^{l}z^{-l})(1-\mathfrak{t}^{2l}z^{-2l})}. 
\end{align}

\subsubsection{$U(3)$}
\label{sec_C_HL_u3}
For $N=3$ the matrix integral \eqref{wilsonsCoulomb} splits into three parts. 
We have 
\begin{align}
\label{C_u3}
&\langle
W_{{\cal R}_1}
W_{{\cal R}_2}
\cdots
W_{{\cal R}_k}
\rangle^{U(3)\text{ ADHM-}[l](C)}(z;\mathfrak{t})
\nonumber\\
&=\frac16 \frac{1}{(1-\mathfrak{t}^2)^3}
\sum_{m_1\in \mathbb{Z}}
\mathfrak{t}^{3l|m_1|}z^{3lm_1}
\oint \prod_{i=1}^{3}\frac{ds_i}{2\pi is_i}
\prod_{i\neq j}^{3}
\frac{1-\frac{s_i}{s_j}}
{1-\mathfrak{t}^2\frac{s_i}{s_j}}
\prod_{j=1}^{k}\chi_{\mathcal{R}_j}(s)
\nonumber\\
&\quad +\frac12 \frac{1}{(1-\mathfrak{t}^2)^3}
\sum_{
\begin{smallmatrix}
m_1,m_2\in \mathbb{Z}\\
(m_1\neq m_2)
\end{smallmatrix}
}
\mathfrak{t}^{2l(|m_1|+|m_2|)}z^{2l(m_1+m_2)}
\oint \prod_{i=1}^{2}\frac{ds_i}{2\pi is_i}
\prod_{i\neq j}^{2}
\frac{1-\frac{s_i}{s_j}}
{1-\mathfrak{t}^2\frac{s_i}{s_j}}\nonumber \\
&\quad\quad\times 
\oint \frac{ds_3}{2\pi is_3}\prod_{j=1}^{k}\chi_{\mathcal{R}_j}(s)
\nonumber\\
&\quad +\frac16 \frac{1}{(1-\mathfrak{t}^2)^3}
\sum_{
\begin{smallmatrix}
m_1,m_2,m_3\in \mathbb{Z}\\
(m_1\neq m_2, m_1\neq m_3, m_2\neq m_3)
\end{smallmatrix}
}
\mathfrak{t}^{l(|m_1|+|m_2|+|m_3|)}z^{l(m_1+m_2+m_3)}
\oint \prod_{i=1}^{3}\frac{ds_i}{2\pi is_i}
\prod_{j=1}^{k}\chi_{\mathcal{R}_j}(s).
\end{align}
For example, the Coulomb index is expanded as
\begin{align}
&
\mathcal{I}^{U(3)\text{ ADHM-}[l](C)}(z;\mathfrak{t})
=\Lambda_{\emptyset}^{(3)}
+\Lambda_{\emptyset}^{(2,1)}
+\Lambda_{\emptyset}^{(1,1,1)}, 
\end{align}
where 
\begin{align}
\Lambda_{\emptyset}^{(3)}&=
\frac{1-\mathfrak{t}^{6l}}
{(1-\mathfrak{t}^2)(1-\mathfrak{t}^4)(1-\mathfrak{t}^6) (1-\mathfrak{t}^{3l}z^{3l}) (1-\mathfrak{t}^{3l}z^{-3l})},\\
\Lambda_{\emptyset}^{(2,1)}&=
\frac{(1-\mathfrak{t}^{2l})(1-\mathfrak{t}^{4l})}
{(1-\mathfrak{t}^2)^2(1-\mathfrak{t}^4)(1-\mathfrak{t}^{l}z^l)(1-\mathfrak{t}^{l}z^{-l}) (1-\mathfrak{t}^{2l}z^{2l}) (1-\mathfrak{t}^{2l}z^{-2l})}
\nonumber\\
&\quad -
\frac{1-\mathfrak{t}^{6l}}
{(1-\mathfrak{t}^2)^2(1-\mathfrak{t}^4)(1-\mathfrak{t}^{3l}z^{3l}) (1-\mathfrak{t}^{3l}z^{-3l})},\\
\Lambda_{\emptyset}^{(1,1,1)}&=
\frac{1}{(1-\mathfrak{t}^2)^3 \prod_{i=1}^3(1-\mathfrak{t}^{li}z^{li})(1-\mathfrak{t}^{li}z^{-li})}\nonumber\\
&\quad \times 
(\mathfrak{t}^{2l}+\mathfrak{t}^{3l}(z^{3l}+z^{-3l})
-\mathfrak{t}^{4l}(z^{2l}+z^{-2l})-\mathfrak{t}^{5l}(z^l+z^{-l}+z^{3l}+z^{-3l})
-\mathfrak{t}^{6l}
\nonumber\\
&\quad +\mathfrak{t}^{7l}(z^{l}+z^{-l})+\mathfrak{t}^{8l}(1+z^{2l}+z^{-2l})-\mathfrak{t}^{12l}
). 
\end{align}
The $2$-point function of the fundamental Wilson lines is given by 
\begin{align}
\langle W_{\ydiagram{1}}W_{\overline{\ydiagram{1}}} \rangle^{U(3)\text{ ADHM-}[l](C)}(z;\mathfrak{t})
&=\Lambda_{\ydiagram{1},\overline{\ydiagram{1}}}^{(3)}
+\Lambda_{\ydiagram{1},\overline{\ydiagram{1}}}^{(2,1)}
+\Lambda_{\ydiagram{1},\overline{\ydiagram{1}}}^{(1,1,1)}, 
\end{align}
where 
\begin{align}
\Lambda_{\ydiagram{1},\overline{\ydiagram{1}}}^{(3)}
&=\frac{1-\mathfrak{t}^{6l}}
{(1-\mathfrak{t}^2)^2 (1-\mathfrak{t}^4) (1-\mathfrak{t}^{3l}z^{3l})(1-\mathfrak{t}^{3l}z^{-3l})},\\
\Lambda_{\ydiagram{1},\overline{\ydiagram{1}}}^{(2,1)}&=
\frac{(2+\mathfrak{t}^2)(1-\mathfrak{t}^{2l})(1-\mathfrak{t}^{4l})}
{(1-\mathfrak{t}^2)^2(1-\mathfrak{t}^4)(1-\mathfrak{t}^{l}z^l)(1-\mathfrak{t}^{l}z^{-l}) (1-\mathfrak{t}^{2l}z^{2l}) (1-\mathfrak{t}^{2l}z^{-2l})}\nonumber\\
&\quad -
\frac{(2+\mathfrak{t}^2)(1-\mathfrak{t}^{6l})}
{(1-\mathfrak{t}^2)^2(1-\mathfrak{t}^4)(1-\mathfrak{t}^{3l}z^{3l}) (1-\mathfrak{t}^{3l}z^{-3l})},\\
\Lambda_{\ydiagram{1},\overline{\ydiagram{1}}}^{(1,1,1)}&=
\frac{3}{(1-\mathfrak{t}^2)^3 \prod_{i=1}^3(1-\mathfrak{t}^{li}z^{li})(1-\mathfrak{t}^{li}z^{-li})}
\nonumber\\
&\quad \times 
(\mathfrak{t}^{2l}+\mathfrak{t}^{3l}(z^{3l}+z^{-3l})
-\mathfrak{t}^{4l}(z^{2l}+z^{-2l})-\mathfrak{t}^{5l}(z^l+z^{-l}+z^{3l}+z^{-3l})
-\mathfrak{t}^{6l}
\nonumber\\
&\quad +\mathfrak{t}^{7l}(z^{l}+z^{-l})+\mathfrak{t}^{8l}(1+z^{2l}+z^{-2l})-\mathfrak{t}^{12l}
). 
\end{align}

Similarly, the $2$-point function of the rank-$2$ symmetric Wilson lines is
\begin{align}
\langle W_{\ydiagram{2}}W_{\overline{\ydiagram{2}}} \rangle^{U(3)\text{ ADHM-}[l](C)}(z;\mathfrak{t})
&=\Lambda_{\ydiagram{2},\overline{\ydiagram{2}}}^{(3)}
+\Lambda_{\ydiagram{2},\overline{\ydiagram{2}}}^{(2,1)}
+\Lambda_{\ydiagram{2},\overline{\ydiagram{2}}}^{(1,1,1)}, 
\end{align}
where 
\begin{align}
\Lambda_{\ydiagram{2},\overline{\ydiagram{2}}}^{(3)}
&=\frac{(1+\mathfrak{t}^4)(1-\mathfrak{t}^{6l})}
{(1-\mathfrak{t}^2)^2 (1-\mathfrak{t}^4) (1-\mathfrak{t}^{3l}z^{3l})(1-\mathfrak{t}^{3l}z^{-3l})},\\
\Lambda_{\ydiagram{2},\overline{\ydiagram{2}}}^{(2,1)}&=
\frac{(3+2\mathfrak{t}^2+\mathfrak{t}^4)(1-\mathfrak{t}^{2l})(1-\mathfrak{t}^{4l})}
{(1-\mathfrak{t}^2)^3(1+\mathfrak{t}^2)(1-\mathfrak{t}^{l}z^l)(1-\mathfrak{t}^{l}z^{-l}) (1-\mathfrak{t}^{2l}z^{2l}) (1-\mathfrak{t}^{2l}z^{-2l})}\nonumber\\
&\quad -
\frac{(3+2\mathfrak{t}^2+\mathfrak{t}^4)(1-\mathfrak{t}^{6l})}
{(1-\mathfrak{t}^2)^3(1+\mathfrak{t}^2)(1-\mathfrak{t}^{3l}z^{3l}) (1-\mathfrak{t}^{3l}z^{-3l})},\\
\Lambda_{\ydiagram{2},\overline{\ydiagram{2}}}^{(1,1,1)}&=
\frac{6}{(1-\mathfrak{t}^2)^3 \prod_{i=1}^3(1-\mathfrak{t}^{li}z^{li})(1-\mathfrak{t}^{li}z^{-li})}
\nonumber\\
&\quad \times 
(\mathfrak{t}^{2l}+\mathfrak{t}^{3l}(z^{3l}+z^{-3l})
-\mathfrak{t}^{4l}(z^{2l}+z^{-2l})-\mathfrak{t}^{5l}(z^l+z^{-l}+z^{3l}+z^{-3l})
-\mathfrak{t}^{6l}
\nonumber\\
&\quad +\mathfrak{t}^{7l}(z^{l}+z^{-l})+\mathfrak{t}^{8l}(1+z^{2l}+z^{-2l})-\mathfrak{t}^{12l}
).
\end{align}

\subsection{Fermi-gas method}
\label{Coulombgeneralformulas}
In the above we have calculated the correlation functions independently for each representation.
However, for the purpose of exact calculation of the correlators of Wilson lines in (anti)symmetric representations, 
it is also useful to consider generating functions for the correlators. 
Consider 
\begin{align}
\label{Coulombgeneratingfcn}
F(a,b,c,d;z;\mathfrak{t})^{U(N)\text{ ADHM-}[l](C)}
&:=\frac{1}{N!}\sum_{{\bm m}\in\mathbb{Z}^N}
\mathfrak{t}^{l \sum_i|m_i|}z^{l\sum_im_i}
\oint
\prod_{i=1}^N
\frac{ds_i}{2\pi is_i}
\frac{
\displaystyle \prod_{\substack{i\neq j\\ (m_i=m_j)}}
\Bigl(1-\frac{s_i}{s_j}\Bigr)}
{\displaystyle \prod_{\substack{i,j\\ (m_i=m_j)}}
\Bigl(1-\mathfrak{t}^2\frac{s_i}{s_j}\Bigr)}\nonumber\\
&\quad \times 
E(a;s)E(b;s^{-1})H(c;s)H(d;s^{-1}), 
\end{align}
which gives a generating function for the correlation functions of 
Wilson lines in the antisymmetric representations 
and those in the symmetric representations as it can be expanded in powers of fugacities $a$, $b$, $c$ and $d$ as
\begin{align}
&F(a,b,c,d;z;\mathfrak{t})^{U(N)\text{ ADHM-}[l](C)}
\nonumber \\
&=
\sum_{k_1, k_2, k_3, k_4 \ge 0}
a^{k_1}b^{k_2}
c^{k_3}d^{k_4}
\langle
W_{(1^{k_1})}
W_{\overline{(1^{k_2})}}
W_{(k_3)}
W_{\overline{(k_4)}}
\rangle^{U(N)\text{ ADHM-}[l](C)}(z;\mathfrak{t}).
\label{FabcdC}
\end{align}

In \cite{Hayashi:2022ldo} it was found that the matrix integral \eqref{wilsonsCoulomb} 
without any insertion can be computed in terms of the Fermi-gas method. 
The result is summarized in terms of the grand canonical partition function as
\begin{align}
\Xi^{(C)}(u)=\sum_{N=0}^\infty u^N\langle 1\rangle^{U(N)\text {ADHM-}[l](C)}(z;\mathfrak{t})=\prod_{m=-\infty}^\infty\biggl(1+\sum_{\nu=1}^\infty u^\nu\mathfrak{t}^{l|m|}z^{lm}\Omega(\nu)\biggr),
\end{align}
where
\begin{align}
\Omega(\nu)=\frac{\mathfrak{t}^{-\nu(\nu-1)}}{\nu!}\oint \prod_{i=1}^\nu
\frac{ds_i}{2\pi is_i}
\det\rho(s_i,s_j)
\end{align}
is identified with the canonical partition function of the free Fermi-gas consisting of $\nu$ particles 
whose one-particle density matrix is given by
\begin{align}
\rho(s,s')=\frac{1}{\frac{s}{s'}-\mathfrak{t}^2}.
\end{align}
Note that the canonical partition function $\Omega(\nu)$ can be evaluated from the spectral traces
\begin{align}
\text{Tr}\rho^n = \oint\prod_{i=1}^n\frac{ds_i}{2\pi is_i}
\rho(s_1,s_2)
\rho(s_2,s_3)
\cdots
\rho(s_n,s_1)
\end{align}
as
\begin{align}
\sum_{\nu=0}^\infty u^\nu\mathfrak{t}^{\nu(\nu-1)}\Omega(\nu)
=\text{det}(1+u\rho)
=\exp\biggl[\sum_{n=1}^\infty\frac{(-1)^{n-1}}{n}u^n\text{Tr}\rho^n\biggr].
\end{align}

These formulae can be extended straightforwardly to the generating function \eqref{Coulombgeneratingfcn}. 
It can be also expressed in terms of Fermi-gas partition functions which are characterized by the density matrix which takes the form
\begin{align}
\rho_f(s,s')=f(s)\rho(s,s'),
\end{align}
where the function $f(s)$ is an extra factor associated with the generating function (\ref{Coulombgeneratingfcn}) and can be expanded into a power sum
\begin{align}
f(s)=\frac{(1+as)(1+bs^{-1})}{(1-cs)(1-ds^{-1})}=\sum_{k=-\infty}^\infty A_ks^k. 
\end{align}
Now we further consider a grand canonical ensemble of the generating function (\ref{Coulombgeneratingfcn}) with respect to $N$, the rank of gauge group
\begin{align}
\Xi^{(C)}_f(u)
&=\sum_{N=0}^\infty u^N
F(a,b,c,d;z;\mathfrak{t})^{U(N)\text{ ADHM-}[l](C)}
\nonumber\\
&=\prod_{m=-\infty}^\infty\biggl(1+\sum_{\nu=1}^\infty u^\nu\mathfrak{t}^{l|m|\nu}z^{lm\nu}\Omega_f(\nu)\biggr). 
\label{XiCf}
\end{align}
The function $\Omega_f(\nu)$ in (\ref{XiCf}) can be viewed as the modified canonical partition function of the Fermi-gas 
whose one-particle density matrix is $\rho_f$. 
It satisfies the relation 
\begin{align}
\sum_{\nu=0}^\infty u^\nu\mathfrak{t}^{\nu(\nu-1)}\Omega_f(\nu)
=\text{det}(1+u\rho_f)
=\exp\biggl[\sum_{n=1}^\infty\frac{(-1)^{n-1}}{n}u^n\text{Tr}\rho_f^n\biggr].
\label{OmegafTrrhofrelation}
\end{align}

To calculate the modified spectral traces $\text{Tr}\rho_f^n$ systematically, 
it is useful to change the integration variable as $s=e^{2\pi i\alpha}$ 
where $\alpha$ is the position space coordinate on a circle.
Since $\rho_f(s,s')$ can be expanded as
\begin{align}
\rho_f(e^{2\pi i\alpha},e^{2\pi i\alpha'})=f(e^{2\pi i\alpha})\frac{1}{e^{2\pi i(\alpha-\alpha')}-\mathfrak{t}^2}
=f(e^{2\pi i\alpha})\sum_{p>0}\mathfrak{t}^{2(p-1)}e^{-2\pi i(\alpha-\alpha')p},
\end{align}
we have $\rho_f(e^{2\pi i\alpha},e^{2\pi i\alpha'})=\langle \alpha|{\hat\rho}_f|\alpha'\rangle$ with
\begin{align}
{\hat\rho}_f=f(e^{2\pi i{\hat\alpha}}){\hat\rho}_1,\quad
{\hat\rho}_1=\theta_\mathbb{Z}({\hat p}+1)\mathfrak{t}^{2(-{\hat p}-1)}.
\end{align}
Here we have introduced the following notation of 1d quantum mechanics:
\begin{align}
&[{\hat \alpha},{\hat p}]=\frac{i}{2\pi},\\
&|\alpha\rangle\text{: position eigenstate},\quad
\langle \alpha|\alpha'\rangle=\delta(\alpha-\alpha'),\quad
1=\int_0^1d\alpha |\alpha\rangle\langle\alpha|,\\
&|p\rangle\!\rangle\text{: momentum eigenstate},\quad
\langle\!\langle p|p'\rangle\!\rangle=\delta_{p,p'},\quad
1=\sum_{p=-\infty}^\infty |p\rangle\!\rangle\langle\!\langle p|,\\
&\langle\alpha|p\rangle\!\rangle=e^{2\pi i\alpha p},\quad
\langle\!\langle p|\alpha\rangle=e^{-2\pi i\alpha p},
\end{align}
and defined $\theta_\mathbb{Z}(p)$ as
\begin{align}
\theta_\mathbb{Z}(x)=
\begin{cases}
1\quad (x\le 0)\\
0\quad (x>0)
\end{cases}.
\end{align}

The calculation of $\text{Tr}\rho_f^n$, or equivalently the trace $\text{tr}{\hat \rho}_f^n$ in 1d quantum mechanics, reduces to that of
$\text{tr}(
e^{2\pi ik_1{\hat \alpha}}
{\hat\rho}_1
e^{2\pi ik_2{\hat \alpha}}
{\hat\rho}_1
\cdots
e^{2\pi ik_n{\hat \alpha}}
{\hat\rho}_1)$, which can be performed explicitly in the momentum basis by using the similarity transformation formula $e^{-2\pi i{\hat \alpha}}{\hat p}e^{2\pi i{\hat\alpha}}={\hat p}+1$. 
Let us define
\begin{align}
K_i:=\sum_{j=1}^ik_j. 
\end{align}
Then we have 
\begin{align}
\text{tr}(
e^{2\pi ik_1{\hat \alpha}}
{\hat\rho}_1
e^{2\pi ik_2{\hat \alpha}}
{\hat\rho}_1
\cdots
e^{2\pi ik_n{\hat \alpha}}
{\hat\rho}_1)
&=\sum_{p=-\infty}^\infty
\prod_{i=1}^n
\theta_\mathbb{Z}\Bigl(p-K_i+1\Bigr)\mathfrak{t}^{2(-p+K_i-1)}\nonumber \\
&=
\frac{\mathfrak{t}^{2\sum_{i=1}^n(K_i-\text{min}\{K_j\}_{j=1}^n)}}{1-\mathfrak{t}^{2n}}
\end{align}
when $K_n=0$, and
\begin{align}
\text{tr}(
e^{2\pi ik_1{\hat \alpha}}
{\hat\rho}_1
e^{2\pi ik_2{\hat \alpha}}
{\hat\rho}_1
\cdots
e^{2\pi ik_n{\hat \alpha}}
{\hat\rho}_1)=0
\end{align}
when $K_n\neq 0$.
Therefore we have
\begin{align}
\text{Tr}\rho^n_f=\sum_{k_1,\cdots,k_{n-1}=-\infty}^\infty
A_{k_1}
A_{k_2}
\cdots
A_{k_{n-1}}
A_{-\sum_{i=1}^{n-1}k_i}
\frac{\mathfrak{t}^{2\sum_{i=1}^n(K_i-\text{min}\{K_j\}_{j=1}^n)}}{1-\mathfrak{t}^{2n}}.
\label{trrhoformula}
\end{align}

Once we obtain $\text{Tr}\rho_f^n$, we can calculate the generating function (\ref{Coulombgeneratingfcn}) by expanding the right-hand side of \eqref{XiCf} in $u$.
To treat the infinite sums 
over $m$ in (\ref{XiCf}) at each order in $u$, it is convenient to organize the expansion as
\begin{align}
\Xi^{(C)}_f(u)
&=\sum_{\nu=0}^\infty u^\nu
\biggl\{\sum_{\substack{\nu_m=0\\ (\sum_m\nu_m=\nu)}}^\infty\biggr\}\mathfrak{t}^{l\sum_m|m|\nu_m}z^{l\sum_m m\nu_m}\biggl(\prod_{m=-\infty}^\infty \Omega_{f}(\nu_m)\biggr).
\end{align}
Note that $\prod_m\Omega_f(\nu_m)$ in the summand depends on the entries of non-zero components $\{\nu_m\}$, 
but does not depend on to which $m$'s these non-zero components are assigned.
By regarding the entries of non-zero components as a Young diagram, we can write the grand canonical correlator $\Xi_f^{(C)}(u)$ as
\begin{align}
\Xi_{f}^{(C)}(u)
&=\sum_{\nu=0}^\infty u^\nu
\sum_{\substack{\lambda\\ (|\lambda|=\nu)}}
\frac{G_\lambda}{\prod_im_i(\lambda)!}
\prod_{i=1}^{\ell(\lambda)} \Omega_{f}(\lambda_i).
\label{CoulombXiinformfactorexpansion}
\end{align}
where $m_i(\lambda)$ is the multiplicity \eqref{milambda},\footnote{
Here we have denoted the dummy indices originating from $\prod_{m=-\infty}^\infty$ in \eqref{XiCf} as $\text{m}_i$ rather than $m_i$ to avoid possible confusion with the multiplicities $m_i(\lambda)$.
}
and we have defined the form factor $G_\lambda$ as
\begin{align}
\frac{G_\lambda}{
\prod_im_i(\lambda)!
}
=\biggl\{\sum_{\nu_\text{m}}\biggr\}
\mathfrak{t}^{l \sum_\text{m}|\text{m}|\nu_\text{m}}
z^{l \sum_\text{m}\text{m}\nu_\text{m}},
\end{align}
where the summation runs over all choices of $\nu_\text{m}$'s whose non-zero components coincide with $\lambda=\{\lambda_1,\lambda_2,\cdots,\lambda_\ell\}$ up to permutations.
For each Young diagram $\lambda$, this summation reduces to a finite dimensional summation over the indices of $\nu_\text{m}$'s.
More concretely, for
\begin{align}
\lambda=\{
\underbrace{x_1,\cdots,x_1}_{k_1},
\underbrace{x_2,\cdots,x_2}_{k_2},
\cdots,
\underbrace{x_n,\cdots,x_n}_{k_n}
\}
\end{align}
we have
\begin{align}
&\frac{G_\lambda}{
\prod_im_i(\lambda)!
}\nonumber \\
&=
\sum_{
\substack{
\text{m}_1^{(1)}<\cdots<\text{m}^{(1)}_{k_1},\\
\text{m}_1^{(2)}<\cdots<\text{m}^{(2)}_{k_2},\cdots,\\
\text{m}_1^{(n)}<\cdots<\text{m}^{(n)}_{k_n}\\
(\text{m}^{(a)}_i\neq \text{m}^{(b)}_j\text{ for all }(a,i)\neq (b,j))
}
}
\mathfrak{t}^{l x_1\sum_{i=1}^{k_1}|\text{m}_i^{(1)}|}
\mathfrak{t}^{l x_2\sum_{i=1}^{k_2}|\text{m}_i^{(2)}|}
\cdots
\mathfrak{t}^{l x_n\sum_{i=1}^{k_n}|\text{m}_i^{(n)}|}
z^{l x_1\sum_{i=1}^{k_1}\text{m}_i^{(1)}}\nonumber \\
&
\quad
\quad
\quad
\quad
\quad
\quad
\quad
\quad
\quad
\quad
\quad
\quad
\times z^{l x_2\sum_{i=1}^{k_2}\text{m}_i^{(2)}}
\cdots
z^{l x_n\sum_{i=1}^{k_n}\text{m}_i^{(n)}}.
\label{Flambdaexplicitsum}
\end{align}
Note that the inequality ($<$) constraints in \eqref{Flambdaexplicitsum} can be cancelled by the factor of the multiplicities $\prod_im_i(\lambda)!$, as we have denoted in the denominator of the left-hand side:
\begin{align}
G_\lambda=\sum_{\substack{-\infty<\text{m}_1,\text{m}_2,\cdots,\text{m}_{\ell(\lambda)}<\infty\\ (\text{m}_i\neq \text{m}_j)}}
\mathfrak{t}^{l\sum_{i=1}^{\ell(\lambda)}|\text{m}_i|\lambda_i}
z^{l\sum_{i=1}^{\ell(\lambda)}\text{m}_i\lambda_i}.
\label{Glambdaexplicisum}
\end{align}
When $\lambda$ is of a single row, $\lambda=(k)$, the form factor $G_\lambda$ is given as
\begin{align}
G_{(k)}={\tilde A}_k,\quad
{\tilde A}_k=\sum_{\text{m}=-\infty}^\infty \mathfrak{t}^{l k|\text{m}|}z^{l k\text{m}}=\frac{1-\mathfrak{t}^{2kl}}{
(1-z^{kl}\mathfrak{t}^{kl})
(1-z^{-kl}\mathfrak{t}^{kl})
},
\label{Atilde}
\end{align}
while $G_\lambda$ for a more general $\lambda$, $\lambda=(k_1,k_2,\cdots,k_\ell)$, can be calculated by the following recursion relation
\begin{align}
G_{(k_1,k_2,\cdots,k_\ell)}=
{\tilde A}_{k_1}
G_{(k_2,\cdots,k_\ell)}
-
\sum_{i=2}^{\ell}
G_{(k_i+k_1,k_2,\cdots,\hat{k_i},\cdots,k_\ell)},
\label{Grecurson}
\end{align}
where each subscript in the second terms $(k_i+k_1,k_2,\cdots,\hat{k_i},\cdots,k_\ell)$ stands for the Young diagram of length $\ell(\lambda)-1$ obtained from $\lambda$ by removing the first row, adding it to the $i$-th row and then reordering the $\ell(\lambda)-1$ elements in non-increasing manner so that the sequence can be interpreted as a Young diagram.
For example, for $\lambda=(4,3,2,1)$ we have
\begin{align}
&(k_2+k_1,{\hat k}_2,k_3,k_4)=(k_2+k_1,k_3,k_4)=(7,2,1),\\
&(k_3+k_1,k_2,{\hat k}_3,k_4)=(k_3+k_1,k_2,k_4)=(6,3,1),\\
&(k_4+k_1,k_2,k_3,{\hat k}_4)=(k_4+k_1,k_2,k_3)=(5,3,2).
\end{align}
The recursion relation \eqref{Grecurson} can be easily shown from the original definition \eqref{Glambdaexplicisum}.
For example, we have
\begin{align}
&G_{\ydiagram{1,1}}={\tilde A}_1^2-{\tilde A}_2,\quad
G_{\ydiagram{2,1}}={\tilde A}_1{\tilde A}_2-{\tilde A}_3,\quad
G_{\ydiagram{1,1,1}}={\tilde A}_1^3-3{\tilde A}_1{\tilde A}_2+2{\tilde A}_3,\quad
G_{\ydiagram{3,1}}={\tilde A}_1{\tilde A}_3-{\tilde A}_4,\nonumber \\
&G_{\ydiagram{2,2}}={\tilde A}_2^2-{\tilde A}_4,\quad
G_{\ydiagram{2,1,1}}={\tilde A}_1^2{\tilde A}_2-2{\tilde A}_1{\tilde A}_3-{\tilde A}_2^2+2{\tilde A}_4,\nonumber \\
&G_{\ydiagram{1,1,1,1}}={\tilde A}_1^4-6{\tilde A}_1^2{\tilde A}_2+8{\tilde A}_1{\tilde A}_3+3{\tilde A}_2^2-6{\tilde A}_4.
\end{align}



Note that if we consider the large $l$ limit and ignore the terms proportional to $\mathfrak{t}^{nl}$ with $n\ge 1$, the grand canonical correlator \eqref{XiCf} simplifies drastically as
\begin{align}
\lim_{l\rightarrow\infty} \Xi^{(C)}_f(u)=1+\sum_{\nu=1}^\infty u^\nu \Omega_f(\nu).
\end{align}
Namely, the generating functions \eqref{FabcdC} are the canonical partition function $\Omega_f(\nu)$ themselves
\begin{align}
F(a,b,c,d;z;\mathfrak{t})^{U(N)\text{ ADHM-}[\infty](C)}=\Omega_f(N).
\label{largelsimplification}
\end{align}

\subsubsection{Antisymmetric representations}
First let us look into the $2$-point functions of Wilson lines in the antisymmetric representations. 
As explained above, these correlation functions can be obtained  from the expansion coefficients of the generating function \eqref{Coulombgeneratingfcn} with $c=d=0$ at $(ab)^k$.
By calculating the canonical partition function $\Omega_{(1+as)(1+bs^{-1})}(\nu)$ for various $\nu$ following the algorithm \eqref{OmegafTrrhofrelation} and \eqref{trrhoformula}, we find the closed-form expression for general $\nu$
\begin{align}
\Omega_{(1+as)(1+bs^{-1})}(\nu)=\sum_{k=0}^\nu \frac{(ab)^k}{(\mathfrak{t}^2;\mathfrak{t}^2)_k(\mathfrak{t}^2;\mathfrak{t}^2)_{\nu-k}}.
\end{align}
Plugging this into the grand canonical correlator $\Xi_{(1+as)(1+bs^{-1})}^{(C)}(u)$ \eqref{CoulombXiinformfactorexpansion} and 
reading off the coefficient of $u^N$ for each $N$, we obtain
\begin{align}
F(a,b,0,0;z;\mathfrak{t})^{U(N)\text{ ADHM-}[l](C)}
=\sum_{k=0}^N(ab)^k
{\cal I}^{U(k)\text{ ADHM-}[l](C)}(z;\mathfrak{t})
{\cal I}^{U(N-k)\text{ ADHM-}[l](C)}(z;\mathfrak{t}).
\end{align}
Namely, we find the following interesting algebraic relation between the Wilson line correlators and the Coulomb indices at different ranks:
\begin{align}
\langle W_{(1^k)}W_{\overline{(1^k)}}\rangle^{U(N)\text{ADHM-}[l](C)}(z;\mathfrak{t})=
{\cal I}^{U(k)\text{ ADHM-}[l](C)}(z;\mathfrak{t})
{\cal I}^{U(N-k)\text{ ADHM-}[l](C)}(z;\mathfrak{t}).
\label{ICantisymW}
\end{align}
Note that the identity is valid for any number of flavors $l\ge 1$ with generic value of the fugacity $z$ ($|z|=1$) for the topological symmetry.
This generalizes the relation (\ref{u2C_fund2pt}) we have found for $N=2$ and $k=1$.

\subsubsection{Symmetric representations}
Next let us consider the $2$-point function of Wilson lines in the symmetric representations.
These are obtained as the expansion coefficients of the generating function \eqref{Coulombgeneratingfcn} with $a=b=0$ at $(cd)^k$,
\begin{align}
F(0,0,c,d;z;\mathfrak{t})^{U(N)\text{ ADHM-}[l](C)}
=\sum_{k=0}^\infty (cd)^k\langle W_{(k)}W_{\overline{(k)}}\rangle^{U(N)\text{ ADHM-}[l](C)}(z;\mathfrak{t}).
\end{align}
By calculating the canonical partition function $\Omega_{\frac{1}{(1-cs)(1-ds^{-1})}}(\nu)$ \eqref{OmegafTrrhofrelation} for various $\nu$ from the method in section \ref{Coulombgeneralformulas}, 
we find the closed-form expression for general $\nu$
\begin{align}
\Omega_{\frac{1}{(1-cs)(1-ds^{-1})}}(\nu)=\frac{1}{
(\mathfrak{t}^2;\mathfrak{t}^2)_{\nu}
(\mathfrak{t}^2;\mathfrak{t}^2)_{\nu-1}
}
\sum_{k=0}^\infty (cd)^k(\mathfrak{t}^{2k+2};\mathfrak{t}^2)_{\nu-1}.
\label{Omeganusym}
\end{align}
The infinite summation over $k$ can be performed explicitly for each $\nu$ as
\begin{align}
&\Omega_{\frac{1}{(1-cs)(1-ds^{-1})}}(1)=\frac{1}{(\mathfrak{t}^2;\mathfrak{t}^2)_1}\frac{1}{1-cd},\\
&\Omega_{\frac{1}{(1-cs)(1-ds^{-1})}}(2)=\frac{1}{(\mathfrak{t}^2;\mathfrak{t}^2)_2(\mathfrak{t}^2;\mathfrak{t}^2)_1}\Bigl(\frac{1}{1-cd}-\frac{\mathfrak{t}^2}{1-\mathfrak{t}^2cd}\Bigr),\\
&\Omega_{\frac{1}{(1-cs)(1-ds^{-1})}}(3)=\frac{1}{(\mathfrak{t}^2;\mathfrak{t}^2)_3(\mathfrak{t}^2;\mathfrak{t}^2)_2}\Bigl(\frac{1}{1-cd}-\frac{\mathfrak{t}^2(1-\mathfrak{t}^4)}{(1-\mathfrak{t}^2)(1-\mathfrak{t}^2cd)}+\frac{\mathfrak{t}^6}{1-\mathfrak{t}^4cd}\Bigr),\\
&\Omega_{\frac{1}{(1-cs)(1-ds^{-1})}}(4)\nonumber \\
&=\frac{1}{(\mathfrak{t}^2;\mathfrak{t}^2)_4(\mathfrak{t}^2;\mathfrak{t}^2)_3}\Bigl(\frac{1}{1-cd}-\frac{\mathfrak{t}^2(1-\mathfrak{t}^6)}{(1-\mathfrak{t}^2)(1-\mathfrak{t}^2cd)}+\frac{\mathfrak{t}^6(1-\mathfrak{t}^6)}{(1-\mathfrak{t}^2)(1-\mathfrak{t}^4cd)}
-\frac{\mathfrak{t}^{12}}{1-\mathfrak{t}^6cd}\Bigr),\\
&\Omega_{\frac{1}{(1-cs)(1-ds^{-1})}}(5)\nonumber \\
&=\frac{1}{(\mathfrak{t}^2;\mathfrak{t}^2)_5(\mathfrak{t}^2;\mathfrak{t}^2)_4}\Bigl(\frac{1}{1-cd}-\frac{\mathfrak{t}^2(1-\mathfrak{t}^8)}{(1-\mathfrak{t}^2)(1-\mathfrak{t}^2cd)}
+\frac{\mathfrak{t}^6(1-\mathfrak{t}^6)(1-\mathfrak{t}^8)}{(1-\mathfrak{t}^2)(1-\mathfrak{t}^4)(1-\mathfrak{t}^4cd)}\nonumber \\
&\quad -\frac{\mathfrak{t}^{12}(1-\mathfrak{t}^8)}{(1-\mathfrak{t}^2)(1-\mathfrak{t}^6cd)}+\frac{\mathfrak{t}^{20}}{1-\mathfrak{t}^8cd}\Bigr),
\end{align}
and so on.
In contrast to the case of the antisymmetric Wilson lines, general expression for the generating function is not available. 
Nevertheless by substituting these $\Omega_{\frac{1}{(1-cs)(1-d^{-1}s)}}(\nu)$ into the grand canonical correlator \eqref{CoulombXiinformfactorexpansion} we obtain 
a closed-form expression for the generating function for each $N$ 
as
\begin{align}
%
&\sum_{k_1,k_2=0}^\infty c^{k_1}d^{k_2}\langle W_{(k_1)}W_{\overline{(k_2)}}\rangle^{U(2)\text{ ADHM-}[l](C)}(z;\mathfrak{t})\nonumber \\
&=\frac{1}{2(1-cd)^2}{\cal I}^{U(1)\text{ADHM-}[l](C)}(z;\mathfrak{t})^2+\frac{1-cd(2+\mathfrak{t}^2)}{2(1-cd)^2(1-cd\mathfrak{t}^2)}{\cal I}^{U(1)\text{ ADHM-}[l](C)}(z^2;\mathfrak{t}^2),\\
&\sum_{k_1,k_2=0}^\infty c^{k_1}d^{k_2}\langle W_{(k_1)}W_{\overline{(k_2)}}\rangle^{U(3)\text{ ADHM-}[l](C)}(z;\mathfrak{t})\nonumber \\
&=\frac{1}{6(1-cd)^3}{\cal I}^{U(1)\text{ ADHM-}[l](C)}(z;\mathfrak{t})^3
+\frac{1-cd(2+\mathfrak{t}^2)}{2(1-cd)^3(1-cd\mathfrak{t}^2)}{\cal I}^{U(1)\text{ADHM-}[l](C)}(z;\mathfrak{t})\nonumber \\
&\quad\times {\cal I}^{U(1)\text{ADHM-}[l](C)}(z^2;\mathfrak{t}^2)
+\frac{1-cd(3+\mathfrak{t}^2+\mathfrak{t}^4)+(cd)^2(3+3\mathfrak{t}^2+3\mathfrak{t}^4+\mathfrak{t}^6)}{3(1-cd)^3(1-cd\mathfrak{t}^2)(1-cd\mathfrak{t}^4)}\nonumber \\
&\quad\times {\cal I}^{U(1)\text{ ADHM-}[l](C)}(z^3;\mathfrak{t}^3),\\
&\sum_{k_1,k_2=0}^\infty c^{k_1}d^{k_2}\langle W_{(k_1)}W_{\overline{(k_2)}}\rangle^{U(4)\text{ ADHM-}[l](C)}(z;\mathfrak{t})\nonumber \\
&=
\frac{1}{24 (1 - cd)^4}{\cal I}^{U(1)\text{ ADHM-}[l](C)}(z;\mathfrak{t})^4
+\frac{(1 - cd (2 + \mathfrak{t}^2))}{4 (1 - cd)^4 (1 - cd \mathfrak{t}^2)}{\cal I}^{U(1)\text{ ADHM-}[l](C)}(z;\mathfrak{t})^2\nonumber \\
&\quad\times {\cal I}^{U(1)\text{ ADHM-}[l](C)}(z^2;\mathfrak{t}^2)
+\frac{(1 - cd (2 + \mathfrak{t}^2))^2}{8 (1 - cd)^4 (1 - cd \mathfrak{t}^2)^2}{\cal I}^{U(1)\text{ ADHM-}[l](C)}(z^2;\mathfrak{t}^2)^2\nonumber \\
&\quad +\frac{1 - cd (3 + \mathfrak{t}^2 + \mathfrak{t}^4) + (cd)^2 (3 + 3 \mathfrak{t}^2 + 3 \mathfrak{t}^4 + \mathfrak{t}^6)}{3 (1 - cd)^4 (1 - cd \mathfrak{t}^2) (1 - cd \mathfrak{t}^4)} {\cal I}^{U(1)\text{ ADHM-}[l](C)}(z;\mathfrak{t})\nonumber \\
&\quad\times {\cal I}^{U(1)\text{ ADHM-}[l](C)}(z^3;\mathfrak{t}^3)
+\frac{1}{4 (1 - cd)^4 (1 - cd \mathfrak{t}^2)^2 (1 - cd \mathfrak{t}^4) (1 - cd \mathfrak{t}^6)}\nonumber \\
&\quad\times (
1 - cd (4 + 2 \mathfrak{t}^2 + \mathfrak{t}^4 + \mathfrak{t}^6)+ (cd)^2 (6 + 8 \mathfrak{t}^2 + 5 \mathfrak{t}^4 + 6 \mathfrak{t}^6 + 2 \mathfrak{t}^8 + \mathfrak{t}^{10})\nonumber \\
&\quad\quad - (cd)^3 (4 + 12 \mathfrak{t}^2 + 10 \mathfrak{t}^4 + 14 \mathfrak{t}^6 + 9 \mathfrak{t}^8 + 5 \mathfrak{t}^{10} + 2 \mathfrak{t}^{12})\nonumber \\
&\quad\quad + (cd)^4 \mathfrak{t}^2 (4 + 4 \mathfrak{t}^2 + 8 \mathfrak{t}^4 + 8 \mathfrak{t}^6 + 6 \mathfrak{t}^8 + 4 \mathfrak{t}^{10} + \mathfrak{t}^{12})
) {\cal I}^{U(1)\text{ ADHM-}[l](C)}(z^4;\mathfrak{t}^4),\\
&\sum_{k_1,k_2=0}^\infty c^{k_1}d^{k_2}\langle W_{(k_1)}W_{\overline{(k_2)}}\rangle^{U(5)\text{ ADHM-}[l](C)}(z;\mathfrak{t})\nonumber \\
&=
\frac{1}{120 (1 - cd)^5}{\cal I}^{U(1)\text{ ADHM-}[l](C)}(z;\mathfrak{t})^5
+\frac{1 - cd (2 + \mathfrak{t}^2)}{12 (1 - cd)^5 (1 - cd \mathfrak{t}^2)}
{\cal I}^{U(1)\text{ ADHM-}[l](C)}(z;\mathfrak{t})^3\nonumber \\
&\quad\times {\cal I}^{U(1)\text{ ADHM-}[l](C)}(z^2;\mathfrak{t}^2)
+\frac{1 - cd (3 + \mathfrak{t}^2 + \mathfrak{t}^4) + (cd)^2 (3 + 3 \mathfrak{t}^2 + 3 \mathfrak{t}^4 + \mathfrak{t}^6)}{6 (1 - cd)^5 (1 - cd \mathfrak{t}^2) (1 - cd \mathfrak{t}^4)}\nonumber \\
&\quad\times {\cal I}^{U(1)\text{ ADHM-}[l](C)}(z;\mathfrak{t})^2{\cal I}^{U(1)\text{ ADHM-}[l](C)}(z^3;\mathfrak{t}^3)
+\frac{1 - cd (2 + \mathfrak{t}^2)}{6 (1 - cd)^5 (1 - cd \mathfrak{t}^2)^2 (1 - cd \mathfrak{t}^4)}\nonumber \\
&\quad\times
(1 - cd (3 + \mathfrak{t}^2 + \mathfrak{t}^4)+ (cd)^2 (3 + 3 \mathfrak{t}^2 + 3 \mathfrak{t}^4 + \mathfrak{t}^6))
{\cal I}^{U(1)\text{ ADHM-}[l](C)}(z^2;\mathfrak{t}^2)\nonumber \\
&\quad\times {\cal I}^{U(1)\text{ ADHM-}[l](C)}(z^3;\mathfrak{t}^3)
+\frac{(1 - cd (2 + \mathfrak{t}^2))^2}{8 (1 - cd)^5 (1 - cd \mathfrak{t}^2)^2}
{\cal I}^{U(1)\text{ ADHM-}[l](C)}(z;\mathfrak{t})\nonumber \\
&\quad\times {\cal I}^{U(1)\text{ ADHM-}[l](C)}(z^2;\mathfrak{t}^2)^2
+\frac{
1
}{4 (1 - cd)^5 (1 - cd \mathfrak{t}^2)^2 (1 - cd \mathfrak{t}^4) (1 - cd \mathfrak{t}^6)}(
1\nonumber \\
&\quad\quad - cd (4 + 2 \mathfrak{t}^2 + \mathfrak{t}^4 + \mathfrak{t}^6)
+ (cd)^2 (6 + 8 \mathfrak{t}^2 + 5 \mathfrak{t}^4 + 6 \mathfrak{t}^6 + 2 \mathfrak{t}^8 + \mathfrak{t}^{10})\nonumber \\
&\quad\quad - (cd)^3 (4 + 12 \mathfrak{t}^2 + 10 \mathfrak{t}^4 + 14 \mathfrak{t}^6 + 9 \mathfrak{t}^8 + 5 \mathfrak{t}^{10} + 2 \mathfrak{t}^{12})\nonumber \\
&\quad\quad + (cd)^4 \mathfrak{t}^2 (4 + 4 \mathfrak{t}^2 + 8 \mathfrak{t}^4 + 8 \mathfrak{t}^6 + 6 \mathfrak{t}^8 + 4 \mathfrak{t}^{10} + \mathfrak{t}^{12})
){\cal I}^{U(1)\text{ ADHM-}[l](C)}(z;\mathfrak{t})\nonumber \\
&\quad \times {\cal I}^{U(1)\text{ ADHM-}[l](C)}(z^4;\mathfrak{t}^4)+\frac{
1
}{5 (1 - cd)^5 (1 - cd \mathfrak{t}^2)^2 (1 - cd \mathfrak{t}^4) (1 - cd \mathfrak{t}^6) (1 - cd \mathfrak{t}^8)}
(
1\nonumber \\
&\quad\quad + cd (-5 - 2 \mathfrak{t}^2 - \mathfrak{t}^4 - \mathfrak{t}^6 - \mathfrak{t}^8)
+ (cd)^2 (10 + 10 \mathfrak{t}^2 + 6 \mathfrak{t}^4 + 7 \mathfrak{t}^6 + 7 \mathfrak{t}^8 + 3 \mathfrak{t}^{10} + \mathfrak{t}^{12} + \mathfrak{t}^{14})\nonumber \\
&\quad\quad + (cd)^3 (-10 - 20 \mathfrak{t}^2 - 15 \mathfrak{t}^4 - 20 \mathfrak{t}^6 - 21 \mathfrak{t}^8 - 16 \mathfrak{t}^{10} - 8 \mathfrak{t}^{12} - 7 \mathfrak{t}^{14} - 2 \mathfrak{t}^{16} - \mathfrak{t}^{18})\nonumber \\
&\quad\quad + (cd)^4 (5 + 20 \mathfrak{t}^2 + 20 \mathfrak{t}^4 + 30 \mathfrak{t}^6 + 35 \mathfrak{t}^8 + 35 \mathfrak{t}^{10} + 25 \mathfrak{t}^{12} + 21 \mathfrak{t}^{14} + 11 \mathfrak{t}^{16} + 6 \mathfrak{t}^{18} + 2 \mathfrak{t}^{20})\nonumber \\
&\quad\quad + (cd)^5 (-5 \mathfrak{t}^2 - 5 \mathfrak{t}^4 - 10 \mathfrak{t}^6 - 15 \mathfrak{t}^8 - 20 \mathfrak{t}^{10} - 20 \mathfrak{t}^{12} - 20 \mathfrak{t}^{14} - 15 \mathfrak{t}^{16} - 10 \mathfrak{t}^{18} - 5 \mathfrak{t}^{20}\nonumber \\
&\quad\quad - \mathfrak{t}^{22})
){\cal I}^{U(1)\text{ ADHM-}[l](C)}(z^5;\mathfrak{t}^5),
\end{align}
and so on, where we have used
${\tilde A}_\alpha=(1-\mathfrak{t}^{2\alpha}){\cal I}^{U(1)\text{ ADHM-}[l](C)}(z^\alpha,\mathfrak{t}^\alpha)$ \eqref{Atilde}.
Here we shall also list the explicit expressions for the correlators for $l=1$:
\begin{align}
%
&\langle W_{\ydiagram{2}}W_{\overline{\ydiagram{2}}}\rangle^{U(2)\text{ ADHM-}[1](C)}(z;\mathfrak{t})
=\frac{1+2(z+z^{-1})\mathfrak{t}+(z+z^{-1})\mathfrak{t}^3-\mathfrak{t}^4}{\prod_\pm (1-z^{\pm 1}\mathfrak{t})(1-z^{\pm 2}\mathfrak{t}^2)},\\
&\langle W_{\ydiagram{2}}W_{\overline{\ydiagram{2}}}\rangle^{U(3)\text{ ADHM-}[1](C)}(z;\mathfrak{t})
=\frac{1+(z+z^{-1})\mathfrak{t}+(z+z^{-1})\mathfrak{t}^3-\mathfrak{t}^4}{\prod_\pm (1-z^{\pm 1}\mathfrak{t})^2(1-z^{\pm 2}\mathfrak{t}^2)},\\
&\langle W_{\ydiagram{2}}W_{\overline{\ydiagram{2}}}\rangle^{U(4)\text{ ADHM-}[1](C)}(z;\mathfrak{t})\nonumber \\
&=
\frac{1
}{
\prod_\pm
(1 - z^{\pm 1}\mathfrak{t})
(1 - z^{\pm 2}\mathfrak{t}^2)^2
(1 - z^{\pm 3}\mathfrak{t}^3)
}
(
1
+ (2z^{-1} + 2 z) \mathfrak{t}
+ (z^{-2} +3 +  z^2) \mathfrak{t}^2\nonumber \\
&\quad + (z^{-3} + 5z^{-1} + 5 z + z^3) \mathfrak{t}^3
+ (3z^{-2} +7 +  3 z^2) \mathfrak{t}^4
+ (2z^{-3} + 5z^{-1} + 5 z + 2 z^3) \mathfrak{t}^5\nonumber \\
&\quad + (z^{-2} +4 +  z^2) \mathfrak{t}^6
+ (z^{-3} + 2z^{-1} + 2 z + z^3) \mathfrak{t}^7
- \mathfrak{t}^{10}
),\\
&\langle W_{\ydiagram{2}}W_{\overline{\ydiagram{2}}}\rangle^{U(5)\text{ ADHM-}[1](C)}(z;\mathfrak{t})\nonumber \\
&=
\frac{1}{\prod_\pm
(1 - z^{\pm 1}\mathfrak{t})^2
(1 - z^{\pm 2}\mathfrak{t}^2)
(1 - z^{\pm 3}\mathfrak{t}^3)
(1 - z^{\pm 4}\mathfrak{t}^4)
}
(
1
+(z^{-1} + z) \mathfrak{t}\nonumber \\
&\quad +(z^{-3} + 4z^{-1} + 4 z + z^3) \mathfrak{t}^3  
+(z^{-2} +2 +  z^2) \mathfrak{t}^4
+(2z^{-3} + 5z^{-1} + 5 z + 2 z^3) \mathfrak{t}^5\nonumber \\
&\quad +(z^{-4} + 2z^{-2} + 3 + 2 z^2 + z^4) \mathfrak{t}^6
+(z^{-3} + 5z^{-1} + 5 z + z^3) \mathfrak{t}^7\nonumber \\
&\quad +(z^{-4} + 2z^{-2} + 2 + 2 z^2 + z^4) \mathfrak{t}^8
+(z^{-3} + 4z^{-1} + 4 z + z^3) \mathfrak{t}^9  
+(-z^{-2} - z^2) \mathfrak{t}^{10}\nonumber \\
&\quad +(z^{-3}+ 2z^{-1} + 2 z + z^3) \mathfrak{t}^{11}
+(- z^{-2}-1  - z^2) \mathfrak{t}^{12}
+(z^{-1} + z) \mathfrak{t}^{13}
- \mathfrak{t}^{14}
),\\
%
%
&\langle W_{\ydiagram{3}}W_{\overline{\ydiagram{3}}}\rangle^{U(2)\text{ ADHM-}[1](C)}(z;\mathfrak{t})\nonumber \\
&=\frac{1
}{\prod_\pm (1-z^{\pm 1}\mathfrak{t})(1-z^{\pm 2}\mathfrak{t}^2)}
(1
+3(z^{-1}+z)\mathfrak{t}
-\mathfrak{t}^2
+2(z+z^{-1})\mathfrak{t}^3
-3\mathfrak{t}^4
+(z+z^{-1})\mathfrak{t}^5\nonumber \\
&\quad
-\mathfrak{t}^6
),\\
&\langle W_{\ydiagram{3}}W_{\overline{\ydiagram{3}}}\rangle^{U(3)\text{ ADHM-}[1](C)}(z;\mathfrak{t})\nonumber \\
&=\frac{1}{
\prod_\pm
(1 - z^{\pm 1}\mathfrak{t})
(1 - z^{\pm 2}\mathfrak{t}^2)
(1 - z^{\pm 3}\mathfrak{t}^3)
}
(
1
 + (3z^{-1} + 3 z)                     \mathfrak{t}
+ (3z^{-2} + 5 + 3 z^2)               \mathfrak{t}^2\nonumber \\
&\quad + (3z^{-3} + 6z^{-1} + 6 z + 3 z^3)   \mathfrak{t}^3  
 + (2z^{-2} +8 +  2 z^2)               \mathfrak{t}^4
+ (4z^{-3} + 3z^{-1} + 3 z + 4 z^3)   \mathfrak{t}^5\nonumber \\
&\quad + (- 2z^{-2} +4- 2 z^2)               \mathfrak{t}^6  
 + (2z^{-3} - z^{-1} - z + 2 z^3)      \mathfrak{t}^7
 + (-2z^{-2} - 2 z^2)                  \mathfrak{t}^8  
 + (z^{-3} + z^3)                      \mathfrak{t}^9\nonumber \\
 &\quad  + (- z^{-2}+1 - z^2)                  \mathfrak{t}^{10}
 + (-z^{-1} - z)                       \mathfrak{t}^{11}
 + \mathfrak{t}^{12}
),\\
&\langle W_{\ydiagram{3}}W_{\overline{\ydiagram{3}}}\rangle^{U(4)\text{ ADHM-}[1](C)}(z;\mathfrak{t})\nonumber \\
&=\frac{1}{
\prod_\pm
(1 - z^{\pm 1}\mathfrak{t})
(1 - z^{\pm 2}\mathfrak{t}^2)^2
(1 - z^{\pm 3}\mathfrak{t}^3)
}
(
1
 + (3z^{-1} + 3 z)                         \mathfrak{t}
+ (2z^{-2} +5 +  2 z^2)                   \mathfrak{t}^2\nonumber \\
&\quad + (3z^{-3} + 9z^{-1} + 9 z + 3 z^3)       \mathfrak{t}^3  
 + (z^{-4} + 5z^{-2} +13 +  5 z^2 + z^4)   \mathfrak{t}^4\nonumber \\
&\quad + (6z^{-3} + 10z^{-1} + 10 z + 6 z^3)     \mathfrak{t}^5  
 + (2z^{-4} + z^{-2} +11 +  z^2 + 2 z^4)   \mathfrak{t}^6\nonumber \\
&\quad + (4z^{-3} + 3z^{-1} + 3 z + 4 z^3)       \mathfrak{t}^7  
 + (z^{-4} - 4z^{-2}+ 2   - 4 z^2 + z^4)   \mathfrak{t}^8\nonumber \\  
&\quad + (z^{-3} - z^{-1} - z + z^3)             \mathfrak{t}^9  
 + (z^{-4} - 2z^{-2}+ 1  - 2 z^2 + z^4)    \mathfrak{t}^{10}
 + (-2z^{-2} - 2 z^2)                      \mathfrak{t}^{12}
+ \mathfrak{t}^{14}
),\\
%
%
&\langle W_{\ydiagram{4}}W_{\overline{\ydiagram{4}}}\rangle^{U(2)\text{ ADHM-}[1](C)}(z;\mathfrak{t})\nonumber \\
&=
\frac{1
}{\prod_\pm (1-z^{\pm 1}\mathfrak{t})(1-z^{\pm 2}\mathfrak{t}^2)}
(1
+4(z+z^{-1})\mathfrak{t}
-2\mathfrak{t}^2
+3(z+z^{-1})\mathfrak{t}^3
-5\mathfrak{t}^4
+2(z+z^{-1})\mathfrak{t}^5\nonumber \\
&\quad -3\mathfrak{t}^6
+(z+z^{-1})\mathfrak{t}^7
-\mathfrak{t}^8
),\\
&\langle W_{\ydiagram{4}}W_{\overline{\ydiagram{4}}}\rangle^{U(3)\text{ ADHM-}[1](C)}(z;\mathfrak{t})\nonumber \\
&=
\frac{1}{
\prod_\pm (1-z^{\pm 1}\mathfrak{t})
(1-z^{\pm 2}\mathfrak{t}^2)
(1-z^{\pm 3}\mathfrak{t}^3)
}
(
1
 + (4z^{-1} + 4 z)                   \mathfrak{t}
+ (4z^{-2} +8 +  4 z^2)             \mathfrak{t}^2\nonumber \\
&\quad + (6z^{-3} + 8z^{-1} + 8 z + 6 z^3) \mathfrak{t}^3  
 + ( z^{-2} +15 + z^2)              \mathfrak{t}^4
+ (9z^{-3} + 3z^{-1} + 3 z + 9 z^3) \mathfrak{t}^5\nonumber \\
&\quad + ( - 7z^{-2}+11 - 7 z^2)           \mathfrak{t}^6  
 + (7z^{-3} - 3z^{-1} - 3 z + 7 z^3) \mathfrak{t}^7
+ (- 8z^{-2}+7  - 8 z^2)            \mathfrak{t}^8\nonumber \\
&\quad + (5z^{-3} - 2z^{-1} - 2 z + 5 z^3) \mathfrak{t}^9  
 + ( - 6z^{-2}+8 - 6 z^2)            \mathfrak{t}^{10}
+ (2z^{-3} - 3z^{-1} - 3 z + 2 z^3) \mathfrak{t}^{11}\nonumber \\
&\quad + ( - 3z^{-2}+6 - 3 z^2)            \mathfrak{t}^{12}
 + (z^{-3} - z^{-1} - z + z^3)     \mathfrak{t}^{13}
+ ( - z^{-2}+3 - z^2)              \mathfrak{t}^{14}\nonumber \\
&\quad + (-z^{-1} - z)                  \mathfrak{t}^{15}
+ \mathfrak{t}^{16}
).
\end{align}
The results for $l\ge 2$ are listed in appendix \ref{app_unflavoredC}.
The $2$-point function of symmetric Wilson lines in the Coulomb limit diverges as the size of the representation $k$ becomes large.
This is in contrast to the behavior in the Higgs limit studied in section \ref{sec_Higgs}.

In the large $l$ limit, on the other hand, we obtain the following unified formula thanks to the simplification \eqref{largelsimplification}
\begin{align}
\langle W_{(k)}W_{\overline{(k)}}\rangle^{U(N)\text{ADHM-}[\infty](C)}(z;\mathfrak{t})=
\frac{(\mathfrak{t}^{2N};\mathfrak{t}^2)_k}{(\mathfrak{t}^2;\mathfrak{t}^2)_N(\mathfrak{t}^2;\mathfrak{t}^2)_k}.
\label{largelantisym}
\end{align}

\subsubsection{Multi-point functions of charged Wilson lines}
\label{sec_chargedCoulomb}
Let us also consider the correlation function of charged Wilson lines
\begin{align}
\langle W_{m_1}W_{m_2}\cdots W_{m_k}\rangle^{U(N)\text{ ADHM-}[l](C)}(z;\mathfrak{t}),
\end{align}
with $m_j\neq 0$ for all $j$ and $\sum_{j=1}^km_j=0$, which is calculated by substituting power sum symmetric functions $p_{m_j}(s)$ \eqref{powersum} for $\chi_{{\cal R}_j}(s)$ in \eqref{wilsonsCoulomb}. 
These correlators are important in the sense that they form a complete basis of the correlators of arbitrary number of Wilson lines in arbitrary representations $\langle
W_{\lambda^{(1)}}
W_{\lambda^{(2)}}
\cdots
\rangle^{U(N)\text{ ADHM-}[l](C)}(z;\mathfrak{t})$ for $N\le k$. 

In the Fermi-gas formalism these correlators can be obtained by inserting $\prod_{j=1}^k\prod_{i=1}^N(1+a_js_i^{m_j})$ instead of $E(a;s)E(b;s^{-1})H(c;s)H(d;s^{-1})$ into the generating function \eqref{Coulombgeneratingfcn} and reading the coefficient of $a_1\cdots a_k$.
Hence the calculation reduces to the calculation of trace $\text{Tr}{\rho}_f^n$ with $f=\prod_{j=1}^k(1+a_js^{m_j})$ to the order $a_1\cdots a_k$.
For simplicity we consider only the cases with $k=2$ and $k=3$, and here we only display the final results for $\text{Tr}\rho^n_f$, the grand canonical correlator $\Xi_f^{(C)}(u)$ and the correlation function $\langle W_{m_1}W_{m_2}\cdots W_{m_k}\rangle^{U(N)\text{ ADHM-}[l](C)}(z;\mathfrak{t})$.
For the detail of the derivation of $\text{Tr}\rho^n_f$ and $\Xi_f^{(C)}(u)$, see appendix \ref{app_chargedCoulombderivation}.

First let us consider the case with $k=2$, where the charges are $(m_1,m_2)=(m_1,-m_1)$.
Without loss of generality we assume $m_1>0$.
In this case we find
\begin{align}
\text{Tr}\rho_{(1+a_1s^{m_1})(1+a_2s^{-m_1})}^n=\frac{1}{1-\mathfrak{t}^{2n}}\biggl(1+a_1a_2\frac{n(1-\mathfrak{t}^{2m_1n})}{1-\mathfrak{t}^{2m_1}}+{\cal O}(a_1^2,a_2^2)\biggr).
\label{Trrhochargedk2}
\end{align}
Plugging this to the grand partition function \eqref{XiCf}, we find
\begin{align}
\Xi_f^{(C)}(u)&=\Xi^{(C)}(u)\biggl[1+a_1a_2\sum_{n=1}^{m_1}
\frac{(\mathfrak{t}^{-2n+2+2m_1};\mathfrak{t}^2)_n}{1-\mathfrak{t}^{2m_1}}
{\cal I}^{U(1)\text{ ADHM-}[l](C)}(z^n;\mathfrak{t}^n)u^n
\biggr]
+{\cal O}(a_1^2,a_2^2).
\label{Xifchargedk2}
\end{align}
Expanding the whole right-hand side in $u$ we obtain
\begin{align}
&\langle W_{m_1}W_{-{m_1}}\rangle^{U(N)\text{ ADHM-}[l](C)}(z;\mathfrak{t})\nonumber \\
&=\sum_{n=1}^{\text{min}(N,m_1)}
\frac{(\mathfrak{t}^{-2n+2+2m_1};\mathfrak{t}^2)_n}{1-\mathfrak{t}^{2m_1}}
{\cal I}^{U(1)\text{ ADHM-}[l](C)}(z^n;\mathfrak{t}^n){\cal I}^{U(N-n)\text{ ADHM-}[l](C)}(z;\mathfrak{t}).
\label{ICchargedW2}
\end{align}

Next we consider the case with $k=3$.
Since the three charges $m_1,m_2$ and $m_3$ add up to zero, the signs of the charges are either $++-$ or $+--$.
Noticing that the generating function for $(m_1,m_2,m_3)$ is identical to that for $(-m_1,-m_2,-m_3)$, we may assume without loss of generality that $m_1,m_2>0$ and $m_3=-m_1-m_2<0$.
In this case we find
\begin{align}
&\text{Tr}\rho_{(1+a_1s^{m_1})(1+a_2s^{m_2})(1+a_3s^{-m_1-m_2})}^n\nonumber \\
&=\frac{1}{1-\mathfrak{t}^{2n}}\Bigl(1+a_1a_2a_3\frac{n(1-\mathfrak{t}^{2nm_1})(1-\mathfrak{t}^{2nm_2})}{(1-\mathfrak{t}^{2m_1})(1-\mathfrak{t}^{2m_2})}+{\cal O}(a_1^2,a_2^2,a_3^2)\Bigr),\label{Trrhochargedk3} \\
&\Xi_f^{(C)}(u)\nonumber \\
&=\Xi^{(C)}(u)
\biggl[1+a_1a_2a_3
\sum_{n=1}^{m_1+m_2}
\frac{
(\mathfrak{t}^{-2n+2+2m_1};\mathfrak{t}^2)_n
+(\mathfrak{t}^{-2n+2+2m_2};\mathfrak{t}^2)_n
-(\mathfrak{t}^{-2n+2+2m_1+2m_2};\mathfrak{t}^2)_n
}{
(1-\mathfrak{t}^{2m_1})
(1-\mathfrak{t}^{2m_2})
}\nonumber \\
&\quad\quad \times {\cal I}^{U(1)\text{ ADHM-}[l](C)}(z^n;\mathfrak{t}^n)u^n\biggr]+{\cal O}(a_1^2,a_2^2,a_3^2),
\label{Xifchargedk3}
\end{align}
and
\begin{align}
&\langle W_{m_1}W_{m_2}W_{-m_1-m_2}\rangle^{U(N)\text{ ADHM-}[l](C)}(z;\mathfrak{t})\nonumber \\
&=
\sum_{n=1}^{\text{min}(N,m_1+m_2)}
\frac{
(\mathfrak{t}^{-2n+2+2m_1};\mathfrak{t}^2)_n
+(\mathfrak{t}^{-2n+2+2m_2};\mathfrak{t}^2)_n
-(\mathfrak{t}^{-2n+2+2m_1+2m_2};\mathfrak{t}^2)_n
}{
(1-\mathfrak{t}^{2m_1})
(1-\mathfrak{t}^{2m_2})
}\nonumber \\
&\quad \times {\cal I}^{U(1)\text{ ADHM-}[l](C)}(z^n;\mathfrak{t}^n)
{\cal I}^{U(N-n)\text{ ADHM-}[l](C)}(z;\mathfrak{t}).
\label{ICchargedW3}
\end{align}


\subsubsection{Large $N$ limit}
Let us study the large $N$ limit of the correlation functions of Wilson lines in the Coulomb limit we have studied above.
We can obtain from the plethystic logarithm \cite{MR1601666} of the large $N$ normalized $2$-point functions of Wilson lines in the representation $\mathcal{R}$ 
the single particle gravity indices that would capture the quantum fluctuation modes on the gravity dual geometry\footnote{
See \cite{Gang:2012yr,Drukker:2015spa,Hatsuda:2023iwi,Hatsuda:2023imp,Hatsuda:2023iof} 
for the relevant analysis of large $N$ line defect correlators in 4d $\mathcal{N}=4$ $U(N)$ SYM theory.
}
$X$
\begin{align}
i_{X}^{(C)}(z;\mathfrak{t})&=
\text{PL}[\langle \mathcal{W}_{\mathcal{R}}\mathcal{W}_{\overline{\mathcal{R}}}\rangle^{U(\infty)\text{ADHM-}[l] (C)}(z;\mathfrak{t}) ]
\nonumber\\
&=\sum_{d\ge1}\frac{\mu(d)}{d}\log [
\langle \mathcal{W}_{\mathcal{R}}\mathcal{W}_{\overline{\mathcal{R}}}\rangle^{U(\infty)\text{ ADHM-}[l] (C)}(z^d;\mathfrak{t}^d)
],
\end{align}
where $\mu(d)$ is the M\"{o}bius function
\begin{align}
\mu(d)&=
\begin{cases}
0&\textrm{if $d$ has repeated primes}\cr
1&\textrm{if $d=1$}\cr
(-1)^k&\textrm{if $d$ is a product of $k$ distinct primes}\cr
\end{cases}. 
\end{align}

\begin{itemize}

\item Antisymmetric representations

In the large $N$ limit $N\rightarrow \infty$ with $k$ being fixed,  
the $2$-point function \eqref{ICantisymW} 
normalized by the Coulomb index ${\cal I}^{U(\infty)\text{ ADHM-}[l](C)}$ (see (\ref{normalize_Wilson}) for the convention) reduces to
\begin{align}
\langle {\cal W}_{(1^k)}{\cal W}_{\overline{(1^k)}}\rangle^{U(\infty)\text{ADHM-}[l](C)}(z;\mathfrak{t})=
{\cal I}^{U(k)\text{ADHM-}[l](C)}(z;\mathfrak{t}).
\end{align}
When $k=1$, the ADHM theory is holographically dual to the M-theory geometry $AdS_4\times S^7/\mathbb{Z}_l$ 
and the Wilson line in the fundamental representation is expected to be holographically dual to an M2-brane wrapping the $AdS_2$ 
in the global $AdS_4$ and the M-theory circle. 
The resulting Coulomb gravity index is
\begin{align}
\label{gindexCl_fund}
i_{\textrm{X}}^{(C)}(z;\mathfrak{t})&=\mathfrak{t}^2+\mathfrak{t}^{l}z^{l}+\mathfrak{t}^{l}z^{-l}-\mathfrak{t}^{2l}. 
\end{align}
It indicates a finite number of quantum fluctuations of the M2-brane on the gravity dual geometry. 
For $l=1$, the Coulomb gravity index gets simplified as
\begin{align}
\label{gindexC_fund}
i_{\textrm{X}}^{(C)}(z;\mathfrak{t})&=\mathfrak{t}z+\mathfrak{t}z^{-1}. 
\end{align}
The gravity index (\ref{gindexC_fund}) involves two bosonic scalar fields. 
They are expected to describe the quantum fluctuations around the motion of the M2-brane wrapping the $AdS_2$. 
They are similar to the quantum fluctuation modes of the fundamental string wrapping $AdS_2$ $\subset$ $AdS_5$ 
which is dual to the fundamental Wilson line in 4d $\mathcal{N}=4$ $U(N)$ SYM theory \cite{Drukker:2000ep}.\footnote{
From the semiclassical analysis of the action of a single M2-brane in \cite{Sakaguchi:2010dg} 
it is argued that the spectrum of the quantum fluctuations around the $AdS_2\times S^1$ solution 
involves an infinite set of $\mathcal{N}=1$ supermultiplets on the $AdS_2$. 
However, it is not clear how it can be verified from the ADHM theory as the BPS index only counts a finite number of excitations.
}

When $l=1$ and $k\rightarrow \infty$ with $\frac{k}{N}<1$, 
the large $N$ normalized $2$-point function is the large $N$ index of the $U(N)$ ADHM theory with one flavor. 
In this case the gravity index is 
\begin{align}
\label{gindexC_brane}
i_{\textrm{X}}^{(C)}(z;\mathfrak{t})&=
\sum_{m=1}^{\infty}\mathfrak{t}^{2m}
+\sum_{m=0}^{\infty}\sum_{k=1}^{\infty}
\left(
\mathfrak{t}^{2m+k}z^k
+\mathfrak{t}^{2m+k}z^{-k}
\right)
\nonumber\\
&=\frac{\mathfrak{t}z+\mathfrak{t}z^{-1}-\mathfrak{t}^2}
{(1-\mathfrak{t}z)(1-\mathfrak{t}z^{-1})}. 
\end{align}
Unlike the case with $k=1$, 
the gravity index (\ref{gindexC_brane}) involves an infinite tower of quantum fluctuation modes. 
Such an infinite tower of excitations is similar to that for the geometry dual to the large (anti)symmetric representation of the Wilson line 
in 4d $\mathcal{N}=4$ $U(N)$ SYM theory. 

\item Symmetric representations

In contrast to the case of antisymmetric representations, even though we have an explicit expression for the canonical partition function
\eqref{Omeganusym} for all $\nu$ and all order in $b,c$, it is still not obvious to write down the correlators $\langle W_{(k)}W_{(\overline{k})}\rangle^{U(N)\text{ ADHM-}[l](C)}$ for general $N,l,$ and $k$.
Nevertheless comparing the closed-form expression for the correlators at different values of $N$ for each $l,k$, we observe that the coefficients of the small $\mathfrak{t}$ expansion 
saturate as $N$ increases. 
For example, for $l=1$, $k\le 3$ and $l=2$, $k=2$ we have
\begin{align}
&\langle {\cal W}_{\ydiagram{2}}{\cal W}_{\overline{\ydiagram{2}}}\rangle^{U(\infty)\text{ADHM-}[1](C)}(z;\mathfrak{t})=
\frac{
1
+(z+z^{-1})\mathfrak{t}
+(-z^2-1-z^{-2})\mathfrak{t}^2
+(z+z^{-1})\mathfrak{t}^3
}{
\prod_\pm
(1-z^{\pm 1}\mathfrak{t})
(1-z^{\pm 2}\mathfrak{t}^2)
},\label{symmetricCoulombl1z1k2largeNconjecture2} \\
&\langle {\cal W}_{\ydiagram{3}}{\cal W}_{\overline{\ydiagram{3}}}\rangle^{U(\infty)\text{ ADHM-}[1](C)}(z;\mathfrak{t})\nonumber \\
&=
\frac{1}{\prod_\pm
(1-z^{\pm 1}\mathfrak{t})
(1-z^{\pm 2}\mathfrak{t}^2)
(1-z^{\pm 3}\mathfrak{t}^3)}
(1
+ (2z^{-1} + 2 z)\mathfrak{t} + (-1 - z^{-2} - z^2) \mathfrak{t}^2\nonumber \\
&\quad + (2z^{-1} + 2 z) \mathfrak{t}^3
+ (1 - 2z^{-4} - 3z^{-2} - 3 z^2 - 2 z^4) \mathfrak{t}^4\nonumber \\
&\quad + (z^{-5} + 4z^{-3} + 2z^{-1} + 2 z + 4 z^3 + z^5) \mathfrak{t}^5
+ (3 - 2z^{-4} - 4z^{-2} - 4 z^2 - 2 z^4) \mathfrak{t}^6\nonumber \\
&\quad + (3z^{-3} + z^{-1} + z + 3 z^3) \mathfrak{t}^7
+ (-z^{-4} - 4z^{-2} - 4 z^2 - z^4) \mathfrak{t}^8\nonumber \\
&\quad + (2z^{-3} + 2z^{-1} + 2 z + 2 z^3) \mathfrak{t}^9    
+ (-z^{-2} - z^2) \mathfrak{t}^{10} 
),\label{symmetricCoulombl1k3largeNconjecture2} \\
&\langle {\cal W}_{\ydiagram{2}}{\cal W}_{\overline{\ydiagram{2}}}\rangle^{U(\infty)\text{ ADHM-}[2](C)}(z;\mathfrak{t})\nonumber \\
&=
\frac{
}{\prod_\pm (1 - z^{\pm 2}\mathfrak{t}^2) (1 - z^{\pm 4}\mathfrak{t}^4)
}
(1
 + (z^{-2}+1+z^2)\mathfrak{t}^2
+ (-z^{-4}+2z^{-2}+2z^2-z^4)\mathfrak{t}^4\nonumber \\
&\quad + (-z^{-4}+z^{-2}-1+z^2-z^4)\mathfrak{t}^6
+ (z^{-2}+1+z^2)\mathfrak{t}^8
).\label{symmetricCoulombl2k2largeNconjecture2}
\end{align}
We have confirmed these expressions at least to the order $\mathfrak{t}^{14}$.
For higher rank cases expressions become rather lengthy. 
We show the large $N$ correlators of the Wilson lines in the rank-$4$ symmetric representations for $l=1$ and rank-$3$ symmetric representations for $l=2$ in the unflavored limit ($z=1$)
\begin{align}
%
&\langle{\cal W}_{\ydiagram{4}}{\cal W}_{\overline{\ydiagram{4}}}\rangle^{U(\infty)\text{ADHM-}[1](C)}(\mathfrak{t})\nonumber \\
&=
\frac{1}{
(1 - \mathfrak{t})^2(1 - \mathfrak{t}^2)^2(1 - \mathfrak{t}^3)^2(1 - \mathfrak{t}^4)^2
}(1 
+ 6 \mathfrak{t} 
- 2 \mathfrak{t}^{2}
 + 10 \mathfrak{t}^{3}
- 13 \mathfrak{t}^{4}
+ 12 \mathfrak{t}^{5}
- 3 \mathfrak{t}^{6}\nonumber \\
&\quad + 24 \mathfrak{t}^{8}
- 32 \mathfrak{t}^{9}
+ 34 \mathfrak{t}^{10}
- 38 \mathfrak{t}^{11}
+ 32 \mathfrak{t}^{12}
- 12 \mathfrak{t}^{13}
+ 7 \mathfrak{t}^{14}+ 6 \mathfrak{t}^{15}
- 18 \mathfrak{t}^{16}
+ 12 \mathfrak{t}^{17}\nonumber \\
&\quad - 6 \mathfrak{t}^{18}
+ 10 \mathfrak{t}^{19}
- 8 \mathfrak{t}^{20}
+ 2 \mathfrak{t}^{21}
),\\
&\langle {\cal W}_{\ydiagram{3}}{\cal W}_{\overline{\ydiagram{3}}}\rangle^{U(\infty)\text{ ADHM-}[2](C)}(\mathfrak{t})\nonumber \\
&=
\frac{
1 + 5 \mathfrak{t}^2 + 6 \mathfrak{t}^4 + 7 \mathfrak{t}^6
+ 7 \mathfrak{t}^8 + 5 \mathfrak{t}^{10} + 7 \mathfrak{t}^{12} + 2 \mathfrak{t}^{14} + 3 \mathfrak{t}^{16} + 5 \mathfrak{t}^{18}
}{
(1 - \mathfrak{t}^2)^2 (1 - \mathfrak{t}^4)^2 (1 - \mathfrak{t}^6)^2}.
\end{align}
We have confirmed these expressions to the order $\mathfrak{t}^{23}$ and $\mathfrak{t}^{20}$ respectively.

On the other hand, in the large $l$ limit we obtain from \eqref{largelantisym} a simple unified formula for arbitrary $k$.
Curiously, in this limit the $2$-point function coincides with the $2$-point function in an antisymmetric representation of the same dimension
\begin{align}
\langle {\cal W}_{(k)}{\cal W}_{(\overline{k})}\rangle^{U(\infty)\text{ ADHM-}[\infty](C)}(z;\mathfrak{t})=\langle {\cal W}_{(1^k)}{\cal W}_{(\overline{1^k})}\rangle^{U(\infty)\text{ ADHM-}[\infty](C)}(z;\mathfrak{t})=\frac{1}{(\mathfrak{t}^2;\mathfrak{t}^2)_k}.
\end{align}

\item Charged Wilson lines

For the $2$-point functions and the $3$-point functions, we find from \eqref{ICchargedW2} and \eqref{ICchargedW3}
\begin{align}
&\langle {\cal W}_{m_1}{\cal W}_{-m_1}\rangle^{U(\infty)\text{ ADHM-}[l](C)}(z;\mathfrak{t})
=\sum_{n=1}^{m_1}\frac{(\mathfrak{t}^{-2n+2+2m_1};\mathfrak{t}^2)_n}{1-\mathfrak{t}^{2m_1}}{\cal I}^{U(1)\text{ ADHM-}[l](C)}(z^n;\mathfrak{t}^n),\nonumber \\
&\langle {\cal W}_{m_1}{\cal W}_{m_2}{\cal W}_{-m_1-m_2}\rangle^{U(\infty)\text{ ADHM-}[l](C)}(z;\mathfrak{t})\nonumber \\
&=\sum_{n=1}^{m_1}\frac{
(\mathfrak{t}^{-2n+2+2m_1};\mathfrak{t}^2)_n
+(\mathfrak{t}^{-2n+2+2m_2};\mathfrak{t}^2)_n
-(\mathfrak{t}^{-2n+2+2m_1+2m_2};\mathfrak{t}^2)_n
}{
(1-\mathfrak{t}^{2m_1})
(1-\mathfrak{t}^{2m_2})
}\nonumber \\
&\quad \times {\cal I}^{U(1)\text{ ADHM-}[l](C)}(z^n;\mathfrak{t}^n).
\end{align}

\end{itemize}

\section{Higgs line defect indices}
\label{sec_Higgs}
Next consider the Higgs limit of the Wilson line defect correlator (\ref{Wilsonfullindex}). 
It is obtained by taking the limit 
\begin{align}
q,t^{-1}\rightarrow 0\text{ with }\mathfrak{t}=q^{\frac{1}{4}}t^{1}\text{ fixed}, 
\label{Higgslimit}
\end{align}
for which we obtain the matrix integral
\begin{align}
\label{Higgs_wilson}
&\langle W_{\mathcal{R}_1}\cdots W_{\mathcal{R}_k}\rangle^{\textrm{$U(N)$ ADHM-$[l] (H)$}}
(x,y_\alpha;\mathfrak{t})\nonumber \\
&=
\frac{1}{N!}\oint \prod_{i=1}^{N}
\frac{ds_{i}}{2\pi is_i}
\frac{
\prod_{i\neq j}(1-\frac{s_i}{s_j})
\prod_{ij}(1-\mathfrak{t}^2\frac{s_i}{s_j})
}
{
\prod_{ij}(1-\mathfrak{t}\frac{s_i}{s_j}x)(1-\mathfrak{t}\frac{s_i}{s_j}x^{-1})
}
\prod_{i=1}^N
\prod_{\alpha=1}^{l}
\frac{1}{(1-\mathfrak{t}s_i y_{\alpha})(1-\mathfrak{t}s_{i}^{-1}y_{\alpha}^{-1})}
\prod_{j=1}^k \chi_{\mathcal{R}_j}(s). 
\end{align}

\subsection{Hall-Littlewood expansions}
In this subsection, we derive the closed-form expressions for the Higgs line defect correlators (\ref{Higgs_wilson}) 
from the orthogonality (\ref{HL_inner}) of the Hall-Littlewood functions. 

\subsubsection{$\mathcal{N}=8$ $U(N)$ SYM}
While the localization formula of the full supersymmetric index leads to divergent series 
for $\mathcal{N}=8$ SYM theories, i.e.~ADHM theory without fundamental hyper, due to the non-positive dimension of the monopole operator, 
the Higgs line defect correlators are well-defined matrix integral with the form
\begin{align}
\label{unN8_wilsons}
&
\langle W_{\mathcal{R}_1}\cdots W_{\mathcal{R}_k}\rangle^{\textrm{$\mathcal{N}=8$ $U(N) (H)$}}(x;\mathfrak{t})
=
\frac{1}{N!}\oint \prod_{i=1}^{N}
\frac{ds_{i}}{2\pi is_i}
\frac{
\prod_{i\neq j}(1-\frac{s_i}{s_j})
\prod_{ij}(1-\mathfrak{t}^2\frac{s_i}{s_j})
}
{
\prod_{ij}(1-\mathfrak{t}\frac{s_i}{s_j}x)(1-\mathfrak{t}\frac{s_i}{s_j}x^{-1})
}
\prod_{j=1}^k \chi_{\mathcal{R}_j}(s). 
\end{align}
While there is no non-trivial $1$-point function, 
the matrix integral (\ref{unN8_wilsons}) is non-trivial even if there is no insertion of the Wilson line. 
It gives rise to the Higgs index 
\begin{align}
\langle  1 \rangle^{\textrm{$\mathcal{N}=8$ $U(N) (H)$}}(x;\mathfrak{t})&=\mathcal{I}^{\textrm{$\mathcal{N}=8$ $U(N) (H)$}}(x;\mathfrak{t}). 
\end{align}

To evaluate the $\mathcal{N}=8$ Higgs correlator (\ref{unN8_wilsons}), 
we start by noticing that the Hall-Littlewood functions satisfy the Cauchy identity
\begin{align}
\label{cauchy}
\sum_{\mu}
b_{\mu}(\mathfrak{t})
P_{\mu}(x;\mathfrak{t})P_{\mu}(y;\mathfrak{t})
&=\prod_{i\ge1}\prod_{j\ge1}
\frac{1-\mathfrak{t}x_i y_j}{1-x_i y_j}, 
\end{align}
where $b_\mu(\mathfrak{t})$ is defined as \eqref{bmut},
we can expand the integrand in the matrix integral (\ref{unN8_wilsons}) 
and then carry out the integration by the orthogonality relation (\ref{HL_inner}). 

Without any insertion of the Wilson lines, we find the Higgs index 
\begin{align}
\label{unN8_index1}
\mathcal{I}^{\textrm{$\mathcal{N}=8$ $U(N) (H)$}}(x;\mathfrak{t})
&=\sum_{\begin{smallmatrix}
\lambda\\
\ell(\lambda)\le N\\
\end{smallmatrix}}
\frac{\mathfrak{t}^{|\lambda|}x^{|\lambda|}}
{(\mathfrak{t}x^{-1};\mathfrak{t}x^{-1})_{N-l(\lambda)}}
=\sum_{k=0}^{N}\frac{\mathfrak{t}^kx^k}
{(\mathfrak{t}x;\mathfrak{t}x)_{k} (\mathfrak{t}x^{-1};\mathfrak{t}x^{-1})_{N-k}}. 
\end{align}

For example, for $N=1$, $2$, $3$ and $4$ we have 
\begin{align}
\label{u1N8_index}
&\mathcal{I}^{\textrm{$\mathcal{N}=8$ $U(1) (H)$}}(x;\mathfrak{t})=\frac{1-\mathfrak{t}^2}{(1-\mathfrak{t}x) (1-\mathfrak{t}x^{-1})}
=1+\sum_{n=1}^{\infty}\mathfrak{t}^n(x^n+x^{-n}), \\
\label{u2N8_index}
&\mathcal{I}^{\textrm{$\mathcal{N}=8$ $U(2) (H)$}}(x;\mathfrak{t})=
\frac{1-\mathfrak{t}^3(x+x^{-1})-\mathfrak{t}^4+\mathfrak{t}^5(x+x^{-1})}
{(1-\mathfrak{t}x)(1-\mathfrak{t}^2x^2)(1-\mathfrak{t}x^{-1})(1-\mathfrak{t}^2x^{-2})}, \\
\label{u3N8_index}
&\mathcal{I}^{\textrm{$\mathcal{N}=8$ $U(3) (H)$}}(x;\mathfrak{t})\nonumber \\
&=
\frac{1}
{(1-\mathfrak{t}x)(1-\mathfrak{t}^2x^2)(1-\mathfrak{t}^3x^3)(1-\mathfrak{t}x^{-1})(1-\mathfrak{t}^2x^{-2})(1-\mathfrak{t}^3x^{-3})}
(1-\mathfrak{t}^4(1+x^2+x^{-2})\nonumber \\
&\quad -\mathfrak{t}^5(x+x^{-1})+\mathfrak{t}^6+\mathfrak{t}^7(x^3+x+x^{-1}+x^{-3})
+\mathfrak{t}^8(x^2+x^{-2})-\mathfrak{t}^9(x^{3}+x^{-3})-\mathfrak{t}^{10}), \\
\label{u4N8_index}
&\mathcal{I}^{\textrm{$\mathcal{N}=8$ $U(4) (H)$}}(x;\mathfrak{t})\nonumber \\
&=\frac{1}{\prod_{i=1}^{4}(1-\mathfrak{t}^ix^i)(1-\mathfrak{t}^ix^{-i})}
(
1+\mathfrak{t}^5(-x^3-x-x^{-1}-x^{-3})+\mathfrak{t}^6(-1-x^2-x^{-2})\nonumber \\
&\quad +\mathfrak{t}^8(1+x^2+x^{-2})
+\mathfrak{t}^9(x^5+x^3+2x+2x^{-1}+x^{-3}+x^{-5})+\mathfrak{t}^{10}(x^4+x^{-4})\nonumber \\
&\quad +\mathfrak{t}^{12}(-x^6-x^4-x^2-x^{-2}-x^{-4}-x^{-6})+\mathfrak{t}^{13}(-x^5-x-x^{-1}-x^{-5})\nonumber \\
&\quad +\mathfrak{t}^{14}(-1+x^{6}+x^{-6})
+\mathfrak{t}^{16}(x^2+x^{-2})
). 
\end{align}

The $2$-point functions of the Wilson lines in the antisymmetric and symmetric representations 
can be obtained by using the Pieri rules \cite{MR1354144,kirillov1998new}
\begin{align}
\label{pieri1}
e_k(s)P_{\mu}(s;\mathfrak{t})
&=\sum_{
\begin{smallmatrix}
\lambda\supset \mu\\
|\lambda/\mu|=k\\
\end{smallmatrix}
}
\prod_{i\ge1}\left[
\begin{matrix}
\mu'_i-\mu'_{i+1}\\
\mu'_i-\lambda'_i\\
\end{matrix}
\right]_{\mathfrak{t}}
P_{\lambda}(s;\mathfrak{t})
\end{align}
and
\begin{align}
\label{pieri2}
h_k(s)P_{\mu}(s;\mathfrak{t})
&=\sum_{
\begin{smallmatrix}
\lambda\supset \mu\\
|\lambda/\mu|=k\\
\end{smallmatrix}
}
\mathfrak{t}^{n(\lambda/\mu)}
\prod_{i\ge1}\left[
\begin{matrix}
\mu'_i-\lambda'_{i+1}\\
\mu'_i-\lambda'_i\\
\end{matrix}
\right]_{\mathfrak{t}}
P_{\lambda}(s;\mathfrak{t}),
\end{align}
where 
$\left[
\begin{matrix}
n\\
k\\
\end{matrix}
\right]_{q}
$
is the $q$-binomial coefficient (\ref{qbincoef}) and $n(\lambda/\mu)$ is defined as \eqref{nlambda/mu}.
From the orthogonality relation (\ref{HL_inner}), 
the Cauchy identity (\ref{cauchy}) and the Pieri rules (\ref{pieri1})-(\ref{pieri2}) 
we find
\begin{align}
&
\langle W_{(1^k)}W_{(\overline{1^k})}\rangle^{\textrm{$\mathcal{N}=8$ $U(N) (H)$}}(x;\mathfrak{t})
\nonumber\\
&=\sum_{\begin{smallmatrix}
\alpha\supset \lambda\\
|\alpha/\lambda|=k\\
\end{smallmatrix}}
\sum_{\begin{smallmatrix}
\lambda\\
\ell(\lambda)\le N\\
\end{smallmatrix}}
\mathfrak{t}^{|\lambda|}x^{|\lambda|}
\frac{\prod_{j\ge1}(\mathfrak{t}x^{-1};\mathfrak{t}x^{-1})_{m_j(\lambda)}}
{(\mathfrak{t}x^{-1};\mathfrak{t}x^{-1})_{N-\ell(\alpha)} \prod_{j\ge1}(\mathfrak{t}x^{-1};\mathfrak{t}x^{-1})_{m_j(\alpha)}}
\prod_{i\ge1}\left[
\begin{matrix}
\alpha'_{i}-\alpha'_{i+1}\\
\alpha'_{i}-\lambda'_{i}\\
\end{matrix}
\right]_{\mathfrak{t}x^{-1}}^2
\end{align}
and
\begin{align}
&
\langle W_{(k)}W_{(\overline{k})}\rangle^{\textrm{$\mathcal{N}=8$ $U(N) (H)$}}(x;\mathfrak{t})
\nonumber\\
&=
\sum_{\begin{smallmatrix}
\alpha\supset \lambda\\
|\alpha/\lambda|=k\\
\end{smallmatrix}}
\sum_{\begin{smallmatrix}
\lambda\\
\ell(\lambda)\le N\\
\end{smallmatrix}}
\mathfrak{t}^{2n(\alpha/\lambda)+|\lambda|}
x^{-2n(\alpha/\lambda)+|\lambda|}
\frac{\prod_{j\ge1}(\mathfrak{t}x^{-1};\mathfrak{t}x^{-1})_{m_j(\lambda)}}
{(\mathfrak{t}x^{-1};\mathfrak{t}x^{-1})_{N-\ell(\alpha)} \prod_{j\ge1}(\mathfrak{t}x^{-1};\mathfrak{t}x^{-1})_{m_j(\alpha)}}\nonumber\\
&\quad \times 
\prod_{i\ge1}\left[
\begin{matrix}
\alpha'_{i}-\lambda'_{i+1}\\
\alpha'_{i}-\lambda'_{i}\\
\end{matrix}
\right]_{\mathfrak{t}x^{-1}}^2. 
\end{align}

For example, for $\mathcal{N}=8$ $U(2)$ SYM theory, the $2$-point function of the Wilson lines in the fundamental representation is 
\begin{align}
\label{u2N8_w1w1}
\langle W_{\ydiagram{1}}W_{\overline{\ydiagram{1}}}\rangle^{\textrm{$\mathcal{N}=8$ $U(2) (H)$}}(x;\mathfrak{t})
&=\frac{1-\mathfrak{t}^2-\mathfrak{t}^3(x+x^{-1})+\mathfrak{t}^4+\mathfrak{t}^5(x+x^{-1})-\mathfrak{t}^6}
{(1-\mathfrak{t}x)^2(1-\mathfrak{t}x^{-1})^2}.
\end{align}

Although the flavored $2$-point functions of the Wilson lines 
in the rank-$k$ symmetric representations can be written as the rational functions, they become expensive as $k$ increases. 
For example, for $k=2$ and $3$ we find
\begin{align}
&
\langle W_{\ydiagram{2}}W_{\overline{\ydiagram{2}}}\rangle^{\textrm{$\mathcal{N}=8$ $U(2) (H)$}}(x;\mathfrak{t})
\nonumber\\
&=\frac{1}{(1-\mathfrak{t}x)(1-\mathfrak{t}^2x^2)(1-\mathfrak{t}x^{-1})(1-\mathfrak{t}^2x^{-2})}
(1+\mathfrak{t}(x+x^{-1})+\mathfrak{t}^2(1+x^2+x^{-2})\nonumber \\
&\quad +\mathfrak{t}^3(-2x-2x^{-1})+\mathfrak{t}^4(-4-2x^2-2x^{-2})
+\mathfrak{t}^5(-x^3-x^{-3})+\mathfrak{t}^6(3+x^{2}+x^{-2})\nonumber \\
&\quad +\mathfrak{t}^7(x^3+2x+2x^{-1}+x^{-3})
-\mathfrak{t}^8+\mathfrak{t}^9(-x-x^{-1})
)
\end{align}
and 
\begin{align}
&
\langle W_{\ydiagram{3}}W_{\overline{\ydiagram{3}}}\rangle^{\textrm{$\mathcal{N}=8$ $U(2) (H)$}}(x;\mathfrak{t})
\nonumber\\
&=\frac{1}{(1-\mathfrak{t}x)^2(1-\mathfrak{t}x^{-1})^2}
(
1+\mathfrak{t}^2(x^2+x^{-2})+\mathfrak{t}^3(-2x-2x^{-1})+\mathfrak{t}^4(-x^2-x^{-2})\nonumber \\
&\quad +\mathfrak{t}^5(-x^3+x+x^{-1}-x^{-3})
+\mathfrak{t}^6(x^2+x^{-2})+\mathfrak{t}^7(x^3+x+x^{-1}+x^{-3})\nonumber \\
&\quad +\mathfrak{t}^8(-1-x^2-x^{-2})
). 
\end{align}

\subsubsection{$\mathcal{N}=8$ grand canonical ensemble}
Let us define the grand canonical index by
\begin{align}
\label{N8_gindex1}
\Xi^{\mathcal{N}=8}(u;x;\mathfrak{t})
&:=\sum_{N=0}^{\infty} \mathcal{I}^{\textrm{$\mathcal{N}=8$ $U(N) (H)$}}(x;\mathfrak{t})u^N. 
\end{align}
It is given by
\begin{align}
\label{N8_gindex2}
\Xi^{\mathcal{N}=8}(u;x;\mathfrak{t})&=
\frac{1}{1-u}
\frac{1}{(u\mathfrak{t}x;\mathfrak{t}x)_{\infty}(u\mathfrak{t}x^{-1};\mathfrak{t}x^{-1})_{\infty}}. 
\end{align}
In the unflavored limit $x\rightarrow 1$, the index (\ref{N8_gindex2}) becomes
\begin{align}
\label{N8_gindex3}
\Xi^{\mathcal{N}=8}(u;\mathfrak{t})&=
\frac{1}{1-u}
\frac{1}{(u\mathfrak{t};\mathfrak{t})_{\infty}^2}. 
\end{align}
The unflavored grand canonical index (\ref{N8_gindex3}) obeys the difference equation
\begin{align}
\Xi^{\mathcal{N}=8}(\mathfrak{t}u;\mathfrak{t})
&=(1-\mathfrak{t}u)(1-u)\Xi^{\mathcal{N}=8}(u;\mathfrak{t}). 
\end{align}
This implies the recursion relation for the unflavored indices
\begin{align}
\mathcal{I}^{\textrm{$\mathcal{N}=8$ $U(N) (H)$}}(\mathfrak{t})
=\frac{1+\mathfrak{t}}{1-\mathfrak{t}^N}\mathcal{I}^{\textrm{$\mathcal{N}=8$ $U(N-1) (H)$}}(\mathfrak{t})
-\frac{\mathfrak{t}}{1-\mathfrak{t}^N}\mathcal{I}^{\textrm{$\mathcal{N}=8$ $U(N-2) (H)$}}(\mathfrak{t}). 
\end{align}

\subsubsection{Large $N$ limit of $\mathcal{N}=8$ $U(N)$ SYM}
In the large $N$ limit, the index becomes
\begin{align}
\label{N8_largen_index1}
\mathcal{I}^{\textrm{$\mathcal{N}=8$ $U(\infty) (H)$}}(x;\mathfrak{t})
&=\prod_{n=1}^{\infty}\frac{1}{(1-\mathfrak{t}^nx^n)(1-\mathfrak{t}^nx^{-n})}. 
\end{align}
The unflavored index is
\begin{align}
\label{N8_largen_index2}
\mathcal{I}^{\textrm{$\mathcal{N}=8$ $U(\infty) (H)$}}(\mathfrak{t})
&=\prod_{n=1}^{\infty}\frac{1}{(1-\mathfrak{t}^n)^2}. 
\end{align}

In the large $N$ limit, the correlation functions of the charged Wilson line operators 
associated to the power sum symmetric functions can be exactly calculated according to scalar product on the ring of polynomials in $\mathfrak{t}$ \cite{MR1354144}. 
The basic ingredient is the normalized $2$-point function of the charged Wilson lines
\begin{align}
\label{N8_largen_wnwn}
\langle \mathcal{W}_{n}\mathcal{W}_{-n}\rangle^{\textrm{$\mathcal{N}=8$ $U(\infty) (H)$}}(x;\mathfrak{t})
&=\frac{n}{(1-\mathfrak{t}x)(1-\mathfrak{t}x^{-1})}. 
\end{align}
The higher-point function of the charged Wilson lines obey the factorization\footnote{
See \cite{Hatsuda:2023iwi,Hatsuda:2023imp} for the factorization of the large Schur line defect correlators of $\mathcal{N}=4$ SYM theory.
}
\begin{align}
\label{N8_largen_fac}
\langle \prod_{j=1}^{k}(\mathcal{W}_{n_j} \mathcal{W}_{-n_j})^{m_j}\rangle^{\textrm{$\mathcal{N}=8$ $U(\infty) (H)$}}(x;\mathfrak{t})
&=\prod_{j=1}^k 
(m_j!)\langle \mathcal{W}_{n_j} \mathcal{W}_{-n_j}\rangle^{\textrm{$\mathcal{N}=8$ $U(\infty) (H)$}}(x;\mathfrak{t})^{m_j}. 
\end{align}
Also it follows that 
\begin{align}
\label{N8_largen_sym1}
\langle \mathcal{W}_{\mu}\mathcal{W}_{\overline{\nu}}\rangle^{\textrm{$\mathcal{N}=8$ $U(\infty) (H)$}}(x;\mathfrak{t})
&=\langle \mathcal{W}_{\mu'}\mathcal{W}_{\overline{\nu'}}\rangle^{\textrm{$\mathcal{N}=8$ $U(\infty) (H)$}}(x;\mathfrak{t}).
\end{align}
From the exact form (\ref{N8_largen_wnwn}) for the $2$-point function, 
the factorization (\ref{N8_largen_fac}) and the well-known relationship between the symmetric functions, 
we can calculate arbitrary large $N$ correlators. 

For example, the large $N$ normalized $2$-point functions of the Wilson lines in the (anti)symmetric representations are given by
\begin{align}
\label{N8_largen_w1w1}
\langle \mathcal{W}_{\ydiagram{1}}\mathcal{W}_{\overline{\ydiagram{1}}}\rangle^{\textrm{$\mathcal{N}=8$ $U(\infty) (H)$}}(x;\mathfrak{t})
&=\frac{1}{(1-\mathfrak{t}x)(1-\mathfrak{t}x^{-1})},\\
\label{N8_largen_wsym2wsym2}
\langle \mathcal{W}_{\ydiagram{2}}\mathcal{W}_{\overline{\ydiagram{2}}}\rangle^{\textrm{$\mathcal{N}=8$ $U(\infty) (H)$}}(x;\mathfrak{t})
&=\langle \mathcal{W}_{\ydiagram{1,1}}\mathcal{W}_{\overline{\ydiagram{1,1}}}\rangle^{\textrm{$\mathcal{N}=8$ $U(\infty) (H)$}}(x;\mathfrak{t})\nonumber \\
&=\frac{1+\mathfrak{t}^2}{(1-\mathfrak{t}x)(1-\mathfrak{t}^2x^2)(1-\mathfrak{t}x^{-1})(1-\mathfrak{t}^2x^{-2})},\\
\label{N8_largen_wsym3wsym3}
\langle \mathcal{W}_{\ydiagram{3}}\mathcal{W}_{\overline{\ydiagram{3}}}\rangle^{\textrm{$\mathcal{N}=8$ $U(\infty) (H)$}}(x;\mathfrak{t})
&=\langle \mathcal{W}_{\ydiagram{1,1,1}}\mathcal{W}_{\overline{\ydiagram{1,1,1}}}\rangle^{\textrm{$\mathcal{N}=8$ $U(\infty) (H)$}}(x;\mathfrak{t})
\nonumber\\
&=\frac{1+\mathfrak{t}^2+\mathfrak{t}^3(x+x^{-1})+\mathfrak{t}^4+\mathfrak{t}^6}
{\prod_{i=1}^3 (1-\mathfrak{t}^ix^i) (1-\mathfrak{t}^ix^{-i})}, \\
\label{N8_largen_wsym4wsym4}
\langle \mathcal{W}_{\ydiagram{4}}\mathcal{W}_{\overline{\ydiagram{4}}}\rangle^{\textrm{$\mathcal{N}=8$ $U(\infty) (H)$}}(x;\mathfrak{t})
&=\langle \mathcal{W}_{\ydiagram{1,1,1,1}}\mathcal{W}_{\overline{\ydiagram{1,1,1,1}}}\rangle^{\textrm{$\mathcal{N}=8$ $U(\infty) (H)$}}(x;\mathfrak{t})
\nonumber\\
&=\frac{1}{\prod_{i=1}^4 (1-\mathfrak{t}^ix^i) (1-\mathfrak{t}^ix^{-i})}
(1+\mathfrak{t}^2+\mathfrak{t}^3(x+x^{-1})\nonumber \\
&\quad +\mathfrak{t}^4(2+x^2+x^{-2})
+\mathfrak{t}^5(x+x^{-1})
+\mathfrak{t}^6(2+x^2+x^{-2})\nonumber \\
&\quad +\mathfrak{t}^7(x+x^{-1})
+\mathfrak{t}^8(2+x^2+x^{-2})+\mathfrak{t}^9(x+x^{-1})+\mathfrak{t}^{10}\nonumber \\
&\quad +\mathfrak{t}^{12}).
\end{align}

In the large representation limit $k\rightarrow\infty$, we obtain
\begin{align}
\label{N8_largen_winfwinf}
\langle \mathcal{W}_{(\infty)}\mathcal{W}_{\overline{(\infty)}}\rangle^{\textrm{$\mathcal{N}=8$ $U(\infty) (H)$}}(x;\mathfrak{t})
&=\langle \mathcal{W}_{(1^{\infty})}\mathcal{W}_{\overline{(1^{\infty})}}\rangle^{\textrm{$\mathcal{N}=8$ $U(\infty) (H)$}}(x;\mathfrak{t})
\nonumber\\
&=\prod_{n=1}^{\infty}
\frac{1}{1-\mathfrak{t}^{2n}}
\prod_{m=0}^{\infty}\prod_{n=1}^{\infty}
\frac{1}{(1-\mathfrak{t}^{2m+n}x^n) (1-\mathfrak{t}^{2m+n}x^{-n})}. 
\end{align}
As we will see in (\ref{uNADHM1_N8}), 
the large $N$ normalized $2$-point functions of the Wilson lines in the rank-$k$ (anti)symmetric representations 
turn out to coincide with the Higgs (or equivalently Coulomb) indices of the $U(k)$ ADHM theory with one flavor. 

\subsubsection{$U(N)$ ADHM with $l=1$}
\label{sec_UNADHMl1H}
The $U(N)$ ADHM theory with $l=1$ flavor can describe the low-energy dynamics of $N$ M2-branes probing the flat space. 
The Higgs limit (\ref{Higgs_wilson}) of the matrix integral reduces to
\begin{align}
\label{unADHM1_wilsons}
&
\langle W_{\mathcal{R}_1}\cdots W_{\mathcal{R}_k}\rangle^{\textrm{$U(N)$ ADHM-$[1] (H)$}}
(x,y_1;\mathfrak{t})
\nonumber\\
&=
\frac{1}{N!}\oint \prod_{i=1}^{N}
\frac{ds_{i}}{2\pi is_i}
\frac{
\prod_{i\neq j}(1-\frac{s_i}{s_j})
\prod_{ij}(1-\mathfrak{t}^2\frac{s_i}{s_j})
}
{
\prod_{ij}(1-\mathfrak{t}\frac{s_i}{s_j}x)(1-\mathfrak{t}\frac{s_i}{s_j}x^{-1})
}
\prod_{i=1}^N
\frac{1}{(1-\mathfrak{t}s_iy_1)(1-\mathfrak{t}s_{i}^{-1}y_1^{-1})}
\prod_{j=1}^k \chi_{\mathcal{R}_j}(s). 
\end{align}

In this case, useful formulae can be obtained by expanding the integrand in terms of the Hall-Littlewood symmetric functions 
by making use of the identity \cite{MR2200851}
\begin{align}
\label{cauchy2}
\sum_{\lambda,\mu}
\mathfrak{t}^{n(\lambda)+n(\mu)-\sum_i \lambda'_i\mu'_i}
P_{\lambda}(x;\mathfrak{t})P_{\mu}(y;\mathfrak{t})
&=
\prod_{i\ge1}
\frac{1}{(1-x_i) (1-y_i)}
\prod_{i,j\ge 1}
\frac{1-x_i y_j}{1-\mathfrak{t}^{-1}x_i y_j}, 
\end{align} 
where $n(\lambda)$ is given as \eqref{nlambda},
and the inner product (\ref{HL_inner}) of the Hall-Littlewood functions. 

Again without insertion of the Wilson lines, one finds the Higgs index for the $U(N)$ ADHM theory with one flavor. 
Since the theory is self-mirror, it is equal to the Coulomb index of the same theory. 
We find that it can be expanded as
\begin{align}
\label{uNADHM1_HL}
\mathcal{I}^{\textrm{$U(N)$ ADHM-$[1] (H)$}}(x;\mathfrak{t})
&=\sum_{\begin{smallmatrix}
\lambda\\
\ell(\lambda)\le N
\end{smallmatrix}}
\frac{\mathfrak{t}^{2|\lambda|+2n(\lambda)-\sum_{i}{\lambda_i'}^2}x^{2n(\lambda)-\sum_{i}{\lambda_i'}^2}}
{(\mathfrak{t}x;\mathfrak{t}x)_{N-\ell(\lambda)}\prod_{j\ge1}(\mathfrak{t}x;\mathfrak{t}x)_{m_j(\lambda)}}. 
\end{align}
Note that the index is independent of the flavor fugacity $y_1$. 

For example, for $N=1$, $2$ and $3$ we have 
\begin{align}
\label{u1ADHM1_index}
\mathcal{I}^{\textrm{$U(1)$ ADHM-$[1] (H)$}}(x;\mathfrak{t})
&=\frac{1}{(1-\mathfrak{t}x)(1-\mathfrak{t}x^{-1})}, \\
\label{u2ADHM1_index}
\mathcal{I}^{\textrm{$U(2)$ ADHM-$[1] (H)$}}(x;\mathfrak{t})
&=\frac{1+\mathfrak{t}^2}
{(1-\mathfrak{t}x)(1-\mathfrak{t}^2x^2)(1-\mathfrak{t}x^{-1})(1-\mathfrak{t}^2x^{-2})}, \\
\label{u3ADHM1_index}
\mathcal{I}^{\textrm{$U(3)$ ADHM-$[1] (H)$}}(x;\mathfrak{t})
&=\frac{1+\mathfrak{t}^2+\mathfrak{t}^3(x+x^{-1})+\mathfrak{t}^4+\mathfrak{t}^6}
{(1-\mathfrak{t}x)(1-\mathfrak{t}^2x^2)(1-\mathfrak{t}^3x^3) (1-\mathfrak{t}x^{-1})(1-\mathfrak{t}^2x^{-2})(1-\mathfrak{t}^3x^{-3})}. 
\end{align}

The Higgs index is related to the $2$-point function of the Wilson lines in the symmetric representations for $\mathcal{N}=8$ SYM theory. 
It follows that it is identified with the large $N$ normalized $2$-point function
\begin{align}
\label{uNADHM1_N8}
\mathcal{I}^{\textrm{$U(N)$ ADHM-$[1] (H)$}}(x;\mathfrak{t})
&=\langle \mathcal{W}_{(N)}\mathcal{W}_{\overline{(N)}}\rangle^{\textrm{$\mathcal{N}=8$ $U(\infty) (H)$}}(x;\mathfrak{t}). 
\end{align}
Alternatively, it can be expanded with respect to the $2$-point functions 
of the symmetric Wilson lines for $\mathcal{N}=8$ SYM theory
\begin{align}
\mathcal{I}^{\textrm{$U(N)$ ADHM-$[1] (H)$}}(x;\mathfrak{t})
&=\sum_{k=0}^{\infty}\langle W_{(k)} W_{(\overline{k})}\rangle^{\textrm{$\mathcal{N}=8$ $U(N) (H)$}}(x;\mathfrak{t})
\mathfrak{t}^{2k}. 
\end{align}

Since the ADHM theories with $l>0$ contain antifundamental (resp.~fundamental) chiral multiplets as constituents of the Higgs branch operators, 
the Wilson line operator transforming in the tensor product of fundamental (resp.~antifundamental) representations can attach the Higgs operators 
so that the $1$-point function does not vanish. 
The $1$-point function of the Wilson line in the rank-$k$ antisymmetric representation is given by
\begin{align}
\label{uNADHM_1ptasym_HL}
&
\langle W_{(1^k)}\rangle^{\textrm{$U(N)$ ADHM-$[1] (H)$}}(x,y_1;\mathfrak{t})
\nonumber\\
&=
y_1^{-k}
\sum_{
\begin{smallmatrix}
\alpha\supset \lambda \\
|\alpha/\lambda|=k\\
\ell(\alpha)\le N\\
\ell(\lambda)\le N\\
\end{smallmatrix}
}
\frac{\mathfrak{t}^{|\alpha|+|\lambda|+n(\alpha)+n(\lambda)-\sum_{i}\alpha'_i\lambda'_i}
x^{n(\alpha)+n(\lambda)-\sum_{i}\alpha'_i\lambda'_i}
}
{(\mathfrak{t}x;\mathfrak{t}x)_{N-\ell(\alpha)} \prod_{j\ge1}(\mathfrak{t}x;\mathfrak{t}x)_{m_j(\alpha)}}
\prod_{i\ge1} 
\left[
\begin{matrix}
\alpha'_i-\alpha'_{i+1}\\
\alpha'_i-\lambda'_{i}
\end{matrix}
\right]_{\mathfrak{t}x}. 
\end{align}
The dependence on the flavor fugacity $y_1$ simply shows up as an overall factor $y_1^{-k}$.
More generally, from the integration \eqref{unADHM1_wilsons} it is easy to see that the correlation function
$
\langle W_{\lambda_1}
\cdots
W_{\lambda_m}
W_{\overline{\rho_1}}
\cdots
W_{\overline{\rho_n}}
\rangle^{U(N)\text{ ADHM-}[1](H)}$
depends on $y_1$ only through the overall factor $y_1^{-\sum_{i=1}^m|\lambda_i|+\sum_{i=1}^n|\rho_i|}$.
For simplicity, we set $y_1$ to unity in the following. 

For example, when $N=1$ and $k=1$ 
the sum is taken over the Young diagrams with a single row. 
One finds
\begin{align}
\label{u1ADHM1w1}
\langle W_{\ydiagram{1}}\rangle^{\textrm{$U(1)$ ADHM-$[1] (H)$}}(x;\mathfrak{t})
&=\frac{\mathfrak{t}}{1-\mathfrak{t}x}
+\frac{\mathfrak{t}^2x^{-1}}{1-\mathfrak{t}x}
+\frac{\mathfrak{t}^3x^{-2}}{1-\mathfrak{t}x}
+\frac{\mathfrak{t}^4x^{-3}}{1-\mathfrak{t}x}+\cdots\nonumber \\
&=\frac{\mathfrak{t}}{(1-\mathfrak{t}x)(1-\mathfrak{t}x^{-1})}. 
\end{align}
For $N=2$ and $k=1$ we have
\begin{align}
\label{u2ADHM1w1}
&
\langle W_{\ydiagram{1}}\rangle^{\textrm{$U(2)$ ADHM-$[1] (H)$}}(x;\mathfrak{t})
\nonumber\\
&=\frac{\mathfrak{t}}{(1-\mathfrak{t}x)^2}
+\frac{2\mathfrak{t}^2x^{-1}}{(1-\mathfrak{t}x)^2}
+\frac{3\mathfrak{t}^3x^{-2}}{(1-\mathfrak{t}x)^2}
+\frac{4\mathfrak{t}^4x^{-3}}{(1-\mathfrak{t}x)^2}
+\frac{5\mathfrak{t}^5x^{-4}}{(1-\mathfrak{t}x)^2}
+\cdots
\nonumber\\
&=\sum_{k=1}^{\infty}
\frac{k\mathfrak{t}^k x^{-k+1}}
{(1-\mathfrak{t}x)^2}
=\frac{\mathfrak{t}}{(1-\mathfrak{t}x)^2(1-\mathfrak{t}x^{-1})^2}.
\end{align}
We observe that the $1$-point function (\ref{u2ADHM1w1}) is factorized as 
\begin{align}
\langle W_{\ydiagram{1}} \rangle^{\textrm{$U(2)$ ADHM-$[1] (H)$}}(x;\mathfrak{t})
=\langle W_{\ydiagram{1}} \rangle^{\textrm{$U(1)$ ADHM-$[1] (H)$}}(x;\mathfrak{t})
\mathcal{I}^{\textrm{$U(1)$ ADHM-$[1] (H)$}}(x;\mathfrak{t}). 
\end{align}
For $N=3$ and $k=1$ we find
\begin{align}
\label{u3ADHM1_w1}
\langle W_{\ydiagram{1}}\rangle^{\textrm{$U(3)$ ADHM-$[1] (H)$}}(x;\mathfrak{t})
&=\frac{\mathfrak{t}+\mathfrak{t}^3}
{(1-\mathfrak{t}x)^2(1-\mathfrak{t}^2x^2)(1-\mathfrak{t}x^{-1})^2(1-\mathfrak{t}^2x^{-2})}. 
\end{align}
We note that the $1$-point function (\ref{u3ADHM1_w1}) is decomposed as
\begin{align}
\langle W_{\ydiagram{1}} \rangle^{\textrm{$U(3)$ ADHM-$[1] (H)$}}(x;\mathfrak{t})
=\langle W_{\ydiagram{1}} \rangle^{\textrm{$U(1)$ ADHM-$[1] (H)$}}(x;\mathfrak{t})
\mathcal{I}^{\textrm{$U(2)$ ADHM-$[1] (H)$}}(x;\mathfrak{t}). 
\end{align}

We can also compute the $1$-point functions of the Wilson lines in the higher rank antisymmetric representations. 
For $N=2$ and $k=2$ we get
\begin{align}
&
\langle W_{\ydiagram{1,1}}\rangle^{\textrm{$U(2)$ ADHM-$[1] (H)$}}(x;\mathfrak{t})
\nonumber\\
&=\frac{\mathfrak{t}^3x}{(\mathfrak{t}x;\mathfrak{t}x)_{2}}
+\frac{\mathfrak{t}x^{-1}}{(1-\mathfrak{t}x)^2}
+\left(
\frac{\mathfrak{t}^4x^{-2}}{(1-\mathfrak{t}x)^2}
+\frac{\mathfrak{t}^5x^{-1}}{(\mathfrak{t}x;\mathfrak{t}x)_2}
\right)
+\frac{2\mathfrak{t}^5x^{-3}}{(1-\mathfrak{t}x)^2}
\nonumber\\
&\quad +\left(
\frac{2\mathfrak{t}^6x^{-4}}{(1-\mathfrak{t}x)^2}
+\frac{\mathfrak{t}^7x^{-3}}{(\mathfrak{t}x;\mathfrak{t}x)_2}
\right)
+\frac{3\mathfrak{t}^7x^{-5}}{(1-\mathfrak{t}x)^2}+\cdots
\nonumber\\
&=\sum_{k=1}^{\infty}
\left[
\left(
\frac{(k-1)\mathfrak{t}^{2k}x^{-2k+2}}{(1-\mathfrak{t}x)^2}
+\frac{\mathfrak{t}^{2k+1}x^{-2k+3}}{(\mathfrak{t}x;\mathfrak{t}x)_2}
\right)
+\frac{k\mathfrak{t}^{2k+1}x^{-2k+1}}{(1-\mathfrak{t}x)^2}
\right]
\nonumber\\
&=\frac{\mathfrak{t}^3(x+x^{-1})}
{(1-\mathfrak{t}x)(1-\mathfrak{t}^2x^2)(1-\mathfrak{t}x^{-1})(1-\mathfrak{t}^2x^{-2})}. 
\label{u2ADHM1_wasym2}
\end{align}
For $N=3$ and $k=2$ we find
\begin{align}
\label{u3ADHM1_wasym2}
\langle W_{\ydiagram{1,1}}\rangle^{\textrm{$U(3)$ ADHM-$[1] (H)$}}(x;\mathfrak{t})
&=\frac{\mathfrak{t}^3(x+x^{-1})}
{(1-\mathfrak{t}x)^2(1-\mathfrak{t}^2x^2)(1-\mathfrak{t}x^{-1})^2(1-\mathfrak{t}^2x^2)}. 
\end{align}
For $N=3$ and $k=3$ we obtain
\begin{align}
\label{u3ADHM1_wasym3}
&
\langle W_{\ydiagram{1,1,1}}\rangle^{\textrm{$U(3)$ ADHM-$[1] (H)$}}(x;\mathfrak{t})
\nonumber\\
&=\frac{\mathfrak{t}^5 (1+\mathfrak{t}(x^3+x+x^{-1}+x^{-3})+\mathfrak{t}^2)}
{(1-\mathfrak{t}x)(1-\mathfrak{t}^2x^2)(1-\mathfrak{t}^3x^3) (1-\mathfrak{t}x^{-1})(1-\mathfrak{t}^2x^{-2})(1-\mathfrak{t}^3x^{-3})}. 
\end{align}

We find that the $1$-point function of the Wilson line in the rank-$k$ antisymmetric representation can be factorized as
\begin{align}
\label{uNADHM_1ptasym_relation}
\langle W_{(1^k)}\rangle^{\textrm{$U(N)$ ADHM-$[1] (H)$}}(x;\mathfrak{t})
=\langle W_{(1^k)}\rangle^{\textrm{$U(k)$ ADHM-$[1] (H)$}}(x;\mathfrak{t})
\mathcal{I}^{\textrm{$U(N-k)$ ADHM-$[1] (H)$}}(x;\mathfrak{t})
\end{align}
for $N>k$.
As we explain in section \ref{sec_topvertex_1ptfcn}, this relation can be derived analytically by using the refined topological vertex.
Hence the $1$-point function of the rank-$k$ antisymmetric Wilson line for the $U(N)$ ADHM theory 
can be determined by that of the lower rank $U(k)$ ADHM theory 
and the Higgs index of the lower rank $U(N-k)$ ADHM theory. 

Similarly, the $1$-point function of the Wilson line in the rank-$k$ symmetric representation is evaluated as
\begin{align}
\label{uNADHM_1ptsym_HL}
&
\langle W_{(k)}\rangle^{\textrm{$U(N)$ ADHM-$[1] (H)$}}(x;\mathfrak{t})
\nonumber\\
&=
\sum_{
\begin{smallmatrix}
\alpha\supset \lambda \\
|\alpha/\lambda|=k\\
\ell(\alpha)\le N\\
\ell(\lambda)\le N\\
\end{smallmatrix}
}
\frac{\mathfrak{t}^{|\alpha|+|\lambda|+n(\alpha/\lambda)+n(\alpha)+n(\lambda)-\sum_{i}\alpha'_i\lambda'_i}
x^{n(\alpha/\lambda)+n(\alpha)+n(\lambda)-\sum_{i}\alpha'_i\lambda'_i}
}
{(\mathfrak{t}x;\mathfrak{t}x)_{N-\ell(\alpha)} \prod_{j\ge1}(\mathfrak{t}x;\mathfrak{t}x)_{m_j(\alpha)}}\nonumber \\
&\quad\times \prod_{i\ge1} 
\left[
\begin{matrix}
\alpha'_i-\lambda'_{i+1}\\
\alpha'_i-\lambda'_{i}
\end{matrix}
\right]_{\mathfrak{t}x}. 
\end{align}
For example, when $N=2$ and $k=2$, we obtain
\begin{align}
\label{u2ADHM1wsym2}
&
\langle W_{\ydiagram{2}}\rangle^{\textrm{$U(2)$ ADHM-$[1] (H)$}}(x;\mathfrak{t})
\nonumber\\
&=
\left(
\frac{\mathfrak{t}^2}{(1-\mathfrak{t}x)^2}+\frac{\mathfrak{t}^4x^2}{(\mathfrak{t}x;\mathfrak{t}x)_2}
\right)
+
\left(
\frac{\mathfrak{t}^3x^{-1}}{(1-\mathfrak{t}x)^2}
+\frac{\mathfrak{t}^3x^{-1}(\mathfrak{t}x;\mathfrak{t}x)_2}{(1-\mathfrak{t}x)^4}
\right)
\nonumber\\
&\quad +
\left(
\frac{4\mathfrak{t}^4x^{-2}}{(1-\mathfrak{t}x)^2}+\frac{\mathfrak{t}^6}{(\mathfrak{t}x;\mathfrak{t}x)_2}
\right)
+
\left(
\frac{4\mathfrak{t}^5x^{-3}}{(1-\mathfrak{t}x)^2}
+\frac{\mathfrak{t}^5x^{-3}(\mathfrak{t}x;\mathfrak{t}x)_2}{(1-\mathfrak{t}x)^4}
\right)+\cdots
\nonumber\\
&=
\sum_{k=0}^{\infty}
\biggl[
\biggl(
\frac{(3k+1)\mathfrak{t}^{2k+2}x^{-2k}}{(1-\mathfrak{t}x)^2}+\frac{\mathfrak{t}^{2k+4}x^{-2k+2}}{(\mathfrak{t}x;\mathfrak{t}x)_2}
\biggr)
+
\biggl(
\frac{(3k+1)\mathfrak{t}^{2k+3}x^{-2k-1}}{(1-\mathfrak{t}x)^2}\nonumber \\
&\quad
+\frac{\mathfrak{t}^{2k+3}x^{-2k-1}(\mathfrak{t}x;\mathfrak{t}x)_2}{(1-\mathfrak{t}x)^4}
\biggr)
\biggr]
\nonumber\\
&=\frac{\mathfrak{t}^2(1+\mathfrak{t}(x+x^{-1})+\mathfrak{t}^2(2+x^{2}+x^{-2})-\mathfrak{t}^4)}
{(1-\mathfrak{t}x)(1-\mathfrak{t}^2x^2)(1-\mathfrak{t}x^{-1})(1-\mathfrak{t}^2x^{-2})}. 
\end{align}

For rank-$3$, $4$ and $5$ symmetric Wilson lines in the $U(2)$ ADHM theory with one flavor we get
\begin{align}
\label{u2ADHM1wsym3}
&\langle W_{\ydiagram{3}} \rangle^{\textrm{$U(2)$ ADHM-$[1] (H)$}}(x;\mathfrak{t})\nonumber \\
&=\frac{\mathfrak{t}^3}
{(1-\mathfrak{t}x)(1-\mathfrak{t}^2x^2)(1-\mathfrak{t}x^{-1})(1-\mathfrak{t}^2x^{-2})}
(
1+\mathfrak{t}(x+x^{-1})+\mathfrak{t}^2(2+x^2+x^{-2})
\nonumber \\
&\quad +\mathfrak{t}^3(x^3+x+x^{-1}+x^{-3})-\mathfrak{t}^4+\mathfrak{t}^5(-x-x^{-1})
), \\
\label{u2ADHM1wsym4}
&\langle W_{\ydiagram{4}} \rangle^{\textrm{$U(2)$ ADHM-$[1] (H)$}}(x;\mathfrak{t})\nonumber \\
&=\frac{\mathfrak{t}^4}
{(1-\mathfrak{t}x)(1-\mathfrak{t}^2x^2)(1-\mathfrak{t}x^{-1})(1-\mathfrak{t}^2x^{-2})}
(
1+\mathfrak{t}(x+x^{-1})+\mathfrak{t}^2(2+x^2+x^{-2})\nonumber \\
&\quad +\mathfrak{t}^3(x^3+x+x^{-1}+x^{-3})
+\mathfrak{t}^4(x^4+x^2+x^{-2}+x^{-4})
+\mathfrak{t}^5(-x-x^{-1})\nonumber \\
&\quad +\mathfrak{t}^6(-1-x^2-x^{-2})
), \\
\label{u2ADHM1wsym5}
&\langle W_{\ydiagram{5}} \rangle^{\textrm{$U(2)$ ADHM-$[1] (H)$}}(x;\mathfrak{t})\nonumber \\
&=\frac{\mathfrak{t}^5}
{(1-\mathfrak{t}x)(1-\mathfrak{t}^2x^2)(1-\mathfrak{t}x^{-1})(1-\mathfrak{t}^2x^{-2})}
(
1+\mathfrak{t}(x+x^{-1})+\mathfrak{t}^2(2+x^2+x^{-2})\nonumber \\
&\quad +\mathfrak{t}^3(x^3+x+x^{-1}+x^{-3})
+\mathfrak{t}^4(x^4+x^2+x^{-2}+x^{-4})
+\mathfrak{t}^5(x^5+x^3+x^{-3}+x^{-5})\nonumber \\
&\quad +\mathfrak{t}^6(-1-x^2-x^{-2})
+\mathfrak{t}^7(-x^3-x-x^{-1}-x^{-3})
). 
\end{align}

It is straightforward to extend the analyses to the multi-point functions. 
The $2$-point function of the Wilson line in the rank-$k$ antisymmetric representation is 
\begin{align}
&
\langle W_{(1^k)} W_{(\overline{1^k})} \rangle^{\textrm{$U(N)$ ADHM-$[1] (H)$}}(x;\mathfrak{t})
\nonumber\\
&=
\sum_{
\begin{smallmatrix}
\alpha\supset \lambda,\mu \\
|\alpha/\lambda|=|\alpha/\mu|=k\\
\ell(\alpha)\le N\\
\ell(\lambda), \ell(\mu)\le N\\
\end{smallmatrix}
}
\frac{\mathfrak{t}^{|\lambda|+|\mu|+n(\lambda)+n(\mu)-\sum_{i}\lambda'_i\mu'_i}
x^{n(\lambda)+n(\mu)-\sum_{i}\lambda'_i\mu'_i}
}
{(\mathfrak{t}x;\mathfrak{t}x)_{N-\ell(\alpha)} \prod_{j\ge1}(\mathfrak{t}x;\mathfrak{t}x)_{m_j(\alpha)}}\nonumber \\
&\quad\times \prod_{i\ge1} 
\left[
\begin{matrix}
\alpha'_i-\alpha'_{i+1}\\
\alpha'_i-\lambda'_{i}
\end{matrix}
\right]_{\mathfrak{t}x}
\left[
\begin{matrix}
\alpha'_i-\alpha'_{i+1}\\
\alpha'_i-\mu'_{i}
\end{matrix}
\right]_{\mathfrak{t}x}.
\end{align}
For $N=2$ and $k=1$ we obtain the $2$-point function of the Wilson lines in the fundamental representation 
\begin{align}
\label{u2ADHM1w1w1withx}
\langle W_{\ydiagram{1}}W_{\overline{\ydiagram{1}}}\rangle^{\textrm{$U(2)$ ADHM-$[1] (H)$}}(\mathfrak{t})
&=\frac{1+\mathfrak{t}^2}{(1-\mathfrak{t}x)^2(1-\mathfrak{t}x^{-1})^2}. 
\end{align}
For $N=3$ and $k=1$ we find the $2$-point function of the Wilson lines in the fundamental representation
\begin{align}
\label{u3ADHM1_w1w1}
\langle W_{\ydiagram{1}} W_{\overline{\ydiagram{1}}} \rangle^{\textrm{$U(3)$ ADHM-$[1] (H)$}}(x;\mathfrak{t})
&=
\frac{1+2\mathfrak{t}^2+\mathfrak{t}^3(x+x^{-1})+\mathfrak{t}^4}
{(1-\mathfrak{t}x)^2(1-\mathfrak{t}^2x^2)(1-\mathfrak{t}x^{-1})^2(1-\mathfrak{t}^2x^{-2})}. 
\end{align}

The $2$-point function of the Wilson line in the rank-$k$ symmetric representation is 
\begin{align}
&
\langle W_{(k)} W_{(\overline{k})} \rangle^{\textrm{$U(N)$ ADHM-$[1] (H)$}}(x;\mathfrak{t})
\nonumber\\
&=
\sum_{
\begin{smallmatrix}
\alpha\supset \lambda,\mu \\
|\alpha/\lambda|=|\alpha/\mu|=k\\
\ell(\alpha)\le N\\
\ell(\lambda), \ell(\mu)\le N\\
\end{smallmatrix}
}
\frac{\mathfrak{t}^{|\lambda|+|\mu|+n(\alpha/\lambda)+n(\alpha/\mu)+n(\lambda)+n(\mu)-\sum_{i}\lambda'_i\mu'_i}
}
{(\mathfrak{t}x;\mathfrak{t}x)_{N-\ell(\alpha)} \prod_{j\ge1}(\mathfrak{t}x;\mathfrak{t}x)_{m_j(\alpha)}}
\nonumber\\
&\quad \times 
x^{n(\alpha/\lambda)+n(\alpha/\mu)+n(\lambda)+n(\mu)-\sum_{i}\lambda'_i\mu'_i}
\prod_{i\ge1} 
\left[
\begin{matrix}
\alpha'_i-\lambda'_{i+1}\\
\alpha'_i-\lambda'_{i}
\end{matrix}
\right]_{\mathfrak{t}x}
\left[
\begin{matrix}
\alpha'_i-\lambda'_{i+1}\\
\alpha'_i-\mu'_{i}
\end{matrix}
\right]_{\mathfrak{t}x}.
\end{align}

For $N=2$ and $k=2$ we find
\begin{align}
\label{u2ADHM1wsym2wsym2withx}
&
\langle W_{\ydiagram{2}}W_{\overline{\ydiagram{2}}}\rangle^{\textrm{$U(2)$ ADHM-$[1] (H)$}}(\mathfrak{t})
\nonumber\\
&=\frac{(1+\mathfrak{t}^2) (1+\mathfrak{t}(x+x^{-1})+\mathfrak{t}^2(2+x^2+x^{-2})+\mathfrak{t}^3(x+x^{-1}))}
{(1-\mathfrak{t}x)(1-\mathfrak{t}^2x^2)(1-\mathfrak{t}x^{-1})(1-\mathfrak{t}^2x^{-2})}.
\end{align}

For $N=2$ and $k\rightarrow \infty$, we find
\begin{align}
\langle W_{(\infty)}W_{(\overline{\infty})}\rangle^{\textrm{$U(2)$ ADHM-$[1]$}}(x;\mathfrak{t})
&=\frac{(1+\mathfrak{t}^2)^2}
{(1-\mathfrak{t}x)^2(1-\mathfrak{t}^2x^2)(1-\mathfrak{t}x^{-1})^2(1-\mathfrak{t}^2x^{-2})}. 
\end{align}
The normalized correlator is
\begin{align}
\langle \mathcal{W}_{(\infty)}\mathcal{W}_{(\overline{\infty})}\rangle^{\textrm{$U(2)$ ADHM-$[1]$}}(x;\mathfrak{t})
&=\frac{1+\mathfrak{t}^2}{(1-\mathfrak{t}x)(1-\mathfrak{t}x^{-1})}. 
\end{align}

For $N=3$ and $k=2$ we get
\begin{align}
\label{u3ADHM1_wsym2wsym2}
&
\langle W_{\ydiagram{2}} W_{\overline{\ydiagram{2}}} \rangle^{\textrm{$U(3)$ ADHM-$[1] (H)$}}(x;\mathfrak{t})
\nonumber\\
&=\frac{1}{(1-\mathfrak{t}x)^2(1-\mathfrak{t}^2x^2) (1-\mathfrak{t}x^{-1})^2(1-\mathfrak{t}^2x^{-2})}
(
1+\mathfrak{t}^2(3+x^2+x^{-2})
+\mathfrak{t}^3(2x+2x^{-1})\nonumber \\
&\quad +\mathfrak{t}^4(4+2x^2+2x^{-2})
+\mathfrak{t}^5(x^3+2x+2x^{-1}+x^{-3})+\mathfrak{t}^6(1+x^2+x^{-2})+\mathfrak{t}^7(-x-x^{-1})\nonumber \\
&\quad -\mathfrak{t}^8
). 
\end{align}

\subsubsection{ADHM with $l=1$ in grand canonical ensemble}
Let us consider the grand canonical index
\begin{align}
\Xi(u;x;\mathfrak{t})&:=\sum_{N=0}^{\infty}
\mathcal{I}^{\textrm{$U(N)$ ADHM-$[1] (H)$}}(x;\mathfrak{t})u^N. 
\label{Xiuxt}
\end{align}
It is given by \cite{Hayashi:2022ldo}
\begin{align}
\Xi(u;x;\mathfrak{t})
&=\prod_{m=0}^{\infty}
\frac{1}{1-u\mathfrak{t}^{2m}}
\prod_{n=1}^{\infty}
\frac{1}{(1-u\mathfrak{t}^{2m+n}x^n)(1-u\mathfrak{t}^{2m+n}x^{-n})}.
\label{Xiuxt2}
\end{align}
It satisfies the difference equation 
\begin{align}
\Xi(u\mathfrak{t}x;x;\mathfrak{t})
&=(u;\mathfrak{t}x^{-1})_{\infty}
\Xi(u;x;\mathfrak{t}).
\end{align}
By expanding the both sides with respect to $u$, we find the recurrence relation
\begin{align}
\mathcal{I}^{\textrm{$U(N)$ ADHM-$[1] (H)$}}(\mathfrak{t};x)
&=\frac{1}{1-\mathfrak{t}^Nx^N}
\sum_{n=0}^{N-1}\frac{\mathfrak{t}^nx^n}{(\mathfrak{t}x^{-1};\mathfrak{t}x^{-1})_{N-n}}
\mathcal{I}^{\textrm{$U(n)$ ADHM-$[1] (H)$}}(\mathfrak{t};x). 
\end{align}
This generalizes the relation for the unflavored indices in \cite{Okazaki:2022sxo}. 

\subsubsection{Large $N$ limit of $U(N)$ ADHM with $l=1$}
In the large $N$ limit $N\rightarrow \infty$, the index is
\begin{align}\label{largeN_Higgsindex}
\mathcal{I}^{\textrm{$U(\infty)$ ADHM-$[1] (H)$}}(x;\mathfrak{t})
&=\prod_{n=1}^{\infty}\frac{1}{1-\mathfrak{t}^{2n}}
\prod_{m=0}^{\infty}\prod_{n=1}^{\infty}
\frac{1}{(1-\mathfrak{t}^{2m+n}x^n)(1-\mathfrak{t}^{2m+n}x^{-n})}, 
\end{align}
which precisely agrees with the large $N$ normalized $2$-point function (\ref{N8_largen_winfwinf}). 

From the algebraic relation (\ref{uNADHM_1ptasym_relation}), the large $N$ limit of the normalized $1$-point function of the Wilson line in the rank-$k$ antisymmetric representation is given by
\begin{align}
\langle \mathcal{W}_{(1^k)} \rangle^{\textrm{$U(\infty)$ ADHM-$[1] (H)$}}(x;\mathfrak{t})
&=\langle W_{(1^k)}\rangle^{\textrm{$U(k)$ ADHM-$[1] (H)$}}(x;\mathfrak{t}),
\label{uNADHM_1ptasym_relation_Ninfty}
\end{align}
where we have set $y_1=1$.

We find that the large $N$ normalized $1$-point function of the Wilson line in the rank-$k$ symmetric representation 
is equal to the Higgs index of the $U(k)$ ADHM theory with one flavor up to the overall factor
\begin{align}
\langle \mathcal{W}_{(k)}\rangle^{\textrm{$U(\infty)$ ADHM-$[1] (H)$}}(x;\mathfrak{t})
&=\mathfrak{t}^k
\mathcal{I}^{\textrm{$U(k)$ ADHM-$[1] (H)$}}(x;\mathfrak{t}). 
\end{align}

We find that the large $N$ normalized $2$-point function of the Wilson lines in the fundamental representation is given by
\begin{align}
\langle \mathcal{W}_{\ydiagram{1}} \mathcal{W}_{\overline{\ydiagram{1}}}\rangle^{\textrm{$U(\infty)$ ADHM-$[1] (H)$}}(x;\mathfrak{t})
&=\frac{1-(x+x^{-1})\mathfrak{t}+2\mathfrak{t}^2}
{(1-\mathfrak{t}x)^2(1-\mathfrak{t}x^{-1})^2}. 
\end{align}
In the unflavored limit we obtain
\begin{align}
\langle \mathcal{W}_{\ydiagram{1}} \mathcal{W}_{\overline{\ydiagram{1}}}\rangle^{\textrm{$U(\infty)$ ADHM-$[1] (H)$}}(\mathfrak{t})
&=\frac{1-2\mathfrak{t}+2\mathfrak{t}^2}
{(1-\mathfrak{t})^4}
=\sum_{n\ge0}C_n \mathfrak{t}^n, 
\end{align}
which is the generating function for the cake number $C_n$. 

Unlike the Coulomb limit, there exists non-trivial $1$-point functions. 
Taking this into account, let us define the connected $2$-point function of the Wilson line operators in the fundamental representation by 
\begin{align}
&
\langle \mathcal{W}_{\ydiagram{1}} \mathcal{W}_{\overline{\ydiagram{1}}}\rangle^{\textrm{$U(\infty)$ ADHM-$[1] (H)$}}_{c}(x;\mathfrak{t})
\nonumber\\
&:=\langle \mathcal{W}_{\ydiagram{1}} \mathcal{W}_{\overline{\ydiagram{1}}}\rangle^{\textrm{$U(\infty)$ ADHM-$[1] (H)$}}(x;\mathfrak{t})
-
\langle \mathcal{W}_{\ydiagram{1}} \rangle^{\textrm{$U(\infty)$ ADHM-$[1] (H)$}}(x;\mathfrak{t})
\langle \mathcal{W}_{\overline{\ydiagram{1}}} \rangle^{\textrm{$U(\infty)$ ADHM-$[1] (H)$}}(x;\mathfrak{t}).
\end{align}
Then we find
\begin{align}
\label{largeN_ADHM1_w1H}
\langle \mathcal{W}_{\ydiagram{1}} \mathcal{W}_{\overline{\ydiagram{1}}}\rangle^{\textrm{$U(\infty)$ ADHM-$[1] (H)$}}_{c}(x;\mathfrak{t})
&=\frac{1}{(1-\mathfrak{t}x)(1-\mathfrak{t}x^{-1})}. 
\end{align}
This precisely agrees with the Higgs index (or equivalently Coulomb index) of the $U(1)$ ADHM theory with $l=1$. 
The corresponding single particle index is 
\begin{align}
i_{X}^{(H)}(x;\mathfrak{t})&=
\text{PL}[\langle \mathcal{W}_{\mathcal{R}}\mathcal{W}_{\overline{\mathcal{R}}}\rangle^{U(\infty)\text{ADHM-}[1] (H)}(x;\mathfrak{t}) ]
\nonumber\\
&=\sum_{d\ge1}\frac{\mu(d)}{d}\log [
\langle \mathcal{W}_{\mathcal{R}}\mathcal{W}_{\overline{\mathcal{R}}}\rangle^{U(\infty)\text{ ADHM-}[1] (H)}(x^d;\mathfrak{t}^d)
]
\nonumber\\
&=\mathfrak{t}x+\mathfrak{t}x^{-1}. 
\end{align}
This is expected to count two bosonic scalar fields in the spectrum of the quantum fluctuations on the gravity dual geometry. 

\subsubsection{$U(N)$ ADHM with $l>1$}
For general $U(N)$ ADHM theory with multiple $l>1$ flavors 
the Higgs limit (\ref{Higgslimit}) of the line defect index takes the form (\ref{Higgs_wilson}). 
One can adopt the same strategy to find the expressions.  
First, the integrand can be expanded with respect to the Hall-Littlewood functions according to the identity (\ref{cauchy2}). 
Repeatedly using the Pieri rule \cite{kirillov1998new}
\begin{align}
P_{\mu}(s;\mathfrak{t})
\prod_{i=1}^{N}\frac{1}{1-zs_i}
&=\sum_{\lambda\supset \mu}
z^{|\lambda|-|\mu|}
\mathfrak{t}^{n(\lambda/\mu)}
\prod_{i\ge1}
\left[
\begin{matrix}
\lambda'_i-\mu'_{i+1}\\
\lambda'_i-\mu'_i\\
\end{matrix}
\right]_{\mathfrak{t}}
P_{\lambda}(s;\mathfrak{t}), 
\end{align}
the integral can be calculated from the orthogonality (\ref{HL_inner}) of the Hall-Littlewood functions. 
Accordingly, we get an exact formula for the Higgs index for the $U(N)$ ADHM theory with $l$ flavors as nested sums\footnote{
Also the Higgs index can be expanded in terms of the modified Hall-Littlewood function as discussed in \cite{Crew:2020psc}. 
}
\begin{align}
\label{uNADHMNf_HL}
&
\mathcal{I}^{\textrm{$U(N)$ ADHM-$[l] (H)$}}(x,y_{\alpha};\mathfrak{t})
\nonumber\\
&=
\sum_{\nu=\lambda^{(l)}\supset \cdots \lambda^{(1)}}
\sum_{\nu=\mu^{(l)}\supset \cdots \mu^{(1)}}
\frac{1}{
(\mathfrak{t}x;\mathfrak{t}x)_{N-\ell(\nu)}
\prod_{j\ge1}(\mathfrak{t}x;\mathfrak{t}x)_{m_j(\nu)}
}
\nonumber\\
&\quad \times 
\mathfrak{t}^{
2|\nu|
+\sum_{\alpha=0}^{l-1} \left[
n(\lambda^{(\alpha+1)}/\lambda^{(\alpha)})
+n(\mu^{(\alpha+1)}/\mu^{(\alpha)})
\right]-\sum_{i\ge1}{\lambda^{(1)}}'_i{\mu^{(1)}}'_i
}
\nonumber\\
&\quad \times 
x^{
\sum_{\alpha=0}^{l-1} \left[
n(\lambda^{(\alpha+1)}/\lambda^{(\alpha)})
+n(\mu^{(\alpha+1)}/\mu^{(\alpha)})
\right]-\sum_{i\ge1}{\lambda^{(1)}}'_i{\mu^{(1)}}'_i
}
\prod_{\alpha=1}^{l}
y_{\alpha}^
{|\lambda^{(\alpha)}|-|\lambda^{(\alpha-1)}| 
-|\mu^{(\alpha)}|+|\mu^{(\alpha-1)}|}
\nonumber\\
&\quad \times 
\prod_{\alpha=0}^{l-1}
\prod_{i\ge1}
\left[
\begin{matrix}
{\lambda^{(\alpha+1)}}'_{i} - {\lambda^{(\alpha)}}'_{i+1} \\
{\lambda^{(\alpha+1)}}'_{i} - {\lambda^{(\alpha)}}'_{i} \\
\end{matrix}
\right]_{\mathfrak{t}x}
\left[
\begin{matrix}
{\mu^{(\alpha+1)}}'_{i} - {\mu^{(\alpha)}}'_{i+1} \\
{\mu^{(\alpha+1)}}'_{i} - {\mu^{(\alpha)}}'_{i} \\
\end{matrix}
\right]_{\mathfrak{t}x}. 
\end{align}
Here the partitions $\lambda^{(\alpha)}$ and $\mu^{(\alpha)}$ 
in the sequence satisfy $\ell(\lambda^{(\alpha)})\le N$ and $\ell(\mu^{(\alpha)})\le N$. 
We have set $\lambda^{(0)}=\emptyset$ and $\mu^{(0)}=\emptyset$. 

For example, for $(N,l)=(1,2)$ and $(2,2)$ we have 
\begin{align}
&\mathcal{I}^{\textrm{$U(1)$ ADHM-$[2] (H)$}}(x,y_{\alpha};\mathfrak{t})
=\frac{1+\mathfrak{t}^2}{(1-\mathfrak{t}x^{\pm})(1-\mathfrak{t}^2y_1^{\pm}y_2^{\mp})},\\ 
&
\mathcal{I}^{\textrm{$U(2)$ ADHM-$[2] (H)$}}(x;\mathfrak{t})
\nonumber\\
&=\frac{1}{(1-\mathfrak{t}x^{\pm})(1-\mathfrak{t}^2x^{\pm2})(1-\mathfrak{t}^2y_1^{\pm}y_2^{\mp})
(1-\mathfrak{t}^3xy_1^{\pm}y_2^{\mp})(1-\mathfrak{t}^3x^{-1}y_1^{\pm}y_2^{\mp})}
(1+2\mathfrak{t}^2\nonumber \\
&\quad +(x+x^{-1})\mathfrak{t}^3+2\mathfrak{t}^4
+(-x^{\pm} y_1^{\pm}y_2^{\mp}-x^{\pm}y_1^{\mp}y_2^{\pm})\mathfrak{t}^5
+(1-y_1^{\pm}y_2^{\mp})\mathfrak{t}^6\nonumber \\
&\quad +(-x^{\pm} y_1^{\pm}y_2^{\mp}-x^{\pm}y_1^{\mp}y_2^{\pm})\mathfrak{t}^7
+(1-y_1^{\pm}y_2^{\mp})\mathfrak{t}^8
+(-x^{\pm} y_1^{\pm}y_2^{\mp}-x^{\pm}y_1^{\mp}y_2^{\pm})\mathfrak{t}^9
+2\mathfrak{t}^{10}\nonumber \\
&\quad +(x+x^{-1})\mathfrak{t}^{11}+2\mathfrak{t}^{12}+\mathfrak{t}^{14}
).\label{u2ADHM2_index}
\end{align}

The $1$-point function of the Wilson line in the rank-$k$ antisymmetric representation can be expanded as
\begin{align}
&
\langle W_{(1^k)}\rangle^{\textrm{$U(N)$ ADHM-$[l] (H)$}}(x,y_{\alpha};\mathfrak{t})
\nonumber\\
&=
\sum_{
\begin{smallmatrix}
\nu\supset \lambda^{(l)}\supset \cdots \lambda^{(1)}\\
|\nu|-|\lambda^{(l)}|=k\\
\end{smallmatrix}
}
\sum_{\nu=\mu^{(l)}\supset \cdots \mu^{(1)}}
\frac{1}{
(\mathfrak{t}x;\mathfrak{t}x)_{N-\ell(\nu)}
\prod_{j\ge1}(\mathfrak{t}x;\mathfrak{t}x)_{m_j(\nu)}
}
\nonumber\\
&\quad \times 
\mathfrak{t}^{
2|\nu|-k
+\sum_{\alpha=0}^{l-1} \left[
n(\lambda^{(\alpha+1)}/\lambda^{(\alpha)})
+n(\mu^{(\alpha+1)}/\mu^{(\alpha)})
\right]-\sum_{i\ge1}{\lambda^{(1)}}'_i{\mu^{(1)}}'_i
}
\nonumber\\
&\quad \times 
x^{
\sum_{\alpha=0}^{l-1} \left[
n(\lambda^{(\alpha+1)}/\lambda^{(\alpha)})
+n(\mu^{(\alpha+1)}/\mu^{(\alpha)})
\right]-\sum_{i\ge1}{\lambda^{(1)}}'_i{\mu^{(1)}}'_i
}
\prod_{\alpha=1}^{l}
y_{\alpha}^
{|\lambda^{(\alpha)}|-|\lambda^{(\alpha-1)}| 
-|\mu^{(\alpha)}|+|\mu^{(\alpha-1)}|}
\nonumber\\
&\quad \times 
\prod_{\alpha=0}^{l-1}
\prod_{i\ge1}
\left[
\begin{matrix}
{\lambda^{(\alpha+1)}}'_{i} - {\lambda^{(\alpha)}}'_{i+1} \\
{\lambda^{(\alpha+1)}}'_{i} - {\lambda^{(\alpha)}}'_{i} \\
\end{matrix}
\right]_{\mathfrak{t}x}
\left[
\begin{matrix}
{\mu^{(\alpha+1)}}'_{i} - {\mu^{(\alpha)}}'_{i+1} \\
{\mu^{(\alpha+1)}}'_{i} - {\mu^{(\alpha)}}'_{i} \\
\end{matrix}
\right]_{\mathfrak{t}x}
\left[
\begin{matrix}
{\lambda^{(l)}}'_{i} - {\lambda^{(l)}}'_{i+1} \\
{\lambda^{(l)}}'_{i} - {\nu'}_{i} \\
\end{matrix}
\right]_{\mathfrak{t}x}. 
\end{align}

\subsubsection{Large $N$ limit of $U(N)$ ADHM with $l>1$}
In the large $N$ limit, we encounter simple expressions of line defect indices. 
We list several results in the following. 
We find that 
the normalized $1$-point function of the charged Wilson line is given by
\begin{align}
\langle \mathcal{W}_{n} \rangle^{\textrm{$U(\infty)$ ADHM-$[l] (H)$}}(x,y_{\alpha};\mathfrak{t})
&=\frac{\sum_{\alpha=1}^{l}y_{\alpha}^{-n}\mathfrak{t}^n}
{(1-\mathfrak{t}^nx^n)(1-\mathfrak{t}^nx^{-n})}. 
\end{align}
The large $N$ limit of the normalized $2$-point function of the Wilson line in the fundamental representation 
and that in the antifundamental representation is
\begin{align}
\langle \mathcal{W}_{\ydiagram{1}} \mathcal{W}_{\overline{\ydiagram{1}}}\rangle^{\textrm{$U(\infty)$ ADHM-$[l] (H)$}}(x,y_{\alpha};\mathfrak{t})
&=\frac{1-(x+x^{-1})\mathfrak{t}+(l+1+\sum_{i<j}^{l} y_i^{\pm}y_j^{\mp})\mathfrak{t}^2}
{(1-\mathfrak{t}x)^2(1-\mathfrak{t}x^{-1})^2}. 
\label{calWcalW1Uinfty-[1]H}
\end{align}

Again the connected $2$-point function of the Wilson line operators in the fundamental representation is given by
\begin{align}
&
\langle \mathcal{W}_{\ydiagram{1}} \mathcal{W}_{\overline{\ydiagram{1}}}\rangle^{\textrm{$U(\infty)$ ADHM-$[l] (H)$}}(x,y_{\alpha};\mathfrak{t})\nonumber \\
&\quad -
\langle \mathcal{W}_{\ydiagram{1}} \rangle^{\textrm{$U(\infty)$ ADHM-$[l] (H)$}}(x,y_{\alpha};\mathfrak{t})
\langle \mathcal{W}_{\overline{\ydiagram{1}}} \rangle^{\textrm{$U(\infty)$ ADHM-$[l] (H)$}}(x,y_{\alpha};\mathfrak{t})
\nonumber\\
&=\frac{1}{(1-\mathfrak{t}x)(1-\mathfrak{t}x^{-1})}. 
\end{align}
While the large $N$ normalized $1$-point function and $2$-point function depend on the flavor fugacities $y_{\alpha}$, the connected $2$-point function turns out to be independent of them. 
Again the corresponding single particle index is
\begin{align}
i_{X}^{(H)}(x;\mathfrak{t})&=
\text{PL}[\langle \mathcal{W}_{\ydiagram{1}}\mathcal{W}_{\overline{\ydiagram{1}}}\rangle^{U(\infty)\text{ADHM-}[l] (H)}(x;\mathfrak{t}) ]
\nonumber\\
&=\sum_{d\ge1}\frac{\mu(d)}{d}\log [
\langle \mathcal{W}_{\ydiagram{1}}\mathcal{W}_{\overline{\ydiagram{1}}}\rangle^{U(\infty)\text{ ADHM-}[l] (H)}(x^d;\mathfrak{t}^d)
]
\nonumber\\
&=\mathfrak{t}x+\mathfrak{t}x^{-1}, 
\end{align}
which is expected to count the two bosonic scalars in the spectrum of the fluctuation modes on the dual geometry.


\subsection{JK residue sums}
\label{sec_JK}

In the previous section we have studied the Higgs line defect indices by using extensively the method of Hall-Littlewood expansion.
While this method is powerful and applicable for Wilson lines in any representations, 
it becomes rather difficult to perform the infinite sums whereas it gives rise to the small $\mathfrak{t}$-expansions.
In this subsection and the next subsection, we explain alternative techniques to evaluate the Higgs indices of the ADHM theory: (i) the direct evaluation of the integrations \eqref{Higgs_wilson} by finite residue sum and (ii) the method of refined topological vertex.
For simplicity, here we consider in method (i) only the correlation functions of Wilson lines in symmetric representations or antisymmetric representations while in method (ii) only the correlation functions of Wilson lines in antisymmetric representations.
While in these 
methods
it is rather non-trivial to extend the calculation to Wilson lines in more general representations, both methods 
can
give the indices as closed-form expressions a priori.

First let us explain the method (i), which allows us to study the correlation functions of Wilson lines in (anti)symmetric representation through the generating function
\begin{align}
&F(a_I,b_I,c_I,d_I;x,y_\alpha;\mathfrak{t})^{U(N)\text{ADHM-}[l](H)}\nonumber \\
&=
\frac{(1-\mathfrak{t})^N}{N!\prod_\pm (1-x^{\pm 1}\mathfrak{t})^N}\int\prod_{i=1}^N\frac{ds_i}{2\pi is_i}
\prod_{i=1}^N
\frac{
\prod_{I=1}^m
(1+a_Is_i)(1+b_Is_i^{-1})
}{
\prod_{I=1}^n
(1-c_Is_i)(1-d_Is_i^{-1})}\nonumber \\
&\quad \times \frac{\prod_{i<j}^N\prod_\pm (1-s_i^{\pm 1}s_j^{\mp 1})(1-\mathfrak{t}^2s_i^{\pm 1}s_j^{\mp 1})}{\prod_{i<j}^N\prod_{\pm,\pm'}(1-\mathfrak{t}(s_is_j^{-1}x^{\pm'1})^{\pm 1})}
\frac{1}{\prod_{i=1}^N\prod_{\alpha=1}^l\prod_\pm (1-\mathfrak{t}(s_iy_\alpha)^{\pm 1})}\nonumber \\
&=\sum_{k_1,k_2,\cdots,k_{2m+2n}\ge 0}
a_1^{k_1}
a_2^{k_2}
\cdots
a_m^{k_m}
b_1^{k_{m+1}}
b_2^{k_{m+2}}
\cdots
b_m^{k_{2m}}
c_1^{k_{2m+1}}
c_2^{k_{2m+2}}
c_n^{k_{2m+n}}
d_1^{k_{2m+n+1}}\nonumber \\
&\quad \times d_2^{k_{2m+n+2}}
\cdots
d_n^{k_{2m+2n}}\nonumber \\
&\quad\times \langle
\prod_{I=1}^m
W_{(1^{k_I})}
W_{(\overline{1^{k_{m+I}}})}
\prod_{I=1}^n
W_{(k_{2m+I})}
W_{(\overline{k_{2m+n+I}})}
\rangle^{U(N)\text{ADHM-}[l](H)}(x,y_\alpha;\mathfrak{t}).
\label{calIHtilde}
\end{align}
For this purpose we consider the following integration
\begin{align}
&{\tilde {\cal I}}^{U(N)\text{-}[l,m](H)}\nonumber \\
&=
\frac{(1-\mathfrak{t}^2)^N}{N!\prod_\pm (1-x^{\pm 1}\mathfrak{t})^N}
\int
\prod_{i=1}^N
\frac{ds_i}{2\pi is_i}
{\cal A}_{l,m}
\prod_{i<j}^N\prod_{\pm,\pm'}\frac{1}{1-\mathfrak{t}(s_is_j^{-1}x^{\pm'1})^{\pm 1}}
\prod_{i=1}^N\prod_{\alpha=1}^l\frac{1}{1-\mathfrak{t}s_i^{-1}y_\alpha^{(2)}},
\label{calIHtilde2}
\end{align}
which we shall call an auxiliary index, where
\begin{align}
{\cal A}_{l,m}=
\prod_{i=1}^N
\frac{
\prod_{I=1}^m(1+a_Is_i)(1+b_Is_i^{-1})
}{
\prod_{\alpha=1}^l(1-\mathfrak{t}s_iy_\alpha^{(1)})}
\prod_{i<j}^N\prod_\pm(1-s_i^{\pm 1}s_j^{\mp 1})(1-\mathfrak{t}^2s_i^{\pm 1}s_j^{\mp 1})
\label{A}
\end{align}
is the ``regular factor'' (although ${\cal A}_{l,m}$ has poles at $s_i=0$ and $s_i=\mathfrak{t}^{-1}(y_\alpha^{(1)})^{-1}$, these poles does not contribute in the calculation with JK residue prescription explained below).

The generating function \eqref{calIHtilde} is given by the auxiliary index
as
\begin{align}
F(a_I,b_I,c_I,d_I;x,y_\alpha;\mathfrak{t})^{U(N)\text{ADHM-}[l](H)}
=
{\tilde {\cal I}}^{U(N)\text{-}[l+n,m](H)}\biggr|_{\substack{
y_\alpha^{(1)}=y_\alpha,y_\alpha^{(2)}=y_\alpha^{-1},\,\,(\alpha\le l)\\
y_{l+I}^{(1)}=\mathfrak{t}^{-1}c_I, y_{l+I}^{(2)}=\mathfrak{t}^{-1}d_I
}
}.
\end{align}
Let us expand the auxiliary index 
\eqref{calIHtilde2} as
\begin{align}
{\tilde {\cal I}}^{U(N)\text{-}[l,m](H)}
\sum_{k_1,\cdots,k_{2m}\ge 0}
\prod_{I=1}^ma_I^{k_I}b_I^{k_{m+I}}
{\tilde {\cal I}}_{k_1,\cdots,k_{2m}}^{U(N)\text{-}[l,m](H)},
\end{align}
Then the expansion coefficients
satisfy the following symmetries
\begin{align}
&{\tilde {\cal I}}_{k_1,\cdots,k_{2m}}^{U(N)\text{-}[l,m](H)}(y_\alpha^{(1)},y_\alpha^{(2)})
=
{\tilde {\cal I}}_{k_{m+1},\cdots,k_{2m},k_1,\cdots,k_m}^{U(N)\text{-}[l,m](H)}(y_\alpha^{(2)},y_\alpha^{(1)}),\\
&{\tilde {\cal I}}_{k_1,\cdots,k_{2m}}^{U(N)\text{-}[l,m](H)}(y_\alpha^{(1)},y_\alpha^{(2)})
=
{\tilde {\cal I}}_{N-k_{m+1},k_2,\cdots k_m,N-k_1,k_{m+2},\cdots,k_{2m}}^{U(N)\text{-}[l,m](H)}(y_\alpha^{(1)},y_\alpha^{(2)}).
\end{align}
This implies that the correlation functions of Wilson lines satisfy the following symmetries
\begin{align}
&\langle
\prod_{I=1}^m
W_{(1^{k_I})}
W_{(\overline{1^{k_{m+I}}})}
\prod_{I=1}^n
W_{(k_{2m+I})}
W_{(\overline{k_{2m+n+I}})}
\rangle^{U(N)\text{ ADHM-}[l](H)}(x,y_\alpha;\mathfrak{t})\nonumber \\
&=
\langle
\prod_{I=1}^m
W_{(1^{k_{m+I}})}
W_{(\overline{1^{k_I}})}
\prod_{I=1}^n
W_{(k_{2m+n+I})}
W_{(\overline{k_{2m+I}})}
\rangle^{U(N)\text{ ADHM-}[l](H)}(x,y_\alpha^{-1};\mathfrak{t}),\nonumber \\
&\langle
\prod_{I=1}^m
W_{(1^{k_I})}
W_{(\overline{1^{k_{m+I}}})}
\prod_{I=1}^n
W_{(k_{2m+I})}
W_{(\overline{k_{2m+n+I}})}
\rangle^{U(N)\text{ ADHM-}[l](H)}(x,y_\alpha;\mathfrak{t})\nonumber \\
&=
\langle
W_{(1^{N-k_{m+1}})}
W_{(\overline{1^{N-k_1}})}
\prod_{I=2}^m
W_{(1^{k_I})}
W_{(\overline{1^{k_I}})}
\prod_{I=1}^n
W_{(k_{2m+n+I})}
W_{(\overline{k_{2m+I}})}
\rangle^{U(N)\text{ ADHM-}[l](H)}(x,y_\alpha;\mathfrak{t}).
\label{Higgssymmetrymanifestinintegral}
\end{align}

First let us consider the case with $m<l$ so that the integrand of the auxiliary index \eqref{calIHtilde2} does not have a pole at $s_i=0$.
In this case, the auxiliary index \eqref{calIHtilde2} can be evaluated by the JK residue prescription \cite{MR1318878,Benini:2013nda,Benini:2013xpa,Hwang:2014uwa} which goes as follows.
First we choose an $N$-dimensional real vector $\eta$ called ``reference vector''.
Here we shall choose it as $\eta=(-1,-1,\cdots,-1)$.
Second, we pick $N$ factors $\{g_a(s_1,\cdots,s_N)\}_{a=1}^N$ from the denominators $1-\mathfrak{t} (s_is_j^{-1}x^{\pm'1})^{\pm 1}$ and $1-\mathfrak{t}s_i^{-1}y_\alpha^{(2)}$ so that it satisfies the following conditions:\footnote{
Here we have excluded the factors $1-\mathfrak{t}s_iy_\alpha^{(1)}$ in the denominator of ${\cal A}_m$ \eqref{A} from the candidates since any choice containing these factors turns out to violate the Condition 2.
}
\begin{itemize}
\item $\text{Condition 1}$: the intersection of $N$ hyperplanes $g_a(s_1,\cdots,s_N)=0$ gives a single point $(s_1^{\{g_a\}}, s_2^{\{g_a\}}, \cdots, s_N^{\{g_a\}})$;
\item $\text{Condition 2}$: define an $N$-dimensional vector $v$ for each factor as
\begin{align}
\frac{1}{g_a(s_1,\cdots,s_N)}=\frac{1}{1-(\cdots)\prod_{i=1}^Ns_i^{\nu_i}}\rightarrow v=(\nu_1,\nu_2,\cdots,\nu_N).
\end{align}
Then we require that the cone $\{\sum_{a=1}^N r_av_a|r_a>0\}$ defined from the vectors $v_a$ associated with the chosen $N$ factors contain the reference vector $\eta$.
\end{itemize}
Now the integration \eqref{calIHtilde2} is evaluated by summing the iterative residue for each choice of the $N$ factors $\{g_a\}$ in the denominator of the integrand of \eqref{calIHtilde2}, as
\begin{align}
{\tilde {\cal I}}^{U(N)\text{-}[l,m](H)}
={\tilde {\cal I}}_{\text{JK}}^{U(N)\text{-}[l,m](H)}\quad (m<l)
\end{align}
with
\begin{align}
&{\tilde {\cal I}}_{\text{JK}}^{U(N)\text{-}[l,m](H)}\nonumber \\
&=
\frac{(1-\mathfrak{t}^2)^N}{N!\prod_\pm(1-x^{\pm 1}\mathfrak{t})^N}
\sum_{\{g_a\}}
\biggl[
\Bigl(\prod_{i=1}^N\frac{1}{s_i}\Bigr)
{\cal A}_{l,m}
\biggl(
\prod_{\substack{\pm,\pm',i<j\\ (\text{not in }\{g_a\})}}
\frac{1}{1-\mathfrak{t}(s_is_j^{-1}x^{\pm' 1})^{\pm 1}}
\biggr)\nonumber \\
&\quad \times \biggl(
\prod_{\substack{\pm,i,\alpha\\ (\text{not in }\{g_a\})}}
\frac{1}{1-\mathfrak{t}s_i^{-1}y_\alpha^{(2)}}
\biggr)
\frac{1}{H}
\biggr]_{s_i\rightarrow s_i^{\{g_a\}}},
\label{JK}
\end{align}
where in the product $\prod_{\pm,\pm',i<j}$ in the first line the index $(\pm,\pm',i,j)$ runs over those for which $1-\mathfrak{t}(s_is_j^{-1}x^{\pm' 1})^{\pm 1}$ does not belong to $\{g_a\}$, while in the product $\prod_{\pm,i,\alpha}$ in the second line the index $(\pm,i,\alpha)$ runs over those for which $1-\mathfrak{t}s_i^{-1}y_\alpha^{(2)}$ does not belong to $\{g_a\}$, and $H$ is given by
\begin{align}
H=\det_{1\le a,j\le N}\Bigl[\frac{\partial g_a(s_1,\cdots,s_N)}{\partial s_j}\Bigr].
\end{align}
Note that even when a choice of $N$ factors $\{g_a\}$ satisfies the above two condition, it does not contribute to the auxiliary index 
\eqref{calIHtilde2} 
if the regular factor \eqref{A} evaluated at the intersection $s_i^{\{g_a\}}$ is zero, namely if $s_i^{\{g_a\}}=s_j^{\{g_a\}}$ or $s_i^{\{g_a\}}=\mathfrak{t}^2s_j^{\{g_a\}}$ for a pair of $(i,j)$.\footnote{
Note however that for $N\ge 4$ there are choices of $\{g_a\}$ where the intersection point satisfies $s_i^{\{g_a\}}=\mathfrak{t}^2s_j^{\{g_a\}}$ but at the same time the denominator has a double pole (see e.g.~\cite{Hwang:2014uwa}).
In these cases the zero in the regular factor ${\cal A}_{l,m}$ \eqref{A} and the double pole cancels with each other and hence the residue can be non-vanishing.
\label{fn_JKpoleats0}
}
As found in \cite{Nekrasov:2002qd,Nekrasov:2003rj,Hwang:2014uwa}, the choices of $N$ factors $\{g_a\}$ with non-vanishing residues are characterized by sets of $l$ Young diagrams $\lambda^{(1)},\lambda^{(2)},\cdots,\lambda^{(l)}$ satisfying $\sum_{\alpha=1}^l|\lambda^{(\alpha)}|=N$, up to the permutation of $(s_1,\cdots,s_N)$ which gives an identical contribution to the auxiliary index.
More concretely, each of $s_1,\cdots,s_N$ corresponds to one of the $N$ boxes in the $l$ Young diagrams which are drawn by associating the factor $1-\mathfrak{t}s_i^{-1}y_\alpha^{(2)}\in \{g_a\}$ with the box of $s_i$ on top-left box in the $\alpha$-th Young diagram and the factor $1-\mathfrak{t}s_is_j^{-1}x^{\pm 1}\in \{g_a\}$ with the box of $s_j$ attached horizontally/vertically to the box $s_i$.
As examples, in Table \ref{JKwlist} we list 
the correspondence between $\{g_a\}$ and $(\lambda^{(1)},\lambda^{(2)})$ for $N=3$ and $l=2$.
\begin{table}
\begin{center}
\begin{tabular}{c|l}
   &\\[-14pt]
$\{g_a\}$&$(\lambda_1,\lambda_2)$\\ \hline
   &\\[-14pt]
$\{1-\mathfrak{t}s_1^{-1}y_1^{(2)},1-\mathfrak{t}s_2^{-1}y_2^{(2)},1-\mathfrak{t}s_1s_3^{-1}x\}$&$(\ydiagram{2},\ydiagram{1})$\\ \hline
   &\\[-14pt]
$\{1-\mathfrak{t}s_1^{-1}y_2^{(2)},1-\mathfrak{t}s_2^{-1}y_1^{(2)},1-\mathfrak{t}s_1s_3^{-1}x\}$&$(\ydiagram{1},\ydiagram{2})$\\ \hline
   &\\[-14pt]
$\{1-\mathfrak{t}s_1^{-1}y_1^{(2)},1-\mathfrak{t}s_2^{-1}y_2^{(2)},1-\mathfrak{t}s_1s_3^{-1}x^{-1}\}$&$(\ydiagram{1,1},\ydiagram{1})$\\ \hline
   &\\[-14pt]
$\{1-\mathfrak{t}s_1^{-1}y_2^{(2)},1-\mathfrak{t}s_2^{-1}y_1^{(2)},1-\mathfrak{t}s_1s_3^{-1}x^{-1}\}$&$(\ydiagram{1},\ydiagram{1,1})$\\ \hline
   &\\[-14pt]
$\{1-\mathfrak{t}s_1^{-1}y_1^{(2)},1-\mathfrak{t}s_1s_2^{-1}x,1-\mathfrak{t}s_1s_3^{-1}x^{-1}\}$&$(\ydiagram{2,1},\emptyset)$\\ \hline
   &\\[-14pt]
$\{1-\mathfrak{t}s_1^{-1}y_1^{(2)},1-\mathfrak{t}s_1s_2^{-1}x,1-\mathfrak{t}s_2s_3^{-1}x\}$&$(\ydiagram{3},\emptyset)$\\ \hline
   &\\[-14pt]
$\{1-\mathfrak{t}s_1^{-1}y_1^{(2)},1-\mathfrak{t}s_1s_2^{-1}x^{-1},1-\mathfrak{t}s_2s_3^{-1}x^{-1}\}$&$(\ydiagram{1,1,1},\emptyset)$\\ \hline
   &\\[-14pt]
$\{1-\mathfrak{t}s_1^{-1}y_2^{(2)},1-\mathfrak{t}s_1s_2^{-1}x,1-\mathfrak{t}s_1s_3^{-1}x^{-1}\}$&$(\emptyset,\ydiagram{2,1})$\\ \hline
   &\\[-14pt]
$\{1-\mathfrak{t}s_1^{-1}y_2^{(2)},1-\mathfrak{t}s_1s_2^{-1}x,1-\mathfrak{t}s_2s_3^{-1}x\}$&$(\emptyset,\ydiagram{3})$\\ \hline
   &\\[-14pt]
$\{1-\mathfrak{t}s_1^{-1}y_2^{(2)},1-\mathfrak{t}s_1s_2^{-1}x^{-1},1-\mathfrak{t}s_2s_3^{-1}x^{-1}\}$&$(\emptyset,\ydiagram{1,1,1})$\\ \hline
\end{tabular}
\caption{
Choice of $N=3$ factors $\{g_a\}$ (up to permutations of $(s_1,s_2,s_3)$) in the denominator of the integrand of the auxiliary index ${\tilde {\cal I}}^{U(3)\text{-}[2,m](H)}$ \eqref{calIHtilde2} with non-vanishing residues, and the pair of Young diagrams associated with each $\{g_a\}$.
}
\label{JKwlist}
\end{center}
\end{table}
Evaluating the residue for each choice of the $N$ intersections $\{g_a\}$, we finally obtain the following formula
\begin{align}
{\tilde {\cal I}}_{\text{JK}}^{U(N)\text{-}[l,m](H)}
&=\frac{1}{\mathfrak{t}^{2Nl}}\sum_{\substack{\lambda^{(1)},\cdots,\lambda^{(l)}\\ (\sum_{\alpha=1}^l |\lambda^{(\alpha)}|=N)}}
\prod_{\alpha,\beta=1}^l
\frac{
1
}{{\cal N}_{\lambda^{(\alpha)},\lambda^{(\beta)}}(\frac{y_\alpha^{(2)}}{y_\beta^{(2)}})}
\frac{
{\cal N}_{\lambda^{(\alpha)},\emptyset}(\mathfrak{t}^2\frac{y^{(2)}_\alpha}{y^{(2)}_\beta})
}{
{\cal N}_{\lambda^{(\alpha)},\emptyset}(\mathfrak{t}^2y^{(2)}_\alpha y_\beta^{(1)})
}\nonumber \\
&\quad\times \prod_{\alpha=1}^l
\prod_{I=1}^m
{\cal N}_{\lambda^{(\alpha)},\emptyset}(-a_I\mathfrak{t}y_\alpha^{(2)})
{\cal N}_{\emptyset,\lambda^{(\alpha)}}(-b_I\mathfrak{t}(y_\alpha^{(2)})^{-1}),
\label{calIHtilde2Nekrasov}
\end{align}
where ${\cal N}_{\lambda,\mu}(u)$
is so-called Nekrasov factor which is defined as \eqref{calNandE}.
Note that when $y_\alpha^{(1)}y_\alpha^{(2)}=1$ for all $\alpha=1,2,\cdots,l$ the formula \eqref{calIHtilde2Nekrasov} reduces to the known formula for the $N$-instanton partition function of the 5d ${\cal N}=1$ $U(l)$ Yang-Mills theory with $m$ fundamental and $m$ anti-fundamental matter fields \cite{Nekrasov:2002qd,Nekrasov:2003rj} (see also \cite{Hwang:2014uwa}).

Here we list some examples of the generating function \eqref{calIHtilde}.
For $l=0$ we have
\begin{align}
%
&\sum_{k_1,k_2\ge 0}c^{k_1}d^{k_2}\langle W_{(k_1)}W_{(\overline{k_2})}\rangle^{{\cal N}=8\text{ }U(2)(H)}(x;\mathfrak{t})\nonumber \\
&=\frac{(1-\mathfrak{t}^2)(1+\mathfrak{t}^2+\mathfrak{t}^3(-x^{-1}-x)+cd\mathfrak{t}^3(-x^{-1}-x+\mathfrak{t}+\mathfrak{t}^3))}{(1-cd)\prod_\pm (1-cdx^{\pm 1}\mathfrak{t})(1-x^{\pm 1}\mathfrak{t})(1-x^{\pm 2}\mathfrak{t}^2)},\label{IHN2l0_cd} \\
&\sum_{k_1,k_2\ge 0}c^{k_1}d^{k_2}\langle W_{(k_1)}W_{(\overline{k_2})}\rangle^{{\cal N}=8\text{ }U(3)(H)}(x;\mathfrak{t})\nonumber \\
&=\frac{1-\mathfrak{t}^2}{
(1 - cd)\prod_\pm
(1-cdx^{\pm 1}\mathfrak{t})
(1-cdx^{\pm 2}\mathfrak{t}^2)
(1-x^{\pm 1}\mathfrak{t})
(1-x^{\pm 2}\mathfrak{t}^2)
(1-x^{\pm 3}\mathfrak{t}^3)
}\nonumber \\
&\quad \times \Bigl[
1
+ \mathfrak{t}^2
+ \mathfrak{t}^4 (-x^{-2} - x^2)
+ \mathfrak{t}^5 (-x^{-1} - x)
+ \mathfrak{t}^6 (-x^{-2}+1 - x^2)
+ \mathfrak{t}^7 (x^{-3} + x^3)
+ \mathfrak{t}^8
\nonumber \\
&\quad +c d \mathfrak{t}^2 [1
+ \mathfrak{t}^2 (-x^{-2}-1 - x^2)
+ \mathfrak{t}^3 (-x^{-3} - x^{-1} - x - x^3)
+ \mathfrak{t}^4 (-x^{-2}+1- x^2)\nonumber \\
&\quad\quad + \mathfrak{t}^5 (x^{-3} + x^{-1} + x + x^3)
+ \mathfrak{t}^6 (x^{-4}+4 + x^4)
+ \mathfrak{t}^7 (x^{-3} + x^{-1} + x + x^3)\nonumber \\
&\quad\quad + \mathfrak{t}^8 (-x^{-2}+1 - x^2)
+ \mathfrak{t}^9 (-x^{-3} - x^{-1} - x - x^3)
+ \mathfrak{t}^{10} (-x^{-2}-1 - x^2)
+ \mathfrak{t}^{12}
]\nonumber \\
&\quad +c^2 d^2 \mathfrak{t}^8 [1
+ \mathfrak{t} (x^{-3}+ x^3)
+ \mathfrak{t}^2 (-x^{-2}+1 - x^2)
+ \mathfrak{t}^3 (-x^{-1} - x)
+ \mathfrak{t}^4 (-x^{-2} - x^2)
+ \mathfrak{t}^6\nonumber \\
&\quad + \mathfrak{t}^8
]
\Bigr]\label{IHN3l0_cd}.
\end{align}
For $l=1$ we have
\begin{align}
&\sum_{k=0}^\infty c^k\langle W_{(k)}\rangle^{U(2)\text{ADHM-}[1](H)}(x;\mathfrak{t})
=\frac{
1+\mathfrak{t}^2-(x+x^{-1})c\mathfrak{t}^4
}
{
(1-c\mathfrak{t})\prod_\pm
(1-x^{\pm 1}\mathfrak{t})
(1-x^{\pm 2}\mathfrak{t}^2)
(1-cx^{\pm 1}\mathfrak{t}^2)
},\nonumber \\
&\sum_{k=0}^\infty c^k\langle W_{(k)}\rangle^{U(3)\text{ ADHM-}[1](H)}(x;\mathfrak{t})\nonumber \\
&=
\frac{1}{
(1-c\mathfrak{t})
\prod_\pm
(1-\mathfrak{t}x^{\pm 1})
(1-\mathfrak{t}^2x^{\pm 2})
(1-\mathfrak{t}^3x^{\pm 3})
(1-cx^{\pm 1}\mathfrak{t}^2)
(1-cx^{\pm 2}\mathfrak{t}^3)
}
(
1+\mathfrak{t}^2\nonumber \\
&\quad +(x+x^{-1})\mathfrak{t}^3+\mathfrak{t}^4+\mathfrak{t}^6\nonumber \\
&\quad +\mathfrak{t}^3(1
-(x^2+1+x^{-2})\mathfrak{t}^2
-(x^3+x+x^{-1}+x^{-3})\mathfrak{t}^3
-(x^2+x^{-2})\mathfrak{t}^4
-(x+x^{-1})\mathfrak{t}^5\nonumber \\
&\quad -(x^2+x^{-2})\mathfrak{t}^6
)c\nonumber \\
&\quad +\mathfrak{t}^{10}(1+(x^3+x+x^{-1}+x^{-3})\mathfrak{t}+\mathfrak{t}^2)c^2
),\label{IHN3l1_c}
\end{align}
where we have set $y_1=1$ since the $y_1$-dependences are trivial, as explained below \eqref{uNADHM_1ptasym_HL} in section \ref{sec_UNADHMl1H}.
For $l=1$ and $l\ge 2$ we have
\begin{align}
\sum_{k=0}^\infty c^k\langle W_{(k)}\rangle^{U(1)\text{ ADHM-}[l](H)}(x,y_\alpha;\mathfrak{t})
=\frac{1}{\prod_\pm(1-\mathfrak{t}x^{\pm 1})}\sum_{\alpha=1}^l\frac{1}{1-\mathfrak{t}y_\alpha^{-1}c}
\prod_{\beta(\neq\alpha)}^l
\frac{1}{1-\frac{y_\alpha}{y_\beta}}
\frac{1}{1-\mathfrak{t}^2\frac{y_\beta}{y_\alpha}}
.
\end{align}
These generating functions precisely reproduce the 1-point functions of Wilson line in symmetric representations obtained by the Hall-Littlewood expansion \eqref{u2ADHM1w1},\eqref{u3ADHM1_w1},\eqref{u2ADHM1wsym2}-\eqref{u2ADHM1wsym5}.


One can also obtain the generating function for the diagonal $2$-point functions
$\langle W_{(1^k)}W_{(\overline{1^k})}\rangle^{U(N)\text{ADHM-}[l](H)}(x,y_\alpha;\mathfrak{t})$
or
$\langle W_{(k)}W_{(\overline{k})}\rangle^{U(N)\text{ADHM-}[l](H)}(x,y_\alpha;\mathfrak{t})$
from the generating function $F(a,b,c,d;x,y_\alpha;\mathfrak{t})^{U(N)\text{ADHM-}[l](H)}$ \eqref{calIHtilde} by collecting only the expansion coefficients of $(ab)^k$'s or $(cd)^k$'s, and then resumming them.
For the diagonal $2$-point correlator of the antisymmetric Wilson lines, this can be performed directly since $F(a,b,0,0;x,y_\alpha;\mathfrak{t})^{U(N)\text{ADHM-}[l](H)}$ is a finite order polynomial in $a,b$.
For the diagonal $2$-point correlator of the symmetric Wilson lines, an efficient way to perform this resummation is to consider the following integration
\begin{align}
\sum_{k=0}^\infty a^k
\langle W_{(k)}W_{(\overline{k})}\rangle^{U(N)\text{ADHM-}[l](H)}(x,y_\alpha;\mathfrak{t})
=
\oint\frac{da'}{2\pi ia'}
F(0,0,a',\frac{a}{a'};x,y_\alpha;\mathfrak{t})^{U(N)\text{ADHM-}[l](H)},
\label{diagonalWWHbyresidue}
\end{align}
which can be evaluated by picking the residues at the poles in $|a'|<1$ under the assumption of $|a|=1$ and $|\mathfrak{t}|<1$, or more explicitly the residues at $a'=a\mathfrak{t},a\mathfrak{t}^2,\cdots$.
For example, we obtain
\begin{align}
&\sum_{k=0}^\infty a^k\langle W_{(k)}W_{(\overline{k})}\rangle^{U(2)\text{ADHM-}[1](H)}(x;\mathfrak{t})
=
\frac{
1 + \mathfrak{t}^2 + a (\mathfrak{t}^2 + \mathfrak{t}^4)
}{
(1 - a)\prod_{\pm}(1-ax^{\pm 1}\mathfrak{t})(1 - x^{\pm 1}\mathfrak{t})(1 - x^{\pm 2}\mathfrak{t}^2)
},\label{IHN2l1_cd} \\
&\sum_{k=0}^\infty a^k\langle W_{(k)}W_{(\overline{k})}\rangle^{U(3)\text{ADHM-}[1](H)}(x;\mathfrak{t})\nonumber \\
&=\frac{1}{
(1 - a)(1 - a\mathfrak{t}^2)\prod_\pm (1 - ax^{\pm 1}\mathfrak{t})(1 - ax^{\pm 2}\mathfrak{t}^2)(1 - x^{\pm 1}\mathfrak{t})(1 - x^{\pm 2}\mathfrak{t}^2)(1 - x^{\pm 3}\mathfrak{t}^3)}\nonumber \\
&\quad\times (
1 + \mathfrak{t}^2 + \mathfrak{t}^3  (x^{-1} + x) +\mathfrak{t}^4 + \mathfrak{t}^6\nonumber \\
&\quad + a \mathfrak{t}^2 [
1
+ \mathfrak{t} (2x^{-1} + 2 x)
+ \mathfrak{t}^2 (x^{-2} + 2 + x^2)
+ \mathfrak{t}^3 (2x^{-1} + 2 x)
+ \mathfrak{t}^4 (x^{-2} + 3 + x^2)\nonumber \\
&\quad\quad + \mathfrak{t}^5 (x^{-1} + x)
- \mathfrak{t}^6 (x^{-2} + x^2)
]\nonumber \\
&\quad + a^2 \mathfrak{t}^6[
  x^{-2} + x^2
 - \mathfrak{t} (x^{-1} + x)
 - \mathfrak{t}^2 (x^{-2}+3  + x^2)
 - \mathfrak{t}^3 (2x^{-1} + 2 x)
 - \mathfrak{t}^4 (x^{-2}+2  + x^2)\nonumber \\
&\quad\quad\quad - \mathfrak{t}^5 (2x^{-1} + 2 x)
 - \mathfrak{t}^6
]\nonumber \\
&\quad + a^3 \mathfrak{t}^8[
-1
 - \mathfrak{t}^2
 - \mathfrak{t}^3 (x^{-1} + x)
 - \mathfrak{t}^4
 - \mathfrak{t}^{6}
]
).\label{IHN3l1_cd}
\end{align}
Again, the expansion coefficients of these generating functions correctly reproduce the results obtained by the Hall-Littlewood expansion \eqref{u2ADHM1w1w1withx}, \eqref{u3ADHM1_w1w1}, \eqref{u2ADHM1wsym2wsym2withx} and \eqref{u3ADHM1_wsym2wsym2}.

Note that the generating functions become complicated quickly as $N$ increases, both for the generic multi-point functions and for the diagonal $2$-point functions.
Nevertheless, we can study the large $N$ behavior for small representations by focusing only on the first few terms of the generating functions expanded in the fugacities.
For example we find
$\langle {\cal W}_{\ydiagram{1}}{\cal W}_{\overline{\ydiagram{1}}}\rangle^{U(\infty)\text{ADHM-}[1](H)}(x;\mathfrak{t})$
as \eqref{calWcalW1Uinfty-[1]H} and
\begin{align}
%
&\langle {\cal W}_{\ydiagram{2}}{\cal W}_{\overline{\ydiagram{2}}}\rangle^{U(\infty)\text{ADHM-}[1](H)}(x;\mathfrak{t})\nonumber \\
&=
\frac{1}{
\prod_\pm (1 - \mathfrak{t}x^{\pm 1})^2 (1 - \mathfrak{t}^2x^{\pm 2})^2
}(
1
+ (-x^{-1} - x)                  \mathfrak{t}
+ (3 - x^{-2} - x^2)             \mathfrak{t}^2\nonumber \\
&\quad + (x^{-3} + x^{-1} + x + x^3)    \mathfrak{t}^3 
+ (4 - 3x^{-2} - 3 x^2)          \mathfrak{t}^4 
+ (-x^{-1} - x)                  \mathfrak{t}^5
+ (5 - 2x^{-2} - 2 x^2)          \mathfrak{t}^6\nonumber \\
&\quad + 3 \mathfrak{t}^8
).
\end{align}

We can further calculate the closed-form expressions for the correlators of Wilson lines in $U(N)$ ADHM theory with $l\ge 2$ flavors, which are hard in the Hall-Littlewood expansion.
We display these results and more results for $l=1$ with  $x=y_\alpha=1$ in appendix \ref{app_unflavoredH}.

\subsubsection{Auxiliary index ${\tilde {\cal I}}^{U(N)\text{-}[l,m](H)}$ with $m=l$}

Lastly let us elaborate the subtlety of our formula \eqref{calIHtilde2Nekrasov} (or \eqref{JK}) for $m\ge l$.
As we mentioned below \eqref{Higgssymmetrymanifestinintegral}, when $m\ge l$ the integrand of the auxiliary index ${\tilde {\cal I}}^{U(N)\text{-}[l,m](H)}$ \eqref{calIHtilde2} has a pole $s_i^{-(m+1-l)}$, hence the residue sum \eqref{calIHtilde2Nekrasov} is insufficient for evaluating ${\tilde {\cal I}}^{U(N)\text{-}[l,m](H)}$.
For the terms which is proportional to $a_{I_1}^Na_{I_2}^N\cdots a_{I_p}^N$ and does not contain $b_{J_1}b_{J_2}\cdots b_{J_q}$
with arbitrary subsets of the indices $\{I_1,\cdots,I_p\},\{J_1,\cdots,J_q\}\subset \{1,\cdots,m\}$, however, the order of the pole at the origin is modified as $s_i^{-(m+1-l-p-q)}$, and hence the residue sum \eqref{calIHtilde2Nekrasov} gives the correct answer if $m<l+p+q$.
In this way we can calculate for example the 1-point functions $\langle W_{(1^k)}\rangle^{U(N)\text{ ADHM-}[1](H)}(x,y_\alpha;\mathfrak{t})$ by \eqref{calIHtilde2Nekrasov}, which precisely reproduces those obtained by the Hall-Littlewood expansion \eqref{u2ADHM1_wasym2},\eqref{u3ADHM1_wasym2} and \eqref{u3ADHM1_wasym3}.
We can also calculate the following mixed generating functions for $l=1$ which are not included in the generating function for the symmetric Wilson lines $F(0,0,c_I,d_I;x;\mathfrak{t})^{U(N)\text{ ADHM-}[1](H)}$:
\begin{align}
%
&\sum_{k\ge 0}c^{k}\langle W_{\ydiagram{1,1}}W_{(k)}\rangle^{U(2)\text{ ADHM-}[1](H)}(x;\mathfrak{t})
=
\frac{\mathfrak{t}^3(x+x^{-1}-\mathfrak{t}^2(1+\mathfrak{t}^2)c)
}{
(1-c\mathfrak{t})\prod_\pm(1-x^{\pm 1}c\mathfrak{t}^2)
(1-x^{\pm 1}\mathfrak{t})(1-x^{\pm 2}\mathfrak{t}^2)
},\\
&\sum_{k\ge 0}d^{k}\langle W_{\ydiagram{1,1}}W_{(\overline{k})}\rangle^{U(2)\text{ ADHM-}[1](H)}(x;\mathfrak{t})\nonumber \\
&=
\frac{
\mathfrak{t}
}{
(1-d\mathfrak{t})\prod_\pm(1-x^{\pm 1}d\mathfrak{t}^2)
(1-x^{\pm 1}\mathfrak{t})(1-x^{\pm 2}\mathfrak{t}^2)
}
(
\mathfrak{t}^2(x+x^{-1})\nonumber \\
&\quad +(1+(x+x^{-1})\mathfrak{t}+\mathfrak{t}^2-(x+x^{-1})\mathfrak{t}^3-(x^2+2+x^{-2})\mathfrak{t}^4)d\nonumber \\
&\quad +(x+x^{-1}-(x+x^{-1})\mathfrak{t}^2-(x^2+2+x^{-2})\mathfrak{t}^3-(x+x^{-1})\mathfrak{t}^4+(x^2+2+x^{-2})\mathfrak{t}^5\nonumber \\
&\quad\quad +(x+x^{-1})\mathfrak{t}^6)d^2\nonumber \\
&\quad +\mathfrak{t}^2(-1+(x+x^{-1})\mathfrak{t}^3+\mathfrak{t}^4-(x+x^{-1})\mathfrak{t}^5)d^3
),
\label{IHN2l1_ac}
\end{align}
where we have set $y_1=1$.

As we have mentioned above, when $l$ is small the JK residue sum \eqref{calIHtilde2Nekrasov} gives incorrect answer to some of the coefficients of $b_I$ as correlators including $W_{(\overline{1^k})}$.
However, we can still obtain the closed-form expression of some of these coefficients indirectly from \eqref{calIHtilde2Nekrasov} by using the symmetry \eqref{Higgssymmetrymanifestinintegral} or noting $W_{(\overline{1^1})}=W_{(\overline{1})}$, which can be calculated through the insertion of $\frac{1}{1-ds_i^{-1}}$ in ${\tilde {\cal I}}^{U(N)\text{-}[l+1,m](H)}$ \eqref{calIHtilde2}.

Interestingly, we find that the correct auxiliary index ${\tilde {\cal I}}^{U(N)\text{-}[l,m](H)}$ is obtained from ${\tilde {\cal I}}^{U(N)\text{-}[m+1,m](H)}={\tilde {\cal I}}^{U(N)\text{-}[m+1,m](H)}_{\text{JK}}$, which can be evaluated by the JK residue formula \eqref{calIHtilde2Nekrasov}, by simply taking the limit $y^{(1)}_\alpha,y^{(2)}_\alpha\rightarrow 0$ for $\alpha=l+1,\cdots,m+1$ in ${\tilde {\cal I}}^{U(N)\text{-}[m+1,m](H)}_{\text{JK}}$.
By assuming this prescription we can also calculate for example the correlators involving $W_{(\overline{1^k})}$ with $k\ge 2$ for the $U(N)$ ADHM theory with $l=1$ flavor.
As a result we obtain
\begin{align}
&\langle W_{\ydiagram{1,1}}W_{\overline{\ydiagram{1,1}}}\rangle^{U(4)\text{ ADHM-}[1](H)}(x;\mathfrak{t})
=
\frac{
(1+\mathfrak{t}^2)
(
1 + 2 \mathfrak{t}^2 + \mathfrak{t}^3 (2x^{-1} + 2 x) + \mathfrak{t}^4 (x^{-2}+3 + x^2)
)
}{
\prod_\pm
(1 - x^{\pm 1} \mathfrak{t})^2
(1 - x^{\pm 2} \mathfrak{t}^2)^2
},\\
%
&\langle W_{\ydiagram{1,1}}W_{\overline{\ydiagram{1,1}}}\rangle^{U(5)\text{ ADHM-}[1](H)}(x;\mathfrak{t})\nonumber \\
&
=
\frac{
1
}{
\prod_\pm (1 - x^{\pm 1} \mathfrak{t})^2
(1 - x^{\pm 2} \mathfrak{t}^2)^2
(1 - x^{\pm 3} \mathfrak{t}^3)
}
(
1 + 3 \mathfrak{t}^2
+ \mathfrak{t}^3 (3x^{-1} + 3 x)
+ \mathfrak{t}^4 (2x^{-2} +7 +  2 x^2)\nonumber \\
&\quad + \mathfrak{t}^5 (x^{-3} + 7x^{-1} + 7 x + x^3)
+ \mathfrak{t}^6 (5x^{-2} +12 +  5 x^2)
+ \mathfrak{t}^7 (2x^{-3} + 9x^{-1} + 9 x + 2 x^3)\nonumber \\
&\quad + \mathfrak{t}^8 (x^{-4} + 5x^{-2} +10 +  5 x^2 + x^4)
+ \mathfrak{t}^9 (x^{-3} + 5x^{-1} + 5 x + x^3)
+ \mathfrak{t}^{10} (x^{-2} + 3 + x^2)
),\\
&\langle W_{\ydiagram{1,1}}W_{\overline{\ydiagram{1,1}}}\rangle^{U(6)\text{ ADHM-}[1](H)}(x;\mathfrak{t})\nonumber \\
&=
\frac{1}{
\prod_\pm (1 - x^{\pm 1} \mathfrak{t})^2
(1 - x^{\pm 2} \mathfrak{t}^2)^2
(1 - x^{\pm 3} \mathfrak{t}^3)
(1 - x^{\pm 4} \mathfrak{t}^4)
}
(
1
+ 3 \mathfrak{t}^2
+ \mathfrak{t}^3 (3x^{-1} + 3 x)\nonumber \\
&\quad + \mathfrak{t}^4 (3x^{-2}+ 8  + 3 x^2)
+ \mathfrak{t}^5 ( 2x^{-3}  + 9x^{-1} + 9 x + 2 x^3)\nonumber \\
&\quad 
+ \mathfrak{t}^6 (x^{-4} + 11x^{-2}+19   + 11 x^2 + x^4)
 + \mathfrak{t}^7 ( 7x^{-3} + 21x^{-1} + 21 x + 7 x^3)\nonumber \\
&\quad 
+ \mathfrak{t}^8 (5x^{-4}+ 22x^{-2} + 33 +  22 x^2 + 5 x^4)\nonumber \\
&\quad
 + \mathfrak{t}^9 ( 2x^{-5} + 14x^{-3} + 32x^{-1} + 32 x + 14 x^3 + 2 x^5) \nonumber \\
&\quad
+ \mathfrak{t}^{10} (x^{-6} + 8x^{-4} + 27x^{-2} +40 +  27 x^2 + 8 x^4 + x^6)\nonumber \\
&\quad + \mathfrak{t}^{11} (2x^{-5} + 14x^{-3} + 31x^{-1} + 31 x + 14 x^3 + 2 x^5) \nonumber \\
&\quad + \mathfrak{t}^{12} (x^{-6} + 6x^{-4} + 20x^{-2} + 31 + 20 x^2 + 6 x^4 + x^6)\nonumber \\
&\quad + \mathfrak{t}^{13} (x^{-5}  + 7x^{-3} + 17x^{-1} + 17 x + 7 x^3 + x^5) 
+ \mathfrak{t}^{14} (2x^{-4} + 8x^{-2} +14 +  8 x^2 + 2 x^4) \nonumber \\
&\quad + \mathfrak{t}^{15} (x^{-3}  + 5x^{-1} + 5 x + x^3)
 + \mathfrak{t}^{16} (x^{-2} +3 +  x^2) 
),\\
&\langle W_{\ydiagram{1,1,1}}W_{\overline{\ydiagram{1,1,1}}}\rangle^{U(6)\text{ ADHM-}[1](H)}(x;\mathfrak{t})\nonumber \\
&=
\frac{1}{
\prod_\pm (1 - x^{\pm 1} \mathfrak{t})^2
(1 - x^{\pm 2} \mathfrak{t}^2)^2
(1 - x^{\pm 3} \mathfrak{t}^3)^2
}
(
1
+ 3 \mathfrak{t}^2
+ \mathfrak{t}^3 (4x^{-1} + 4 x) 
+ \mathfrak{t}^4 (3x^{-2} +9  +  3 x^2)\nonumber \\
&\quad + \mathfrak{t}^5 ( 2x^{-3} + 12x^{-1} + 12 x + 2 x^3)
+ \mathfrak{t}^6 (x^{-4}  + 12x^{-2} +26 +  12 x^2 + x^4)\nonumber \\
&\quad + \mathfrak{t}^7 ( 8x^{-3} + 28x^{-1} + 28 x + 8 x^3)
 + \mathfrak{t}^8 (5x^{-4} + 24x^{-2} +43 +  24 x^2 + 5 x^4) \nonumber \\
&\quad
+ \mathfrak{t}^9 ( 2x^{-5} + 16x^{-3} + 42x^{-1} + 42 x + 16 x^3 + 2 x^5)\nonumber \\
&\quad
+ \mathfrak{t}^{10} (x^{-6} + 7x^{-4} + 28x^{-2} + 47 + 28 x^2 + 7 x^4 + x^6)\nonumber \\
&\quad
+ \mathfrak{t}^{11} (2x^{-5} + 14x^{-3} + 32x^{-1} + 32 x + 14 x^3 + 2 x^5)\nonumber \\
&\quad + \mathfrak{t}^{12} (x^{-6} + 5x^{-4} + 16x^{-2} + 27 + 16 x^2 + 5 x^4 + x^6) 
+ \mathfrak{t}^{13} (4x^{-3} + 10x^{-1} + 10 x + 4 x^3) \nonumber \\
&\quad + \mathfrak{t}^{14} (x^{-2} +4 +  x^2) 
).
\end{align}
See appendix \ref{app_unflavoredH} for more results in the unflavored limit.

Moreover, we observe for $l=m=1$ that the correct correlation functions and those obtained by naively applying \eqref{calIHtilde2Nekrasov} are related by a simple coefficient in the grand canonical sum in the rank $N$. 
Namely, if we define the grand canonical sum $\Xi_{\text{small }\mathfrak{t}}^{(l=m=1)}(u)$ for the generating function for ${\tilde {\cal I}}^{U(N)\text{-}[1,1](H)}$ with $m=1$ whose expansion coefficients are the correct ones (namely those obtained by the small $\mathfrak{t}$ expansion of \eqref{calIHtilde2}) and the grand canonical sum $\Xi_{\text{JK}}^{(l=m=1)}(u)$ of the JK residue sum
\eqref{calIHtilde2Nekrasov}
with $l=m=1$
\begin{align}
\Xi_{\text{small }\mathfrak{t}}^{(l=m=1)}(u)=\sum_{N=0}^\infty u^N{\tilde {\cal I}}^{U(N)\text{-}[1,1](H)},\quad
\Xi_{\text{JK}}^{(l=m=1)}(u)
=\sum_{N=0}^\infty u^N
{\tilde {\cal I}}^{U(N)\text{-}[1,1](H)}_{\text{JK}},
\end{align}
then we conjecture that
\begin{align}
\frac{\Xi_{\text{small }\mathfrak{t}}^{(l=m=1)}(u)}{\Xi_{\text{JK}}^{(l=m=1)}(u)}
=\prod_{n_1,n_2=0}^\infty\frac{1+b(y_1^{(2)})^{-1}x^{n_1-n_2}\mathfrak{t}^{n_1+n_2+1}u}{1+b(y_1^{(2)})^{-1}x^{n_1-n_2}\mathfrak{t}^{n_1+n_2-1}u}
=\prod_{n=-\infty}^\infty\frac{1}{1+b(y_1^{(2)})^{-1}x^n\mathfrak{t}^{|n|-1}u}.
\label{Xil1m1ratio}
\end{align}
We have checked this formula for all of the expansion coefficients of $a_1,b_1,c_1$ and $d_1$ in $N\le 4$ by comparing either the closed-form expressions of the coefficients or their small $\mathfrak{t}$ expansion to the order ${\cal O}(\mathfrak{t}^6)$.
We observe that the formula \eqref{Xil1m1ratio} also generalizes to $l=m\ge 2$ as
\begin{align}
\frac{\Xi_{\text{small }\mathfrak{t}}^{(l=m)}(u)}{\Xi_{\text{JK}}^{(l=m)}(u)}
&=\prod_{n_1,n_2=0}^\infty\frac{1-(\prod_{\alpha=1}^l(-\frac{b_\alpha}{y_\alpha^{(2)}}))x^{n_1-n_2}\mathfrak{t}^{n_1+n_2+2-l}u}{1-(\prod_{\alpha=1}^l(-\frac{b_\alpha}{y_\alpha^{(2)}}))x^{n_1-n_2}\mathfrak{t}^{n_1+n_2-l}u}\nonumber \\
&=\prod_{n=-\infty}^\infty\frac{1}{1-(\prod_{\alpha=1}^l(-\frac{b_\alpha}{y_\alpha^{(2)}}))x^n\mathfrak{t}^{|n|-l}u},\label{Xilratio}
\end{align}
with
\begin{align}
\Xi_{\text{small }\mathfrak{t}}^{(l=m)}(u)=\sum_{N=0}^\infty u^N{\tilde {\cal I}}^{U(N)\text{-}[l,l](H)},\quad
\Xi_{\text{JK}}^{(l=m)}(u)
=\sum_{N=0}^\infty u^N
{\tilde {\cal I}}^{U(N)\text{-}[l,l](H)}_{\text{JK}}.
\end{align}

\subsection{Refined topological vertex}
\label{sec_topvertex}

The (refined) topological vertex 
can also compute the Higgs limit of the correlation functions of Wilson lines in the antisymmetric representations in the 3d $U(N)$ ADHM theory. The refined topological vertex has already been used to calculate the Nekrasov partition functions of 5d $\mathcal{N}=1$ $U(l)$ gauge theories with flavors \cite{Iqbal:2003ix,Iqbal:2003zz,Eguchi:2003sj,Taki:2007dh,Awata:2008ed}. As mentioned in section \ref{sec_JK}, the $N$-instanton partition function is related to a generating function for the correlators of Wilson lines in the antisymmetric representations in the 3d $U(N)$ ADHM theory. Hence it is possible to use the refined topological vertex to calculate such Wilson line defect correlation functions. The advantage of using the refined topological vertex is that one can obtain the exact 5d Nekrasov partition functions by summing up all the instantons when $l=1$. 

Before starting the computation of the refined topological vertex, we first identify a slight difference between the instanton partition function and the Higgs limit of the Wilson line correlator. The $N$-instanton partition function of the 5d $\mathcal{N}=1$ $U(l)$ gauge theory with $N_f$ flavors and the Chern-Simons level $\kappa$ satisfying $2\left|\kappa\right| + N_f \leq 2$ is given by 
\begin{equation}\label{instanton}
\begin{split}
Z_N^{l, N_f, \kappa} =& \frac{1}{N!}\int\prod_{i=1}^N\frac{d\phi_i}{2\pi i}\frac{\left(\prod_{i<j}^N\prod_{\pm}2\sinh\left(\frac{\pm\left(\phi_i - \phi_j\right)}{2}\right)\right)\left(\prod_{i=1}^N\prod_{j=1}^N2\sinh\left(\frac{\phi_i - \phi_j + 2\epsilon_+}{2}\right)\right)}{\prod_{i=1}^N\prod_{j=1}^N2\sinh\left(\frac{\phi_i - \phi_j + \epsilon_1}{2}\right)2\sinh\left(\frac{\phi_i - \phi_j + \epsilon_2}{2}\right)}\cr
&\hspace{2.5cm} \times \frac{\prod_{i=1}^N\prod_{I=1}^{N_f}2\sinh\left(\frac{\phi_i + m_{I}}{2}\right)}{\prod_{i=1}^N\prod_{\alpha=1}^l\prod_{\pm}2\sinh\left(\frac{\pm \left(\phi_i - a_{\alpha}\right) + \epsilon_+}{2}\right)}\;\exp\left(\kappa \sum_{i=1}^N\phi_i\right),
\end{split}
\end{equation}
where $\epsilon_+ = \frac{\epsilon_1+\epsilon_2}{2}$ and $\epsilon_1, \epsilon_2$ are the $\Omega$-deformation parameters and $a_{\alpha}$'s are Coulomb branch moduli.
The $N$-instanton partition function \eqref{instanton} can be rewritten as
\begin{equation}\label{instanton2}
\begin{split}
Z_N^{l, N_f, \kappa}&=\bt^{lN}\left(-1\right)^{AN}\left(\prod_{I=1}^AM_I^{-\frac{1}{2}}\prod_{J=A+1}^{N_f}M_J^{\frac{1}{2}}\right)^{N}\cr
&\qquad\times\int\prod_{i=1}^N\frac{ds_i}{2\pi is_i}\mathcal{Z}_N\prod_{j=1}^Ns_j^{\kappa + \frac{N_f}{2} - A}\left(\prod_{I'=1}^A\left(1-s_jM_{I'}\right)\prod_{J'=A+1}^{N_f}\left(1-s_j^{-1}M_{J'}^{-1}\right)\right),
\end{split}
\end{equation}
where $A$ is an integer satisfying $0\leq A \leq N_f$ and
\begin{equation}
\mathcal{Z}_N = \frac{\left(1-\bt^2\right)^N}{N!\prod_{\pm}\left(1-x^{\pm 1}\bt\right)^N}\frac{\prod_{i<j}^N\prod_{\pm}\left(1-\left(s_is_j^{-1}\right)^{\pm 1}\right)\left(1-\bt^2\left(s_is_j^{-1}\right)^{\pm 1}\right)}{\prod_{i<j}^N\prod_{\pm, \pm'}\left(1-\bt\left(s_is_j^{-1}x^{\pm' 1}\right)^{\pm 1}\right)\prod_{i=1}^N\prod_{\alpha=1}^l\prod_{\pm}\left(1-\bt \left(s_iy_{\alpha}\right)^{\pm 1}\right)}.
\end{equation}
Here we have defined $\bt = e^{-\frac{\epsilon_1+\epsilon_2}{2}}, x=e^{-\frac{\epsilon_1-\epsilon_2}{2}}$ and $s_i = e^{\phi_i}, y_{\alpha}=e^{-a_{\alpha}}, M_I=e^{m_I}$ for $i=1, \cdots, N, \alpha=1, \cdots, l, I=1, \cdots, N_f$. When we set $\kappa = A-\frac{N_f}{2}$, \eqref{instanton2} becomes
\begin{equation}\label{instanton3}
\begin{split}
&Z_N^{l, N_f, \kappa = A-\frac{N_f}{2}}\bt^{-lN}\prod_{I=1}^A\left(-M_I^{\frac{1}{2}}\right)^N\prod_{J=A+1}^{N_f}M_J^{-\frac{N}{2}}\cr
&=\int\prod_{i=1}^N\frac{ds_i}{2\pi i s_i}\mathcal{Z}_N\prod_{j=1}^N\left(\prod_{I'=1}^A\left(1-s_jM_{I'}\right)\prod_{J'=A+1}^{N_f}\left(1-s_j^{-1}M_{J'}^{-1}\right)\right)\cr
&=\sum_{k_1, k_2, \cdots, k_{N_f}\geq 0}\left(-M_1\right)^{k_1}\cdots\left(-M_A\right)^{k_A}\left(-M_{A+1}\right)^{-k_{A+1}}\cdots\left(-M_{N_f}\right)^{-k_{N_f}}\cr
&\hspace{8cm}\times \Braket{\prod_{I=1}^AW_{(1^{k_I})}\prod_{J=A+1}^{N_f}W_{(\overline{1^{k_J}})}}^{U(N)\;\text{ADHM}-[1](H)}.
\end{split}
\end{equation}
Namely the leftmost side of \eqref{instanton3} gives rise to the generating function for the correlation functions of Wilson lines in the antisymmetric representations. 

Having identified the relation between the instanton partition function and the Higgs limit of correlators of Wilson lines in the antisymmetric representations, we turn to the computation of the instanton partition function \eqref{instanton} from the refined topological vertex. The relation between the topological string partition function and the Nekrasov partition function comes from the fact that 5d $\mathcal{N}=1$ supersymmetric field theories can be realized from M-theory on non-compact Calabi-Yau threefolds. When the Calabi-Yau threefolds are toric, one can obtain the Nekrasov partition functions of the 5d theories by applying the refined topological vertex to the toric diagrams of the Calabi-Yau threefolds. The formulae of the refined topological vertex are summarized in appendix \ref{app:top}.

\subsubsection{1-point function in $U(N)$ ADHM with $l=1$}
\label{sec_topvertex_1ptfcn}

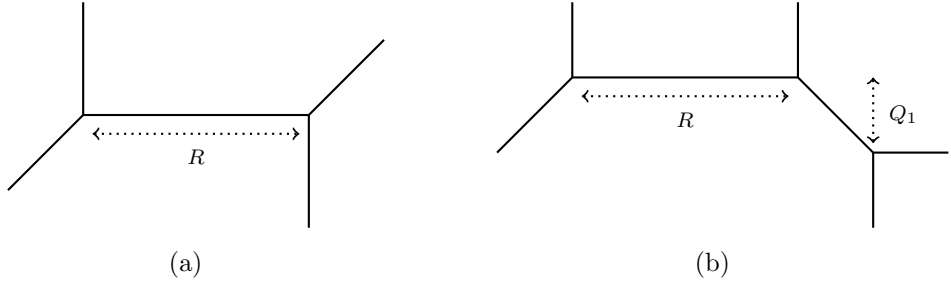
\begin{figure}
\centering
\subfigure[]{\label{fig:Nf0}
\begin{scriptsize}
\begin{tikzpicture}[scale=0.5]
\draw[thick] (-3,0)--(3,0) {};
\draw[thick] (-3,0)--(-3,3) {};
\draw[thick] (-3,0)--(-5,-2) {};
\draw[thick] (3,0)--(3,-3) {};
\draw[thick] (3,0)--(5,2) {};
\draw[arrows=<->, dotted, thick] (-3+1/4,-1/2)--(3-1/4,-1/2) {};
\node[label=below:{$R$}]  at (0,-1/2) {};
\end{tikzpicture}
\end{scriptsize}}\hspace{1cm}
\subfigure[]{\label{fig:Nf1}
\begin{scriptsize}
\begin{tikzpicture}[scale=0.5]
\draw[thick] (-3,0)--(3,0) {};
\draw[thick] (-3,0)--(-3,2) {};
\draw[thick] (-3,0)--(-5,-2) {};
\draw[thick] (3,0)--(3,2) {};
\draw[thick] (3,0)--(5,-2) {};
\draw[thick] (5,-2)--(5,-4) {};
\draw[thick] (5,-2)--(7,-2) {};
\draw[arrows=<->, dotted, thick] (-3+1/4,-1/2)--(3-1/4,-1/2) {};
\draw[arrows=<->, dotted, thick] (5,-2+1/4)--(5,0) {};
\node[label=below:{$R$}]  at (0,-1/2) {};
\node[label=right:{$Q_1$}]  at (5,-1) {};
\end{tikzpicture}
\end{scriptsize}}
\caption{(a) The toric diagram for the 5d $U(1)$ gauge theory with $N_f = 0, \kappa = 0$. (b) The toric diagram for the 5d $U(1)$ gauge theory with $N_f = 1, \kappa = \frac{1}{2}$.}
\label{fig:toric1}
\end{figure}

We first consider the cases with $N_f - A < l$ in order to avoid poles at $s_i=0$ for some $i$. Furthermore, we focus on $l=1$ since we can sum over all the instantons. More specifically, the cases we first consider are $\left(l, N_f, \kappa, A\right) = \left(1, 0, 0, 0, 0\right), \left(1, 1, \frac{1}{2}, 1\right)$.
For these choices, the 5d partition function $Z_N^{l=1, N_f,\kappa=A-\frac{N_f}{2}}$ \eqref{instanton3} give the generating function of the Higgs indices ${\cal I}^{U(N)\text{ ADHM-}[1](H)}$ and the generating function of the 1-point functions $\langle W_{(1^k)}\rangle^{U(N)\text{ ADHM-}[1](H)}$ respectively.
The toric diagram which realizes the 5d $\mathcal{N}=1$ $U(1)$ gauge theory with $N_f$ flavors and the Chern-Simons level $\kappa$ satisfying $\left(N_f, \kappa\right) = \left(0, 0\right), \left(1, \frac{1}{2}\right)$ are depicted in Figure \ref{fig:toric1}. 
When we choose the horizontal direction as the preferred direction and assign $q$ to the upper vertical external legs, the application of the refined topological vertex to the toric diagrams in Figure \ref{fig:toric1} yields
\begin{align}
Z^{l=1,N_f=0, \kappa=0} &=\sum_{N=0}^{\infty}\tu^NZ_N^{l=1,N_f=0, \kappa=0},\\
Z^{l=1,N_f=1, \kappa=\frac{1}{2}} &=Z_{\text{pert}}^{l=1,N_f=1, \kappa=\frac{1}{2}}\sum_{N=0}^{\infty}\left(-\tu\right)^NZ_N^{l=1,N_f=1, \kappa=\frac{1}{2}},
\end{align}
where
\begin{align}
Z_N^{l=1,N_f, \kappa}&=(-1)^N\sum_{\substack{\lambda^{(1)}\\\left|\lambda^{(1)}\right| = N}}\prod_{s\in\lambda^{(1)}}\frac{e^{\kappa\left(E_{10}(s) - \epsilon_+\right)}\prod_{I=1}^{N_f}2\sinh\left(\frac{E_{10}(s) - \epsilon_+ +m_I}{2}\right)}{2\sinh\left(\frac{E_{11}(s)}{2}\right)2\sinh\left(\frac{E_{11}(s) - 2\epsilon_+}{2}\right)},\label{instanton_top1}\\
Z_{\text{pert}}^{l=1,N_f=1, \kappa = \frac{1}{2}} &= \text{PE}\left[-\frac{y_1 M_1^{-1}t^{\frac{1}{2}}q^{\frac{1}{2}}}{(1-t)(1-q)}\right] \qquad \left(t = e^{\epsilon_1}, q = e^{-\epsilon_2}\right),
\end{align}
and
\be\label{Es2}
E_{\alpha\beta}(s) = a_{\alpha} - a_{\beta} - \epsilon_1\text{arm}_{\lambda^{(a)}}(s) + \epsilon_2\left(\text{leg}_{\lambda^{(\beta)}}(s) + 1\right),
\ee
with $\lambda^{(0)} = \emptyset, a_0 = 0$.\footnote{In the case of $\left(N_f, \kappa\right) = \left(0, 0\right)$, the different assignment of $q$ and $t$ gives the same result. In the case of $\left(N_f, \kappa\right) = \left(1, \frac{1}{2}\right)$, it turns out that the different assignment of $q$ and $t$ gives rise to the same instanton partition function $Z_N^{N=1, \kappa=\frac{1}{2}}$ by a rescaling of $y_1$.} 
The notation of \eqref{Es2} is slightly different from that of \eqref{Es} and the subscripts of $E_{\alpha\beta}(s)$ are labels for the Coulomb branch moduli as well as the Young diagrams. $\text{PE}\left[f(x_1, \cdots, x_k)\right]$ represents the plethystic expontential, 
\be
\text{PE}\left[\pm f(x_1, \cdots, x_k)\right] = \exp\left(\pm\sum_{n=1}^{\infty}\frac{f(x_1^n, \cdots, x_k^n)}{n}\right).
\ee
In the calculations of the refined topological vertex, we defined the K\"ahler parameters as
\be
R = \tu M_1^{\frac{1}{2}}y_1^{-1}, \quad Q_1 = y_1M_1^{-1}
\ee
for $\left(N_f, \kappa\right) = \left(1, \frac{1}{2}\right)$ and $R=\tu$ for $\left(N_f, \kappa\right)=(0, 0)$. The partition function \eqref{instanton_top1} is expected to reproduce \eqref{instanton} and we have checked the equality until the order $\bt^8$ from the 1-instanton to 5-instantons. We will turn off $y_1$ hereafter since it can be absorbed by the rescaling of $M_1$ and $\tu$. 

We can also apply the refined topological vertex to the toric diagrams in Figure \ref{fig:toric1} with the preferred direction changed. When we choose the vertical direction as the preferred direction, we can explicitly sum over the Young diagrams and obtain
\begin{align}
Z^{l=1,N_f=0, \kappa=0} &=\text{PE}\left[-\frac{\tu t^{\frac{1}{2}}q^{\frac{1}{2}}}{(1-t)(1-q)}\right],\label{PE_Nf0}\\
Z^{l=1,N_f=1, \kappa=\frac{1}{2}} &=\text{PE}\left[\frac{\tu M_1^{\frac{1}{2}}q}{(1-t)(1-q)} - \frac{\tu M_1^{-\frac{1}{2}}t^{\frac{1}{2}}q^{\frac{1}{2}}}{(1-t)(1-q)} - \frac{M_1^{-1}t^{\frac{1}{2}}q^{\frac{1}{2}}}{(1-t)(1-q)}\right].\label{PE_Nf1}
\end{align}
It has been conjectured that the partition functions computed by the refined topological vertex are independent of the choice of the preferred direction \cite{Iqbal:2007ii,Iqbal:2008ra,Awata:2009yc} and \eqref{instanton_top1} are expected to agree with \eqref{PE_Nf0} and \eqref{PE_Nf1}. Since \eqref{instanton_top1} is equal to the $N$-instanton partition function \eqref{instanton}, the relation \eqref{PE_Nf0} and \eqref{PE_Nf1} imply
\begin{equation}\label{PE_Nf0_1}
\text{PE}\left[\frac{\tu\bt}{\left(1-x^{-1}\bt\right)\left(1-x\bt\right)}\right] = 1 + \sum_{N=1}^{\infty}\tu^N\bt^N\mathcal{I}^{U(N)\text{ ADHM-}[1](H)},
\end{equation}
and 
\begin{equation}\label{PE_Nf1_1}
\begin{split}
&\text{PE}\left[-\frac{\tu\bt^2M_1^{\frac{1}{2}}}{\left(1-x^{-1}\bt\right)\left(1-x\bt\right)} + \frac{\tu M_1^{-\frac{1}{2}}\bt}{\left(1-x^{-1}\bt\right)\left(1-x\bt\right)}\right]\cr
&\hspace{4cm} = 1 + \sum_{N=1}^{\infty}\left(-\tu\right)^N\bt^N\left(-M_1^{\frac{1}{2}}\right)^{-N}\left(\sum_{k=0}^N\left(-M_1\right)^k\Braket{W_{(1^k)}}^{U(N)\text{ ADHM-}[1](H)}\right).
\end{split}
\end{equation}
The relation \eqref{PE_Nf1_1} further implies
\begin{equation}\label{PE_Nf1_3}
\text{PE}\left[-\frac{u\bt M_1}{\left(1-x^{-1}\bt\right)\left(1-x\bt\right)}\right]=1 + \sum_{N=1}^{\infty}u^N\left(-M_1\right)^N\Braket{W_{(1^N)}}^{U(N)\text{ ADHM-}[1](H)},
\end{equation}
where $u=\tu \mathfrak{t}M_1^{-\frac{1}{2}}$.
Then substituting \eqref{PE_Nf1_3} and \eqref{PE_Nf0_1} into \eqref{PE_Nf1_1} gives
\begin{equation}\label{PE_Nf1_4}
\begin{split}
&\left(1 + \sum_{N'=1}^{\infty}u^{N'}\left(-M_1\right)^{N'}\Braket{W_{(1^{N'})}}^{U(N')\text{ ADHM-}[1](H)}\right)\left( 1 + \sum_{N''=1}^{\infty}u^{N''}\mathcal{I}^{U(N'')\;\text{ADHM}-[1](H)}\right)\cr
 &\hspace{6cm}=1 + \sum_{N=1}^{\infty}u^N\left(\sum_{k=0}^N\left(-M_1\right)^k\Braket{W_{(1^k)}}^{U(N)\text{ ADHM-}[1](H)}\right).
\end{split}
\end{equation}
The order $u^NM_1^k$ of \eqref{PE_Nf1_4} yields
\begin{equation}\label{1point_relation}
\Braket{W_{(1^{k})}}^{U(k)\text{ ADHM-}[1](H)}\mathcal{I}^{U(N-k)\text{ ADHM-}[1](H)} = \Braket{W_{(1^k)}}^{U(N)\text{ ADHM-}[1](H)}
\end{equation}
which reproduces \eqref{uNADHM_1ptasym_relation}.

Note that the relation \eqref{PE_Nf1_3} can also be written as
\begin{align}
\Xi(-\mathfrak{t}u;x;\mathfrak{t})\Bigl(1+\sum_{N=1}^\infty u^N\langle W_{(1^N)}\rangle^{U(N)\text{ ADHM-}[1](H)}(x;\mathfrak{t})\Bigr)=1,
\label{Higgsl1_1ptfcngeneratingfcn}
\end{align}
where $\Xi(u;x;\mathfrak{t})=\sum_{N=0}^\infty {\cal I}^{U(N)\text{ ADHM-}[1](H)}(x;\mathfrak{t})u^N$ is the grand canonical index \eqref{Xiuxt}.
By expanding this relation in $u$ we find that the 1-point functions $\langle W_{(1^k)}\rangle^{U(N)\text{ ADHM-}[1](H)}$ with $k=N$ can be calculated recursively in $N$ as
\begin{align}
&\langle W_{(1^N)}\rangle^{U(N)\text{ ADHM-}[1](H)}(x;\mathfrak{t})\nonumber \\
&=-\sum_{n=1}^N (-\mathfrak{t})^n{\cal I}^{U(n)\text{ ADHM-}[1](H)}(x;\mathfrak{t})
\langle W_{(1^{N-n})}\rangle^{U(N-n)\text{ ADHM-}[1](H)}(x;\mathfrak{t})+\delta_{N,0},
\end{align}
where $\langle W_{(1^0)}\rangle^{U(0)\text{ADHM-}[1](H)}(x;\mathfrak{t})=1$.
This relation together with the relation \eqref{uNADHM_1ptasym_relation} allows us to express the 1-point functions $\langle W_{(1^k)}\rangle^{U(N)\text{ ADHM-}[1](H)}$ for all $N$ and $k$ as polynomials in the Higgs indices ${\cal I}^{U(n)\text{ ADHM-}[1](H)}$ whose coefficients are polynomial in $\mathfrak{t}$.

\subsubsection{2-point function in $U(N)$ ADHM with $l=1$}

So far, we have focused on the cases where the integrand \eqref{instanton} does not have poles at $s_i = 0$ since the refined topological vertex computations reproduce the results obtained by using the JK residues. In section \ref{sec_JK}, the ratio between the correct generating function for $\tilde{\mathcal{I}}^{U(N)\text{-}[1,1](H)}$ and the one evaluated from the JK residues has been conjectured in \eqref{Xil1m1ratio} and the relation implies
\be\label{ratioNf2}\begin{split}
&\sum_{N=0}^{\infty}u^N\sum_{k_1, k_2 \geq 0}\left(-M_1\right)^{k_1}\left(-M_2\right)^{-k_2}\Braket{W_{(1^{k_1})}W_{(\overline{1^{k_2}})}}^{U(N)\text{ ADHM-}[1](H)}\cr
=&\prod_{n_1, n_2=0}^{\infty}\frac{1-M_2^{-1}x^{n_1-n_2}\bt^{n_1+n_2+1}u}{1-M_2^{-1}x^{n_1-n_2}\bt^{n_1+n_2-1}u}\sum_{N=0}^{\infty}u^N\sum_{k_1, k_2 \geq 0}\left(-M_1\right)^{k_1}\left(-M_2\right)^{-k_2}\Braket{W_{(1^{k_1})}W_{(\overline{1^{k_2}})}}^{U(N)\text{ ADHM-}[1](H)}_{\text{JK}}\cr
=&\text{PE}\left[-\frac{\bt M_2^{-1}u}{\left(1-x\bt\right)\left(1-x^{-1}\bt\right)} + \frac{\bt^{-1} M_2^{-1}u}{\left(1-x\bt\right)\left(1-x^{-1}\bt\right)}\right]\cr
&\hspace{4cm}\times\sum_{N=0}^{\infty}u^N\sum_{k_1, k_2 \geq 0}\left(-M_1\right)^{k_1}\left(-M_2\right)^{-k_2}\Braket{W_{(1^{k_1})}W_{(\overline{1^{k_2}})}}^{U(N)\text{ ADHM-}[1](H)}_{\text{JK}}.
\end{split}\ee
$\Braket{W_{(1^{k_1})}W_{(\overline{1^{k_2}})}}^{U(N)\text{ ADHM-}[1](H)}_{\text{JK}}$ is the 2-point function whose integral is evaluated by the JK residues.

\begin{figure}[t]
\centering
\begin{scriptsize}
\begin{tikzpicture}[scale=0.5]
\draw[thick] (-3,0)--(3,0) {};
\draw[thick] (-3,0)--(-3,2) {};
\draw[thick] (-3,0)--(-5,-2) {};
\draw[thick] (-5,-2)--(-7,-2) {};
\draw[thick] (-5,-2)--(-5,-4) {};
\draw[arrows=<->, dotted, thick] (-5,-2+1/4)--(-5,0) {};
\draw[thick] (3,0)--(3,2) {};
\draw[thick] (3,0)--(5,-2) {};
\draw[thick] (5,-2)--(5,-4) {};
\draw[thick] (5,-2)--(7,-2) {};
\draw[arrows=<->, dotted, thick] (-3+1/4,-1/2)--(3-1/4,-1/2) {};
\draw[arrows=<->, dotted, thick] (5,-2+1/4)--(5,0) {};
\node[label=below:{$R$}]  at (0,-1/2) {};
\node[label=right:{$P_1$}]  at (5,-1) {};
\node[label=left:{$Q_1$}]  at (-5,-1) {};
\end{tikzpicture}
\end{scriptsize}
\caption{The toric diagram for the 5d $U(1)$ gauge theory with $N_f = 2, \kappa = 0$.}
\label{fig:Nf2}
\end{figure}
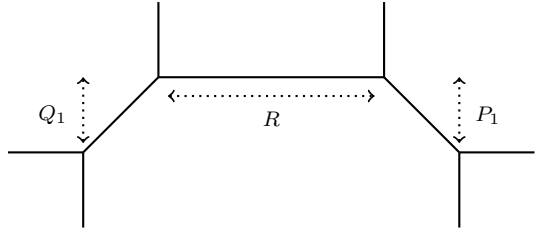

From \eqref{instanton3}, the 2-point function $\Braket{W_{(1^{k_1})}W_{(\overline{1^{k_2}})}}^{U(N)\text{ ADHM-}[1](H)}_{\text{JK}}$ can be computed by applying the refined topological vertex to a toric diagram which realizes the 5d $\mathcal{N}=1$ $U(1)$ gauge theory with $2$ flavors and $\kappa = 0$. Such a toric diagram is depicted in Figure \ref{fig:Nf2}. 
The application of the refined topological vertex to the toric diagram in Figure \ref{fig:Nf2} yields
\be\label{instanton4}\begin{split}
&\sum_{N=0}^{\infty}\tu^N\sum_{\substack{\lambda^{(1)}\\\left|\lambda^{(1)}\right| = N}}\prod_{s\in\lambda^{(1)}}\frac{\prod_{I=1}^{2}2\sinh\left(\frac{E_{10}(s) - \epsilon_+ +m_I}{2}\right)}{2\sinh\left(\frac{E_{11}(s)}{2}\right)2\sinh\left(\frac{E_{11}(s) - 2\epsilon_+}{2}\right)}\cr
= &\text{PE}\left[\frac{\tu y_1M_1^{-\frac{1}{2}}M_2^{-\frac{1}{2}}t}{(1-t)(1-q)} + \frac{\tu y_1^{-1}M_1^{\frac{1}{2}}M_2^{\frac{1}{2}}q}{(1-t)(1-q)} - \frac{\tu M_1^{\frac{1}{2}}M_2^{-\frac{1}{2}}t^{\frac{1}{2}}q^{\frac{1}{2}}}{(1-t)(1-q)} -\frac{\tu M_1^{-\frac{1}{2}}M_2^{\frac{1}{2}}t^{\frac{1}{2}}q^{\frac{1}{2}}}{(1-t)(1-q)}\right],
\end{split}\ee
where we assigned 
\be
R = \tu y_1^{-1}M_1^{\frac{1}{2}}M_2^{\frac{1}{2}}, \quad Q_1 = y_1M_1^{-1}, \quad P_1 = y_1M_2^{-1}.
\ee
The left-hand/right-hand side of \eqref{instanton4} is obtained by choosing the preferred direction in the horizontal/vertical direction respectively in Figure \ref{fig:Nf2}. To obtain the left-hand side of \eqref{instanton4}, we assigned $t$ to the lower parallel external legs.\footnote{Again, the different assignment of $q$ and $t$ gives the same result by a rescaling of $y_1$.} We will set $y_1 = 1$ since it can be again absorbed by rescaling $\tu, M_1$ and $M_2$. Then the relation \eqref{instanton3} gives
\be\label{2ptPE1}\begin{split}
&\sum_{N=0}^{\infty}\left(-\tu\right)^N\bt^{N}\left(-M_1^{-\frac{1}{2}}\right)^NM_2^{\frac{N}{2}}\sum_{k_1, k_2 \geq 0}\left(-M_1\right)^{k_1}\left(-M_2\right)^{-k_2}\Braket{W_{(1^{k_1})}W_{(\overline{1^{k_2}})}}^{U(N)\;\text{ADHM}-[1](H)}_{\text{JK}} \cr
= &\text{PE}\left[-\frac{\tu M_1^{-\frac{1}{2}}M_2^{-\frac{1}{2}}}{\left(1-x\bt\right)\left(1-x^{-1}\bt\right)} - \frac{\tu\bt^2M_1^{\frac{1}{2}}M_2^{\frac{1}{2}}}{\left(1-x\bt\right)\left(1-x^{-1}\bt\right)} + \frac{\tu\bt M_1^{\frac{1}{2}}M_2^{-\frac{1}{2}}}{\left(1-x\bt\right)\left(1-x^{-1}\bt\right)} + \frac{\tu\bt M_1^{-\frac{1}{2}}M_2^{\frac{1}{2}}}{\left(1-x\bt\right)\left(1-x^{-1}\bt\right)}\right].
\end{split}\ee
When we define $u= \tu\bt M_1^{-\frac{1}{2}}M_2^{\frac{1}{2}}$, \eqref{2ptPE1} further becomes
\be\begin{split}
&\sum_{N=0}^{\infty}u^N\sum_{k_1, k_2 \geq 0}\left(-M_1\right)^{k_1}\left(-M_2\right)^{-k_2}\Braket{W_{(1^{k_1})}W_{(\overline{1^{k_2}})}}^{U(N)\text{ ADHM-}[1](H)}_{\text{JK}} \cr
= &\text{PE}\left[-\frac{u\bt^{-1}M_2^{-1}}{\left(1-x\bt\right)\left(1-x^{-1}\bt\right)} - \frac{u\bt M_1}{\left(1-x\bt\right)\left(1-x^{-1}\bt\right)} + \frac{u M_1M_2^{-1}}{\left(1-x\bt\right)\left(1-x^{-1}\bt\right)} + \frac{u}{\left(1-x\bt\right)\left(1-x^{-1}\bt\right)}\right].
\end{split}\ee
Then combining \eqref{2ptPE1} with \eqref{ratioNf2} implies
\be\label{2ptPE2}\begin{split}
&\sum_{N=0}^{\infty}u^N\sum_{k_1, k_2 \geq 0}\left(-M_1\right)^{k_1}\left(-M_2\right)^{-k_2}\Braket{W_{(1^{k_1})}W_{(\overline{1^{k_2}})}}^{U(N)\text{ ADHM-}[1](H)}\cr
=&\text{PE}\left[- \frac{u\bt M_1}{\left(1-x\bt\right)\left(1-x^{-1}\bt\right)} - \frac{u\bt M_2^{-1}}{\left(1-x\bt\right)\left(1-x^{-1}\bt\right)} + \frac{uM_1M_2^{-1}}{\left(1-x\bt\right)\left(1-x^{-1}\bt\right)} + \frac{u}{\left(1-x\bt\right)\left(1-x^{-1}\bt\right)}\right].
\end{split}\ee

In terms of the grand canonical index $\Xi(u;x;\mathfrak{t})$, the relation \eqref{2ptPE2} is rephrased as
\begin{align}
&\Xi(u;x;\mathfrak{t})\Xi(M_1M_2^{-1}u;x;\mathfrak{t})\nonumber \\
&=\Xi(\mathfrak{t}M_1u;x;\mathfrak{t})\Xi(\mathfrak{t}M_2^{-1}u;x;\mathfrak{t})\nonumber \\
&\quad\times \Bigl(
\sum_{N=0}^\infty u^N\sum_{k_1,k_2\ge 0}(-M_1)^{k_1}(-M_2)^{-k_2}\langle
W_{(1^{k_1})}
W_{(\overline{1^{k_2}})}\rangle^{U(N)\text{ ADHM-}[1](H)}(x;\mathfrak{t})
\Bigr).
\label{Higgsl1_2ptfcngeneratingfcn}
\end{align}
Expanding in $u,-M_1,(-M_2)^{-1}$, we obtain the following algebraic relation among the 2-point functions.
\begin{align}
&\sum_{\substack{N_1,N_2\\
0\le N_1\le \text{min}(k_1,N-k_2)\\
0\le N_2\le \text{min}(N-k_1,k_2)\\
N_1+N_2\le N
}}
(-\mathfrak{t})^{N_1+N_2}
\langle
W_{(1^{k_1-N_1})}
W_{(\overline{1^{k_2-N_2}})}
\rangle^{U(N-N_1-N_2)\text{ ADHM-}[1](H)}(x;\mathfrak{t})\nonumber \\
&\quad\quad\quad\quad\times {\cal I}^{U(N_1)\text{ ADHM-}[1](H)}(x;\mathfrak{t})
{\cal I}^{U(N_2)\text{ ADHM-}[1](H)}(x;\mathfrak{t})\nonumber \\
&=\delta_{k_1,k_2}
{\cal I}^{U(k_1)\text{ ADHM-}[1](H)}(x;\mathfrak{t})
{\cal I}^{U(N-k_1)\text{ ADHM-}[1](H)}(x;\mathfrak{t}).
\end{align}
These relations also allow us to determine $\langle W_{(1^{k_1})}W_{(\overline{1^{k_2}})}\rangle^{U(N)\text{ ADHM-}[1](H)}$ recursively in $N,k_1,k_2$ and express in terms of ${\cal I}^{U(N)\text{ ADHM-}[1](H)}$.
For example, we find
\begin{align}
&\langle W_{\ydiagram{1}}W_{\overline{\ydiagram{1}}}\rangle^{U(2)\text{ ADHM-}[1](H)}(x;\mathfrak{t})=(1+\mathfrak{t}^2){\cal I}^{U(1)\text{ ADHM-}[1](H)}(x;\mathfrak{t})^2,\\
&\langle W_{\ydiagram{1}}W_{\overline{\ydiagram{1}}}\rangle^{U(3)\text{ ADHM-}[1](H)}(x;\mathfrak{t})\nonumber \\
&=
{\cal I}^{U(1)\text{ ADHM-}[1](H)}(x;\mathfrak{t})
{\cal I}^{U(2)\text{ ADHM-}[1](H)}(x;\mathfrak{t})
+
\mathfrak{t}^2{\cal I}^{U(1)\text{ ADHM-}[1](H)}(x;\mathfrak{t})^3,\\
&\langle W_{\ydiagram{1,1}}W_{\overline{\ydiagram{1}}}\rangle^{U(3)\text{ ADHM-}[1](H)}(x;\mathfrak{t})\nonumber \\
&=
-\mathfrak{t}^3{\cal I}^{U(1)\text{ ADHM-}[1](H)}(x;\mathfrak{t})
{\cal I}^{U(2)\text{ ADHM-}[1](H)}(x;\mathfrak{t})
+
(\mathfrak{t}+\mathfrak{t}^3){\cal I}^{U(1)\text{ ADHM-}[1](H)}(x;\mathfrak{t})^3,\\
&\langle W_{\ydiagram{1}}W_{\overline{\ydiagram{1}}}\rangle^{U(4)\text{ ADHM-}[1](H)}(x;\mathfrak{t})\nonumber \\
&=
{\cal I}^{U(1)\text{ ADHM-}[1](H)}(x;\mathfrak{t})
{\cal I}^{U(3)\text{ ADHM-}[1](H)}(x;\mathfrak{t})\nonumber \\
&\quad +
\mathfrak{t}^2{\cal I}^{U(1)\text{ ADHM-}[1](H)}(x;\mathfrak{t})^2
{\cal I}^{U(2)\text{ ADHM-}[1](H)}(x;\mathfrak{t}),\\
&\langle W_{\ydiagram{1,1}}W_{\overline{\ydiagram{1}}}\rangle^{U(4)\text{ ADHM-}[1](H)}(x;\mathfrak{t})\nonumber \\
&=
(\mathfrak{t}-\mathfrak{t}^3)
{\cal I}^{U(1)\text{ ADHM-}[1](H)}(x;\mathfrak{t})^2
{\cal I}^{U(2)\text{ ADHM-}[1](H)}(x;\mathfrak{t})
+
\mathfrak{t}^3
{\cal I}^{U(1)\text{ ADHM-}[1](H)}(x;\mathfrak{t})^4,\\
&\langle W_{\ydiagram{1,1,1}}W_{\overline{\ydiagram{1}}}\rangle^{U(4)\text{ ADHM-}[1](H)}(x;\mathfrak{t})\nonumber \\
&=
\mathfrak{t}^4
{\cal I}^{U(1)\text{ ADHM-}[1](H)}(x;\mathfrak{t})
{\cal I}^{U(3)\text{ ADHM-}[1](H)}(x;\mathfrak{t})\nonumber \\
&\quad +
(-\mathfrak{t}^2-2\mathfrak{t}^4)
{\cal I}^{U(1)\text{ ADHM-}[1](H)}(x;\mathfrak{t})^2
{\cal I}^{U(2)\text{ ADHM-}[1](H)}(x;\mathfrak{t})\nonumber \\
&\quad +
(\mathfrak{t}^2+\mathfrak{t}^4)
{\cal I}^{U(1)\text{ ADHM-}[1](H)}(x;\mathfrak{t})^4,\\
&\langle W_{\ydiagram{1,1}}W_{\overline{\ydiagram{1,1}}}\rangle^{U(4)\text{ ADHM-}[1](H)}(x;\mathfrak{t})\nonumber \\
&=
(1+\mathfrak{t}^4)
{\cal I}^{U(2)\text{ ADHM-}[1](H)}(x;\mathfrak{t})^2
-2\mathfrak{t}^4
{\cal I}^{U(1)\text{ ADHM-}[1](H)}(x;\mathfrak{t})^2
{\cal I}^{U(2)\text{ ADHM-}[1](H)}(x;\mathfrak{t})\nonumber \\
&\quad +
(\mathfrak{t}^2+\mathfrak{t}^4)
{\cal I}^{U(1)\text{ ADHM-}[1](H)}(x;\mathfrak{t})^4.
\end{align}

It is also interesting to see a relation between the factor \eqref{Xil1m1ratio} and the toric diagram in Figure \ref{fig:Nf2}. Note that the factor \eqref{Xil1m1ratio} can be expressed as
\be\label{extral1}
\text{PE}\left[-\frac{\bt M_2^{-1}u}{\left(1-x\bt\right)\left(1-x^{-1}\bt\right)} + \frac{\bt^{-1} M_2^{-1}u}{\left(1-x\bt\right)\left(1-x^{-1}\bt\right)}\right] = \frac{\prod_{i, j=1}^{\infty}\left(1 - RQ_1P_1q^{i-1}t^j\right)}{\prod_{i, j=1}^{\infty}\left(1 - RQ_1P_1q^i t^{j-1}\right)}.
\ee
$RQ_1P_1$ is associated with the length between the lower external legs of the toric diagram. The contributions of M2-branes wrapped on a two-cycle between parallel external legs are called the extra factor, which one removes to obtain the partition function of the UV complete 5d theory realized on the toric diagram \cite{Bergman:2013ala,Bao:2013pwa,Hayashi:2013qwa,Bergman:2013aca}. In the case of the diagram in Figure \ref{fig:Nf2}, the contribution associated with the parallel external legs extending in the lower direction, to which we assigned $t$ in the refined topological vertex computation, is given by
\be\label{extra}
Z^{\|}_{\text{lower}} = \prod_{i, j=1}^{\infty}\left(1 - RQ_1P_1q^{i-1}t^j\right)^{-1}.
\ee
Hence the numerator of \eqref{extral1} cancels the extra factor given by \eqref{extra}. Namely, we may interpret the effect of the factor \eqref{extral1} as canceling an extra factor associated with parallel external legs 
whose contribution is given by the form $\prod_{i, j=1}^{\infty}\left(1- Qq^{i-1}t^j\right)^{-1}$ and then inserting another factor which is the inverse of the removed extra factor 
with $Q$ replaced with $Qqt^{-1}$. 

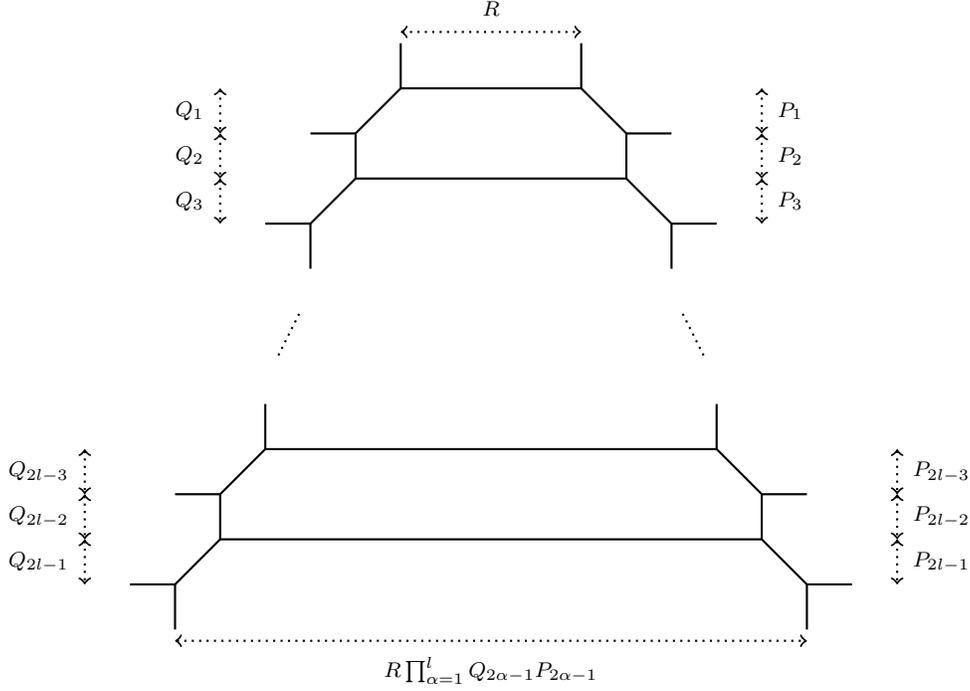
\begin{figure}[t]
\centering
\begin{scriptsize}
\begin{tikzpicture}[scale=0.6]
\draw[thick] (-2,5)--(-2,6);
\draw[thick] (2,5)--(2,6);
\draw[thick] (-2,5)--(2,5);
\draw[thick] (-2,5)--(-2-1,5-1);
\draw[thick] (2,5)--(2+1,5-1);
\draw[thick] (-3,4)--(-3,3);
\draw[thick] (-3,4)--(-4,4);
\draw[thick] (3,4)--(3,3);
\draw[thick] (3,4)--(4,4);
\draw[thick] (-3,3)--(3,3);
\draw[thick] (-3,3)--(-3-1,3-1);
\draw[thick] (3,3)--(3+1,3-1);
\draw[thick] (-4,2)--(-4,1);
\draw[thick] (4,2)--(4,1);
\draw[thick] (4,2)--(5,2);
\draw[thick] (-4,2)--(-5,2);
\draw[thick, dotted] (-4-1/4,0)--(-5+1/4,-1);
\draw[thick, dotted] (4+1/4,0)--(5-1/4,-1);
\draw[thick] (-5,-2)--(-5,-3);
\draw[thick] (5,-2)--(5,-3);
\draw[thick] (-5,-3)--(5,-3);
\draw[thick] (-5,-3)--(-6,-4);
\draw[thick] (5,-3)--(6,-4);
\draw[thick] (6,-4)--(6,-5);
\draw[thick] (-6,-4)--(-6,-5);
\draw[thick] (6,-4)--(7,-4);
\draw[thick] (-6,-4)--(-7,-4);
\draw[thick] (-6,-5)--(6,-5);
\draw[thick] (-6,-5)--(-7,-6);
\draw[thick] (6,-5)--(7,-6);
\draw[thick] (7,-6)--(7,-7);
\draw[thick] (-7,-6)--(-7,-7);
\draw[thick] (7,-6)--(8,-6);
\draw[thick] (-7,-6)--(-8,-6);
\node[label=above:{$R$}]  at (0,6+1/4) {};
\draw[arrows=<->, thick, dotted] (-2,6+1/4)--(2,6+1/4);
\draw[arrows=<->, thick, dotted] (-6,5)--(-6,4);
\draw[arrows=<->, thick, dotted] (-6,4)--(-6,3);
\draw[arrows=<->, thick, dotted] (-6,3)--(-6,2);
\node[label=left:{$Q_1$}]  at (-6,4+1/2) {};
\node[label=left:{$Q_2$}]  at (-6,3+1/2) {};
\node[label=left:{$Q_3$}]  at (-6,2+1/2) {};
\draw[arrows=<->, thick, dotted] (-9,-3)--(-9,-4);
\draw[arrows=<->, thick, dotted] (-9,-4)--(-9,-5);
\draw[arrows=<->, thick, dotted] (-9,-5)--(-9,-6);
\node[label=left:{$Q_{2l-3}$}]  at (-9,-4+1/2) {};
\node[label=left:{$Q_{2l-2}$}]  at (-9,-5+1/2) {};
\node[label=left:{$Q_{2l-1}$}]  at (-9,-6+1/2) {};
\draw[arrows=<->, thick, dotted] (6,5)--(6,4);
\draw[arrows=<->, thick, dotted] (6,4)--(6,3);
\draw[arrows=<->, thick, dotted] (6,3)--(6,2);
\node[label=right:{$P_1$}]  at (6,4+1/2) {};
\node[label=right:{$P_2$}]  at (6,3+1/2) {};
\node[label=right:{$P_3$}]  at (6,2+1/2) {};
\draw[arrows=<->, thick, dotted] (9,-3)--(9,-4);
\draw[arrows=<->, thick, dotted] (9,-4)--(9,-5);
\draw[arrows=<->, thick, dotted] (9,-5)--(9,-6);
\node[label=right:{$P_{2l-3}$}]  at (9,-4+1/2) {};
\node[label=right:{$P_{2l-2}$}]  at (9,-5+1/2) {};
\node[label=right:{$P_{2l-1}$}]  at (9,-6+1/2) {};
\draw[arrows=<->, thick, dotted] (-7,-7-1/4)--(7,-7-1/4);
\node[label=below:{$R\prod_{\alpha=1}^lQ_{2\alpha-1}P_{2\alpha-1}$}]  at (0,-7-1/4) {};
\end{tikzpicture}
\end{scriptsize}
\caption{The toric diagram for the 5d $U(l)$ gauge theory with $N_f = 2l, \kappa = 0$.}
\label{fig:2Nf}
\end{figure}
Based on this observation, we can infer the ratio between $\Xi_{\text{small $\bt$}}^{(l=m)}(u)$ and $\Xi_{\text{JK}}^{(l=m)}(u)$ for general $l$ from the associated toric diagram. For that we consider the case $\left(l, N_f, \kappa, A\right) = \left(l, 2l, 0, l\right)$. The toric diagram which realizes the 5d theory is depicted in Figure \ref{fig:2Nf}. We introduce the gauge theory parameters as 
\begin{align}
Q_{2\alpha-1} &= y_{\alpha}M_{\alpha}^{-1}, \quad Q_{2\beta} = y_{\beta+1}^{-1}M_{\beta}, \quad P_{2\alpha-1} = y_{\alpha}M_{\alpha+l}^{-1}, \quad P_{2\beta} = y_{\beta+1}^{-1}M_{\beta+l}\\
R &=\tu\prod_{\alpha=1}^lQ_{2\alpha-1}^{-\frac{1}{2}}P_{2\alpha-1}^{-\frac{1}{2}} = \tu\prod_{\alpha=1}^l y_{\alpha}^{-1}M_{\alpha}^{\frac{1}{2}}M_{\alpha+l}^{\frac{1}{2}},
\end{align}
for $\alpha=1, 2, \cdots, l$ and $\beta= 1, 2, \cdots, l-1$, and assign $t$ to the lower parallel external legs. Then the factor corresponding to the ratio between $\Xi_{\text{small $\bt$}}^{(l=m)}(u)$ and $\Xi_{\text{JK}}^{(l=m)}(u)$ may be inferred as
\be\label{extral}
\begin{split}
&\frac{\prod_{i, j=1}^{\infty}\left(1 - R\left(\prod_{\alpha=1}^lQ_{2\alpha-1}P_{2\alpha-1}\right)q^{i-1}t^j\right)}{\prod_{i, j=1}^{\infty}\left(1 - R\left(\prod_{\alpha=1}^lQ_{2\alpha-1}P_{2\alpha-1}\right)q^i t^{j-1}\right)} \cr
&\hspace{3cm}=\frac{\prod_{i, j=1}^{\infty}\left(1 - \tu\left(\prod_{\alpha=1}^l y_{\alpha}M_{\alpha}^{-\frac{1}{2}}M_{\alpha+l}^{-\frac{1}{2}}\right)q^{i-1}t^j\right)}{\prod_{i, j=1}^{\infty}\left(1 - \tu\left(\prod_{\alpha=1}^l y_{\alpha}M_{\alpha}^{-\frac{1}{2}}M_{\alpha+l}^{-\frac{1}{2}}\right)q^i t^{j-1}\right)}\cr
&\hspace{3cm}=\text{PE}\left[\frac{\tu\prod_{\alpha=1}^ly_{\alpha}M_{\alpha}^{-\frac{1}{2}}M_{\alpha+l}^{-\frac{1}{2}}}{\left(1-x\bt\right)\left(1-x^{-1}\bt\right)} -\frac{\tu\bt^2\prod_{\alpha=1}^ly_{\alpha}M_{\alpha}^{-\frac{1}{2}}M_{\alpha+l}^{-\frac{1}{2}}}{\left(1-x\bt\right)\left(1-x^{-1}\bt\right)}  \right].
\end{split}
\ee
We can rewrite the expression as
\be\label{PE.general}\begin{split}
&\text{PE}\left[\frac{\tu\prod_{\alpha=1}^ly_{\alpha}M_{\alpha}^{-\frac{1}{2}}M_{\alpha+l}^{-\frac{1}{2}}}{\left(1-x\bt\right)\left(1-x^{-1}\bt\right)} -\frac{\tu\bt^2\prod_{\alpha=1}^ly_{\alpha}M_{\alpha}^{-\frac{1}{2}}M_{\alpha+l}^{-\frac{1}{2}}}{\left(1-x\bt\right)\left(1-x^{-1}\bt\right)}  \right] \cr
&\hspace{6cm}= \text{PE}\left[\frac{u\bt^{-l}\prod_{\alpha=1}^ly_{\alpha}M_{\alpha+l}^{-1}}{\left(1-x\bt\right)\left(1-x^{-1}\bt\right)} -\frac{u\bt^{2-l}\prod_{\alpha=1}^ly_{\alpha}M_{\alpha+l}^{-1}}{\left(1-x\bt\right)\left(1-x^{-1}\bt\right)}  \right],
\end{split}\ee
with the identification $\tu = u\bt^{-l}\prod_{\alpha=1}^lM_{\alpha}^{\frac{1}{2}}M_{\alpha+l}^{-\frac{1}{2}}$.
We expect that this identification between $u$ and ${\widetilde u}$ for general $l$ also follows from the relation \eqref{instanton3} between the 5d instanton partition function and the auxiliary index, similarly to the case $l=1$.
\eqref{PE.general} reproduces \eqref{Xilratio} by further identifying the parameters as $-b_{\alpha} = M_{\alpha+l}^{-1}, y_{\alpha} = \left(y_{\alpha}^{(2)}\right)^{-1}$.

\subsubsection{Large $N$ limit of $U(N)$ ADHM with $l=1$}
\label{topvertexlargeN}
Lastly let us study the correlation functions in the large $N$ limit.
For this purpose let us first recall that the large $N$ limit of the Higgs index \eqref{largeN_Higgsindex} is obtained from the grand canonical index $\Xi(u;x;\mathfrak{t})$ \eqref{Xiuxt2} as
\begin{align}
{\cal I}^{U(\infty)\text{ ADHM-}[1](H)}=-\text{Res}[\Xi(u;x;\mathfrak{t}),u\rightarrow 1].
\end{align}
This is because the relation between the grand canonical index and the Higgs index \eqref{Xiuxt} is inverted as
\begin{align}
{\cal I}^{U(N)\text{ ADHM-}[1](H)}(x;\mathfrak{t})=\oint_{|u|=\epsilon} \frac{du}{2\pi iu}u^{-N}\Xi(u;x;\mathfrak{t})
=-\sum_{w\,(|w|>\epsilon)}
\text{Res}[u^{-N}\Xi(u;x;\mathfrak{t}),u\rightarrow w],
\end{align}
where $\epsilon$ is a sufficiently small positive number so that $\Xi(u;x;\mathfrak{t})$ does not have poles in $|u|<\epsilon$.
The grand canonical index $\Xi(u;x;\mathfrak{t})$ has poles at $u=1$, $u=\mathfrak{t}^{-2m}$ ($m\ge 1$) and $u=x^{\pm n}\mathfrak{t}^{-2m-n}$ ($m\ge 0,n\ge 1$), where the residues at the poles in the latter two groups only give non-perturbative corrections ${\cal O}(\mathfrak{t}^{2mN})$ and ${\cal O}(\mathfrak{t}^{(2m+n)N})$ in the large $N$ limit.
Hence we find that the Higgs index in the large $N$ limit is given by the residue at $u=1$ as \eqref{largeN_Higgsindex}.

In the same way, we obtain the generating function of the 2-point functions in the large $N$ limit from their grand canonical sum \eqref{Higgsl1_2ptfcngeneratingfcn}
\begin{align}
&\sum_{N=0}^\infty u^N\sum_{k_1,k_2\ge 0}(-M_1)^{k_1}(-M_2)^{-k_2}\langle
W_{(1^{k_1})}
W_{(\overline{1^{k_2}})}\rangle^{U(N)\text{ ADHM-}[1](H)}(x;\mathfrak{t})\nonumber \\
&=
\frac{
\Xi(u;x;\mathfrak{t})\Xi(M_1M_2^{-1}u;x;\mathfrak{t})
}{
\Xi(\mathfrak{t}M_1u;x;\mathfrak{t})\Xi(\mathfrak{t}M_2^{-1}u;x;\mathfrak{t})}
\end{align}
by picking the residue of the right-hand side at the pole $u=1$ as
\begin{align}
&\sum_{k_1,k_2\ge 0}(-M_1)^{k_1}(-M_2)^{-k_2}\langle
W_{(1^{k_1})}
W_{(\overline{1^{k_2}})}\rangle^{U(\infty)\text{ ADHM-}[1](H)}(x;\mathfrak{t})\nonumber \\
&=
-
\text{Res}\biggl[
\frac{
\Xi(u;x;\mathfrak{t})\Xi(M_1M_2^{-1}u;x;\mathfrak{t})
}{
\Xi(\mathfrak{t}M_1u;x;\mathfrak{t})\Xi(\mathfrak{t}M_2^{-1}u;x;\mathfrak{t})}
,u\rightarrow 1
\biggr]\nonumber \\
&={\cal I}^{U(\infty)\text{ ADHM-}[1](H)}(x;\mathfrak{t})
\frac{\Xi(M_1M_2^{-1};x;\mathfrak{t})}{\Xi(\mathfrak{t}M_1;x;\mathfrak{t})\Xi(\mathfrak{t}M_2^{-1};x;\mathfrak{t})}
.
\label{Higgsl1_2ptfcngeneratingfcn_largeN}
\end{align}

For the 1-point functions we set $-M_2^{-1}=0$ in \eqref{Higgsl1_2ptfcngeneratingfcn_largeN} and find
\begin{align}
\sum_{k\ge 0}(-M_1)^{k}\langle
{\cal W}_{(1^{k})}
\rangle^{U(\infty)\text{ ADHM-}[1](H)}(x;\mathfrak{t})
=
\frac{
1
}{
\Xi(\mathfrak{t}M_1;x;\mathfrak{t})},
\end{align}
Since the right-hand side is related to the generating function of $\langle W_{(1^k)}\rangle^{U(k)\text{ ADHM-}[1](H)}$ due to the relation \eqref{Higgsl1_1ptfcngeneratingfcn}, we can also rephrase this result as
\begin{align}
\langle{\cal W}_{(1^{k})}
\rangle^{U(\infty)\text{ ADHM-}[1](H)}(x;\mathfrak{t})
=
\langle W_{(1^{k})}
\rangle^{U(k)\text{ ADHM-}[1](H)}(x;\mathfrak{t}),
\end{align}
which reproduces \eqref{uNADHM_1ptasym_relation_Ninfty}.

For the generic two-point functions, by expanding the right-hand side of \eqref{Higgsl1_2ptfcngeneratingfcn_largeN} in $-M_1$ and $-M_2^{-1}$ we find
\begin{align}
\label{g_adhml1_2pt_asym_H}
&\sum_{k_1, k_2 \geq 0}\left(-M_1\right)^{k_1}\left(-M_2\right)^{-k_2}\Braket{\mathcal{W}_{(1^{k_1})}\mathcal{W}_{(\overline{1^{k_2}})}}^{U(\infty)\;\text{ADHM}-[1](H)}(x;\mathfrak{t})
\nonumber\\
&=
\left(1 + \sum_{k_1'=1}^{\infty}\left(-M_1\right)^{k_1'}\Braket{W_{(1^{k_1'})}}^{U(k_1')\;\text{ADHM}-[1](H)}(x;\mathfrak{t})\right)\nonumber \\
&\quad\times \left(1 + \sum_{k_2'=1}^{\infty}\left(-M_2\right)^{-k_2'}\Braket{W_{(1^{k_2'})}}^{U(k_2')\;\text{ADHM}-[1](H)}(x;\mathfrak{t})\right)\nonumber \\
&\quad\times \left(1 + \sum_{N=1}^{\infty}\left(M_1M_2^{-1}\right)^{N}\mathcal{I}^{U(N)\;\text{ADHM}-[1](H)}(x;\mathfrak{t})\right).
\end{align}
In particular, for the diagonal 2-point functions $k_1=k_2$ we find
\begin{align}
\label{largeN_ADHM1_wasymkH}
&
\langle {\cal W}_{(1^k)} {\cal W}_{\overline{(1^k)}}\rangle^{U(\infty)\text{ ADHM-}[1](H)}(x;\mathfrak{t})
\nonumber\\
&=
\mathcal{I}^{U(k)\text{ ADHM-}[1](H)}(x;\mathfrak{t})
+
\sum_{l=1}^{k}
\langle W_{(1^l)} \rangle^{U(k)\text{ ADHM-}[1](H)}(x;\mathfrak{t})
\langle W_{(1^l)} \rangle^{U(l)\text{ ADHM-}[1](H)}(x;\mathfrak{t}). 
\end{align}
Unlike the Coulomb limit of 
the 
large $N$ normalized $2$-point function, 
it contains the terms contributed from the $1$-point functions. 
It coincides with the Higgs or equivalently Coulomb index of the $U(k)$ ADHM theory with one flavor after subtracting them. 

The expansion coefficients stabilize for the infinite rank $k$ of the antisymmetric representation. The finite $k$ corrections appear at the order $\mathfrak{t}^{k+1}$ in the large $N$ limit. 
In the large representation limit $k\rightarrow \infty$, we get 
\begin{align}
\label{largeN_ADHM1_wasymlargekH}
&
\langle {\cal W}_{(1^{\infty})} {\cal W}_{\overline{(1^{\infty})}}\rangle^{U(\infty)\text{ ADHM-}[1](H)}(x;\mathfrak{t})
\nonumber\\
&=
\mathcal{I}^{U(\infty)\text{ ADHM-}[1](H)}(x;\mathfrak{t})
\left(
1+
\sum_{l=1}^{\infty}
{\langle \mathcal{W}_{(1^l)} \rangle^{U(\infty)\text{ ADHM-}[1](H)}(x;\mathfrak{t})}^2
\right). 
\end{align}
We see that the Higgs limit of the large $N$ normalized $2$-point function of the antisymmetric Wilson lines 
is factorized into the large $N$ Higgs index of the $U(N)$ ADHM theory 
and a sum of the squares of the large $N$ normalized $1$-point functions. 

Similarly, we conjecture that 
the large representation limit of the large $N$ normalized $2$-point function of the symmetric Wilson lines is given by
\begin{align}
\label{largeN_ADHM1_wsymlargekH}
&
\langle {\cal W}_{(\infty)} {\cal W}_{\overline{(\infty)}}\rangle^{U(\infty)\text{ ADHM-}[1](H)}(x;\mathfrak{t})
\nonumber\\
&=
\mathcal{I}^{U(\infty)\text{ ADHM-}[1](H)}(x;\mathfrak{t})
\left(
1+
\sum_{l=1}^{\infty}
{\langle \mathcal{W}_{(l)} \rangle^{U(\infty)\text{ ADHM-}[1](H)}(x;\mathfrak{t})}^2
\right). 
\end{align}
We have checked the relation (\ref{largeN_ADHM1_wsymlargekH}) up to the order $\mathfrak{t}^8$ by means of the Hall-Littlewood expansion. 
The expressions (\ref{largeN_ADHM1_wasymlargekH}) and (\ref{largeN_ADHM1_wsymlargekH}) which contain the large $N$ Higgs indices indicate 
that the corresponding single particle gravity indices  
that encode the spectra of the excitations on the gravity dual geometries involve an infinite tower of fluctuation modes.

\section*{Acknowledgement}
The authors would like to thank Nick Dorey, Hai Lin and Futoshi Yagi for useful discussions and comments. 
The work of H.H. is supported in part by JSPS KAKENHI Grand Number JP23K03396. The work of T.O.~was supported by the Startup Funding no.~4007012317 of the Southeast University. 
Part of the results in this paper were obtained by using the high performance computing facility provided by Yukawa Institute for Theoretical Physics (Sushiki server).

\appendix

\section{$q$-factorial and $q$-analogs}
\label{app_qpoch}
We define the $q$-factorial by
\begin{align}
(x;q)_n=\prod_{j=0}^{n-1}(1-xq^j).
\end{align}
We define the $q$-binomial coefficient as
\begin{align}
\left[
\begin{matrix}
n\\
k\\
\end{matrix}
\right]_{q}
&:=\frac{(q;q)_{n}}{(q;q)_k(q;q)_{n-k}}.
\label{qbincoef}
\end{align}

Here we list the formulae involving $q$-factorial which we use in the main text.
\begin{align}
&\sum_{n=0}^\infty \frac{x^n}{(q;q)_n}=\frac{1}{(x;q)_\infty},
\label{qPochhammerformula1} \\
&\sum_{k=0}^n
\begin{bmatrix}
n\\
k
\end{bmatrix}_q(-z)^jq^{\frac{j(j-1)}{2}}=(z;q)_n.\label{qbincoefformula1}
\end{align}

\section{Notation of Young diagram}
\label{app_Young}
In this appendix we summarize our notation related to the Young diagrams.
We denote a Young diagram as $\lambda=(\lambda_1,\lambda_2,\cdots,\lambda_{\ell(\lambda)})$, where $\lambda_1\ge \lambda_2\ge \cdots \ge \lambda_{\ell(\lambda)}$.
We denote the transpose of a Young diagram $\lambda$ as $\lambda'=(\lambda'_1,\lambda'_2,\cdots,\lambda'_{\ell(\lambda')})$.

For a Young diagram $\lambda$ we define the following numbers and a polynomial in $\mathfrak{t}$.
\begin{align}
m_i(\lambda)&=\#\{j|\lambda_j=i,1\le j\le \ell(\lambda)\},\label{milambda} \\
n(\lambda)&=\sum_{i=1}^{\ell(\lambda)}(i-1)\lambda_i,\label{nlambda} \\
n(\lambda/\mu)&=\sum_{i\ge1}
\left(
\begin{matrix}
\lambda'_i-\mu'_i\\
2\\
\end{matrix}
\right),\label{nlambda/mu} \\
b_{\lambda}(\mathfrak{t})&=\prod_{j\ge 1}(\mathfrak{t};\mathfrak{t})_{m_j(\lambda)}.\label{bmut}
\end{align}
Here $m_i(\lambda)$ is called multiplicity of the Young diagram $\lambda$. 

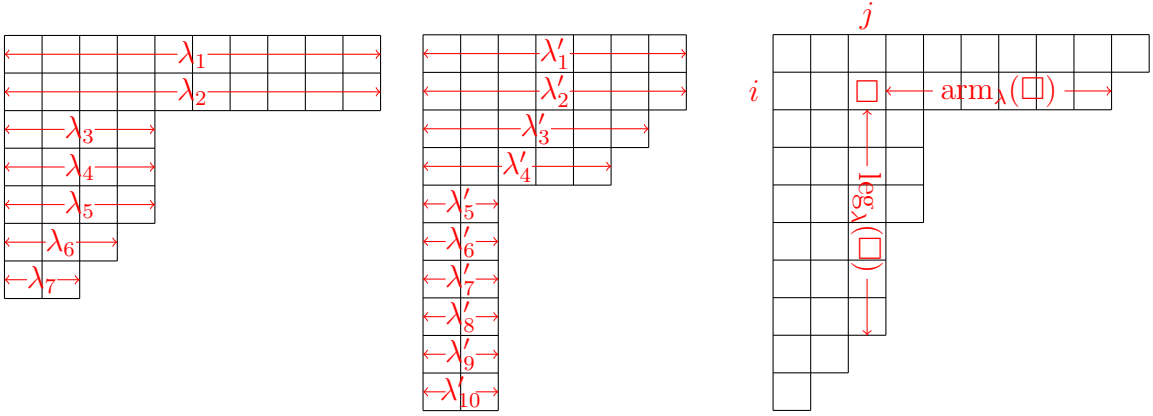
\begin{figure}
\begin{center}
\begin{tikzpicture}[scale=0.25]
\draw (0,50) -- (20,50);
\draw (0,48) -- (20,48);
\draw (0,46) -- (20,46);
\draw (0,44) -- (8,44);
\draw (0,42) -- (8,42);
\draw (0,40) -- (8,40);
\draw (0,38) -- (6,38);
\draw (0,36) -- (4,36);
\draw (0,50) -- (0,36);
\draw (2,50) -- (2,36);
\draw (4,50) -- (4,36);
\draw (6,50) -- (6,38);
\draw (8,50) -- (8,40);
\draw (10,50) -- (10,46);
\draw (12,50) -- (12,46);
\draw (14,50) -- (14,46);
\draw (16,50) -- (16,46);
\draw (18,50) -- (18,46);
\draw (20,50) -- (20,46);
\draw [red,<-] (0,49) -- (9.2,49);
\draw [red,->] (10.8,49) -- (20,49);
\node at (10,49) {$\textcolor{red}{\lambda_1}$};
\draw [red,<-] (0,47) -- (9.2,47);
\draw [red,->] (10.8,47) -- (20,47);
\node at (10,47) {$\textcolor{red}{\lambda_2}$};
\draw [red,<-] (0,45) -- (3.2,45);
\draw [red,->] (4.8,45) -- (8,45);
\node at (4,45) {$\textcolor{red}{\lambda_3}$};
\draw [red,<-] (0,43) -- (3.2,43);
\draw [red,->] (4.8,43) -- (8,43);
\node at (4,43) {$\textcolor{red}{\lambda_4}$};
\draw [red,<-] (0,41) -- (3.2,41);
\draw [red,->] (4.8,41) -- (8,41);
\node at (4,41) {$\textcolor{red}{\lambda_5}$};
\draw [red,<-] (0,39) -- (2.2,39);
\draw [red,->] (3.8,39) -- (6,39);
\node at (3,39) {$\textcolor{red}{\lambda_6}$};
\draw [red,<-] (0,37) -- (1.2,37);
\draw [red,->] (2.8,37) -- (4,37);
\node at (2,37) {$\textcolor{red}{\lambda_7}$};
\end{tikzpicture}
\,\,
\mbox{\raisebox{-1.52cm}{
\begin{tikzpicture}[scale=0.25]
\draw (0,50) -- (14,50);
\draw (0,48) -- (14,48);
\draw (0,46) -- (14,46);
\draw (0,44) -- (12,44);
\draw (0,42) -- (10,42);
\draw (0,40) -- (4,40);
\draw (0,38) -- (4,38);
\draw (0,36) -- (4,36);
\draw (0,34) -- (4,34);
\draw (0,32) -- (4,32);
\draw (0,30) -- (4,30);
\draw (0,50) -- (0,30);
\draw (2,50) -- (2,30);
\draw (4,50) -- (4,30);
\draw (6,50) -- (6,42);
\draw (8,50) -- (8,42);
\draw (10,50) -- (10,42);
\draw (12,50) -- (12,44);
\draw (14,50) -- (14,46);
\draw [red,<-] (0,49) -- (6.2,49);
\draw [red,->] (7.8,49) -- (14,49);
\node at (7,49) {$\textcolor{red}{\lambda'_1}$};
\draw [red,<-] (0,47) -- (6.2,47);
\draw [red,->] (7.8,47) -- (14,47);
\node at (7,47) {$\textcolor{red}{\lambda'_2}$};
\draw [red,<-] (0,45) -- (5.2,45);
\draw [red,->] (6.8,45) -- (12,45);
\node at (6,45) {$\textcolor{red}{\lambda'_3}$};
\draw [red,<-] (0,43) -- (4.2,43);
\draw [red,->] (5.8,43) -- (10,43);
\node at (5,43) {$\textcolor{red}{\lambda'_4}$};
\draw [red,<-] (0,41) -- (1.2,41);
\draw [red,->] (2.8,41) -- (4,41);
\node at (2,41) {$\textcolor{red}{\lambda'_5}$};
\draw [red,<-] (0,39) -- (1.2,39);
\draw [red,->] (2.8,39) -- (4,39);
\node at (2,39) {$\textcolor{red}{\lambda'_6}$};
\draw [red,<-] (0,37) -- (1.2,37);
\draw [red,->] (2.8,37) -- (4,37);
\node at (2,37) {$\textcolor{red}{\lambda'_7}$};
\draw [red,<-] (0,35) -- (1.2,35);
\draw [red,->] (2.8,35) -- (4,35);
\node at (2,35) {$\textcolor{red}{\lambda'_8}$};
\draw [red,<-] (0,33) -- (1.2,33);
\draw [red,->] (2.8,33) -- (4,33);
\node at (2,33) {$\textcolor{red}{\lambda'_9}$};
\draw [red,<-] (0,31) -- (1.2,31);
\draw [red,->] (2.8,31) -- (4,31);
\node at (2,31) {$\textcolor{red}{\lambda'_{10}}$};
\end{tikzpicture}
}}
\,\,
\mbox{\raisebox{-1.43cm}{
\begin{tikzpicture}[scale=0.25]
\draw (0,50) -- (20,50);
\draw (0,48) -- (20,48);
\draw (0,46) -- (18,46);
\draw (0,44) -- (8,44);
\draw (0,42) -- (8,42);
\draw (0,40) -- (8,40);
\draw (0,38) -- (6,38);
\draw (0,36) -- (6,36);
\draw (0,34) -- (6,34);
\draw (0,32) -- (4,32);
\draw (0,30) -- (2,30);
\draw (0,50) -- (0,30);
\draw (2,50) -- (2,30);
\draw (4,50) -- (4,32);
\draw (6,50) -- (6,34);
\draw (8,50) -- (8,40);
\draw (10,50) -- (10,46);
\draw (12,50) -- (12,46);
\draw (14,50) -- (14,46);
\draw (16,50) -- (16,46);
\draw (18,50) -- (18,46);
\draw (20,50) -- (20,48);
\node at (-1,47) {$\textcolor{red}{i}$};
\node at (5,51) {$\textcolor{red}{j}$};
\node at (5,47) {$\textcolor{red}{\Box}$};
\draw [red,<-] (6,47) -- (8.5,47);
\draw [red,->] (15.5,47) -- (18,47);
\node at (12,47) {$\textcolor{red}{\text{arm}_\lambda(\Box)}$};
\draw [red,<-] (5,46) -- (5,43);
\draw [red,->] (5,37) -- (5,34);
\node at (5,40) {\rotatebox{270}{$\textcolor{red}{\text{leg}_\lambda(\Box)}$}};
\end{tikzpicture}
}}
\caption{
Left and center: an example of a pair of Yound diagram $\lambda=(\lambda_1,\cdots,\lambda_{\ell(\lambda)})$ and its transpose $\lambda'=(\lambda'_1,\cdots,\lambda'_{\ell(\lambda')})$.
In the figure $\ell(\lambda)=7$, $\lambda=(10,10,4,4,4,3,2)$, $\ell(\lambda')=10$ and $\lambda'=(7,7,6,5,2,2,2,2,2,2)$.
Right: arm/leg length of a box in a Young diagram.
In the figure $\Box=(i,j)=(2,3)$, $\text{arm}_\lambda(\Box)=6$ and $\text{leg}_\lambda(\Box)=6$.
}
\label{Frobenius}
\end{center}
\end{figure}

Each box in (or outside) a Young diagram is labeled by its vertical position $i$ and horizontal position $j$ from the top-left corner as $\Box=(i,j)$, for which we define the arm length $\text{arm}_\lambda(\Box)$ and the leg length $\text{leg}_\lambda(\Box)$ as
\begin{align}
\Box=(i,j)\rightarrow
\text{arm}_\lambda(\Box)=\begin{cases}
\lambda_i-j&\quad (i\le \ell(\lambda))\\
-j&\quad (i>\ell(\lambda))
\end{cases},\quad
\text{leg}_\lambda(\Box)=\begin{cases}
\lambda'_j-i&\quad (j\le \ell(\lambda'))\\
-i&\quad (j>\ell(\lambda'))
\end{cases}.
\label{armlegAppB}
\end{align}
See figure \ref{Frobenius} for graphical explanations of these notions.
Note that the arm/leg length are also defined for a box outside the Young diagram $\lambda$ through the second lines of \eqref{armlegAppB}.

Under these notations we also define the following functions which we have used in section \ref{sec_JK} and \ref{sec_topvertex}.
\begin{align}
E_{\lambda\mu}(a,s)&=a-\epsilon_1\text{arm}_\lambda(s)+\epsilon_2(\text{leg}_\mu(s)+1),\label{Es}\\
{\cal N}_{\lambda,\mu}(u)
&=\prod_{\Box\in\lambda}(1-e^{E_{\lambda\mu}(\log u,\Box)})
\prod_{\Box'\in\mu}(1-e^{-E_{\mu\lambda}(-\log u-\epsilon_1-\epsilon_2,\Box')})\nonumber \\
&=
\prod_{\Box\in\lambda}(1-u(x\mathfrak{t})^{\text{arm}_\lambda(\Box)}(x^{-1}\mathfrak{t})^{-\text{leg}_\mu(\Box)-1})
\prod_{\Box'\in\mu}(1-u(x\mathfrak{t})^{-\text{arm}_\mu(\Box')-1}(x^{-1}\mathfrak{t})^{\text{leg}_\lambda(\Box')}),
\label{calNandE}
\end{align}
where $e^{\epsilon_1}=x^{-1}\mathfrak{t}^{-1}$ and $e^{\epsilon_2}=x\mathfrak{t}^{-1}$.
Here ${\cal N}_{\lambda,\mu}(u)$ is so called Nekrasov factor.
Note that ${\cal N}_{\lambda,\emptyset}(u)$ and ${\cal N}_{\emptyset,\lambda}(u)$ can also be written as
\begin{align}
{\cal N}_{\lambda,\emptyset}(u)=\prod_{(i,j)\in\lambda}(1-u(x\mathfrak{t})^{j-1}(x^{-1}\mathfrak{t})^{i-1}),\quad
{\cal N}_{\emptyset,\lambda}(u)=\prod_{(i,j)\in\lambda}(1-u(x\mathfrak{t})^{-j}(x^{-1}\mathfrak{t})^{-i}).
\end{align}

\section{Derivations of Coulomb limit of the charged Wilson line indices}
\label{app_chargedCoulombderivation}
In this appendix we display the detailed derivation of the traces $\text{Tr}\rho_f^n$ \eqref{Trrhochargedk2},\eqref{Trrhochargedk3} and the grand canonical correlator $\Xi_f^{(C)}(u)$ \eqref{Xifchargedk2},\eqref{Xifchargedk3} for $f=(1+a_1s^{m_1})(1+a_2s^{-m_1})$ and $f=(1+a_1s^{m_1})(1+a_2s^{m_2})(1+a_3s^{-m_1-m_2})$, to the order $a_1a_2$ and $a_1a_2a_3$ respectively.
From these grand partition functions $\Xi_f^{(C)}(u)$ we can read off the correlators of the charged Wilson lines $\langle W_{m_1} W_{-m_1}\rangle^{U(N)\text{ ADHM-}[l](C)}$ \eqref{ICchargedW2} and $\langle W_{m_1}W_{m_2}W_{-m_1-m_2}\rangle^{U(N)\text{ ADHM-}[l](C)}$ \eqref{ICchargedW3} we have considered in section \ref{sec_chargedCoulomb}, as the expansion coefficients of $u^N$.
As explained in section \ref{sec_chargedCoulomb}, we assume without loss of generality $m_1>0$ in the former case and $m_1,m_2>0$ in the latter case.

First let us consider $f=(1+a_1s^{m_1})(1+a_2s^{-m_1})$.
In this case $\text{Tr}\rho_f^n$ is given by $\text{Tr}\rho_f^n=\text{tr}
(
(1+a_1e^{2\pi im_1{\hat \alpha}})(1+a_2e^{-2\pi im_1{\hat \alpha}}){\hat \rho}_1
\cdots
(1+a_1e^{2\pi im_1{\hat \alpha}})(1+a_2e^{-2\pi im_1{\hat \alpha}}){\hat \rho}_1
)$, which can be evaluated by expanding each $(1+a_1e^{2\pi im_1{\hat\alpha}}) (1+a_2e^{-2\pi im_1{\hat\alpha}})$ and applying the formula \eqref{trrhoformula} to each term.
Since we are interested only in the coefficient of $a_1a_2$, we have only to consider the terms with one $e^{2\pi im_1{\hat\alpha}}$ and one $e^{-2\pi im_1{\hat\alpha}}$ inserted among $n$ ${\hat\rho}_1$.
Due to the cyclic symmetry of the trace, it is sufficient to consider the insertion of $e^{2\pi im_1{\hat\alpha}}$ in front of the first ${\hat\rho}_1$ the insertion of $e^{-2\pi im_1{\hat\alpha}}$ in front of $i$-th ${\hat\rho}_1$ ($i=1,2,\cdots,n$)
\begin{align}
\text{tr}
e^{2\pi im_1{\hat\alpha}}
{\hat\rho}_1^{i-1}
e^{-2\pi im_1{\hat\alpha}}
{\hat\rho}_1^{n-i+1}
=\frac{\mathfrak{t}^{2(i-1)m_1}}{1-\mathfrak{t}^{2n}},
\end{align}
where we have used the formula \eqref{trrhoformula}.
By summing over $i$ and multiplying an overall $n$ to take care of the cyclic images we find that the total coefficient of $a_1a_2$ is $\frac{n}{1-\mathfrak{t}^{2n}}\frac{1-\mathfrak{t}^{2nm_1}}{1-\mathfrak{t}^{2m_1}}$.
Namely, we have
\begin{align}
\text{Tr}\rho_{(1+a_1s^{m_1})(1+a_2s^{-m_1})}^n=\frac{1}{1-\mathfrak{t}^{2n}}\biggl(1+a_1a_2\frac{n(1-\mathfrak{t}^{2m_1n})}{1-\mathfrak{t}^{2m_1}}+\cdots\biggr),
\end{align}
which is \eqref{Trrhochargedk2}.
Plugging this to the grand partition function \eqref{XiCf} and using the formula \eqref{qPochhammerformula1} repeatedly, we find
\begin{align}
\Xi_f^{(C)}(u)&=\Xi^{(C)}(u)\biggl[1+a_1a_2
\sum_{m=-\infty}^\infty
\sum_{n=1}^\infty
\Bigl(\prod_{j=1}^n\frac{1}{1-\mathfrak{t}^{-2j}\mathfrak{t}^{l|m|}z^{lm}u}\Bigr)
(\mathfrak{t}^{l|m|}z^{lm}u)^n
\mathfrak{t}^{-n^2+n}\frac{(-1)^{n-1}}{1-\mathfrak{t}^{2n}}\nonumber \\
&\quad \times \frac{1-\mathfrak{t}^{2nm_1}}{1-\mathfrak{t}^{2m_1}}\biggr]+{\cal O}(a_1^2,a_2^2),
\label{sumnunclear}
\end{align}
By expanding the summation over $n$ in \eqref{sumnunclear} in terms of $\mathfrak{t}^{l|m|}z^{lm}u$ we observe
\begin{align}
&\sum_{n=1}^\infty
\Bigl(\prod_{j=1}^n\frac{1}{1-\mathfrak{t}^{-2j}\mathfrak{t}^{l|m|}z^{lm}u}\Bigr)
(\mathfrak{t}^{l|m|}z^{lm}u)^n
\mathfrak{t}^{-n^2+n}\frac{(-1)^{n-1}}{1-\mathfrak{t}^{2n}}
\frac{1-\mathfrak{t}^{2nm_1}}{1-\mathfrak{t}^{2m_1}}\nonumber \\
&=\sum_{n=1}^\infty\frac{(\mathfrak{t}^{l|m|}z^{lm}u)^n}{1-\mathfrak{t}^{2n}}\sum_{p=0}^n(-1)^{p-1}\mathfrak{t}^{-2np+p^2+p}
\left[
\begin{matrix}
n\\
p\\
\end{matrix}
\right]_{\mathfrak{t}^2}
\Bigl(\frac{1-\mathfrak{t}^{2pm_1}}{1-\mathfrak{t}^{2m_1}}-1\Bigr),
\label{binomialobservedformulaC}
\end{align}
where $\left[
\begin{matrix}
n\\
p\\
\end{matrix}
\right]_{\mathfrak{t}^2}$ is the $q$-binomial coefficient \eqref{qbincoef}.
Plugging this into \eqref{sumnunclear}, the summation over $m$ can be performed explicitly for each $\mathfrak{t}^{l|m|n}z^{lmn}$ as
\begin{align}
\sum_{m=-\infty}^\infty \mathfrak{t}^{l|m|n}z^{lmn}
=(1-\mathfrak{t}^{2n}){\cal I}^{U(1)\text{ ADHM-}[l](C)}(z^n;\mathfrak{t}^n),
\end{align}
where ${\cal I}^{U(1)\text{ ADHM-}[l](C)}(z;\mathfrak{t})=\frac{1-\mathfrak{t}^{2l}}{(1-\mathfrak{t}^2)\prod_{\pm }(1-\mathfrak{t}^lz^{\pm l})}$.
We can also perform the summation over $p$ in \eqref{binomialobservedformulaC} by using the formula
\begin{align}
\sum_{p=0}^n(-1)^{p-1}\mathfrak{t}^{-2np+p^2+p}
\left[
\begin{matrix}
n\\
p\\
\end{matrix}
\right]_{\mathfrak{t}^2}
\Bigl(\frac{1-\mathfrak{t}^{2pm_1}}{1-\mathfrak{t}^{2m_1}}-1\Bigr)
=\frac{(\mathfrak{t}^{-2n+2+2m_1};\mathfrak{t}^2)_n}{1-\mathfrak{t}^{2m_1}}
\end{align}
Note that this vanishes for $n>m_1$.
Putting the results together,
we finally obtain $\Xi_f^{(C)}(u)$ as \eqref{Xifchargedk2}.


Next let us consider $f=(1+a_1s^{m_1})(1+a_2s^{m_2})(1+a_3s^{-m_1-m_2})$.
Again, to obtain the coefficient of $a_1a_2a_3$ in $\text{Tr}\rho_{(1+a_1s^{m_1})(1+a_2s^{m_2})(1+a_3s^{-m_1-m_2})}^n$ it is sufficient to consider $\text{tr}{\hat\rho}_1^n$ with the following two types of insertions:
(i)
$\text{tr}
e^{2\pi im_1{\hat\alpha}}
{\hat\rho}^{i-1}
e^{2\pi im_2{\hat\alpha}}
{\hat\rho}^{j}
e^{2\pi i(-m_1-m_2){\hat\alpha}}
{\hat\rho}^{n+1-i-j}
$ ($1\le i\le n$ and $0\le j\le n-i$) and
(ii)
$\text{tr}
e^{2\pi im_2{\hat\alpha}}
{\hat\rho}^{i-1}
e^{2\pi im_1{\hat\alpha}}
{\hat\rho}^{j}
e^{2\pi i(-m_1-m_2){\hat\alpha}}
{\hat\rho}^{n+1-i-j}
$ ($2\le i\le n$ and $0\le j\le n-i$).
Here we have removed $i=1$ from the case (ii) to avoid an over-counting.
These contributions are evaluated by using the formula \eqref{trrhoformula} as
\begin{align}
&n\sum_{i=1}^n\sum_{j=0}^{n-i}
\text{tr}
e^{2\pi im_1{\hat\alpha}}
{\hat\rho}^{i-1}
e^{2\pi im_2{\hat\alpha}}
{\hat\rho}^{j}
e^{2\pi i(-m_1-m_2){\hat\alpha}}
{\hat\rho}^{n+1-i-j}
=
n\sum_{i=1}^n\sum_{j=0}^{n-i}
\frac{\mathfrak{t}^{2(i-1)m_1+2j(m_1+m_2)}}{1-\mathfrak{t}^{2n}}\nonumber \\
&\quad =\frac{n(1
-\mathfrak{t}^{2m_2}
-\mathfrak{t}^{2m_1n}
-\mathfrak{t}^{2(m_1+m_2)(n+1)}
+\mathfrak{t}^{2m_1(n+1)+2m_2}
+\mathfrak{t}^{2m_1n+2m_2(n+1)}
}{
(1-\mathfrak{t}^{2n})
(1-\mathfrak{t}^{2m_1})
(1-\mathfrak{t}^{2m_2})
(1-\mathfrak{t}^{2m_1+2m_2})
},\\
&n\sum_{i=2}^n\sum_{j=0}^{n-i}
\text{tr}
e^{2\pi im_2{\hat\alpha}}
{\hat\rho}^{i-1}
e^{2\pi im_1{\hat\alpha}}
{\hat\rho}^{j}
e^{2\pi i(-m_1-m_2){\hat\alpha}}
{\hat\rho}^{n+1-i-j}
=
n\sum_{i=2}^n\sum_{j=0}^{n-i}
\frac{\mathfrak{t}^{2(i-1)m_2+2j(m_1+m_2)}}{1-\mathfrak{t}^{2n}},\nonumber \\
&\quad =\frac{n(\mathfrak{t}^{2m_2}
-\mathfrak{t}^{2(m_1+m_2)}
-\mathfrak{t}^{2m_2n}
+\mathfrak{t}^{2(m_1+m_2)n}
+\mathfrak{t}^{2m_1+2m_2(n+1)}
-\mathfrak{t}^{2m_1n+2m_2(n+1)}
}{
(1-\mathfrak{t}^{2n})
(1-\mathfrak{t}^{2m_1})
(1-\mathfrak{t}^{2m_2})
(1-\mathfrak{t}^{2m_1+2m_2})
}.
\end{align}
Hence we find \eqref{Trrhochargedk3}
\begin{align}
\text{Tr}\rho_{(1+a_1s^{m_1})(1+a_2s^{m_2})(1+a_3s^{-m_1-m_2})}^n=\frac{1}{1-\mathfrak{t}^{2n}}\Bigl(1+a_1a_2a_3\frac{n(1-\mathfrak{t}^{2nm_1})(1-\mathfrak{t}^{2nm_2})}{(1-\mathfrak{t}^{2m_1})(1-\mathfrak{t}^{2m_2})}+\cdots\Bigr).
\end{align}

Plugging this to the grand partition function we obtain
\begin{align}
&\Xi_f^{(C)}(u)=\Xi^{(C)}(u)\biggl[1+a_1a_2a_3
\sum_{n=1}^{N}
{\cal I}^{U(1)\text{ ADHM-}[l](C)}(z^n;\mathfrak{t}^n)u^n
\sum_{p=0}^n(-1)^{p-1}\mathfrak{t}^{-2np+p^2+p}
\left[
\begin{matrix}
n\\
p\\
\end{matrix}
\right]_{\mathfrak{t}^2}\nonumber \\
&\quad
\times \frac{1-\mathfrak{t}^{2pm_1}}{1-\mathfrak{t}^{2m_1}}
\frac{1-\mathfrak{t}^{2pm_2}}{1-\mathfrak{t}^{2m_2}}
\biggr]+{\cal O}(a_1^2,a_2^2,a_3^2).
\end{align}
Again the summation over $p$ can be performed by using the formula \eqref{qbincoefformula1} as
\begin{align}
&\sum_{p=0}^n(-1)^{p-1}\mathfrak{t}^{-2np+p^2+p}
\left[
\begin{matrix}
n\\
p\\
\end{matrix}
\right]_{\mathfrak{t}^2}
\Bigl(\frac{1-\mathfrak{t}^{2pm_1}}{1-\mathfrak{t}^{2m_1}}
\frac{1-\mathfrak{t}^{2pm_2}}{1-\mathfrak{t}^{2m_2}}-1\Bigr)\nonumber \\
&=
\frac{
(\mathfrak{t}^{-2n+2+2m_1};\mathfrak{t}^2)_n
+(\mathfrak{t}^{-2n+2+2m_2};\mathfrak{t}^2)_n
-(\mathfrak{t}^{-2n+2+2m_1+2m_2};\mathfrak{t}^2)_n
}{
(1-\mathfrak{t}^{2m_1})
(1-\mathfrak{t}^{2m_2})
},
\end{align}
which vanishes for $n>m_1+m_2$.
Hence we finally obtain $\Xi_f^{(C)}(u)$ as \eqref{Xifchargedk2}.


\section{Formalism of the refined topological vertex}
\label{app:top}

\begin{figure}
\centering
\begin{scriptsize}
\begin{tikzpicture}[scale=0.5]
\draw[arrows=->,thick] (0,0)--(3,0) {};
\draw[arrows=->,thick] (0,0)--(0,3) {};
\draw[arrows=->,thick] (0,0)--(-2,-2) {};
\node[label=above:{$\nu$}]  at (3,0) {};
\node[label=right:{$\mu$}]  at (0,3) {};
\node[label=left:{$q$}]  at (0,3) {};
\node[label=above:{$\lambda$}]  at (-2,-2) {};
\node[label=right:{$t$}]  at (-2,-2) {};
\node[] at (2,0) {$\|$};
\end{tikzpicture}
\end{scriptsize}
\caption{A vertex with three legs. $\lambda, \mu, \nu$ are Young diagrams.}
\label{fig:vertex}
\end{figure}
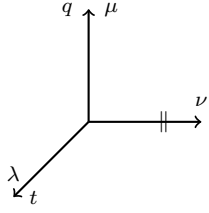

We here summarize the formalism of the refined topological vertex. For a given toric diagram, we first assign an orientation and a Young diagram for each internal line of the diagram. For external legs, we put trivial Young diagrams. We also choose a preferred direction. Then we cut the internal lines and decompose the diagram into vertices with three legs. One of the three legs should be in the preferred direction. For the other two legs, we assign $t$ for one leg and $q$ for the other leg. The assignment of $t, q$ should be chosen so that both $t$ and $q$ are assigned for each internal line in the non-preferred direction after gluing the vertices. A vertex with the outward orientations for the three legs is depicted in Figure \ref{fig:vertex}. 
The preferred direction is chosen in the horizontal direction and it is represented by $\|$ in the figure. For each vertex given in Figure \ref{fig:vertex}, we assign
\be
C_{\lambda\mu\nu}(t, q) = t^{-\frac{\|\mu'\|^2}{2}}q^{\frac{\|\mu\|^2 + \|\nu\|^2}{2}}\widetilde{Z}_{\nu}(t, q)\sum_{\eta}\left(\frac{q}{t}\right)^{\frac{|\eta|+|\lambda|-|\mu|}{2}}s_{\lambda'/\eta}\left(t^{-\rho}q^{-\nu}\right)s_{\mu/\eta}\left(t^{-\nu'}q^{-\rho}\right),
\ee
where 
\begin{align}
\|\lambda\|^2 = \sum_i\lambda_i^2, \quad \left|\lambda\right| = \sum_i\lambda_i,
\end{align}
for a Young diagram $\lambda=\left(\lambda_1, \lambda_2, \cdots\right)$. $\widetilde{Z}_{\lambda}(t, q)$ is defined as
\be
\widetilde{Z}_{\lambda}(t, q) = \prod_{(i, j)\in\lambda}\left(1 - q^{\text{arm}_{\lambda}(i, j)}t^{\text{leg}_{\lambda}(i, j) + 1}\right)^{-1}
\ee
with \eqref{armlegAppB}.
$s_{\mu/\nu}(x_i)$ is the skew Schur function and we also defined
\be
s_{\mu/\nu}\left(t^{-\rho}q^{-\lambda}\right) = s_{\mu/\nu}\left(t^{\frac{1}{2}}q^{-\lambda_1}, t^{\frac{3}{2}}q^{-\lambda_2},  t^{\frac{5}{2}}q^{-\lambda_3}, \cdots\right).
\ee
When the orientation assigned to a leg is inward, the corresponding Young diagram needs to be transposed. 

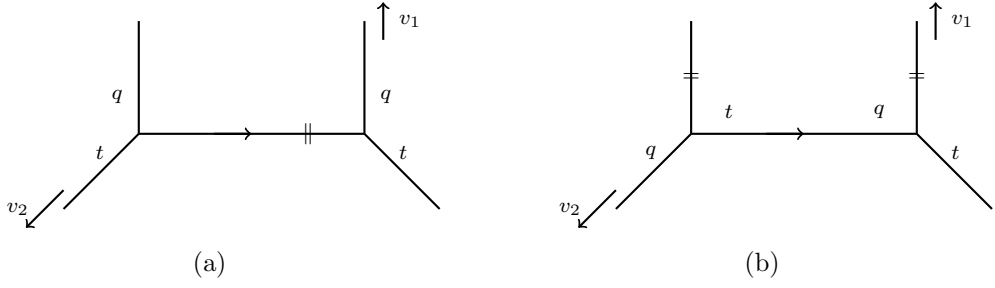
\begin{figure}
\centering
\subfigure[]{\label{fig:glue1}
\begin{scriptsize}
\begin{tikzpicture}[scale=0.5]
\draw[thick] (-3,0)--(3,0) {};
\draw[thick] (-3,0)--(-3,3) {};
\draw[thick] (-3,0)--(-5,-2) {};
\draw[thick] (3,0)--(3,3) {};
\draw[thick] (3,0)--(5,-2) {};
\node[]  at (1.5,0) {$\|$};
\draw[arrows=->, thick] (-1,0)--(0,0) {};
\node[label=right:{$q$}]  at (3,1) {};
\node[label=left:{$q$}]  at (-3,1) {};
\node[label=right:{$t$}]  at (4-1/2,-1/2) {};
\node[label=left:{$t$}]  at (-4+1/2,-1/2) {};
\draw[arrows=->, thick] (3.5,2.5)--(3.5,3.5) {};
\node[label=right:{$v_1$}]  at (3.5,3) {};
\draw[arrows=->, thick] (-5,-1.5)--(-6,-2.5) {};
\node[label=left:{$v_2$}]  at (-5.5,-2) {};
\end{tikzpicture}
\end{scriptsize}}\hspace{1cm}
\subfigure[]{\label{fig:glue2}
\begin{scriptsize}
\begin{tikzpicture}[scale=0.5]
\draw[thick] (-3,0)--(3,0) {};
\draw[thick] (-3,0)--(-3,3) {};
\draw[thick] (-3,0)--(-5,-2) {};
\draw[thick] (3,0)--(3,3) {};
\draw[thick] (3,0)--(5,-2) {};
\node[]  at (3,1.5) {$=$};
\node[]  at (-3,1.5) {$=$};
\draw[arrows=->, thick] (-1,0)--(0,0) {};
\node[label=above:{$q$}]  at (2,0) {};
\node[label=above:{$t$}]  at (-2,0) {};
\node[label=right:{$t$}]  at (4-1/2,-1/2) {};
\node[label=left:{$q$}]  at (-4+1/2,-1/2) {};
\draw[arrows=->, thick] (3.5,2.5)--(3.5,3.5) {};
\node[label=right:{$v_1$}]  at (3.5,3) {};
\draw[arrows=->, thick] (-5,-1.5)--(-6,-2.5) {};
\node[label=left:{$v_2$}]  at (-5.5,-2) {};
\end{tikzpicture}
\end{scriptsize}}
\caption{(a) Gluing along the preferred direction. (b) Gluing along a non-preferred direction. The vertical direction is the preferred direction in this figure.}
\label{fig:glue}
\end{figure}
In order to recover the original diagram one needs to glue the vertices. The gluing is done by summing over all the Young diagrams associated to the glued legs with a weight factor inserted for each glued leg. For the gluing along the preferred direction depicted in Figure \ref{fig:glue1}, the weight is given by
\be\label{weight1}
\left(-Q\right)^{|\nu|}f_{\nu}(t, q)^n \qquad \left(n=\det\left(v_1, v_2\right)\right)
\ee
where 
\be
f_{\nu}(t, q) = \left(-1\right)^{|\nu|}t^{-\frac{\|\nu'\|}{2}}q^{\frac{\|\nu\|}{2}}.
\ee
On the other hand, the weight for the gluing along the non-preferred direction depicted in Figure \ref{fig:glue2} is given by
\be\label{weight2}
\left(-Q\right)^{|\nu|}\widetilde{f}_{\nu}(t, q)^n \qquad \left(n=\det\left(v_1, v_2\right)\right)
\ee
where 
\be
\widetilde{f}_{\nu}(t, q) = \left(-1\right)^{|\nu|}q^{-\frac{\|\nu'\|}{2}}t^{\frac{\|\nu\|}{2}}\left(\frac{t}{q}\right)^{\frac{|\nu|}{2}}.
\ee
For both \eqref{weight1} and \eqref{weight2}, $Q$ is given by $Q=e^{-\text{vol}(C)}$ where $\text{vol}(C)$ is the volume of the curve $C$ associated with the glued leg. For the sum over Young diagrams, one may use the following identities
\begin{align}
\sum_{\mu}s_{\mu/\eta_1}(x_i)s_{\mu/\eta_2}(y_i) &= \prod_{i, j}\left(1-x_iy_i\right)^{-1}\sum_{\tau}s_{\eta_2/\tau}(x_i)s_{\eta_1/\tau}(y_i),\\
\sum_{\mu}s_{\mu'/\eta_1}(x_i)s_{\mu/\eta_2}(y_i) &= \prod_{i, j}\left(1+x_iy_i\right)\sum_{\tau}s_{\eta_2'/\tau}(x_i)s_{\eta_1'/\tau'}(y_i).
\end{align}

\section{Closed-form expressions of line defect indices in unflavored limit}
\label{app_closedform}
In this appendix we list closed-form expressions for the unflavored line defect indices.

\subsection{Unflavored Coulomb line defect indices}
\label{app_unflavoredC}

\subsubsection{$U(2)$}
{\fontsize{9pt}{1pt}\selectfont
\begin{align}
&\langle W_{\ydiagram{2}}W_{\overline{\ydiagram{2}}}\rangle^{U(2)\text{ ADHM-}[1](C)}(\mathfrak{t})
=
\frac{1 + 4 \mathfrak{t} + 2 \mathfrak{t}^3 - \mathfrak{t}^4}{(1 - \mathfrak{t})^2 (1 - \mathfrak{t}^2)^2},\\
&\langle W_{\ydiagram{3}}W_{\overline{\ydiagram{3}}}\rangle^{U(2)\text{ ADHM-}[1](C)}(\mathfrak{t})
=
\frac{1 + 6 \mathfrak{t} - \mathfrak{t}^2 + 4 \mathfrak{t}^3 - 3 \mathfrak{t}^4 + 2 \mathfrak{t}^5 - \mathfrak{t}^6}{(1 - \mathfrak{t})^2 (1 - \mathfrak{t}^2)^2},\\
&\langle W_{\ydiagram{4}}W_{\overline{\ydiagram{4}}}\rangle^{U(2)\text{ ADHM-}[1](C)}(\mathfrak{t})
=
\frac{1 + 8 \mathfrak{t} - 2 \mathfrak{t}^2 + 6 \mathfrak{t}^3 - 5 \mathfrak{t}^4 + 4 \mathfrak{t}^5 - 3 \mathfrak{t}^6 + 2 \mathfrak{t}^7 - \mathfrak{t}^8}{(1 - \mathfrak{t})^2 (1 - \mathfrak{t}^2)^2},\\
&\langle W_{\ydiagram{5}}W_{\overline{\ydiagram{5}}}\rangle^{U(2)\text{ ADHM-}[1](C)}(\mathfrak{t})
=
\frac{1 + 10 \mathfrak{t} - 3 \mathfrak{t}^2 + 8 \mathfrak{t}^3 - 7 \mathfrak{t}^4 + 6 \mathfrak{t}^5 - 5 \mathfrak{t}^6 + 4 \mathfrak{t}^7 - 3 \mathfrak{t}^8 + 2 \mathfrak{t}^9 - \mathfrak{t}^{10}}{(1 - \mathfrak{t})^2 (1 - \mathfrak{t}^2)^2},\\
&\langle W_{\ydiagram{6}}W_{\overline{\ydiagram{6}}}\rangle^{U(2)\text{ ADHM-}[1](C)}(\mathfrak{t})\nonumber \\
&=
\frac{
1 + 12 \mathfrak{t} - 4 \mathfrak{t}^2 + 10 \mathfrak{t}^3 - 9 \mathfrak{t}^4 + 8 \mathfrak{t}^5 - 7 \mathfrak{t}^6 + 6 \mathfrak{t}^7 - 5 \mathfrak{t}^8 + 4 \mathfrak{t}^9 - 3 \mathfrak{t}^{10} + 2 \mathfrak{t}^{11} - \mathfrak{t}^{12}
}{(1 - \mathfrak{t})^2 (1 - \mathfrak{t}^2)^2},\\
%
&\langle W_{\ydiagram{2}}W_{\overline{\ydiagram{2}}}\rangle^{U(2)\text{ ADHM-}[2](C)}(\mathfrak{t})
=
\frac{1 + 6 \mathfrak{t}^2 + 10 \mathfrak{t}^4 + 5 \mathfrak{t}^6 + 3 \mathfrak{t}^8 - \mathfrak{t}^{10}}{(1 - \mathfrak{t}^2)^2 (1 - \mathfrak{t}^4)^2},\\
&\langle W_{\ydiagram{3}}W_{\overline{\ydiagram{3}}}\rangle^{U(2)\text{ ADHM-}[2](C)}(\mathfrak{t})
=
\frac{1 + 6 \mathfrak{t}^2 + 2 \mathfrak{t}^6 - \mathfrak{t}^8}{(1 - \mathfrak{t}^2)^4},\\
&\langle W_{\ydiagram{4}}W_{\overline{\ydiagram{4}}}\rangle^{U(2)\text{ ADHM-}[2](C)}(\mathfrak{t})
=
\frac{1 + 10 \mathfrak{t}^2 + 16 \mathfrak{t}^4 + 10 \mathfrak{t}^6 + 5 \mathfrak{t}^8 - \mathfrak{t}^{10} - \mathfrak{t}^{14}}{(1 - \mathfrak{t}^2)^2 (1 - \mathfrak{t}^4)^2},\\
&\langle W_{\ydiagram{5}}W_{\overline{\ydiagram{5}}}\rangle^{U(2)\text{ ADHM-}[2](C)}(\mathfrak{t})
=
\frac{1 + 10 \mathfrak{t}^2 - 2 \mathfrak{t}^4 + 6 \mathfrak{t}^6 - 4 \mathfrak{t}^8 + 2 \mathfrak{t}^{10} - \mathfrak{t}^{12}}{(1 - \mathfrak{t}^2)^4},\\
&\langle W_{\ydiagram{6}}W_{\overline{\ydiagram{6}}}\rangle^{U(2)\text{ ADHM-}[2](C)}(\mathfrak{t})
=
\frac{1 + 14 \mathfrak{t}^2 + 22 \mathfrak{t}^4 + 14 \mathfrak{t}^6 + 7 \mathfrak{t}^8 - \mathfrak{t}^{14} - \mathfrak{t}^{18}}{(1 - \mathfrak{t}^2)^2 (1 - \mathfrak{t}^4)^2},\\
&\langle W_{\ydiagram{2}}W_{\overline{\ydiagram{2}}}\rangle^{U(2)\text{ ADHM-}[3](C)}(\mathfrak{t})
=
\frac{
(1 + \mathfrak{t}) (1 + \mathfrak{t}^2) (1 - \mathfrak{t} + 5 \mathfrak{t}^3 - 3 \mathfrak{t}^4 - 2 \mathfrak{t}^5 + 3 \mathfrak{t}^6 + \mathfrak{t}^7 - \mathfrak{t}^8)
}{
(1 - \mathfrak{t}^2)^2 (1 - \mathfrak{t}^3) (1 - \mathfrak{t}^6)
},\\
&\langle W_{\ydiagram{3}}W_{\overline{\ydiagram{3}}}\rangle^{U(2)\text{ ADHM-}[3](C)}(\mathfrak{t})
=
\frac{
(1 + \mathfrak{t}) (1 - \mathfrak{t} + \mathfrak{t}^2 + 6 \mathfrak{t}^3 - 5 \mathfrak{t}^4 + 5 \mathfrak{t}^5 - \mathfrak{t}^7 + \mathfrak{t}^8 + 2 \mathfrak{t}^9 - \mathfrak{t}^{10} + \mathfrak{t}^{11} - \mathfrak{t}^{12})
}{
(1 - \mathfrak{t}^2)^2 (1 - \mathfrak{t}^3) (1 - \mathfrak{t}^6)
},\\
&\langle W_{\ydiagram{4}}W_{\overline{\ydiagram{4}}}\rangle^{U(2)\text{ ADHM-}[3](C)}(\mathfrak{t})\nonumber \\
&=
\frac{(1 + \mathfrak{t})
(1 - \mathfrak{t} + \mathfrak{t}^2 + 8 \mathfrak{t}^3 - 7 \mathfrak{t}^4 + 7 \mathfrak{t}^5 - \mathfrak{t}^6 + \mathfrak{t}^8 + 3 \mathfrak{t}^9 - 3 \mathfrak{t}^{10} + 2 \mathfrak{t}^{11} - \mathfrak{t}^{12} + \mathfrak{t}^{13} - \mathfrak{t}^{14})
}{
(1 - \mathfrak{t}^2)^2 (1 - \mathfrak{t}^3) (1 - \mathfrak{t}^6)},\\
&\langle W_{\ydiagram{5}}W_{\overline{\ydiagram{5}}}\rangle^{U(2)\text{ ADHM-}[3](C)}(\mathfrak{t})\nonumber \\
&=
\frac{
(1 + \mathfrak{t})
(1 - \mathfrak{t} + \mathfrak{t}^{2} + 10 \mathfrak{t}^{3} - 9 \mathfrak{t}^{4} + 9 \mathfrak{t}^{5} - 2 \mathfrak{t}^{6} + \mathfrak{t}^{7} + 5 \mathfrak{t}^{9} - 4 \mathfrak{t}^{10} + 3 \mathfrak{t}^{11} - 3 \mathfrak{t}^{12} + 2 \mathfrak{t}^{13} - \mathfrak{t}^{14} + \mathfrak{t}^{15} - \mathfrak{t}^{16})
}{
(1 - \mathfrak{t}^2)^2 (1 - \mathfrak{t}^3) (1 - \mathfrak{t}^6)
},\\
&\langle W_{\ydiagram{6}}W_{\overline{\ydiagram{6}}}\rangle^{U(2)\text{ ADHM-}[3](C)}(\mathfrak{t})\nonumber \\
&=
\frac{
1 + \mathfrak{t}
}{
(1 - \mathfrak{t}^2)^2 (1 - \mathfrak{t}^3) (1 - \mathfrak{t}^6)
}
(1 - \mathfrak{t} + \mathfrak{t}^{2} + 12 \mathfrak{t}^{3} - 11 \mathfrak{t}^{4} + 11 \mathfrak{t}^{5} - 3 \mathfrak{t}^{6} + 2 \mathfrak{t}^{7} - \mathfrak{t}^{8} + 7 \mathfrak{t}^{9} - 6 \mathfrak{t}^{10} + 5 \mathfrak{t}^{11} - 4 \mathfrak{t}^{12} + 3 \mathfrak{t}^{13}\nonumber \\
&\quad - 3 \mathfrak{t}^{14} + 2 \mathfrak{t}^{15} - \mathfrak{t}^{16} + \mathfrak{t}^{17} - \mathfrak{t}^{18}),\\
%
%
&\langle W_{\ydiagram{2}}W_{\overline{\ydiagram{2}}}\rangle^{U(2)\text{ ADHM-}[4](C)}(\mathfrak{t})
=
\frac{
1 + 6 \mathfrak{t}^4 - \mathfrak{t}^6 + 4 \mathfrak{t}^8 + 3 \mathfrak{t}^{12} - \mathfrak{t}^{14}
}{(1 - \mathfrak{t}^2)^2 (1 - \mathfrak{t}^4) (1 - \mathfrak{t}^8)},\\
&\langle W_{\ydiagram{3}}W_{\overline{\ydiagram{3}}}\rangle^{U(2)\text{ ADHM-}[4](C)}(\mathfrak{t})
=
\frac{
1 + 8 \mathfrak{t}^4 + 4 \mathfrak{t}^8 + 4 \mathfrak{t}^{12} - \mathfrak{t}^{16}
}{(1 - \mathfrak{t}^2)^2 (1 - \mathfrak{t}^4) (1 - \mathfrak{t}^8)},\\
&\langle W_{\ydiagram{4}}W_{\overline{\ydiagram{4}}}\rangle^{U(2)\text{ ADHM-}[4](C)}(\mathfrak{t})
=
\frac{
1 + 10 \mathfrak{t}^4 + 6 \mathfrak{t}^8 - \mathfrak{t}^{10} + 5 \mathfrak{t}^{12} - \mathfrak{t}^{18}
}{(1 - \mathfrak{t}^2)^2 (1 - \mathfrak{t}^4) (1 - \mathfrak{t}^8)
},\\
&\langle W_{\ydiagram{5}}W_{\overline{\ydiagram{5}}}\rangle^{U(2)\text{ ADHM-}[4](C)}(\mathfrak{t})
=
\frac{
1 + 12 \mathfrak{t}^4 + 7 \mathfrak{t}^8 + 5 \mathfrak{t}^{12} - \mathfrak{t}^{20}
}{
(1 - \mathfrak{t}^2)^2 (1 - \mathfrak{t}^4) (1 - \mathfrak{t}^8)},\\
&\langle W_{\ydiagram{6}}W_{\overline{\ydiagram{6}}}\rangle^{U(2)\text{ ADHM-}[4](C)}(\mathfrak{t})
=
\frac{
1 + 14 \mathfrak{t}^4 + 8 \mathfrak{t}^8 + 7 \mathfrak{t}^{12} - \mathfrak{t}^{14} - \mathfrak{t}^{22}
}{
(1 - \mathfrak{t}^2)^2 (1 - \mathfrak{t}^4) (1 - \mathfrak{t}^8)}.
\end{align}
}

\subsubsection{$U(3)$}
{\fontsize{9pt}{1pt}\selectfont
\begin{align}
&\langle W_{\ydiagram{2}}W_{\overline{\ydiagram{2}}}\rangle^{U(3)\text{ ADHM-}[1](C)}(\mathfrak{t})
=
\frac{1 + 2 \mathfrak{t} + 2 \mathfrak{t}^3 - \mathfrak{t}^4}{(1 - \mathfrak{t})^4 (1 - \mathfrak{t}^2)^2},\\
&\langle W_{\ydiagram{3}}W_{\overline{\ydiagram{3}}}\rangle^{U(3)\text{ ADHM-}[1](C)}(\mathfrak{t})
=
\frac{1 + 6 \mathfrak{t} + 11 \mathfrak{t}^2 + 18 \mathfrak{t}^3 + 12 \mathfrak{t}^4 + 14 \mathfrak{t}^5 + 2 \mathfrak{t}^7 - 4 \mathfrak{t}^8 + 2 \mathfrak{t}^9 - \mathfrak{t}^{10} - 2 \mathfrak{t}^{11} + \mathfrak{t}^{12}}{(1 - \mathfrak{t})^2 (1 - \mathfrak{t}^2)^2 (1 - \mathfrak{t}^3)^2},\\
&\langle W_{\ydiagram{4}}W_{\overline{\ydiagram{4}}}\rangle^{U(3)\text{ ADHM-}[1](C)}(\mathfrak{t})\nonumber \\
&=
\frac{
1 + 7 \mathfrak{t} + 8 \mathfrak{t}^2 + 13 \mathfrak{t}^3 - 4 \mathfrak{t}^4 + 15 \mathfrak{t}^5 - 14 \mathfrak{t}^6 + 7 \mathfrak{t}^7 - 2 \mathfrak{t}^8 + \mathfrak{t}^9 - 3 \mathfrak{t}^{10} + 3 \mathfrak{t}^{12} - 3 \mathfrak{t}^{13} + \mathfrak{t}^{14}
}{
(1 - \mathfrak{t})^3 (1 - \mathfrak{t}^2)^2 (1 - \mathfrak{t}^3)},\\
&\langle W_{\ydiagram{5}}W_{\overline{\ydiagram{5}}}\rangle^{U(3)\text{ ADHM-}[1](C)}(\mathfrak{t})\nonumber \\
&=
\frac{
1 + 9 \mathfrak{t} + 12 \mathfrak{t}^2 + 19 \mathfrak{t}^3 - 8 \mathfrak{t}^4 + 25 \mathfrak{t}^5 - 24 \mathfrak{t}^6 + 15 \mathfrak{t}^7 - 8 \mathfrak{t}^8 + 7 \mathfrak{t}^9 - 8 \mathfrak{t}^{10} + \mathfrak{t}^{11} + 4 \mathfrak{t}^{12} - 3 \mathfrak{t}^{13} - \mathfrak{t}^{14} + 3 \mathfrak{t}^{16} - 3 \mathfrak{t}^{17} + \mathfrak{t}^{18}
}{
(1 - \mathfrak{t})^3 (1 - \mathfrak{t}^2)^2 (1 - \mathfrak{t}^3)
},\\
&\langle W_{\ydiagram{6}}W_{\overline{\ydiagram{6}}}\rangle^{U(3)\text{ ADHM-}[1](C)}(\mathfrak{t})\nonumber \\
&=
\frac{1}{(1 - \mathfrak{t})^2 (1 - \mathfrak{t}^2)^2 (1 - \mathfrak{t}^3)^2}
(1 + 12 \mathfrak{t} + 29 \mathfrak{t}^2 + 54 \mathfrak{t}^3 + 30 \mathfrak{t}^4 + 50 \mathfrak{t}^5 - 12 \mathfrak{t}^6 + 26 \mathfrak{t}^7 - 27 \mathfrak{t}^8 + 24 \mathfrak{t}^9 - 17 \mathfrak{t}^{10} + 6 \mathfrak{t}^{11}\nonumber \\
&\quad - 8 \mathfrak{t}^{12} + 6 \mathfrak{t}^{13} - 3 \mathfrak{t}^{14} - 4 \mathfrak{t}^{15} + \mathfrak{t}^{16} + 2 \mathfrak{t}^{18} - 4 \mathfrak{t}^{19} + 2 \mathfrak{t}^{20} + \mathfrak{t}^{22} - 2 \mathfrak{t}^{23} + \mathfrak{t}^{24}),\\
%
%
&\langle W_{\ydiagram{2}}W_{\overline{\ydiagram{2}}}\rangle^{U(3)\text{ ADHM-}[2](C)}(\mathfrak{t})
=
\frac{1 + 3 \mathfrak{t}^2 + 8 \mathfrak{t}^4 + 2 \mathfrak{t}^6 + 3 \mathfrak{t}^8 - \mathfrak{t}^{10}}{(1 - \mathfrak{t}^2)^5 (1 - \mathfrak{t}^4)},\\
&\langle W_{\ydiagram{3}}W_{\overline{\ydiagram{3}}}\rangle^{U(3)\text{ ADHM-}[2](C)}(\mathfrak{t})
=
\frac{1 + 7 \mathfrak{t}^2 + 25 \mathfrak{t}^4 + 50 \mathfrak{t}^6 + 65 \mathfrak{t}^8 + 54 \mathfrak{t}^{10} + 34 \mathfrak{t}^{12} + 7 \mathfrak{t}^{14} + 2 \mathfrak{t}^{16} - 3 \mathfrak{t}^{18} - 3 \mathfrak{t}^{20} + \mathfrak{t}^{22}}{(1 - \mathfrak{t}^2)^3 (1 - \mathfrak{t}^4) (1 - \mathfrak{t}^6)^2},\\
&\langle W_{\ydiagram{4}}W_{\overline{\ydiagram{4}}}\rangle^{U(3)\text{ ADHM-}[2](C)}(\mathfrak{t})
=
\frac{1 + 8 \mathfrak{t}^2 + 25 \mathfrak{t}^4 + 41 \mathfrak{t}^6 + 34 \mathfrak{t}^8 + 10 \mathfrak{t}^{10} + 10 \mathfrak{t}^{12}
- 6 \mathfrak{t}^{14} - 2 \mathfrak{t}^{16} - 2 \mathfrak{t}^{20} + \mathfrak{t}^{22}}{(1 - \mathfrak{t}^2)^4 (1 - \mathfrak{t}^4) (1 - \mathfrak{t}^6)},\\
&\langle W_{\ydiagram{5}}W_{\overline{\ydiagram{5}}}\rangle^{U(3)\text{ ADHM-}[2](C)}(\mathfrak{t})\nonumber \\
&=
\frac{
1 + 10 \mathfrak{t}^2 + 33 \mathfrak{t}^4 + 59 \mathfrak{t}^6 + 48 \mathfrak{t}^8 + 17 \mathfrak{t}^{10} + 14 \mathfrak{t}^{12} - 7 \mathfrak{t}^{14} - 2 \mathfrak{t}^{16} - 4 \mathfrak{t}^{18} - 2 \mathfrak{t}^{20} + 2 \mathfrak{t}^{22} - 2 \mathfrak{t}^{24} + \mathfrak{t}^{26}
}{(1 - \mathfrak{t}^2)^4 (1 - \mathfrak{t}^4) (1 - \mathfrak{t}^6)},\\
&\langle W_{\ydiagram{6}}W_{\overline{\ydiagram{6}}}\rangle^{U(3)\text{ ADHM-}[2](C)}(\mathfrak{t})\nonumber \\
&=
\frac{
1
}{(1 - \mathfrak{t}^2)^3 (1 - \mathfrak{t}^4) (1 - \mathfrak{t}^6)^2}
(1 + 13 \mathfrak{t}^2 + 55 \mathfrak{t}^4 + 134 \mathfrak{t}^6 + 186 \mathfrak{t}^8 + 168 \mathfrak{t}^{10} + 110 \mathfrak{t}^{12} + 37 \mathfrak{t}^{14} + 10 \mathfrak{t}^{16} - 16 \mathfrak{t}^{18} - 13 \mathfrak{t}^{20}\nonumber \\
&\quad - 8 \mathfrak{t}^{22} - 4 \mathfrak{t}^{24} - 2 \mathfrak{t}^{26} + \mathfrak{t}^{30} - \mathfrak{t}^{32} + \mathfrak{t}^{34}),\\
&\langle W_{\ydiagram{2}}W_{\overline{\ydiagram{2}}}\rangle^{U(3)\text{ ADHM-}[3](C)}(\mathfrak{t})
=
\frac{
1 - \mathfrak{t} + 3 \mathfrak{t}^3 - \mathfrak{t}^4 + 2 \mathfrak{t}^5 - \mathfrak{t}^6 - \mathfrak{t}^7 + 2 \mathfrak{t}^8 + \mathfrak{t}^9 - \mathfrak{t}^{10}}{(1 - \mathfrak{t}) (1 - \mathfrak{t}^2)^2 (1 - \mathfrak{t}^3)^3},\\
&\langle W_{\ydiagram{3}}W_{\overline{\ydiagram{3}}}\rangle^{U(3)\text{ ADHM-}[3](C)}(\mathfrak{t})\nonumber \\
&=
\frac{
1 - \mathfrak{t} + 6 \mathfrak{t}^{3} - 4 \mathfrak{t}^{4} + 4 \mathfrak{t}^{5} + 6 \mathfrak{t}^{6} - 4 \mathfrak{t}^{7} + 4 \mathfrak{t}^{8} + 7 \mathfrak{t}^{9} - 6 \mathfrak{t}^{10} + 9 \mathfrak{t}^{11} - 2 \mathfrak{t}^{12} - 4 \mathfrak{t}^{13} + 2 \mathfrak{t}^{14} + 4 \mathfrak{t}^{15} - 2 \mathfrak{t}^{18} - \mathfrak{t}^{19} + \mathfrak{t}^{20}
}{
(1 - \mathfrak{t}) (1 - \mathfrak{t}^2)^2 (1 - \mathfrak{t}^3)^2 (1 - \mathfrak{t}^9)
},\\
&\langle W_{\ydiagram{4}}W_{\overline{\ydiagram{4}}}\rangle^{U(3)\text{ ADHM-}[3](C)}(\mathfrak{t})\nonumber \\
&=
\frac{1}{
(1 - \mathfrak{t}) (1 - \mathfrak{t}^2)^2 (1 - \mathfrak{t}^3)^2 (1 - \mathfrak{t}^9)}
(1 - \mathfrak{t} + 8 \mathfrak{t}^{3} - 6 \mathfrak{t}^{4} + 6 \mathfrak{t}^{5} + 9 \mathfrak{t}^{6} - 5 \mathfrak{t}^{7} + 6 \mathfrak{t}^{8} + 13 \mathfrak{t}^{9} - 13 \mathfrak{t}^{10} + 16 \mathfrak{t}^{11} - 6 \mathfrak{t}^{12}\nonumber \\
&\quad - 2 \mathfrak{t}^{13} + \mathfrak{t}^{14} + 8 \mathfrak{t}^{15} - 3 \mathfrak{t}^{16} - 2 \mathfrak{t}^{18} + 2 \mathfrak{t}^{20} - 2 \mathfrak{t}^{21} - \mathfrak{t}^{23} + \mathfrak{t}^{24}),\\
&\langle W_{\ydiagram{5}}W_{\overline{\ydiagram{5}}}\rangle^{U(3)\text{ ADHM-}[3](C)}(\mathfrak{t})\nonumber \\
&=
\frac{1}{
(1 - \mathfrak{t}) (1 - \mathfrak{t}^2)^2 (1 - \mathfrak{t}^3)^2 (1 - \mathfrak{t}^9)}
(1 - \mathfrak{t} + 10 \mathfrak{t}^{3} - 8 \mathfrak{t}^{4} + 8 \mathfrak{t}^{5} + 13 \mathfrak{t}^{6} - 7 \mathfrak{t}^{7} + 8 \mathfrak{t}^{8} + 21 \mathfrak{t}^{9} - 20 \mathfrak{t}^{10} + 25 \mathfrak{t}^{11} - 14 \mathfrak{t}^{12}\nonumber \\
&\quad + 2 \mathfrak{t}^{13} - 2 \mathfrak{t}^{14} + 17 \mathfrak{t}^{15} - 9 \mathfrak{t}^{16} + 2 \mathfrak{t}^{17} - 5 \mathfrak{t}^{18} + 2 \mathfrak{t}^{19} + 3 \mathfrak{t}^{20} - 4 \mathfrak{t}^{21} - 2 \mathfrak{t}^{23} + 4 \mathfrak{t}^{24} - 2 \mathfrak{t}^{25} - \mathfrak{t}^{27} + \mathfrak{t}^{28}),\\
&\langle W_{\ydiagram{6}}W_{\overline{\ydiagram{6}}}\rangle^{U(3)\text{ ADHM-}[3](C)}(\mathfrak{t})\nonumber \\
&=
\frac{1}{
(1 - \mathfrak{t}) (1 - \mathfrak{t}^2)^2 (1 - \mathfrak{t}^3)^2 (1 - \mathfrak{t}^9)}
(1 - \mathfrak{t} + 12 \mathfrak{t}^{3} - 10 \mathfrak{t}^{4} + 10 \mathfrak{t}^{5} + 18 \mathfrak{t}^{6} - 10 \mathfrak{t}^{7} + 11 \mathfrak{t}^{8} + 30 \mathfrak{t}^{9} - 29 \mathfrak{t}^{10} + 36 \mathfrak{t}^{11}\nonumber \\
&\quad - 22 \mathfrak{t}^{12} + 8 \mathfrak{t}^{13} - 9 \mathfrak{t}^{14} + 29 \mathfrak{t}^{15} - 18 \mathfrak{t}^{16} + 9 \mathfrak{t}^{17} - 11 \mathfrak{t}^{18} + 6 \mathfrak{t}^{19} - 4 \mathfrak{t}^{21} + \mathfrak{t}^{22} - 2 \mathfrak{t}^{23} + 5 \mathfrak{t}^{24} - 6 \mathfrak{t}^{25} + 2 \mathfrak{t}^{26} - 2 \mathfrak{t}^{27}\nonumber \\
&\quad + 4 \mathfrak{t}^{28} - 2 \mathfrak{t}^{29} - \mathfrak{t}^{31} + \mathfrak{t}^{32}),\\
%
%
&\langle W_{\ydiagram{2}}W_{\overline{\ydiagram{2}}}\rangle^{U(3)\text{ ADHM-}[4](C)}(\mathfrak{t})
=
\frac{
1 - \mathfrak{t}^2 + 5 \mathfrak{t}^4 - \mathfrak{t}^6 + 3 \mathfrak{t}^8 - \mathfrak{t}^{10} + 3 \mathfrak{t}^{12} - \mathfrak{t}^{14}
}{
(1 - \mathfrak{t}^2)^3 (1 - \mathfrak{t}^4)^3},\\
&\langle W_{\ydiagram{3}}W_{\overline{\ydiagram{3}}}\rangle^{U(3)\text{ ADHM-}[4](C)}(\mathfrak{t})\nonumber \\
&=
\frac{
1 + 8 \mathfrak{t}^4 + 6 \mathfrak{t}^6 + 21 \mathfrak{t}^8 + 9 \mathfrak{t}^{10} + 29 \mathfrak{t}^{12} + 10 \mathfrak{t}^{14} + 20 \mathfrak{t}^{16} + 3 \mathfrak{t}^{18} + 11 \mathfrak{t}^{20} + 3 \mathfrak{t}^{22} + 4 \mathfrak{t}^{24} - 4 \mathfrak{t}^{26} - 2 \mathfrak{t}^{28} + \mathfrak{t}^{30}
}{
(1 - \mathfrak{t}^2)^2 (1 - \mathfrak{t}^4)^2 (1 - \mathfrak{t}^6) (1 - \mathfrak{t}^{12})
},\\
&\langle W_{\ydiagram{4}}W_{\overline{\ydiagram{4}}}\rangle^{U(3)\text{ ADHM-}[4](C)}(\mathfrak{t})\nonumber \\
&=
\frac{
1 - \mathfrak{t}^2 + 10 \mathfrak{t}^4 - \mathfrak{t}^6 + 22 \mathfrak{t}^8 - 5 \mathfrak{t}^{10} + 28 \mathfrak{t}^{12} - 5 \mathfrak{t}^{14} + 7 \mathfrak{t}^{16} - 2 \mathfrak{t}^{18} + 12 \mathfrak{t}^{20} - 3 \mathfrak{t}^{22} - \mathfrak{t}^{24} - 2 \mathfrak{t}^{26} - \mathfrak{t}^{28} + \mathfrak{t}^{30}
}{
(1 - \mathfrak{t}^2)^3 (1 - \mathfrak{t}^4)^2 (1 - \mathfrak{t}^{12})},\\
&\langle W_{\ydiagram{5}}W_{\overline{\ydiagram{5}}}\rangle^{U(3)\text{ ADHM-}[4](C)}(\mathfrak{t})\nonumber \\
&=
\frac{
1 - \mathfrak{t}^2 + 12 \mathfrak{t}^4 - \mathfrak{t}^6 + 30 \mathfrak{t}^8 - 4 \mathfrak{t}^{10} + 39 \mathfrak{t}^{12} - 6 \mathfrak{t}^{14} + 9 \mathfrak{t}^{16} - 3 \mathfrak{t}^{18} + 15 \mathfrak{t}^{20} - \mathfrak{t}^{22} - 2 \mathfrak{t}^{24} - 3 \mathfrak{t}^{26} - \mathfrak{t}^{28} - \mathfrak{t}^{32} + \mathfrak{t}^{34}
}{(1 - \mathfrak{t}^2)^3 (1 - \mathfrak{t}^4)^2 (1 - \mathfrak{t}^{12})},\\
&\langle W_{\ydiagram{6}}W_{\overline{\ydiagram{6}}}\rangle^{U(3)\text{ ADHM-}[4](C)}(\mathfrak{t})\nonumber \\
&=
\frac{
1
}{
(1 - \mathfrak{t}^2)^2 (1 - \mathfrak{t}^4)^2 (1 - \mathfrak{t}^6) (1 - \mathfrak{t}^{12})}
(1 + 14 \mathfrak{t}^4 + 12 \mathfrak{t}^6 + 52 \mathfrak{t}^8 + 34 \mathfrak{t}^{10} + 89 \mathfrak{t}^{12} + 43 \mathfrak{t}^{14} + 57 \mathfrak{t}^{16} + 28 \mathfrak{t}^{20} + 14 \mathfrak{t}^{22}\nonumber \\
&\quad + 18 \mathfrak{t}^{24} - 9 \mathfrak{t}^{26} - 7 \mathfrak{t}^{28} - 5 \mathfrak{t}^{30} - 4 \mathfrak{t}^{32} - 2 \mathfrak{t}^{34} + \mathfrak{t}^{42}).
\end{align}
}

\subsubsection{$U(4)$}
{\fontsize{9pt}{1pt}\selectfont
\begin{align}
&\langle W_{\ydiagram{2}}W_{\overline{\ydiagram{2}}}\rangle^{U(4)\text{ ADHM-}[1](C)}(\mathfrak{t})
=
\frac{1 + 4 \mathfrak{t} + 5 \mathfrak{t}^2 + 12 \mathfrak{t}^3 + 13 \mathfrak{t}^4 + 14 \mathfrak{t}^5 + 6 \mathfrak{t}^6 + 6 \mathfrak{t}^7 - \mathfrak{t}^{10}}{(1 - \mathfrak{t})^2 (1 - \mathfrak{t}^2)^4 (1 - \mathfrak{t}^3)^2},\\
%
&\langle W_{\ydiagram{3}}W_{\overline{\ydiagram{3}}}\rangle^{U(4)\text{ ADHM-}[1](C)}(\mathfrak{t})\nonumber \\
&=
\frac{
1 + 6 \mathfrak{t} + 9 \mathfrak{t}^2 + 24 \mathfrak{t}^3 + 25 \mathfrak{t}^4 + 32 \mathfrak{t}^5 + 17 \mathfrak{t}^6 + 14 \mathfrak{t}^7 - 4 \mathfrak{t}^8 - \mathfrak{t}^{10} - 4 \mathfrak{t}^{12} + \mathfrak{t}^{14}
}{(1 - \mathfrak{t})^2 (1 - \mathfrak{t}^2)^4 (1 - \mathfrak{t}^3)^2},\\
&\langle W_{\ydiagram{4}}W_{\overline{\ydiagram{4}}}\rangle^{U(4)\text{ ADHM-}[1](C)}(\mathfrak{t})\nonumber \\
&=
\frac{1}{(1 - \mathfrak{t})^2 (1 - \mathfrak{t}^2)^2 (1 - \mathfrak{t}^3)^2 (1 - \mathfrak{t}^4)^2}
(1 + 8 \mathfrak{t} + 16 \mathfrak{t}^2 + 56 \mathfrak{t}^3 + 74 \mathfrak{t}^4 + 144 \mathfrak{t}^5 + 141 \mathfrak{t}^6 + 178 \mathfrak{t}^7 + 118 \mathfrak{t}^8 + 104 \mathfrak{t}^9\nonumber \\
&\quad + 30 \mathfrak{t}^{10} + 18 \mathfrak{t}^{11} - 15 \mathfrak{t}^{12} - 4 \mathfrak{t}^{13} - 16 \mathfrak{t}^{14} - 4 \mathfrak{t}^{15} - 11 \mathfrak{t}^{16} + 2 \mathfrak{t}^{18} + 2 \mathfrak{t}^{19} - 4 \mathfrak{t}^{20} + \mathfrak{t}^{22} + 2 \mathfrak{t}^{23} - \mathfrak{t}^{24}),\\
&\langle W_{\ydiagram{5}}W_{\overline{\ydiagram{5}}}\rangle^{U(4)\text{ ADHM-}[1](C)}(\mathfrak{t})\nonumber \\
&=
\frac{1}{(1 - \mathfrak{t})^4 (1 - \mathfrak{t}^2)^2 (1 - \mathfrak{t}^3)^2}
(1 + 8 \mathfrak{t} + 3 \mathfrak{t}^2 + 46 \mathfrak{t}^3 - 20 \mathfrak{t}^4 + 80 \mathfrak{t}^5 - 68 \mathfrak{t}^6 + 92 \mathfrak{t}^7 - 104 \mathfrak{t}^8 + 100 \mathfrak{t}^9 - 96 \mathfrak{t}^{10} + 88 \mathfrak{t}^{11}\nonumber \\
&\quad - 81 \mathfrak{t}^{12} + 66 \mathfrak{t}^{13} - 62 \mathfrak{t}^{14} + 56 \mathfrak{t}^{15} - 40 \mathfrak{t}^{16} + 28 \mathfrak{t}^{17} - 27 \mathfrak{t}^{18} + 22 \mathfrak{t}^{19} - 11 \mathfrak{t}^{20} + 6 \mathfrak{t}^{21} - 6 \mathfrak{t}^{22} + 4 \mathfrak{t}^{23} - \mathfrak{t}^{24}),\\
&\langle W_{\ydiagram{6}}W_{\overline{\ydiagram{6}}}\rangle^{U(4)\text{ ADHM-}[1](C)}(\mathfrak{t})\nonumber \\
&=
\frac{1}{(1 - \mathfrak{t})^2 (1 - \mathfrak{t}^2)^2 (1 - \mathfrak{t}^3)^2 (1 - \mathfrak{t}^4)^2}
(1 + 12 \mathfrak{t} + 29 \mathfrak{t}^2 + 108 \mathfrak{t}^3 + 172 \mathfrak{t}^4 + 302 \mathfrak{t}^5 + 388 \mathfrak{t}^6 + 368 \mathfrak{t}^7 + 413 \mathfrak{t}^8 + 156 \mathfrak{t}^9\nonumber \\
&\quad + 233 \mathfrak{t}^{10} - 72 \mathfrak{t}^{11} + 104 \mathfrak{t}^{12} - 114 \mathfrak{t}^{13} + 41 \mathfrak{t}^{14} - 84 \mathfrak{t}^{15} + 9 \mathfrak{t}^{16} - 40 \mathfrak{t}^{17} + 10 \mathfrak{t}^{18} - 12 \mathfrak{t}^{19} - 11 \mathfrak{t}^{20} - 2 \mathfrak{t}^{22} + 10 \mathfrak{t}^{23}\nonumber \\
&\quad - 13 \mathfrak{t}^{24} + 10 \mathfrak{t}^{25} - 7 \mathfrak{t}^{26} + 10 \mathfrak{t}^{27} - 12 \mathfrak{t}^{28} + 8 \mathfrak{t}^{29} - 3 \mathfrak{t}^{30} + 6 \mathfrak{t}^{31} - 6 \mathfrak{t}^{32} + 2 \mathfrak{t}^{33} - \mathfrak{t}^{34} + 2 \mathfrak{t}^{35} - \mathfrak{t}^{36}),\\
%
%
&\langle W_{\ydiagram{2}}W_{\overline{\ydiagram{2}}}\rangle^{U(4)\text{ ADHM-}[2](C)}(\mathfrak{t})\nonumber \\
&=
\frac{
1 + 6 \mathfrak{t}^2 + 20 \mathfrak{t}^4 + 56 \mathfrak{t}^6 + 118 \mathfrak{t}^8 + 173 \mathfrak{t}^{10} + 201 \mathfrak{t}^{12} + 178 \mathfrak{t}^{14} + 127 \mathfrak{t}^{16} + 58 \mathfrak{t}^{18} + 21 \mathfrak{t}^{20} + 2 \mathfrak{t}^{22} - \mathfrak{t}^{26}
}{(1 - \mathfrak{t}^2)^2 (1 - \mathfrak{t}^4)^4 (1 - \mathfrak{t}^6)^2},\\
&\langle W_{\ydiagram{3}}W_{\overline{\ydiagram{3}}}\rangle^{U(4)\text{ ADHM-}[2](C)}(\mathfrak{t})\nonumber \\
&=
\frac{
1 + 6 \mathfrak{t}^2 + 17 \mathfrak{t}^4 + 56 \mathfrak{t}^6 + 91 \mathfrak{t}^8 + 119 \mathfrak{t}^{10} + 100 \mathfrak{t}^{12} + 71 \mathfrak{t}^{14} + 21 \mathfrak{t}^{16} + 8 \mathfrak{t}^{18} - 5 \mathfrak{t}^{20} - 5 \mathfrak{t}^{22} - \mathfrak{t}^{24} + \mathfrak{t}^{26}
}{(1 - \mathfrak{t}^2)^4 (1 - \mathfrak{t}^4)^2 (1 - \mathfrak{t}^6)^2},\\
&\langle W_{\ydiagram{4}}W_{\overline{\ydiagram{4}}}\rangle^{U(4)\text{ ADHM-}[2](C)}(\mathfrak{t})\nonumber \\
&=
\frac{1}{(1 - \mathfrak{t}^2)^2 (1 - \mathfrak{t}^4)^2 (1 - \mathfrak{t}^6)^2 (1 - \mathfrak{t}^8)^2}
(1 + 10 \mathfrak{t}^2 + 43 \mathfrak{t}^4 + 163 \mathfrak{t}^6 + 437 \mathfrak{t}^8 + 915 \mathfrak{t}^{10} + 1538 \mathfrak{t}^{12} + 2120 \mathfrak{t}^{14} + 2438 \mathfrak{t}^{16}\nonumber \\
&\quad + 2311 \mathfrak{t}^{18} + 1843 \mathfrak{t}^{20} + 1140 \mathfrak{t}^{22} + 561 \mathfrak{t}^{24} + 123 \mathfrak{t}^{26} - 43 \mathfrak{t}^{28} - 92 \mathfrak{t}^{30} - 53 \mathfrak{t}^{32} - 25 \mathfrak{t}^{34} + 5 \mathfrak{t}^{38} + 4 \mathfrak{t}^{40} + 2 \mathfrak{t}^{42} - \mathfrak{t}^{44}),\\
&\langle W_{\ydiagram{5}}W_{\overline{\ydiagram{5}}}\rangle^{U(4)\text{ ADHM-}[2](C)}(\mathfrak{t})\nonumber \\
&=
\frac{1}{(1 - \mathfrak{t}^2)^4 (1 - \mathfrak{t}^4) (1 - \mathfrak{t}^6)^2 (1 - \mathfrak{t}^8)}
(1 + 10 \mathfrak{t}^2 + 33 \mathfrak{t}^4 + 134 \mathfrak{t}^6 + 272 \mathfrak{t}^8 + 484 \mathfrak{t}^{10} + 569 \mathfrak{t}^{12} + 609 \mathfrak{t}^{14} + 388 \mathfrak{t}^{16}\nonumber \\
&\quad + 258 \mathfrak{t}^{18} + 29 \mathfrak{t}^{20} + 10 \mathfrak{t}^{22} - 58 \mathfrak{t}^{24} - 31 \mathfrak{t}^{26} - 25 \mathfrak{t}^{28} - \mathfrak{t}^{30} + 3 \mathfrak{t}^{32} + 2 \mathfrak{t}^{36} + 2 \mathfrak{t}^{40} - \mathfrak{t}^{42}),\\
&\langle W_{\ydiagram{6}}W_{\overline{\ydiagram{6}}}\rangle^{U(4)\text{ ADHM-}[2](C)}(\mathfrak{t})\nonumber \\
&=
\frac{1}{(1 - \mathfrak{t}^2)^2 (1 - \mathfrak{t}^4)^3 (1 - \mathfrak{t}^6)^2 (1 - \mathfrak{t}^8)}
(1 + 14 \mathfrak{t}^2 + 67 \mathfrak{t}^4 + 275 \mathfrak{t}^6 + 787 \mathfrak{t}^8 + 1659 \mathfrak{t}^{10} + 2664 \mathfrak{t}^{12} + 3385 \mathfrak{t}^{14} + 3389 \mathfrak{t}^{16}\nonumber \\
&\quad + 2615 \mathfrak{t}^{18} + 1522 \mathfrak{t}^{20} + 539 \mathfrak{t}^{22} - \mathfrak{t}^{24} - 257 \mathfrak{t}^{26} - 274 \mathfrak{t}^{28} - 195 \mathfrak{t}^{30} - 75 \mathfrak{t}^{32} - 21 \mathfrak{t}^{34} + 4 \mathfrak{t}^{36} + 10 \mathfrak{t}^{38} + 10 \mathfrak{t}^{40} + 6 \mathfrak{t}^{42}\nonumber \\
&\quad + 2 \mathfrak{t}^{44} + 2 \mathfrak{t}^{46} + \mathfrak{t}^{48} - \mathfrak{t}^{52}),\\
&\langle W_{\ydiagram{2}}W_{\overline{\ydiagram{2}}}\rangle^{U(4)\text{ ADHM-}[3](C)}(\mathfrak{t})\nonumber \\
&=
\frac{1}{(1 - \mathfrak{t})^2 (1 - \mathfrak{t}^2)^2 (1 - \mathfrak{t}^3) (1 - \mathfrak{t}^6)^2 (1 - \mathfrak{t}^9)}
(1 - 2 \mathfrak{t} + \mathfrak{t}^{2} + 5 \mathfrak{t}^{3} - 8 \mathfrak{t}^{4} + 5 \mathfrak{t}^{5} + 5 \mathfrak{t}^{6} - 8 \mathfrak{t}^{7} + 11 \mathfrak{t}^{8} - 9 \mathfrak{t}^{10} + 16 \mathfrak{t}^{11}\nonumber \\
&\quad - 6 \mathfrak{t}^{12} - \mathfrak{t}^{13} + 13 \mathfrak{t}^{14} - 10 \mathfrak{t}^{15} + 2 \mathfrak{t}^{16} + 9 \mathfrak{t}^{17} - 9 \mathfrak{t}^{18} + 5 \mathfrak{t}^{19} + 3 \mathfrak{t}^{20} - 5 \mathfrak{t}^{21} + \mathfrak{t}^{22} + 2 \mathfrak{t}^{23} - \mathfrak{t}^{24}),\\
&\langle W_{\ydiagram{3}}W_{\overline{\ydiagram{3}}}\rangle^{U(4)\text{ ADHM-}[3](C)}(\mathfrak{t})\nonumber \\
&=
\frac{1}{(1 - \mathfrak{t})^2 (1 - \mathfrak{t}^2)^2 (1 - \mathfrak{t}^3) (1 - \mathfrak{t}^6)^2 (1 - \mathfrak{t}^9)}
(1 - 2 \mathfrak{t} + \mathfrak{t}^{2} + 7 \mathfrak{t}^{3} - 12 \mathfrak{t}^{4} + 9 \mathfrak{t}^{5} + 8 \mathfrak{t}^{6} - 14 \mathfrak{t}^{7} + 17 \mathfrak{t}^{8} + 7 \mathfrak{t}^{9} - 19 \mathfrak{t}^{10}\nonumber \\
&\quad + 30 \mathfrak{t}^{11} - 10 \mathfrak{t}^{12} - 5 \mathfrak{t}^{13} + 31 \mathfrak{t}^{14} - 23 \mathfrak{t}^{15} + 3 \mathfrak{t}^{16} + 21 \mathfrak{t}^{17} - 18 \mathfrak{t}^{18} + 6 \mathfrak{t}^{19} + 7 \mathfrak{t}^{20} - 5 \mathfrak{t}^{21} - \mathfrak{t}^{22} + 4 \mathfrak{t}^{23} - 5 \mathfrak{t}^{24} + 3 \mathfrak{t}^{25}\nonumber \\
&\quad - 2 \mathfrak{t}^{27} + \mathfrak{t}^{28}),\\
&\langle W_{\ydiagram{4}}W_{\overline{\ydiagram{4}}}\rangle^{U(4)\text{ ADHM-}[3](C)}(\mathfrak{t})\nonumber \\
&=
\frac{1}{(1 - \mathfrak{t})^2 (1 - \mathfrak{t}^2) (1 - \mathfrak{t}^3) (1 - \mathfrak{t}^4) (1 - \mathfrak{t}^6) (1 - \mathfrak{t}^9) (1 - \mathfrak{t}^{12})}
 (1 - 2 \mathfrak{t} + 2 \mathfrak{t}^{2} + 7 \mathfrak{t}^{3} - 15 \mathfrak{t}^{4} + 22 \mathfrak{t}^{5} - 4 \mathfrak{t}^{6} - 9 \mathfrak{t}^{7} + 39 \mathfrak{t}^{8}\nonumber \\
&\quad + \mathfrak{t}^{9} - 25 \mathfrak{t}^{10} + 93 \mathfrak{t}^{11} - 55 \mathfrak{t}^{12} + 36 \mathfrak{t}^{13} + 81 \mathfrak{t}^{14} - 64 \mathfrak{t}^{15} + 57 \mathfrak{t}^{16} + 60 \mathfrak{t}^{17} - 62 \mathfrak{t}^{18} + 75 \mathfrak{t}^{19} + 37 \mathfrak{t}^{20} - 68 \mathfrak{t}^{21} + 84 \mathfrak{t}^{22}\nonumber \\
&\quad - 13 \mathfrak{t}^{23} - 30 \mathfrak{t}^{24} + 59 \mathfrak{t}^{25} - 32 \mathfrak{t}^{26} + 6 \mathfrak{t}^{27} + 4 \mathfrak{t}^{28} + 7 \mathfrak{t}^{29} - 17 \mathfrak{t}^{30} + 17 \mathfrak{t}^{31} - 14 \mathfrak{t}^{32} + 3 \mathfrak{t}^{33} + \mathfrak{t}^{34} - 5 \mathfrak{t}^{35} + 4 \mathfrak{t}^{36} - 2 \mathfrak{t}^{37}\nonumber \\
&\quad + \mathfrak{t}^{38} - \mathfrak{t}^{39} + 2 \mathfrak{t}^{41} - \mathfrak{t}^{42}),\\
&\langle W_{\ydiagram{5}}W_{\overline{\ydiagram{5}}}\rangle^{U(4)\text{ ADHM-}[3](C)}(\mathfrak{t})\nonumber \\
&=
\frac{1 + \mathfrak{t}^2}{(1 - \mathfrak{t})^4 (1 - \mathfrak{t}^3) (1 - \mathfrak{t}^6) (1 - \mathfrak{t}^9) (1 - \mathfrak{t}^{12})}
 (1 - 4 \mathfrak{t} + 7 \mathfrak{t}^{2} + 3 \mathfrak{t}^{3} - 33 \mathfrak{t}^{4} + 67 \mathfrak{t}^{5} - 65 \mathfrak{t}^{6} + 29 \mathfrak{t}^{7} + 9 \mathfrak{t}^{8} + 22 \mathfrak{t}^{9} - 132 \mathfrak{t}^{10}\nonumber \\
&\quad + 280 \mathfrak{t}^{11} - 352 \mathfrak{t}^{12} + 333 \mathfrak{t}^{13} - 255 \mathfrak{t}^{14} + 218 \mathfrak{t}^{15} - 277 \mathfrak{t}^{16} + 429 \mathfrak{t}^{17} - 523 \mathfrak{t}^{18} + 480 \mathfrak{t}^{19} - 350 \mathfrak{t}^{20} + 245 \mathfrak{t}^{21} - 198 \mathfrak{t}^{22}\nonumber \\
&\quad + 183 \mathfrak{t}^{23} - 159 \mathfrak{t}^{24} + 118 \mathfrak{t}^{25} - 71 \mathfrak{t}^{26} + 32 \mathfrak{t}^{27} - 15 \mathfrak{t}^{28} + 18 \mathfrak{t}^{29} - 23 \mathfrak{t}^{30} + 21 \mathfrak{t}^{31} - 23 \mathfrak{t}^{32} + 27 \mathfrak{t}^{33} - 26 \mathfrak{t}^{34} + 23 \mathfrak{t}^{35}\nonumber \\
&\quad - 22 \mathfrak{t}^{36} + 19 \mathfrak{t}^{37} - 13 \mathfrak{t}^{38} + 9 \mathfrak{t}^{39} - 7 \mathfrak{t}^{40} + 4 \mathfrak{t}^{41} - \mathfrak{t}^{42}),\\
&\langle W_{\ydiagram{6}}W_{\overline{\ydiagram{6}}}\rangle^{U(4)\text{ ADHM-}[3](C)}(\mathfrak{t})\nonumber \\
&=
\frac{1}{(1 - \mathfrak{t})^2 (1 - \mathfrak{t}^2) (1 - \mathfrak{t}^3) (1 - \mathfrak{t}^4) (1 - \mathfrak{t}^6) (1 - \mathfrak{t}^9) (1 - \mathfrak{t}^{12})}
 (1 - 2 \mathfrak{t} + 2 \mathfrak{t}^{2} + 11 \mathfrak{t}^{3} - 23 \mathfrak{t}^{4} + 34 \mathfrak{t}^{5} - 3 \mathfrak{t}^{6} - 19 \mathfrak{t}^{7}\nonumber \\
&\quad + 73 \mathfrak{t}^{8} + 8 \mathfrak{t}^{9} - 55 \mathfrak{t}^{10} + 201 \mathfrak{t}^{11} - 104 \mathfrak{t}^{12} + 66 \mathfrak{t}^{13} + 209 \mathfrak{t}^{14} - 164 \mathfrak{t}^{15} + 162 \mathfrak{t}^{16} + 110 \mathfrak{t}^{17} - 72 \mathfrak{t}^{18} + 80 \mathfrak{t}^{19} + 232 \mathfrak{t}^{20}\nonumber \\
&\quad - 310 \mathfrak{t}^{21} + 346 \mathfrak{t}^{22} - 158 \mathfrak{t}^{23} + 37 \mathfrak{t}^{24} + 44 \mathfrak{t}^{25} + 2 \mathfrak{t}^{26} - 29 \mathfrak{t}^{27} + 21 \mathfrak{t}^{28} + 26 \mathfrak{t}^{29} - 67 \mathfrak{t}^{30} + 65 \mathfrak{t}^{31} - 54 \mathfrak{t}^{32} + 22 \mathfrak{t}^{33}\nonumber \\
&\quad - 13 \mathfrak{t}^{34} - 8 \mathfrak{t}^{35} + \mathfrak{t}^{36} - 2 \mathfrak{t}^{37} + 2 \mathfrak{t}^{38} - 5 \mathfrak{t}^{39} + 2 \mathfrak{t}^{40} + 4 \mathfrak{t}^{41} - 3 \mathfrak{t}^{42} + 2 \mathfrak{t}^{43} - 5 \mathfrak{t}^{44} + 8 \mathfrak{t}^{45} - 6 \mathfrak{t}^{46} + 5 \mathfrak{t}^{47} - 5 \mathfrak{t}^{48} + 6 \mathfrak{t}^{49}\nonumber \\
&\quad - 5 \mathfrak{t}^{50} + 3 \mathfrak{t}^{51} - 2 \mathfrak{t}^{52} + 2 \mathfrak{t}^{53} - \mathfrak{t}^{54}),\\
%
%
&\langle W_{\ydiagram{2}}W_{\overline{\ydiagram{2}}}\rangle^{U(4)\text{ ADHM-}[4](C)}(\mathfrak{t})\nonumber \\
&=
\frac{1}{(1 - \mathfrak{t}^2)^3 (1 - \mathfrak{t}^4) (1 - \mathfrak{t}^6) (1 - \mathfrak{t}^8)^2 (1 - \mathfrak{t}^{12})}
(1 - \mathfrak{t}^2 + 7 \mathfrak{t}^4 - 3 \mathfrak{t}^6 + 19 \mathfrak{t}^8 - \mathfrak{t}^{10} + 42 \mathfrak{t}^{12} - 3 \mathfrak{t}^{14} + 60 \mathfrak{t}^{16} - 5 \mathfrak{t}^{18}\nonumber \\
&\quad + 65 \mathfrak{t}^{20} - 7 \mathfrak{t}^{22} + 46 \mathfrak{t}^{24} - 7 \mathfrak{t}^{26} + 27 \mathfrak{t}^{28} - 8 \mathfrak{t}^{30} + 10 \mathfrak{t}^{32} - 4 \mathfrak{t}^{34} + 3 \mathfrak{t}^{36} - \mathfrak{t}^{38}),\\
&\langle W_{\ydiagram{3}}W_{\overline{\ydiagram{3}}}\rangle^{U(4)\text{ ADHM-}[4](C)}(\mathfrak{t})\nonumber \\
&=
\frac{1}{(1 - \mathfrak{t}^2)^3 (1 - \mathfrak{t}^4) (1 - \mathfrak{t}^6) (1 - \mathfrak{t}^8)^2 (1 - \mathfrak{t}^{12})}
(1 - \mathfrak{t}^2 + 9 \mathfrak{t}^4 - 2 \mathfrak{t}^6 + 28 \mathfrak{t}^8 + 5 \mathfrak{t}^{10} + 73 \mathfrak{t}^{12} + 11 \mathfrak{t}^{14} + 108 \mathfrak{t}^{16} + 10 \mathfrak{t}^{18}\nonumber\\
&\quad + 113 \mathfrak{t}^{20} + 6 \mathfrak{t}^{22} + 76 \mathfrak{t}^{24} + 45 \mathfrak{t}^{28} - 7 \mathfrak{t}^{30} + 13 \mathfrak{t}^{32} - 7 \mathfrak{t}^{34} - 2 \mathfrak{t}^{40} + \mathfrak{t}^{42}),\\
&\langle W_{\ydiagram{4}}W_{\overline{\ydiagram{4}}}\rangle^{U(4)\text{ ADHM-}[4](C)}(\mathfrak{t})\nonumber \\
&=
\frac{1}{(1 - \mathfrak{t}^2)^3 (1 - \mathfrak{t}^4) (1 - \mathfrak{t}^6) (1 - \mathfrak{t}^8) (1 - \mathfrak{t}^{12}) (1 - \mathfrak{t}^{16})}
(1 - \mathfrak{t}^2 + 11 \mathfrak{t}^4 - 2 \mathfrak{t}^6 + 41 \mathfrak{t}^8 + 9 \mathfrak{t}^{10} + 126 \mathfrak{t}^{12} + 29 \mathfrak{t}^{14} + 231 \mathfrak{t}^{16}\nonumber \\
&\quad + 41 \mathfrak{t}^{18} + 317 \mathfrak{t}^{20} + 37 \mathfrak{t}^{22} + 337 \mathfrak{t}^{24} + 21 \mathfrak{t}^{26} + 283 \mathfrak{t}^{28} - 7 \mathfrak{t}^{30} + 169 \mathfrak{t}^{32} - 24 \mathfrak{t}^{34} + 84 \mathfrak{t}^{36} - 22 \mathfrak{t}^{38} + 30 \mathfrak{t}^{40} - 23 \mathfrak{t}^{42}\nonumber \\
&\quad - 8 \mathfrak{t}^{46} - 4 \mathfrak{t}^{48} + 3 \mathfrak{t}^{50} - \mathfrak{t}^{52} + 3 \mathfrak{t}^{54} - \mathfrak{t}^{56}),\\
&\langle W_{\ydiagram{5}}W_{\overline{\ydiagram{5}}}\rangle^{U(4)\text{ ADHM-}[4](C)}(\mathfrak{t})\nonumber \\
&=
\frac{1}{(1 - \mathfrak{t}^2)^3 (1 - \mathfrak{t}^4) (1 - \mathfrak{t}^6) (1 - \mathfrak{t}^8) (1 - \mathfrak{t}^{12}) (1 - \mathfrak{t}^{16})}
(1 - \mathfrak{t}^2 + 13 \mathfrak{t}^4 - 2 \mathfrak{t}^6 + 53 \mathfrak{t}^8 + 17 \mathfrak{t}^{10} + 175 \mathfrak{t}^{12} + 59 \mathfrak{t}^{14}\nonumber \\
&\quad + 350 \mathfrak{t}^{16} + 104 \mathfrak{t}^{18} + 479 \mathfrak{t}^{20} + 94 \mathfrak{t}^{22} + 520 \mathfrak{t}^{24} + 93 \mathfrak{t}^{26} + 423 \mathfrak{t}^{28} + 17 \mathfrak{t}^{30} + 243 \mathfrak{t}^{32} - 7 \mathfrak{t}^{34} + 111 \mathfrak{t}^{36} - 26 \mathfrak{t}^{38}\nonumber \\
&\quad + 35 \mathfrak{t}^{40} - 23 \mathfrak{t}^{42} - 15 \mathfrak{t}^{44} - 19 \mathfrak{t}^{46} - 18 \mathfrak{t}^{48} + 8 \mathfrak{t}^{50} - 3 \mathfrak{t}^{52} + 6 \mathfrak{t}^{54} + \mathfrak{t}^{58} + \mathfrak{t}^{60} - \mathfrak{t}^{62}),\\
&\langle W_{\ydiagram{6}}W_{\overline{\ydiagram{6}}}\rangle^{U(4)\text{ ADHM-}[4](C)}(\mathfrak{t})\nonumber \\
&=
\frac{1}{(1 - \mathfrak{t}^2)^3 (1 - \mathfrak{t}^4) (1 - \mathfrak{t}^6) (1 - \mathfrak{t}^8) (1 - \mathfrak{t}^{12}) (1 - \mathfrak{t}^{16})}
(1 - \mathfrak{t}^2 + 15 \mathfrak{t}^4 - 2 \mathfrak{t}^6 + 66 \mathfrak{t}^8 + 25 \mathfrak{t}^{10} + 233 \mathfrak{t}^{12} + 93 \mathfrak{t}^{14}\nonumber \\
&\quad + 501 \mathfrak{t}^{16} + 187 \mathfrak{t}^{18} + 708 \mathfrak{t}^{20} + 163 \mathfrak{t}^{22} + 787 \mathfrak{t}^{24} + 170 \mathfrak{t}^{26} + 663 \mathfrak{t}^{28} + 15 \mathfrak{t}^{30} + 375 \mathfrak{t}^{32} - 26 \mathfrak{t}^{34} + 189 \mathfrak{t}^{36} - 61 \mathfrak{t}^{38}\nonumber \\
&\quad + 63 \mathfrak{t}^{40} - 49 \mathfrak{t}^{42} - 12 \mathfrak{t}^{44} - 50 \mathfrak{t}^{46} - 32 \mathfrak{t}^{48} - \mathfrak{t}^{50} - 3 \mathfrak{t}^{52} + 8 \mathfrak{t}^{54} + 6 \mathfrak{t}^{58} + 2 \mathfrak{t}^{62} - \mathfrak{t}^{64} + \mathfrak{t}^{66} - \mathfrak{t}^{68}).
\end{align}
}

\subsubsection{$U(5)$}
{\fontsize{9pt}{1pt}\selectfont
\begin{align}
&\langle W_{\ydiagram{2}}W_{\overline{\ydiagram{2}}}\rangle^{U(5)\text{ ADHM-}[1](C)}(\mathfrak{t})\nonumber \\
&=
\frac{
1 + 2 \mathfrak{t} + 10 \mathfrak{t}^3 + 4 \mathfrak{t}^4 + 14 \mathfrak{t}^5 + 9 \mathfrak{t}^6 + 12 \mathfrak{t}^7 + 8 \mathfrak{t}^8 + 10 \mathfrak{t}^9 - 2 \mathfrak{t}^{10} + 6 \mathfrak{t}^{11} - 3 \mathfrak{t}^{12} + 2 \mathfrak{t}^{13} - \mathfrak{t}^{14}
}{(1 - \mathfrak{t})^4 (1 - \mathfrak{t}^2)^2 (1 - \mathfrak{t}^3)^2 (1 - \mathfrak{t}^4)^2},\\
&\langle W_{\ydiagram{3}}W_{\overline{\ydiagram{3}}}\rangle^{U(5)\text{ ADHM-}[1](C)}(\mathfrak{t})\nonumber \\
&=
\frac{1}{(1 - \mathfrak{t})^4 (1 - \mathfrak{t}^2)^2 (1 - \mathfrak{t}^3)^2 (1 - \mathfrak{t}^4)^2}
(1 + 4 \mathfrak{t} + 20 \mathfrak{t}^3 + 7 \mathfrak{t}^4 + 38 \mathfrak{t}^5 + 15 \mathfrak{t}^6 + 44 \mathfrak{t}^7 + 8 \mathfrak{t}^8 + 36 \mathfrak{t}^9 - 13 \mathfrak{t}^{10} + 22 \mathfrak{t}^{11}\nonumber \\
&\quad - 16 \mathfrak{t}^{12} + 8 \mathfrak{t}^{13} - 5 \mathfrak{t}^{14} - 2 \mathfrak{t}^{15} + 2 \mathfrak{t}^{16} - 2 \mathfrak{t}^{17} + \mathfrak{t}^{18}),\\
&\langle W_{\ydiagram{4}}W_{\overline{\ydiagram{4}}}\rangle^{U(5)\text{ ADHM-}[1](C)}(\mathfrak{t})\nonumber \\
&=
\frac{1}{(1 - \mathfrak{t})^4 (1 - \mathfrak{t}^2)^2 (1 - \mathfrak{t}^3)^2 (1 - \mathfrak{t}^4)^2}
(1 + 6 \mathfrak{t} + \mathfrak{t}^2 + 32 \mathfrak{t}^3 + 16 \mathfrak{t}^4 + 70 \mathfrak{t}^5 + 35 \mathfrak{t}^6 + 94 \mathfrak{t}^7 + 20 \mathfrak{t}^8 + 80 \mathfrak{t}^9 - 25 \mathfrak{t}^{10}\nonumber \\
&\quad + 44 \mathfrak{t}^{11} - 36 \mathfrak{t}^{12} + 20 \mathfrak{t}^{13} - 18 \mathfrak{t}^{14} - 4 \mathfrak{t}^{16} - 6 \mathfrak{t}^{17} + 10 \mathfrak{t}^{18} - 4 \mathfrak{t}^{19} - 2 \mathfrak{t}^{21} + \mathfrak{t}^{22} + 2 \mathfrak{t}^{23} - \mathfrak{t}^{24}),\\
%
&\langle W_{\ydiagram{5}}W_{\overline{\ydiagram{5}}}\rangle^{U(5)\text{ ADHM-}[1](C)}(\mathfrak{t})\nonumber \\
&=
\frac{1}{(1 - \mathfrak{t})^2 (1 - \mathfrak{t}^2)^2 (1 - \mathfrak{t}^3)^2 (1 - \mathfrak{t}^4)^2 (1 - \mathfrak{t}^5)^2}
(1 + 10 \mathfrak{t} + 22 \mathfrak{t}^2 + 80 \mathfrak{t}^3 + 171 \mathfrak{t}^4 + 370 \mathfrak{t}^5 + 633 \mathfrak{t}^6 + 1044 \mathfrak{t}^7 + 1433 \mathfrak{t}^8\nonumber \\
&\quad + 1882 \mathfrak{t}^9 + 2102 \mathfrak{t}^{10} + 2226 \mathfrak{t}^{11} + 2008 \mathfrak{t}^{12} + 1696 \mathfrak{t}^{13} + 1161 \mathfrak{t}^{14} + 724 \mathfrak{t}^{15} + 277 \mathfrak{t}^{16} + 24 \mathfrak{t}^{17} - 143 \mathfrak{t}^{18} - 176 \mathfrak{t}^{19}\nonumber \\
&\quad - 177 \mathfrak{t}^{20} - 124 \mathfrak{t}^{21} - 74 \mathfrak{t}^{22} - 42 \mathfrak{t}^{23} - 21 \mathfrak{t}^{24} - 4 \mathfrak{t}^{25} - 5 \mathfrak{t}^{26} + 2 \mathfrak{t}^{27} + 10 \mathfrak{t}^{29} + 7 \mathfrak{t}^{30} + 4 \mathfrak{t}^{31} - 3 \mathfrak{t}^{32} - 2 \mathfrak{t}^{34} + 4 \mathfrak{t}^{35}\nonumber \\
&\quad + 2 \mathfrak{t}^{36} - \mathfrak{t}^{38} - 2 \mathfrak{t}^{39} + \mathfrak{t}^{40}),\\
&\langle W_{\ydiagram{6}}W_{\overline{\ydiagram{6}}}\rangle^{U(5)\text{ ADHM-}[1](C)}(\mathfrak{t})\nonumber \\
&=
\frac{1}{(1 - \mathfrak{t})^3 (1 - \mathfrak{t}^2)^2 (1 - \mathfrak{t}^3)^2 (1 - \mathfrak{t}^4)^2 (1 - \mathfrak{t}^5)}
(1 + 11 \mathfrak{t} + 17 \mathfrak{t}^2 + 79 \mathfrak{t}^3 + 139 \mathfrak{t}^4 + 296 \mathfrak{t}^5 + 447 \mathfrak{t}^6 + 661 \mathfrak{t}^7 + 782 \mathfrak{t}^8\nonumber \\
&\quad + 876 \mathfrak{t}^9 + 790 \mathfrak{t}^{10} + 647 \mathfrak{t}^{11} + 461 \mathfrak{t}^{12} + 170 \mathfrak{t}^{13} + 131 \mathfrak{t}^{14} - 141 \mathfrak{t}^{15} + 23 \mathfrak{t}^{16} - 199 \mathfrak{t}^{17} + 15 \mathfrak{t}^{18} - 124 \mathfrak{t}^{19} - 15 \mathfrak{t}^{20}\nonumber \\
&\quad - 19 \mathfrak{t}^{21} - 12 \mathfrak{t}^{22} + 18 \mathfrak{t}^{23} - 39 \mathfrak{t}^{24} + 18 \mathfrak{t}^{25} - 11 \mathfrak{t}^{26} + 24 \mathfrak{t}^{27} - 18 \mathfrak{t}^{28} + 17 \mathfrak{t}^{29} - 12 \mathfrak{t}^{30} + 9 \mathfrak{t}^{31} - 9 \mathfrak{t}^{32} + 7 \mathfrak{t}^{33} - 2 \mathfrak{t}^{34}\nonumber \\
&\quad + 7 \mathfrak{t}^{35} - 6 \mathfrak{t}^{36} + 2 \mathfrak{t}^{37} - 2 \mathfrak{t}^{38} - \mathfrak{t}^{39} + 4 \mathfrak{t}^{40} - 3 \mathfrak{t}^{41} + 3 \mathfrak{t}^{42} - 3 \mathfrak{t}^{43} + \mathfrak{t}^{44}),\\
%
%
&\langle W_{\ydiagram{2}}W_{\overline{\ydiagram{2}}}\rangle^{U(5)\text{ ADHM-}[2](C)}(\mathfrak{t})\nonumber \\
&=
\frac{1}{(1 - \mathfrak{t}^2)^5 (1 - \mathfrak{t}^4) (1 - \mathfrak{t}^6)^2 (1 - \mathfrak{t}^8)^2}
(1 + 3 \mathfrak{t}^2 + 8 \mathfrak{t}^4 + 22 \mathfrak{t}^6 + 58 \mathfrak{t}^8 + 78 \mathfrak{t}^{10} + 145 \mathfrak{t}^{12} + 160 \mathfrak{t}^{14} + 192 \mathfrak{t}^{16} + 170 \mathfrak{t}^{18}\nonumber \\
&\quad + 140 \mathfrak{t}^{20} + 86 \mathfrak{t}^{22} + 62 \mathfrak{t}^{24} + 14 \mathfrak{t}^{26} + 15 \mathfrak{t}^{28} - 4 \mathfrak{t}^{30} + 3 \mathfrak{t}^{32} - \mathfrak{t}^{34}),\\
&\langle W_{\ydiagram{3}}W_{\overline{\ydiagram{3}}}\rangle^{U(5)\text{ ADHM-}[2](C)}(\mathfrak{t})\nonumber \\
&=
\frac{1}{(1 - \mathfrak{t}^2)^5 (1 - \mathfrak{t}^4) (1 - \mathfrak{t}^6)^2 (1 - \mathfrak{t}^8)^2}
(1 + 5 \mathfrak{t}^2 + 12 \mathfrak{t}^4 + 44 \mathfrak{t}^6 + 106 \mathfrak{t}^8 + 191 \mathfrak{t}^{10} + 315 \mathfrak{t}^{12} + 408 \mathfrak{t}^{14} + 466 \mathfrak{t}^{16}\nonumber \\
&\quad + 422 \mathfrak{t}^{18} + 362 \mathfrak{t}^{20} + 197 \mathfrak{t}^{22} + 138 \mathfrak{t}^{24} + 27 \mathfrak{t}^{26} + 5 \mathfrak{t}^{28} - 2 \mathfrak{t}^{30} - 11 \mathfrak{t}^{32} + 3 \mathfrak{t}^{34} - 2 \mathfrak{t}^{36} + \mathfrak{t}^{38}),\\
&\langle W_{\ydiagram{4}}W_{\overline{\ydiagram{4}}}\rangle^{U(5)\text{ ADHM-}[2](C)}(\mathfrak{t})\nonumber \\
&=
\frac{1}{(1 - \mathfrak{t}^2)^5 (1 - \mathfrak{t}^4) (1 - \mathfrak{t}^6)^2 (1 - \mathfrak{t}^8)^2}
(1 + 7 \mathfrak{t}^2 + 17 \mathfrak{t}^4 + 70 \mathfrak{t}^6 + 177 \mathfrak{t}^8 + 354 \mathfrak{t}^{10} + 613 \mathfrak{t}^{12} + 832 \mathfrak{t}^{14} + 985 \mathfrak{t}^{16}\nonumber \\
&\quad + 914 \mathfrak{t}^{18} + 765 \mathfrak{t}^{20} + 436 \mathfrak{t}^{22} + 260 \mathfrak{t}^{24} + 28 \mathfrak{t}^{26} - 6 \mathfrak{t}^{28} - 48 \mathfrak{t}^{30} - 24 \mathfrak{t}^{32} - 9 \mathfrak{t}^{34} - 4 \mathfrak{t}^{36} + 6 \mathfrak{t}^{38} + \mathfrak{t}^{40} + 2 \mathfrak{t}^{42} - \mathfrak{t}^{44}),\\
&\langle W_{\ydiagram{5}}W_{\overline{\ydiagram{5}}}\rangle^{U(5)\text{ ADHM-}[2](C)}(\mathfrak{t})\nonumber \\
&=
\frac{1}{(1 - \mathfrak{t}^2)^3 (1 - \mathfrak{t}^4) (1 - \mathfrak{t}^6)^2 (1 - \mathfrak{t}^8)^2 (1 - \mathfrak{t}^{10})^2}
(1 + 11 \mathfrak{t}^2 + 44 \mathfrak{t}^4 + 178 \mathfrak{t}^6 + 580 \mathfrak{t}^8 + 1564 \mathfrak{t}^{10} + 3572 \mathfrak{t}^{12} + 7053 \mathfrak{t}^{14}\nonumber \\
&\quad + 12154 \mathfrak{t}^{16} + 18504 \mathfrak{t}^{18} + 25159 \mathfrak{t}^{20} + 30609 \mathfrak{t}^{22} + 33507 \mathfrak{t}^{24} + 32888 \mathfrak{t}^{26} + 28902 \mathfrak{t}^{28} + 22427 \mathfrak{t}^{30} + 15181 \mathfrak{t}^{32}\nonumber \\
&\quad + 8465 \mathfrak{t}^{34} + 3526 \mathfrak{t}^{36} + 489 \mathfrak{t}^{38} - 802 \mathfrak{t}^{40} - 1049 \mathfrak{t}^{42} - 740 \mathfrak{t}^{44} - 382 \mathfrak{t}^{46} - 101 \mathfrak{t}^{48} + 39 \mathfrak{t}^{50} + 56 \mathfrak{t}^{52} + 60 \mathfrak{t}^{54} + 17 \mathfrak{t}^{56}\nonumber \\
&\quad + 13 \mathfrak{t}^{58} - 2 \mathfrak{t}^{62} - \mathfrak{t}^{64} - 3 \mathfrak{t}^{66} + \mathfrak{t}^{68}),\\
&\langle W_{\ydiagram{6}}W_{\overline{\ydiagram{6}}}\rangle^{U(5)\text{ ADHM-}[2](C)}(\mathfrak{t})\nonumber \\
&=
\frac{1}{(1 - \mathfrak{t}^2)^4 (1 - \mathfrak{t}^4) (1 - \mathfrak{t}^6)^2 (1 - \mathfrak{t}^8)^2 (1 - \mathfrak{t}^{10})}
(1 + 12 \mathfrak{t}^2 + 42 \mathfrak{t}^4 + 179 \mathfrak{t}^6 + 561 \mathfrak{t}^8 + 1444 \mathfrak{t}^{10} + 3069 \mathfrak{t}^{12} + 5541 \mathfrak{t}^{14}\nonumber \\
&\quad + 8513 \mathfrak{t}^{16} + 11270 \mathfrak{t}^{18} + 13040 \mathfrak{t}^{20} + 13021 \mathfrak{t}^{22} + 11317 \mathfrak{t}^{24} + 8242 \mathfrak{t}^{26} + 4903 \mathfrak{t}^{28} + 1966 \mathfrak{t}^{30} + 195 \mathfrak{t}^{32} - 793 \mathfrak{t}^{34}\nonumber \\
&\quad - 829 \mathfrak{t}^{36} - 705 \mathfrak{t}^{38} - 374 \mathfrak{t}^{40} - 135 \mathfrak{t}^{42} - 2 \mathfrak{t}^{44} + 67 \mathfrak{t}^{46} + 37 \mathfrak{t}^{48} + 41 \mathfrak{t}^{50} + 8 \mathfrak{t}^{52} + 13 \mathfrak{t}^{54} - 2 \mathfrak{t}^{56} + \mathfrak{t}^{58} - 2 \mathfrak{t}^{62} - 2 \mathfrak{t}^{66}\nonumber \\
&\quad + \mathfrak{t}^{68}),\\
&\langle W_{\ydiagram{2}}W_{\overline{\ydiagram{2}}}\rangle^{U(5)\text{ ADHM-}[3](C)}(\mathfrak{t})\nonumber \\
&=
\frac{1}{
(1 - \mathfrak{t})^2
(1 - \mathfrak{t}^2)^2
(1 - \mathfrak{t}^3)^3
(1 - \mathfrak{t}^4)
(1 - \mathfrak{t}^9)
(1 - \mathfrak{t}^{12})
}
(1 - 2 \mathfrak{t} + \mathfrak{t}^{2} + 3 \mathfrak{t}^{3} - 5 \mathfrak{t}^{4} + 5 \mathfrak{t}^{5} - 3 \mathfrak{t}^{6} - \mathfrak{t}^{7} + 9 \mathfrak{t}^{8} - 6 \mathfrak{t}^{10} + 6 \mathfrak{t}^{11}\nonumber \\
&\quad - 8 \mathfrak{t}^{12} + 7 \mathfrak{t}^{13} + 13 \mathfrak{t}^{14} - 4 \mathfrak{t}^{15} - \mathfrak{t}^{16} - \mathfrak{t}^{17} - 8 \mathfrak{t}^{18} + 16 \mathfrak{t}^{19} + 4 \mathfrak{t}^{20} - 4 \mathfrak{t}^{21} + 2 \mathfrak{t}^{22} - 5 \mathfrak{t}^{23} - 3 \mathfrak{t}^{24} + 12 \mathfrak{t}^{25} - 4 \mathfrak{t}^{26} + \mathfrak{t}^{27}\nonumber \\
&\quad - 3 \mathfrak{t}^{29} + \mathfrak{t}^{30} + 2 \mathfrak{t}^{31} - \mathfrak{t}^{32}),\\
&\langle W_{\ydiagram{3}}W_{\overline{\ydiagram{3}}}\rangle^{U(5)\text{ ADHM-}[3](C)}(\mathfrak{t})\nonumber \\
&=
\frac{1}{
(1 - \mathfrak{t})^2
(1 - \mathfrak{t}^2)^2
(1 - \mathfrak{t}^3)^3
(1 - \mathfrak{t}^4)
(1 - \mathfrak{t}^9)
(1 - \mathfrak{t}^{12})
}
 (1 - 2 \mathfrak{t} + \mathfrak{t}^{2} + 5 \mathfrak{t}^{3} - 9 \mathfrak{t}^{4} + 9 \mathfrak{t}^{5} - 4 \mathfrak{t}^{6} - \mathfrak{t}^{7} + 11 \mathfrak{t}^{8} + 5 \mathfrak{t}^{9}\nonumber \\
&\quad - 14 \mathfrak{t}^{10} + 22 \mathfrak{t}^{11} - 19 \mathfrak{t}^{12} + 11 \mathfrak{t}^{13} + 19 \mathfrak{t}^{14} - 2 \mathfrak{t}^{15} + 3 \mathfrak{t}^{16} + 11 \mathfrak{t}^{17} - 30 \mathfrak{t}^{18} + 28 \mathfrak{t}^{19} + 6 \mathfrak{t}^{20} + 2 \mathfrak{t}^{21} + 10 \mathfrak{t}^{22} - 7 \mathfrak{t}^{23}\nonumber \\
&\quad - 21 \mathfrak{t}^{24} + 24 \mathfrak{t}^{25} - 6 \mathfrak{t}^{26} + 9 \mathfrak{t}^{27} - 7 \mathfrak{t}^{29} - 4 \mathfrak{t}^{30} + 6 \mathfrak{t}^{31} - \mathfrak{t}^{32} + \mathfrak{t}^{33} - 2 \mathfrak{t}^{35} + \mathfrak{t}^{36}),\\
&\langle W_{\ydiagram{4}}W_{\overline{\ydiagram{4}}}\rangle^{U(5)\text{ ADHM-}[3](C)}(\mathfrak{t})\nonumber \\
&=
\frac{1}{
(1 - \mathfrak{t})^2
(1 - \mathfrak{t}^2)^2
(1 - \mathfrak{t}^3)^3
(1 - \mathfrak{t}^4)
(1 - \mathfrak{t}^9)
(1 - \mathfrak{t}^{12})
}
(1 - 2 \mathfrak{t} + \mathfrak{t}^{2} + 7 \mathfrak{t}^{3} - 13 \mathfrak{t}^{4} + 13 \mathfrak{t}^{5} - 5 \mathfrak{t}^{6} - \mathfrak{t}^{7} + 16 \mathfrak{t}^{8} + 10 \mathfrak{t}^{9}\nonumber \\
&\quad - 28 \mathfrak{t}^{10} + 45 \mathfrak{t}^{11} - 31 \mathfrak{t}^{12} + 25 \mathfrak{t}^{13} + 31 \mathfrak{t}^{14} - 9 \mathfrak{t}^{15} - 2 \mathfrak{t}^{16} + 34 \mathfrak{t}^{17} - 44 \mathfrak{t}^{18} + 53 \mathfrak{t}^{19} + 9 \mathfrak{t}^{20} - 15 \mathfrak{t}^{21} + 20 \mathfrak{t}^{22} - 3 \mathfrak{t}^{23}\nonumber \\
&\quad - 30 \mathfrak{t}^{24} + 51 \mathfrak{t}^{25} - 26 \mathfrak{t}^{26} + 11 \mathfrak{t}^{27} - 5 \mathfrak{t}^{28} + 3 \mathfrak{t}^{29} - 8 \mathfrak{t}^{30} + 12 \mathfrak{t}^{31} - 7 \mathfrak{t}^{32} - 3 \mathfrak{t}^{33} + 4 \mathfrak{t}^{34} - 6 \mathfrak{t}^{35} + 6 \mathfrak{t}^{36} - 2 \mathfrak{t}^{37} - \mathfrak{t}^{39}\nonumber \\
&\quad + 2 \mathfrak{t}^{41} - \mathfrak{t}^{42}),\\
&\langle W_{\ydiagram{5}}W_{\overline{\ydiagram{5}}}\rangle^{U(5)\text{ ADHM-}[3](C)}(\mathfrak{t})\nonumber \\
&=
\frac{1}{(1 - \mathfrak{t}) (1 - \mathfrak{t}^2)^2 (1 - \mathfrak{t}^3)^2 (1 - \mathfrak{t}^4) (1 - \mathfrak{t}^5) (1 - \mathfrak{t}^9) (1 - \mathfrak{t}^{12}) (1 - \mathfrak{t}^{15})}
 (1 - \mathfrak{t} + 10 \mathfrak{t}^{3} - 9 \mathfrak{t}^{4} + 8 \mathfrak{t}^{5} + 15 \mathfrak{t}^{6} - 8 \mathfrak{t}^{7} + 23 \mathfrak{t}^{8}\nonumber \\
&\quad + 63 \mathfrak{t}^{9} - 19 \mathfrak{t}^{10} + 89 \mathfrak{t}^{11} + 87 \mathfrak{t}^{12} + 19 \mathfrak{t}^{13} + 169 \mathfrak{t}^{14} + 191 \mathfrak{t}^{15} + 56 \mathfrak{t}^{16} + 302 \mathfrak{t}^{17} + 216 \mathfrak{t}^{18} + 114 \mathfrak{t}^{19} + 415 \mathfrak{t}^{20} + 270 \mathfrak{t}^{21}\nonumber \\
&\quad + 176 \mathfrak{t}^{22} + 493 \mathfrak{t}^{23} + 207 \mathfrak{t}^{24} + 198 \mathfrak{t}^{25} + 468 \mathfrak{t}^{26} + 185 \mathfrak{t}^{27} + 173 \mathfrak{t}^{28} + 440 \mathfrak{t}^{29} + 34 \mathfrak{t}^{30} + 149 \mathfrak{t}^{31} + 303 \mathfrak{t}^{32} - 6 \mathfrak{t}^{33}\nonumber \\
&\quad + 107 \mathfrak{t}^{34} + 170 \mathfrak{t}^{35} - 78 \mathfrak{t}^{36} + 22 \mathfrak{t}^{37} + 97 \mathfrak{t}^{38} - 71 \mathfrak{t}^{39} + 28 \mathfrak{t}^{40} + 48 \mathfrak{t}^{41} - 77 \mathfrak{t}^{42} + 9 \mathfrak{t}^{43} - 13 \mathfrak{t}^{44} - 14 \mathfrak{t}^{45} - 9 \mathfrak{t}^{46}\nonumber \\
&\quad + 10 \mathfrak{t}^{47} - 26 \mathfrak{t}^{48} - 2 \mathfrak{t}^{49} - 3 \mathfrak{t}^{50} - 3 \mathfrak{t}^{51} + 6 \mathfrak{t}^{52} + \mathfrak{t}^{53} + 3 \mathfrak{t}^{54} + 2 \mathfrak{t}^{55} + 4 \mathfrak{t}^{56} - \mathfrak{t}^{57} - 3 \mathfrak{t}^{58} + \mathfrak{t}^{60} + 2 \mathfrak{t}^{61} + \mathfrak{t}^{62} - 2 \mathfrak{t}^{64}\nonumber \\
&\quad - \mathfrak{t}^{65} + \mathfrak{t}^{66}),\\
&\langle W_{\ydiagram{6}}W_{\overline{\ydiagram{6}}}\rangle^{U(5)\text{ ADHM-}[3](C)}(\mathfrak{t})\nonumber \\
&=
\frac{1}{(1 - \mathfrak{t})^2 (1 - \mathfrak{t}^2)^2 (1 - \mathfrak{t}^3)^2 (1 - \mathfrak{t}^4) (1 - \mathfrak{t}^9) (1 - \mathfrak{t}^{12}) (1 - \mathfrak{t}^{15})}
(1 - 2 \mathfrak{t} + \mathfrak{t}^{2} + 12 \mathfrak{t}^{3} - 23 \mathfrak{t}^{4} + 22 \mathfrak{t}^{5} + 8 \mathfrak{t}^{6} - 30 \mathfrak{t}^{7}\nonumber \\
&\quad + 55 \mathfrak{t}^{8} + 29 \mathfrak{t}^{9} - 87 \mathfrak{t}^{10} + 159 \mathfrak{t}^{11} - 24 \mathfrak{t}^{12} - 47 \mathfrak{t}^{13} + 264 \mathfrak{t}^{14} - 69 \mathfrak{t}^{15} - 13 \mathfrak{t}^{16} + 332 \mathfrak{t}^{17} - 133 \mathfrak{t}^{18} + 64 \mathfrak{t}^{19} + 463 \mathfrak{t}^{20}\nonumber \\
&\quad - 298 \mathfrak{t}^{21} + 251 \mathfrak{t}^{22} + 317 \mathfrak{t}^{23} - 322 \mathfrak{t}^{24} + 368 \mathfrak{t}^{25} + 211 \mathfrak{t}^{26} - 269 \mathfrak{t}^{27} + 332 \mathfrak{t}^{28} + 103 \mathfrak{t}^{29} - 258 \mathfrak{t}^{30} + 316 \mathfrak{t}^{31} + 73 \mathfrak{t}^{32}\nonumber \\
&\quad - 284 \mathfrak{t}^{33} + 361 \mathfrak{t}^{34} - 229 \mathfrak{t}^{35} + 14 \mathfrak{t}^{36} + 90 \mathfrak{t}^{37} - 12 \mathfrak{t}^{38} - 51 \mathfrak{t}^{39} + 33 \mathfrak{t}^{40} - 14 \mathfrak{t}^{41} - 62 \mathfrak{t}^{42} + 81 \mathfrak{t}^{43} - 66 \mathfrak{t}^{44} + 32 \mathfrak{t}^{45}\nonumber \\
&\quad - 24 \mathfrak{t}^{46} + 6 \mathfrak{t}^{47} - 15 \mathfrak{t}^{48} + 14 \mathfrak{t}^{49} - 14 \mathfrak{t}^{50} + 11 \mathfrak{t}^{51} - 3 \mathfrak{t}^{52} + 11 \mathfrak{t}^{53} - 6 \mathfrak{t}^{54} + \mathfrak{t}^{55} - 4 \mathfrak{t}^{56} + 3 \mathfrak{t}^{57} + \mathfrak{t}^{59} + 2 \mathfrak{t}^{60} - \mathfrak{t}^{61}\nonumber \\
&\quad - 2 \mathfrak{t}^{63} + 3 \mathfrak{t}^{66} - 2 \mathfrak{t}^{67} + \mathfrak{t}^{68} - 2 \mathfrak{t}^{69} + \mathfrak{t}^{70}),\\
%
%
&\langle W_{\ydiagram{2}}W_{\overline{\ydiagram{2}}}\rangle^{U(5)\text{ ADHM-}[4](C)}(\mathfrak{t})\nonumber \\
&=
\frac{1}{(1 - \mathfrak{t}^2)^4 (1 - \mathfrak{t}^4)^3 (1 - \mathfrak{t}^6) (1 - \mathfrak{t}^{12}) (1 - \mathfrak{t}^{16})}
(1 - 2 \mathfrak{t}^2 + 6 \mathfrak{t}^4 - 6 \mathfrak{t}^6 + 10 \mathfrak{t}^8 - 3 \mathfrak{t}^{10} + 23 \mathfrak{t}^{12} - 17 \mathfrak{t}^{14} + 39 \mathfrak{t}^{16} - 17 \mathfrak{t}^{18}\nonumber \\
&\quad + 49 \mathfrak{t}^{20} - 22 \mathfrak{t}^{22} + 52 \mathfrak{t}^{24} - 27 \mathfrak{t}^{26} + 46 \mathfrak{t}^{28} - 14 \mathfrak{t}^{30} + 25 \mathfrak{t}^{32} - 16 \mathfrak{t}^{34} + 20 \mathfrak{t}^{36} - 6 \mathfrak{t}^{38} + 5 \mathfrak{t}^{40} - 5 \mathfrak{t}^{42} + 4 \mathfrak{t}^{44} - \mathfrak{t}^{46}),\\
&\langle W_{\ydiagram{3}}W_{\overline{\ydiagram{3}}}\rangle^{U(5)\text{ ADHM-}[4](C)}(\mathfrak{t})\nonumber \\
&=
\frac{1}{(1 - \mathfrak{t}^2)^4 (1 - \mathfrak{t}^4)^3 (1 - \mathfrak{t}^6) (1 - \mathfrak{t}^{12}) (1 - \mathfrak{t}^{16})}
(1 - 2 \mathfrak{t}^2 + 8 \mathfrak{t}^4 - 7 \mathfrak{t}^6 + 14 \mathfrak{t}^8 - 2 \mathfrak{t}^{10} + 44 \mathfrak{t}^{12} - 22 \mathfrak{t}^{14} + 69 \mathfrak{t}^{16} - 16 \mathfrak{t}^{18}\nonumber \\
&\quad + 96 \mathfrak{t}^{20} - 31 \mathfrak{t}^{22} + 97 \mathfrak{t}^{24} - 31 \mathfrak{t}^{26} + 87 \mathfrak{t}^{28} - 19 \mathfrak{t}^{30} + 46 \mathfrak{t}^{32} - 26 \mathfrak{t}^{34} + 35 \mathfrak{t}^{36} - 4 \mathfrak{t}^{38} + 4 \mathfrak{t}^{40} - 10 \mathfrak{t}^{42} + 6 \mathfrak{t}^{44} + \mathfrak{t}^{46}\nonumber \\
&\quad - 3 \mathfrak{t}^{48} + \mathfrak{t}^{50}),\\
&\langle W_{\ydiagram{4}}W_{\overline{\ydiagram{4}}}\rangle^{U(5)\text{ ADHM-}[4](C)}(\mathfrak{t})\nonumber \\
&=
\frac{1}{(1 - \mathfrak{t}^2)^4 (1 - \mathfrak{t}^4)^3 (1 - \mathfrak{t}^6) (1 - \mathfrak{t}^{12}) (1 - \mathfrak{t}^{16})}
(1 - 2 \mathfrak{t}^2 + 10 \mathfrak{t}^4 - 9 \mathfrak{t}^6 + 22 \mathfrak{t}^8 - 3 \mathfrak{t}^{10} + 69 \mathfrak{t}^{12} - 21 \mathfrak{t}^{14} + 125 \mathfrak{t}^{16}\nonumber \\
&\quad - 20 \mathfrak{t}^{18} + 180 \mathfrak{t}^{20} - 42 \mathfrak{t}^{22} + 184 \mathfrak{t}^{24} - 49 \mathfrak{t}^{26} + 176 \mathfrak{t}^{28} - 42 \mathfrak{t}^{30} + 90 \mathfrak{t}^{32} - 44 \mathfrak{t}^{34} + 66 \mathfrak{t}^{36} - 18 \mathfrak{t}^{38} + 21 \mathfrak{t}^{40} - 24 \mathfrak{t}^{42}\nonumber \\
&\quad + 5 \mathfrak{t}^{44} - 3 \mathfrak{t}^{46} - 2 \mathfrak{t}^{48} + 2 \mathfrak{t}^{50} - 2 \mathfrak{t}^{52} + 3 \mathfrak{t}^{54} - \mathfrak{t}^{56}),\\
&\langle W_{\ydiagram{5}}W_{\overline{\ydiagram{5}}}\rangle^{U(5)\text{ ADHM-}[4](C)}(\mathfrak{t})\nonumber \\
&=
\frac{1}{(1 - \mathfrak{t}^2)^3 (1 - \mathfrak{t}^4)^2 (1 - \mathfrak{t}^6) (1 - \mathfrak{t}^{10}) (1 - \mathfrak{t}^{12}) (1 - \mathfrak{t}^{16}) (1 - \mathfrak{t}^{20})}
(1 - \mathfrak{t}^2 + 12 \mathfrak{t}^4 - \mathfrak{t}^6 + 42 \mathfrak{t}^8 + 27 \mathfrak{t}^{10} + 169 \mathfrak{t}^{12}\nonumber \\
&\quad + 123 \mathfrak{t}^{14} + 480 \mathfrak{t}^{16} + 405 \mathfrak{t}^{18} + 1051 \mathfrak{t}^{20} + 842 \mathfrak{t}^{22} + 1812 \mathfrak{t}^{24} + 1365 \mathfrak{t}^{26} + 2595 \mathfrak{t}^{28} + 1799 \mathfrak{t}^{30} + 3112 \mathfrak{t}^{32} + 1975 \mathfrak{t}^{34}\nonumber \\
&\quad + 3220 \mathfrak{t}^{36} + 1794 \mathfrak{t}^{38} + 2840 \mathfrak{t}^{40} + 1379 \mathfrak{t}^{42} + 2134 \mathfrak{t}^{44} + 815 \mathfrak{t}^{46} + 1303 \mathfrak{t}^{48} + 305 \mathfrak{t}^{50} + 634 \mathfrak{t}^{52} + 25 \mathfrak{t}^{54} + 214 \mathfrak{t}^{56}\nonumber \\
&\quad - 74 \mathfrak{t}^{58} + 26 \mathfrak{t}^{60} - 90 \mathfrak{t}^{62} - 50 \mathfrak{t}^{64} - 44 \mathfrak{t}^{66} - 25 \mathfrak{t}^{68} - \mathfrak{t}^{70} - \mathfrak{t}^{72} + 25 \mathfrak{t}^{74} + 5 \mathfrak{t}^{78} - 2 \mathfrak{t}^{80} + 2 \mathfrak{t}^{82} - 3 \mathfrak{t}^{86} + \mathfrak{t}^{88}),\\
&\langle W_{\ydiagram{6}}W_{\overline{\ydiagram{6}}}\rangle^{U(5)\text{ ADHM-}[4](C)}(\mathfrak{t})\nonumber \\
&=
\frac{1}{(1 - \mathfrak{t}^2)^4 (1 - \mathfrak{t}^4)^2 (1 - \mathfrak{t}^6) (1 - \mathfrak{t}^{12}) (1 - \mathfrak{t}^{16}) (1 - \mathfrak{t}^{20})}
(1 - 2 \mathfrak{t}^2 + 15 \mathfrak{t}^4 - 15 \mathfrak{t}^6 + 54 \mathfrak{t}^8 - 15 \mathfrak{t}^{10} + 187 \mathfrak{t}^{12} - 32 \mathfrak{t}^{14}\nonumber \\
&\quad + 491 \mathfrak{t}^{16} + 8 \mathfrak{t}^{18} + 949 \mathfrak{t}^{20} - 22 \mathfrak{t}^{22} + 1459 \mathfrak{t}^{24} - 35 \mathfrak{t}^{26} + 1903 \mathfrak{t}^{28} - 140 \mathfrak{t}^{30} + 2004 \mathfrak{t}^{32} - 222 \mathfrak{t}^{34} + 1873 \mathfrak{t}^{36} - 309 \mathfrak{t}^{38}\nonumber \\
&\quad + 1472 \mathfrak{t}^{40} - 330 \mathfrak{t}^{42} + 916 \mathfrak{t}^{44} - 338 \mathfrak{t}^{46} + 455 \mathfrak{t}^{48} - 217 \mathfrak{t}^{50} + 195 \mathfrak{t}^{52} - 150 \mathfrak{t}^{54} + 36 \mathfrak{t}^{56} - 60 \mathfrak{t}^{58} - 22 \mathfrak{t}^{60} - 31 \mathfrak{t}^{62}\nonumber \\
&\quad - 51 \mathfrak{t}^{64} + 38 \mathfrak{t}^{66} - 5 \mathfrak{t}^{68} + 20 \mathfrak{t}^{70} - 8 \mathfrak{t}^{72} + 12 \mathfrak{t}^{74} - 2 \mathfrak{t}^{76} - \mathfrak{t}^{78} - 2 \mathfrak{t}^{80} + 3 \mathfrak{t}^{82} - \mathfrak{t}^{84} - 2 \mathfrak{t}^{86} + \mathfrak{t}^{88}).
\end{align}
}

\subsection{Unflavored Higgs indices}
\label{app_unflavoredH}
In this subsection we list the Higgs indices of $U(N)$ ADHM theory with $l$ flavors in the unflavored limit.
For $l\ge 1$, the numerator of the Higgs indices are palindromic polynomials, which we abbreviate with ``$\cdots$''.

\subsubsection{$U(1)$}
\begin{align}
\mathcal{I}^{U(1)\text{ ADHM-}[l](H)}(\mathfrak{t})
&=\frac{1+\mathfrak{t}}{1-\mathfrak{t}} {}_2F_1(l,l;1;\mathfrak{t}^2). 
\end{align}


\subsubsection{$U(2)$}
{\fontsize{9pt}{1pt}\selectfont
\begin{align}
&\mathcal{I}^{{\cal N}=8\text{ }U(2)(H)}(\mathfrak{t})=\frac{1+\mathfrak{t}+2\mathfrak{t}^2}{(1-\mathfrak{t})(1-\mathfrak{t}^2)},\label{u2N8_index2} \\
&\mathcal{I}^{\textrm{$U(2)$ ADHM-$[1] (H)$}}(\mathfrak{t})
=\frac{1+\mathfrak{t}^2}
{(1-\mathfrak{t})^2(1-\mathfrak{t}^2)^2},\label{u2ADHM1_index2} \\
&{\cal I}^{U(2)\text{ ADHM-}[2]\,(H)}(\mathfrak{t})=\frac{1+\mathfrak{t}+3\mathfrak{t}^2+6\mathfrak{t}^3+8\mathfrak{t}^4+6\mathfrak{t}^5+\cdots+\mathfrak{t}^{10}}{(1-\mathfrak{t})(1-\mathfrak{t}^2)^4(1-\mathfrak{t}^3)^3}, \\
&{\cal I}^{U(2)\text{ ADHM-}[3]\,(H)}(\mathfrak{t})=
\frac{
1 + \mathfrak{t} + 6 \mathfrak{t}^2 + 17 \mathfrak{t}^3 + 31 \mathfrak{t}^4 + 52 \mathfrak{t}^5 + 92 \mathfrak{t}^6 + 110 \mathfrak{t}^7 + 112 \mathfrak{t}^8 + \cdots + \mathfrak{t}^{16}
}{(1-\mathfrak{t})(1-\mathfrak{t}^2)^6(1-\mathfrak{t}^3)^5}, \\
&{\cal I}^{U(2)\text{ ADHM-}[4]\,(H)}(\mathfrak{t})\nonumber \\
&=
\frac{
1 + \mathfrak{t} + 11 \mathfrak{t}^2 + 34 \mathfrak{t}^3 + 88 \mathfrak{t}^4 + 216 \mathfrak{t}^5 + 473 \mathfrak{t}^6 + 797 \mathfrak{t}^7  + 1243 \mathfrak{t}^8 + 1738 \mathfrak{t}^9 + 2080 \mathfrak{t}^{10} + 2152 \mathfrak{t}^{11}+\cdots+\mathfrak{t}^{22}
}{
(1-\mathfrak{t})(1-\mathfrak{t}^2)^8(1-\mathfrak{t}^3)^7
}, \\
&{\cal I}^{U(2)\text{ ADHM-}[5](H)}(\mathfrak{t})\nonumber \\
&=\frac{1
}{
(1 + \mathfrak{t})(1 - \mathfrak{t}^2)^{11}(1 - \mathfrak{t}^3)^9
}
(
1 + 3 \mathfrak{t} + 21 \mathfrak{t}^2 + 94 \mathfrak{t}^3 + 341 \mathfrak{t}^4 + 1099 \mathfrak{t}^5 + 3137 \mathfrak{t}^6 + 7624 \mathfrak{t}^7 + 16442 \mathfrak{t}^8 + 31830 \mathfrak{t}^9\nonumber \\
&\quad + 55082 \mathfrak{t}^{10} + 85360 \mathfrak{t}^{11} + 120008 \mathfrak{t}^{12} + 153060 \mathfrak{t}^{13} + 176628 \mathfrak{t}^{14} + 184960 \mathfrak{t}^{15} + \cdots + \mathfrak{t}^{30}
), \\
%
&{\cal I}^{U(2)\text{ ADHM-}[6](H)}(\mathfrak{t})\nonumber \\
&=\frac{1
}{
(1 + \mathfrak{t})^3(1 - \mathfrak{t}^2)^{13}(1 - \mathfrak{t}^3)^{11}
}
(
1 + 5 \mathfrak{t} + 37 \mathfrak{t}^{2} + 204 \mathfrak{t}^{3} + 947 \mathfrak{t}^{4} + 3819 \mathfrak{t}^{5} + 13587 \mathfrak{t}^{6} + 42180 \mathfrak{t}^{7} + 116511 \mathfrak{t}^{8}\nonumber \\
&\quad + 289075 \mathfrak{t}^{9} + 647517 \mathfrak{t}^{10} + 1314730 \mathfrak{t}^{11} + 2435034 \mathfrak{t}^{12} + 4128428 \mathfrak{t}^{13} + 6422514 \mathfrak{t}^{14} + 9189070 \mathfrak{t}^{15} + 12121994 \mathfrak{t}^{16}\nonumber \\
&\quad + 14760964 \mathfrak{t}^{17} + 16603650 \mathfrak{t}^{18} + 17264534 \mathfrak{t}^{19} +\cdots+\mathfrak{t}^{38}).
\end{align}
}

\subsubsection{$U(3)$}
{\fontsize{9pt}{1pt}\selectfont
\begin{align}
&\mathcal{I}^{\textrm{$\mathcal{N}=8$ $U(3) (H)$}}(\mathfrak{t})
=\frac{1+\mathfrak{t}+2\mathfrak{t}^2+3\mathfrak{t}^3+\mathfrak{t}^4}
{(1-\mathfrak{t})(1-\mathfrak{t}^2)(1-\mathfrak{t}^3)}, \label{u3N8_index2}\\
&\mathcal{I}^{\textrm{$U(3)$ ADHM-$[1] (H)$}}(\mathfrak{t})
=\frac{1+\mathfrak{t}^2+2\mathfrak{t}^3+\mathfrak{t}^4+\mathfrak{t}^6}
{(1-\mathfrak{t})^2(1-\mathfrak{t}^2)^2(1-\mathfrak{t}^3)^2},\\
&{\cal I}^{U(3)\text{ ADHM-}[2]\,(H)}(\mathfrak{t})=
\frac{1 + 3 \mathfrak{t}^2 + 6 \mathfrak{t}^3 + 12 \mathfrak{t}^4 + 16 \mathfrak{t}^5 + 31 \mathfrak{t}^6 + 36 \mathfrak{t}^7 + 55 \mathfrak{t}^8 + 54 \mathfrak{t}^9 + 60 \mathfrak{t}^{10} + \cdots + \mathfrak{t}^{20}}
{(1 - \mathfrak{t})^2(1 - \mathfrak{t}^2)^3(1 - \mathfrak{t}^3)^4(1 - \mathfrak{t}^4)^3
},\\
&{\cal I}^{U(3)\text{ ADHM-}[3]\,(H)}(\mathfrak{t})\nonumber \\
&=
\frac{1}{
(1 - \mathfrak{t})^2(1 - \mathfrak{t}^2)^5(1 - \mathfrak{t}^3)^6(1 - \mathfrak{t}^4)^5
}
(
1 + 6 \mathfrak{t}^2 + 14 \mathfrak{t}^3 + 40 \mathfrak{t}^4 + 82 \mathfrak{t}^5 + 213 \mathfrak{t}^6 + 388 \mathfrak{t}^7 + 772 \mathfrak{t}^8 + 1260 \mathfrak{t}^9 + 2079 \mathfrak{t}^{10}\nonumber \\
&\quad + 2986 \mathfrak{t}^{11} + 4226 \mathfrak{t}^{12} + 5226 \mathfrak{t}^{13} + 6384 \mathfrak{t}^{14} + 6940 \mathfrak{t}^{15} + 7334 \mathfrak{t}^{16}+\cdots+\mathfrak{t}^{32}), \\
&{\cal I}^{U(3)\text{ ADHM-}[4](H)}(\mathfrak{t})\nonumber \\
&=\frac{1}{
(1 - \mathfrak{t})(1 - \mathfrak{t}^2)^7(1 - \mathfrak{t}^3)^9(1 - \mathfrak{t}^4)^7
}
(
1 + \mathfrak{t} + 12 \mathfrak{t}^2 + 37 \mathfrak{t}^3 + 141 \mathfrak{t}^4 + 414 \mathfrak{t}^5 + 1272 \mathfrak{t}^6 + 3302 \mathfrak{t}^7+ 8390 \mathfrak{t}^8 + 19116 \mathfrak{t}^9\nonumber \\
&\quad + 41629 \mathfrak{t}^{10} + 83351 \mathfrak{t}^{11} + 158324 \mathfrak{t}^{12} + 279775 \mathfrak{t}^{13} + 468974 \mathfrak{t}^{14} + 736957 \mathfrak{t}^{15} + 1099119 \mathfrak{t}^{16} + 1544172 \mathfrak{t}^{17}\nonumber \\
&\quad + 2061438 \mathfrak{t}^{18} + 2600638 \mathfrak{t}^{19} + 3121068 \mathfrak{t}^{20} + 3546026 \mathfrak{t}^{21} + 3834944 \mathfrak{t}^{22} + 3930198 \mathfrak{t}^{23} + \cdots + \mathfrak{t}^{46}),\\
&{\cal I}^{U(3)\text{ ADHM-}[5](H)}(\mathfrak{t})\nonumber \\
&=\frac{1
}{
(1 - \mathfrak{t}^2)^9 (1 - \mathfrak{t}^3)^{12} (1 - \mathfrak{t}^4)^9
}
(1 + 2 \mathfrak{t} + 21 \mathfrak{t}^{2} + 80 \mathfrak{t}^{3} + 373 \mathfrak{t}^{4} + 1384 \mathfrak{t}^{5} + 5090 \mathfrak{t}^{6} + 16624 \mathfrak{t}^{7} + 51704 \mathfrak{t}^{8} + 147710 \mathfrak{t}^{9}\nonumber \\
&\quad + 398449 \mathfrak{t}^{10} + 1002184 \mathfrak{t}^{11} + 2380125 \mathfrak{t}^{12} + 5315986 \mathfrak{t}^{13} + 11242817 \mathfrak{t}^{14} + 22490428 \mathfrak{t}^{15} + 42734482 \mathfrak{t}^{16}\nonumber \\
&\quad + 77129082 \mathfrak{t}^{17} + 132600175 \mathfrak{t}^{18} + 217226432 \mathfrak{t}^{19} + 339776481 \mathfrak{t}^{20} + 507652538 \mathfrak{t}^{21} + 725554206 \mathfrak{t}^{22}\nonumber \\
&\quad + 992336730 \mathfrak{t}^{23} + 1300184884 \mathfrak{t}^{24} + 1632358790 \mathfrak{t}^{25} + 1965329545 \mathfrak{t}^{26} + 2269415812 \mathfrak{t}^{27} + 2514755940 \mathfrak{t}^{28}\nonumber \\
&\quad + 2674032378 \mathfrak{t}^{29} + 2729508494 \mathfrak{t}^{30} + \cdots + \mathfrak{t}^{60}).
\end{align}
}

\subsubsection{$U(4)$}
{\fontsize{9pt}{1pt}\selectfont
\begin{align}
&\mathcal{I}^{\textrm{$\mathcal{N}=8$ $U(4) (H)$}}(\mathfrak{t})
=\frac{1+\mathfrak{t}+2\mathfrak{t}^2+3\mathfrak{t}^3+5\mathfrak{t}^4+2\mathfrak{t}^5+2\mathfrak{t}^6}
{(1-\mathfrak{t})(1-\mathfrak{t}^2)(1-\mathfrak{t}^3)(1-\mathfrak{t}^4)},\label{u4N8_index2} \\
&{\cal I}^{U(4)\text{ ADHM-}[1]\,(H)}(\mathfrak{t})=
\frac{1 + \mathfrak{t}^2 + 2 \mathfrak{t}^3 + 4 \mathfrak{t}^4 + 2 \mathfrak{t}^5 + 4 \mathfrak{t}^6 + 2 \mathfrak{t}^7 + 4 \mathfrak{t}^8 + 2 \mathfrak{t}^9 + \mathfrak{t}^{10} + \mathfrak{t}^{12}}{(1 - \mathfrak{t})^2 (1 - \mathfrak{t}^4)^2 (1 - \mathfrak{t}^2)^2 (1 - \mathfrak{t}^3)^2}, \\
&{\cal I}^{U(4)\text{ ADHM-}[2](H)}(\mathfrak{t})\nonumber \\
&=
\frac{1}{
(1 - \mathfrak{t}) (1 - \mathfrak{t}^2)^4 (1 - \mathfrak{t}^3)^4 (1 - \mathfrak{t}^4)^4 (1 - \mathfrak{t}^5)^3
}
(
1 + \mathfrak{t} + 3 \mathfrak{t}^{2} + 9 \mathfrak{t}^{3} + 22 \mathfrak{t}^{4} + 43 \mathfrak{t}^{5} + 85 \mathfrak{t}^{6} + 153 \mathfrak{t}^{7} + 273 \mathfrak{t}^{8} + 440 \mathfrak{t}^{9}\nonumber \\
&\quad + 680 \mathfrak{t}^{10} + 982 \mathfrak{t}^{11} + 1364 \mathfrak{t}^{12} + 1778 \mathfrak{t}^{13} + 2225 \mathfrak{t}^{14} + 2633 \mathfrak{t}^{15} + 2981 \mathfrak{t}^{16} + 3187 \mathfrak{t}^{17} + 3274 \mathfrak{t}^{18} + \cdots + \mathfrak{t}^{36}),\\
&{\cal I}^{U(4)\text{ ADHM-}[3](H)}(\mathfrak{t})\nonumber \\
&=
\frac{
1
}{
(1 - \mathfrak{t}) (1 - \mathfrak{t}^2)^6 (1 - \mathfrak{t}^3)^6 (1 - \mathfrak{t}^4)^6 (1 - \mathfrak{t}^5)^5
}
(
1 + \mathfrak{t} + 6 \mathfrak{t}^{2} + 20 \mathfrak{t}^{3} + 58 \mathfrak{t}^{4} + 155 \mathfrak{t}^{5} + 407 \mathfrak{t}^{6} + 984 \mathfrak{t}^{7} + 2293 \mathfrak{t}^{8}\nonumber \\
&\quad + 5007 \mathfrak{t}^{9} + 10462 \mathfrak{t}^{10} + 20786 \mathfrak{t}^{11} + 39504 \mathfrak{t}^{12} + 71651 \mathfrak{t}^{13} + 124558 \mathfrak{t}^{14} + 207293 \mathfrak{t}^{15} + 331114 \mathfrak{t}^{16} + 507632 \mathfrak{t}^{17}\nonumber \\
&\quad + 748125 \mathfrak{t}^{18} + 1059928 \mathfrak{t}^{19} + 1445460 \mathfrak{t}^{20} + 1897583 \mathfrak{t}^{21} + 2400256 \mathfrak{t}^{22} + 2925807 \mathfrak{t}^{23} + 3438988 \mathfrak{t}^{24} + 3897930 \mathfrak{t}^{25}\nonumber \\
&\quad + 4262524 \mathfrak{t}^{26} + 4496723 \mathfrak{t}^{27} + 4577852 \mathfrak{t}^{28}+\cdots +\mathfrak{t}^{56}),\\
&{\cal I}^{U(4)\text{ ADHM-}[4](H)}(\mathfrak{t})\nonumber \\
&=
\frac{1}{
(1 + \mathfrak{t})^3 (1 - \mathfrak{t}^2)^5 (1 - \mathfrak{t}^3)^{12} (1 - \mathfrak{t}^4)^8 (1 - \mathfrak{t}^5)^7
}
(
1 + 5 \mathfrak{t} + 25 \mathfrak{t}^{2} + 107 \mathfrak{t}^{3} + 427 \mathfrak{t}^{4} + 1564 \mathfrak{t}^{5} + 5390 \mathfrak{t}^{6} + 17494 \mathfrak{t}^{7}\nonumber \\
&\quad + 53903 \mathfrak{t}^{8} + 157963 \mathfrak{t}^{9} + 441717 \mathfrak{t}^{10} + 1180555 \mathfrak{t}^{11} + 3021599 \mathfrak{t}^{12} + 7416302 \mathfrak{t}^{13} + 17480180 \mathfrak{t}^{14} + 39610858 \mathfrak{t}^{15}\nonumber \\
&\quad + 86394700 \mathfrak{t}^{16} + 181555520 \mathfrak{t}^{17} + 367969527 \mathfrak{t}^{18} + 719933931 \mathfrak{t}^{19} + 1360920322 \mathfrak{t}^{20} + 2487642612 \mathfrak{t}^{21}\nonumber \\
&\quad + 4400438535 \mathfrak{t}^{22} + 7538275660 \mathfrak{t}^{23} + 12514477648 \mathfrak{t}^{24} + 20146251681 \mathfrak{t}^{25} + 31468375511 \mathfrak{t}^{26} + 47719035584 \mathfrak{t}^{27}\nonumber \\
&\quad + 70285764876 \mathfrak{t}^{28} + 100601604158 \mathfrak{t}^{29} + 139988082494 \mathfrak{t}^{30} + 189450691312 \mathfrak{t}^{31} + 249445133688 \mathfrak{t}^{32}\nonumber \\
&\quad + 319644300984 \mathfrak{t}^{33} + 398746337151 \mathfrak{t}^{34} + 484367021263 \mathfrak{t}^{35} + 573055921814 \mathfrak{t}^{36} + 660460479592 \mathfrak{t}^{37}\nonumber \\
&\quad + 741640891925 \mathfrak{t}^{38} + 811510310462 \mathfrak{t}^{39} + 865349932560 \mathfrak{t}^{40} + 899327677397 \mathfrak{t}^{41} + 910943197426 \mathfrak{t}^{42}\nonumber \\
&\quad + \cdots + \mathfrak{t}^{84}).
\end{align}
}

\subsubsection{$U(5)$}
{\fontsize{9pt}{1pt}\selectfont
\begin{align}
&{\cal I}^{{\cal N}=8\text{ }U(5)(H)}(\mathfrak{t})
=
\frac{
1 + 2 \mathfrak{t}^2 + \mathfrak{t}^3 + 4 \mathfrak{t}^4 + 3 \mathfrak{t}^5 + 2 \mathfrak{t}^6 + 2 \mathfrak{t}^7 + \mathfrak{t}^8
}{
(1 - \mathfrak{t})^2 (1 - \mathfrak{t}^3) (1 - \mathfrak{t}^4) (1 - \mathfrak{t}^5)
},\\
&{\cal I}^{U(5)\text{ ADHM-}[1](H)}(\mathfrak{t})
=
\frac{
1+\mathfrak{t}^{2}+2 \mathfrak{t}^{3}+4 \mathfrak{t}^{4}+6 \mathfrak{t}^{5}+7 \mathfrak{t}^{6}+8 \mathfrak{t}^{7}+12 \mathfrak{t}^{8} +12 \mathfrak{t}^{9}+14 \mathfrak{t}^{10}+\cdots+\mathfrak{t}^{20}
}{
(1-\mathfrak{t})^2
(1-\mathfrak{t}^2)^2
(1-\mathfrak{t}^3)^2
(1-\mathfrak{t}^4)^2
(1-\mathfrak{t}^5)^2
},\\
&{\cal I}^{U(5)\text{ ADHM-}[2](H)}(\mathfrak{t})\nonumber \\
&=
\frac{1}{
(1 - \mathfrak{t})^2 (1 - \mathfrak{t}^2)^4 (1 - \mathfrak{t}^3)^4 (1 - \mathfrak{t}^4)^3 (1 - \mathfrak{t}^5)^4 (1 - \mathfrak{t}^6)^3
}
(
1 
+ 2 \mathfrak{t}^{2}
+ 6 \mathfrak{t}^{3}
+ 14 \mathfrak{t}^{4}
+ 26 \mathfrak{t}^{5}
+ 59 \mathfrak{t}^{6}
+ 108 \mathfrak{t}^{7}
+ 216 \mathfrak{t}^{8}\nonumber \\
&\quad + 382 \mathfrak{t}^{9}
+ 669 \mathfrak{t}^{10}
+ 1090 \mathfrak{t}^{11}
+ 1788 \mathfrak{t}^{12}
+ 2718 \mathfrak{t}^{13}
+ 4080 \mathfrak{t}^{14}
+ 5844 \mathfrak{t}^{15}
+ 8166 \mathfrak{t}^{16}
+ 10902 \mathfrak{t}^{17}
+ 14271 \mathfrak{t}^{18}\nonumber \\
&\quad + 17886 \mathfrak{t}^{19}
+ 21899 \mathfrak{t}^{20}
+ 25824 \mathfrak{t}^{21}
+ 29701 \mathfrak{t}^{22}
+ 32898 \mathfrak{t}^{23}
+ 35621 \mathfrak{t}^{24}
+ 37152 \mathfrak{t}^{25}
+ 37792 \mathfrak{t}^{26}
+\cdots+\mathfrak{t}^{52}
),\\
&{\cal I}^{U(5)\text{ ADHM-}[3](H)}(\mathfrak{t})\nonumber \\
&=
\frac{1}{
(1 - \mathfrak{t})^2 (1 - \mathfrak{t}^2)^3 (1 - \mathfrak{t}^3)^6(1 - \mathfrak{t}^4)^8 (1 - \mathfrak{t}^5)^6 (1 - \mathfrak{t}^6)^5
}
(
1 + 8 \mathfrak{t}^{2} + 14 \mathfrak{t}^{3} + 57 \mathfrak{t}^{4} + 144 \mathfrak{t}^{5} + 428 \mathfrak{t}^{6} + 1050 \mathfrak{t}^{7} + 2767 \mathfrak{t}^{8}\nonumber \\
&\quad  + 6494 \mathfrak{t}^{9} + 15455 \mathfrak{t}^{10} + 34436 \mathfrak{t}^{11} + 75777 \mathfrak{t}^{12} + 159098 \mathfrak{t}^{13} + 327449 \mathfrak{t}^{14} + 648564 \mathfrak{t}^{15} + 1254726 \mathfrak{t}^{16} + 2349032 \mathfrak{t}^{17}\nonumber \\
&\quad  + 4289709 \mathfrak{t}^{18} + 7604480 \mathfrak{t}^{19} + 13149870 \mathfrak{t}^{20} + 22118900 \mathfrak{t}^{21} + 36305505 \mathfrak{t}^{22} + 58057680 \mathfrak{t}^{23} + 90643611 \mathfrak{t}^{24}\nonumber \\
&\quad  + 138043192 \mathfrak{t}^{25} + 205374524 \mathfrak{t}^{26} + 298328112 \mathfrak{t}^{27} + 423592293 \mathfrak{t}^{28} + 587714724 \mathfrak{t}^{29} + 797485928 \mathfrak{t}^{30}\nonumber \\
&\quad  + 1058110306 \mathfrak{t}^{31} + 1373683837 \mathfrak{t}^{32} + 1744737596 \mathfrak{t}^{33} + 2169208793 \mathfrak{t}^{34} + 2639700608 \mathfrak{t}^{35} + 3145467255 \mathfrak{t}^{36}\nonumber \\
&\quad  + 3669881102 \mathfrak{t}^{37} + 4193883413 \mathfrak{t}^{38} + 4693901906 \mathfrak{t}^{39} + 5146817398 \mathfrak{t}^{40} + 5528151672 \mathfrak{t}^{41} + 5817914594 \mathfrak{t}^{42}\nonumber \\
&\quad  + 5998467330 \mathfrak{t}^{43} + 6060236338 \mathfrak{t}^{44} + \cdots + \mathfrak{t}^{88}),\\
&{\cal I}^{U(5)\text{ ADHM-}[4](H)}(\mathfrak{t})\nonumber \\
&=
\frac{1}{
(1 - \mathfrak{t})^2 (1 - \mathfrak{t}^2)^4 (1 - \mathfrak{t}^3)^8
(1 - \mathfrak{t}^4)^{11}
(1 - \mathfrak{t}^5)^8 (1 - \mathfrak{t}^6)^7
}
(1 + 14 \mathfrak{t}^{2} + 26 \mathfrak{t}^{3} + 144 \mathfrak{t}^{4} + 422 \mathfrak{t}^{5} + 1525 \mathfrak{t}^{6} + 4552 \mathfrak{t}^{7}\nonumber \\
&\quad + 14320 \mathfrak{t}^{8} + 40898 \mathfrak{t}^{9} + 117057 \mathfrak{t}^{10} + 316842 \mathfrak{t}^{11} + 842382 \mathfrak{t}^{12} + 2152660 \mathfrak{t}^{13} + 5371784 \mathfrak{t}^{14} + 12971888 \mathfrak{t}^{15}\nonumber \\
&\quad + 30541134 \mathfrak{t}^{16} + 69840248 \mathfrak{t}^{17} + 155701665 \mathfrak{t}^{18} + 337909034 \mathfrak{t}^{19} + 715371143 \mathfrak{t}^{20} + 1476631912 \mathfrak{t}^{21}\nonumber \\
&\quad + 2975725193 \mathfrak{t}^{22} + 5854314772 \mathfrak{t}^{23} + 11254266633 \mathfrak{t}^{24} + 21144255650 \mathfrak{t}^{25} + 38850833773 \mathfrak{t}^{26} + 69832405852 \mathfrak{t}^{27}\nonumber \\
&\quad + 122857970873 \mathfrak{t}^{28} + 211626350724 \mathfrak{t}^{29} + 357074135539 \mathfrak{t}^{30} + 590338971892 \mathfrak{t}^{31} + 956693584086 \mathfrak{t}^{32}\nonumber \\
&\quad + 1520196834136 \mathfrak{t}^{33} + 2369377032433 \mathfrak{t}^{34} + 3623221733944 \mathfrak{t}^{35} + 5437702724926 \mathfrak{t}^{36} + 8011339379254 \mathfrak{t}^{37}\nonumber \\
&\quad + 11589946780455 \mathfrak{t}^{38} + 16468097235248 \mathfrak{t}^{39} + 22987646520104 \mathfrak{t}^{40} + 31529957469668 \mathfrak{t}^{41}\nonumber \\
&\quad + 42503114080404 \mathfrak{t}^{42} + 56320324713258 \mathfrak{t}^{43} + 73372891064703 \mathfrak{t}^{44} + 93994170471124 \mathfrak{t}^{45}\nonumber \\
&\quad + 118421152614268 \mathfrak{t}^{46} + 146750638794106 \mathfrak{t}^{47} + 178900384837948 \mathfrak{t}^{48} + 214572126197044 \mathfrak{t}^{49}\nonumber \\
&\quad + 253229805508166 \mathfrak{t}^{50} + 294087984821878 \mathfrak{t}^{51} + 336124698825151 \mathfrak{t}^{52} + 378109217680110 \mathfrak{t}^{53}\nonumber \\
&\quad + 418657557076879 \mathfrak{t}^{54} + 456299760943124 \mathfrak{t}^{55} + 489569121896438 \mathfrak{t}^{56} + 517091081311966 \mathfrak{t}^{57}\nonumber \\
&\quad + 537680248696289 \mathfrak{t}^{58} + 550419745121320 \mathfrak{t}^{59} + 554732530432476 \mathfrak{t}^{60} + \cdots + \mathfrak{t}^{120}).
\end{align}
}

\subsubsection{$U(6)$}
{\fontsize{9pt}{1pt}\selectfont
\begin{align}
&{\cal I}^{{\cal N}=8\text{ }U(6)(H)}(\mathfrak{t})
=
\frac{
1 + \mathfrak{t} + 2 \mathfrak{t}^{2} + 3 \mathfrak{t}^{3} + 5 \mathfrak{t}^{4} + 7 \mathfrak{t}^{5} + 11 \mathfrak{t}^{6} + 8 \mathfrak{t}^{7} + 9 \mathfrak{t}^{8} + 7 \mathfrak{t}^{9} + 6 \mathfrak{t}^{10} + 2 \mathfrak{t}^{11} + 2 \mathfrak{t}^{12}
}{
(1 - \mathfrak{t}) (1 - \mathfrak{t}^2) (1 - \mathfrak{t}^3) (1 - \mathfrak{t}^4) (1 - \mathfrak{t}^5) (1 - \mathfrak{t}^6)
},\\
&{\cal I}^{U(6)\text{ ADHM-}[1](H)}(\mathfrak{t})\nonumber \\
&=
\frac{
1
}{
(1-\mathfrak{t})^2
(1-\mathfrak{t}^2)^3
(1-\mathfrak{t}^3)^2
(1-\mathfrak{t}^4)
(1-\mathfrak{t}^5)^2
(1-\mathfrak{t}^6)^2
}
(1+2 \mathfrak{t}^{3}+4 \mathfrak{t}^{4}+4 \mathfrak{t}^{5}+8 \mathfrak{t}^{6}+8 \mathfrak{t}^{7} +13 \mathfrak{t}^{8}+18 \mathfrak{t}^{9}+24 \mathfrak{t}^{10}\nonumber \\
&\quad +22 \mathfrak{t}^{11}+31 \mathfrak{t}^{12}+30 \mathfrak{t}^{13}+30 \mathfrak{t}^{14}+\cdots+\mathfrak{t}^{28}
),\\
&{\cal I}^{U(6)\text{ ADHM-}[2](H)}(\mathfrak{t})\nonumber \\
&=
\frac{1}{
(1 - \mathfrak{t}^2)^4 (1 - \mathfrak{t}^3)^4 (1 - \mathfrak{t}^4)^4 (1 - \mathfrak{t}^5)^5 (1 - \mathfrak{t}^6)^4 (1 - \mathfrak{t}^7)^3
}(
1
+ 2 \mathfrak{t}
+ 5 \mathfrak{t}^{2}
+ 14 \mathfrak{t}^{3}
+ 36 \mathfrak{t}^{4}
+ 83 \mathfrak{t}^{5}
+ 193 \mathfrak{t}^{6}
+ 422 \mathfrak{t}^{7}\nonumber \\
&\quad + 892 \mathfrak{t}^{8}
+ 1821 \mathfrak{t}^{9}
+ 3620 \mathfrak{t}^{10}
+ 6955 \mathfrak{t}^{11}
+ 13017 \mathfrak{t}^{12}
+ 23649 \mathfrak{t}^{13}
+ 41856 \mathfrak{t}^{14}
+ 72130 \mathfrak{t}^{15}
+ 121233 \mathfrak{t}^{16} + 198686 \mathfrak{t}^{17}\nonumber \\
&\quad + 317998 \mathfrak{t}^{18}
+ 496951 \mathfrak{t}^{19}
+ 759026 \mathfrak{t}^{20}
+ 1133300 \mathfrak{t}^{21}
+ 1655290 \mathfrak{t}^{22}
+ 2365512 \mathfrak{t}^{23}
+ 3309485 \mathfrak{t}^{24} + 4533761 \mathfrak{t}^{25}\nonumber \\
&\quad + 6084418 \mathfrak{t}^{26}
+ 8000798 \mathfrak{t}^{27}
+ 10312362 \mathfrak{t}^{28}
+ 13030773 \mathfrak{t}^{29}
+ 16147551 \mathfrak{t}^{30}
+ 19625914 \mathfrak{t}^{31} + 23401717 \mathfrak{t}^{32}\nonumber \\
&\quad + 27378910 \mathfrak{t}^{33}
+ 31435436 \mathfrak{t}^{34}
+ 35423981 \mathfrak{t}^{35}
+ 39184907 \mathfrak{t}^{36}
+ 42550833 \mathfrak{t}^{37}
+ 45364374 \mathfrak{t}^{38} + 47484587 \mathfrak{t}^{39}\nonumber \\
&\quad + 48803727 \mathfrak{t}^{40}
+ 49250804 \mathfrak{t}^{41}
+ \cdots + \mathfrak{t}^{82}
),\\
&{\cal I}^{U(6)\text{ ADHM-}[3](H)}(\mathfrak{t})\nonumber \\
&=
\frac{1}{
(1 - \mathfrak{t}^2)^4 (1 - \mathfrak{t}^3)^6 (1 - \mathfrak{t}^4)^8 (1 - \mathfrak{t}^5)^7 (1 - \mathfrak{t}^6)^6 (1 - \mathfrak{t}^7)^5
}
(
1 + 2 \mathfrak{t} + 10 \mathfrak{t}^{2} + 32 \mathfrak{t}^{3} + 103 \mathfrak{t}^{4} + 303 \mathfrak{t}^{5} + 880 \mathfrak{t}^{6} + 2403 \mathfrak{t}^{7}\nonumber \\
&\quad + 6406 \mathfrak{t}^{8} + 16451 \mathfrak{t}^{9} + 41104 \mathfrak{t}^{10} + 99689 \mathfrak{t}^{11} + 235704 \mathfrak{t}^{12} + 542806 \mathfrak{t}^{13} + 1220385 \mathfrak{t}^{14} + 2678842 \mathfrak{t}^{15} + 5747852 \mathfrak{t}^{16}\nonumber \\
&\quad + 12059561 \mathfrak{t}^{17} + 24760541 \mathfrak{t}^{18} + 49768790 \mathfrak{t}^{19} + 97988297 \mathfrak{t}^{20} + 189050446 \mathfrak{t}^{21} + 357574789 \mathfrak{t}^{22} + 663285154 \mathfrak{t}^{23}\nonumber \\
&\quad + 1207114077 \mathfrak{t}^{24} + 2156043656 \mathfrak{t}^{25} + 3780758446 \mathfrak{t}^{26} + 6511008636 \mathfrak{t}^{27} + 11015424083 \mathfrak{t}^{28} + 18313164789 \mathfrak{t}^{29}\nonumber \\
&\quad + 29926658258 \mathfrak{t}^{30} + 48084095959 \mathfrak{t}^{31} + 75981153517 \mathfrak{t}^{32} + 118107659469 \mathfrak{t}^{33} + 180642989031 \mathfrak{t}^{34}\nonumber \\
&\quad + 271914798581 \mathfrak{t}^{35} + 402909461928 \mathfrak{t}^{36} + 587806581955 \mathfrak{t}^{37} + 844499953747 \mathfrak{t}^{38} + 1195045759756 \mathfrak{t}^{39}\nonumber \\
&\quad + 1665969699878 \mathfrak{t}^{40} + 2288343601620 \mathfrak{t}^{41} + 3097543188313 \mathfrak{t}^{42} + 4132588331998 \mathfrak{t}^{43} + 5434990864127 \mathfrak{t}^{44}\nonumber \\
&\quad + 7047045391784 \mathfrak{t}^{45} + 9009554147554 \mathfrak{t}^{46} + 11359010757358 \mathfrak{t}^{47} + 14124352066722 \mathfrak{t}^{48} + 17323432506807 \mathfrak{t}^{49}\nonumber \\
&\quad + 20959468679725 \mathfrak{t}^{50} + 25017729216583 \mathfrak{t}^{51} + 29462814279862 \mathfrak{t}^{52} + 34236842330972 \mathfrak{t}^{53}\nonumber \\
&\quad + 39258874231657 \mathfrak{t}^{54} + 44425799988391 \mathfrak{t}^{55} + 49614858084543 \mathfrak{t}^{56} + 54687780147279 \mathfrak{t}^{57}\nonumber \\
&\quad + 59496456465722 \mathfrak{t}^{58} + 63889811892733 \mathfrak{t}^{59} + 67721499280858 \mathfrak{t}^{60} + 70857850726009 \mathfrak{t}^{61}\nonumber \\
&\quad + 73185524398898 \mathfrak{t}^{62} + 74618214302346 \mathfrak{t}^{63} + 75101907322384 \mathfrak{t}^{64} + \cdots \mathfrak{t}^{128}
).
\end{align}
}

\subsubsection{$U(7)$}
{\fontsize{9pt}{1pt}\selectfont
\begin{align}
&{\cal I}^{{\cal N}=8\text{ }U(7)(H)}(\mathfrak{t})
=
\frac{
(1 + \mathfrak{t} + \mathfrak{t}^2 + \mathfrak{t}^4) (1 - \mathfrak{t} + 2 \mathfrak{t}^2 + \mathfrak{t}^4 + 3 \mathfrak{t}^5 + 2 \mathfrak{t}^6 + 2 \mathfrak{t}^7 + 2 \mathfrak{t}^8 + \mathfrak{t}^9 + 2 \mathfrak{t}^{10} + \mathfrak{t}^{11})
}{
(1 - \mathfrak{t})^2 (1 - \mathfrak{t}^3) (1 - \mathfrak{t}^4) (1 - \mathfrak{t}^5) (1 - \mathfrak{t}^6) (1 - \mathfrak{t}^7)
},\\
&{\cal I}^{U(7)\text{ ADHM-}[1](H)}(\mathfrak{t})\nonumber \\
&=
\frac{1}
{
(1-\mathfrak{t})^2
(1-\mathfrak{t}^2)^2
(1-\mathfrak{t}^3)^2
(1-\mathfrak{t}^4)^2
(1-\mathfrak{t}^5)^2
(1-\mathfrak{t}^6)^2
(1-\mathfrak{t}^7)^2
}
(
1+\mathfrak{t}^{2}+2 \mathfrak{t}^{3}+4 \mathfrak{t}^{4}+6 \mathfrak{t}^{5} +12 \mathfrak{t}^{6}+18 \mathfrak{t}^{7}+26 \mathfrak{t}^{8}\nonumber \\
&\quad +38 \mathfrak{t}^{9}+57 \mathfrak{t}^{10}+76 \mathfrak{t}^{11}+107 \mathfrak{t}^{12}+132 \mathfrak{t}^{13}+170 \mathfrak{t}^{14}+204 \mathfrak{t}^{15}+241 \mathfrak{t}^{16}+272 \mathfrak{t}^{17}+308 \mathfrak{t}^{18}+326 \mathfrak{t}^{19}+345 \mathfrak{t}^{20}\nonumber \\
&\quad +348 \mathfrak{t}^{21}+\cdots+\mathfrak{t}^{42}
),\\
&{\cal I}^{U(7)\text{ ADHM-}[2](H)}(\mathfrak{t})\nonumber \\
&=
\frac{1}{
(1 - \mathfrak{t}) (1 - \mathfrak{t}^2)^3 (1 - \mathfrak{t}^3)^4 (1 - \mathfrak{t}^4)^4 (1 - \mathfrak{t}^5)^5 (1 - \mathfrak{t}^6)^4 (1 - \mathfrak{t}^7)^4 (1 - \mathfrak{t}^8)^3
}
(
1 
+ \mathfrak{t} 
+ 4 \mathfrak{t}^{2}
+ 10 \mathfrak{t}^{3}
 + 26 \mathfrak{t}^{4}
+ 57 \mathfrak{t}^{5}
+ 136 \mathfrak{t}^{6}\nonumber \\
&\quad + 293 \mathfrak{t}^{7}
+ 636 \mathfrak{t}^{8}
+ 1308 \mathfrak{t}^{9}
+ 2671 \mathfrak{t}^{10}
+ 5262 \mathfrak{t}^{11}
+ 10214 \mathfrak{t}^{12}
+ 19257 \mathfrak{t}^{13}
+ 35708 \mathfrak{t}^{14} + 64621 \mathfrak{t}^{15}
+ 114866 \mathfrak{t}^{16}\nonumber \\
&\quad + 199808 \mathfrak{t}^{17}
+ 341432 \mathfrak{t}^{18}
+ 571933 \mathfrak{t}^{19}
+ 941371 \mathfrak{t}^{20}
+ 1520806 \mathfrak{t}^{21}
+ 2415178 \mathfrak{t}^{22} + 3768282 \mathfrak{t}^{23}
+ 5782441 \mathfrak{t}^{24}\nonumber \\
&\quad + 8724301 \mathfrak{t}^{25}
+ 12952022 \mathfrak{t}^{26}
+ 18918304 \mathfrak{t}^{27}
+ 27202887 \mathfrak{t}^{28}
+ 38506000 \mathfrak{t}^{29} + 53681013 \mathfrak{t}^{30}
+ 73706472 \mathfrak{t}^{31}\nonumber \\
&\quad + 99711033 \mathfrak{t}^{32}
+ 132910518 \mathfrak{t}^{33}
+ 174616789 \mathfrak{t}^{34}
+ 226128108 \mathfrak{t}^{35}
+ 288718943 \mathfrak{t}^{36} + 363479944 \mathfrak{t}^{37}\nonumber \\
&\quad + 451298138 \mathfrak{t}^{38}
+ 552656797 \mathfrak{t}^{39}
+ 667629271 \mathfrak{t}^{40}
+ 795666608 \mathfrak{t}^{41}
+ 935642392 \mathfrak{t}^{42} + 1085664220 \mathfrak{t}^{43}\nonumber \\
&\quad + 1243210254 \mathfrak{t}^{44}
+ 1404999634 \mathfrak{t}^{45}
+ 1567247336 \mathfrak{t}^{46}
+ 1725609652 \mathfrak{t}^{47}
+ 1875555249 \mathfrak{t}^{48} + 2012369797 \mathfrak{t}^{49}\nonumber \\
&\quad + 2131606970 \mathfrak{t}^{50}
+ 2229104646 \mathfrak{t}^{51}
+ 2301447720 \mathfrak{t}^{52}
+ 2345932985 \mathfrak{t}^{53}
+ 2360969518 \mathfrak{t}^{54} + \cdots + \mathfrak{t}^{108}
),\\
&{\cal I}^{U(7)\text{ ADHM-}[3](H)}(\mathfrak{t})\nonumber \\
&=
\frac{1}{
(1 - \mathfrak{t}^2)^5 (1 - \mathfrak{t}^3)^6 (1 - \mathfrak{t}^4)^6 (1 - \mathfrak{t}^5)^8 (1 - \mathfrak{t}^6)^6 (1 - \mathfrak{t}^7)^6 (1 - \mathfrak{t}^8)^5
}
(
1 + 2 \mathfrak{t} + 9 \mathfrak{t}^{2} + 30 \mathfrak{t}^{3} + 95 \mathfrak{t}^{4} + 274 \mathfrak{t}^{5} + 793 \mathfrak{t}^{6}\nonumber \\
&\quad + 2158 \mathfrak{t}^{7} + 5755 \mathfrak{t}^{8} + 14816 \mathfrak{t}^{9} + 37314 \mathfrak{t}^{10} + 91484 \mathfrak{t}^{11} + 219635 \mathfrak{t}^{12} + 515496 \mathfrak{t}^{13} + 1186318 \mathfrak{t}^{14} + 2675776 \mathfrak{t}^{15}\nonumber \\
&\quad + 5924159 \mathfrak{t}^{16} + 12875700 \mathfrak{t}^{17} + 27496124 \mathfrak{t}^{18} + 57706270 \mathfrak{t}^{19} + 119092408 \mathfrak{t}^{20} + 241750692 \mathfrak{t}^{21} + 482905002 \mathfrak{t}^{22}\nonumber \\
&\quad + 949467474 \mathfrak{t}^{23} + 1838110265 \mathfrak{t}^{24} + 3504654844 \mathfrak{t}^{25} + 6583041855 \mathfrak{t}^{26} + 12184806532 \mathfrak{t}^{27}\nonumber \\
&\quad + 22229596004 \mathfrak{t}^{28} + 39981798842 \mathfrak{t}^{29} + 70910510194 \mathfrak{t}^{30} + 124041657510 \mathfrak{t}^{31} + 214054755605 \mathfrak{t}^{32}\nonumber \\
&\quad + 364475536128 \mathfrak{t}^{33} + 612467788824 \mathfrak{t}^{34} + 1015898476656 \mathfrak{t}^{35} + 1663600805614 \mathfrak{t}^{36} + 2690021707948 \mathfrak{t}^{37}\nonumber \\
&\quad + 4295807559681 \mathfrak{t}^{38} + 6776224797996 \mathfrak{t}^{39} + 10559773176694 \mathfrak{t}^{40} + 16259696112820 \mathfrak{t}^{41} + 24741540966872 \mathfrak{t}^{42}\nonumber \\
&\quad + 37210082150702 \mathfrak{t}^{43} + 55319126202045 \mathfrak{t}^{44} + 81307356034014 \mathfrak{t}^{45} + 118162976340427 \mathfrak{t}^{46}\nonumber \\
&\quad + 169818581888452 \mathfrak{t}^{47} + 241376255366926 \mathfrak{t}^{48} + 339360127940272 \mathfrak{t}^{49} + 471990910765072 \mathfrak{t}^{50}\nonumber \\
&\quad + 649472488754466 \mathfrak{t}^{51} + 884276769733050 \mathfrak{t}^{52} + 1191407283017910 \mathfrak{t}^{53} + 1588617838525357 \mathfrak{t}^{54}\nonumber \\
&\quad + 2096556786474320 \mathfrak{t}^{55} + 2738804771196409 \mathfrak{t}^{56} + 3541770109040044 \mathfrak{t}^{57} + 4534407373638730 \mathfrak{t}^{58}\nonumber \\
&\quad + 5747725524761090 \mathfrak{t}^{59} + 7214060057868625 \mathfrak{t}^{60} + 8966090884896262 \mathfrak{t}^{61} + 11035603909769281 \mathfrak{t}^{62}\nonumber \\
&\quad + 13452008091919766 \mathfrak{t}^{63} + 16240643120958910 \mathfrak{t}^{64} + 19420930413328940 \mathfrak{t}^{65} + 23004446285931830 \mathfrak{t}^{66}\nonumber \\
&\quad + 26993011489566990 \mathfrak{t}^{67} + 31376913190041974 \mathfrak{t}^{68} + 36133380556367648 \mathfrak{t}^{69} + 41225444884032641 \mathfrak{t}^{70}\nonumber \\
&\quad + 46601302662074454 \mathfrak{t}^{71} + 52194291979728887 \mathfrak{t}^{72} + 57923559194503618 \mathfrak{t}^{73} + 63695466736967801 \mathfrak{t}^{74}\nonumber \\
&\quad + 69405741706630596 \mathfrak{t}^{75} + 74942327640642557 \mathfrak{t}^{76} + 80188843099333230 \mathfrak{t}^{77} + 85028515076245926 \mathfrak{t}^{78}\nonumber \\
&\quad + 89348403757204890 \mathfrak{t}^{79} + 93043716233404999 \mathfrak{t}^{80} + 96021976949488836 \mathfrak{t}^{81} + 98206832251509621 \mathfrak{t}^{82}\nonumber \\
&\quad + 99541266723291972 \mathfrak{t}^{83} + 99990050596853638 \mathfrak{t}^{84} + \cdots + \mathfrak{t}^{168}
).
\end{align}
}

\subsubsection{$U(8)$}
{\fontsize{9pt}{1pt}\selectfont
\begin{align}
&{\cal I}^{{\cal N}=8\text{ }U(8)(H)}(\mathfrak{t})\nonumber \\
&=
\frac{
1
}{
(1 - \mathfrak{t}) (1 - \mathfrak{t}^2) (1 - \mathfrak{t}^3) (1 - \mathfrak{t}^4) (1 - \mathfrak{t}^5) (1 - \mathfrak{t}^6) (1 - \mathfrak{t}^7) (1 - \mathfrak{t}^8)
}(
1 + \mathfrak{t} + 2 \mathfrak{t}^{2} + 3 \mathfrak{t}^{3} + 5 \mathfrak{t}^{4} + 7 \mathfrak{t}^{5} + 11 \mathfrak{t}^{6} + 15 \mathfrak{t}^{7}\nonumber \\
&\quad  + 22 \mathfrak{t}^{8} + 21 \mathfrak{t}^{9} + 25 \mathfrak{t}^{10} + 25 \mathfrak{t}^{11} + 27 \mathfrak{t}^{12} + 23 \mathfrak{t}^{13} + 22 \mathfrak{t}^{14} + 15 \mathfrak{t}^{15} + 13 \mathfrak{t}^{16} + 8 \mathfrak{t}^{17} + 6 \mathfrak{t}^{18} + 2 \mathfrak{t}^{19} + 2 \mathfrak{t}^{20}
),\\
&{\cal I}^{U(8)\text{ ADHM-}[1](H)}(\mathfrak{t})\nonumber \\
&=
\frac{
1}{
(1-\mathfrak{t})^2
(1-\mathfrak{t}^2)^2
(1-\mathfrak{t}^3)^2
(1-\mathfrak{t}^4)^2
(1-\mathfrak{t}^5)^2
(1-\mathfrak{t}^6)^2
(1-\mathfrak{t}^7)^2
(1-\mathfrak{t}^8)^2
}(
1+\mathfrak{t}^{2}+2 \mathfrak{t}^{3}+4 \mathfrak{t}^{4} +6 \mathfrak{t}^{5}+12 \mathfrak{t}^{6}+18 \mathfrak{t}^{7}\nonumber \\
&\quad +33 \mathfrak{t}^{8}+44 \mathfrak{t}^{9}+72 \mathfrak{t}^{10}+102 \mathfrak{t}^{11}+156 \mathfrak{t}^{12}+208 \mathfrak{t}^{13}+298 \mathfrak{t}^{14}+386 \mathfrak{t}^{15}+520 \mathfrak{t}^{16}+646 \mathfrak{t}^{17}+827 \mathfrak{t}^{18}+990 \mathfrak{t}^{19}\nonumber \\
&\quad +1210 \mathfrak{t}^{20}+1390 \mathfrak{t}^{21}+1616 \mathfrak{t}^{22}+1788 \mathfrak{t}^{23}+1996 \mathfrak{t}^{24}+2110 \mathfrak{t}^{25}+2254 \mathfrak{t}^{26}+2294 \mathfrak{t}^{27}+2352 \mathfrak{t}^{28}+\cdots +\mathfrak{t}^{56}
),\\
&{\cal I}^{U(8)\text{ ADHM-}[2](H)}(\mathfrak{t})\nonumber \\
&=
\frac{1}{
(1 - \mathfrak{t}) (1 - \mathfrak{t}^2)^3 (1 - \mathfrak{t}^3)^3 (1 - \mathfrak{t}^4)^4 (1 - \mathfrak{t}^5)^5 (1 - \mathfrak{t}^6)^5 (1 - \mathfrak{t}^7)^4 (1 - \mathfrak{t}^8)^4 (1 - \mathfrak{t}^9)^3
}
(
1 
+ \mathfrak{t}
+ 4 \mathfrak{t}^{2}
+ 11 \mathfrak{t}^{3}
+ 27 \mathfrak{t}^{4}\nonumber \\
&\quad + 61 \mathfrak{t}^{5}
+ 146 \mathfrak{t}^{6}
+ 319 \mathfrak{t}^{7}
+ 701 \mathfrak{t}^{8}
+ 1479 \mathfrak{t}^{9}
+ 3074 \mathfrak{t}^{10}
+ 6227 \mathfrak{t}^{11}
+ 12433 \mathfrak{t}^{12}
+ 24238 \mathfrak{t}^{13}
+ 46559 \mathfrak{t}^{14} + 87755 \mathfrak{t}^{15}\nonumber \\
&\quad + 162791 \mathfrak{t}^{16}
+ 296862 \mathfrak{t}^{17}
+ 533245 \mathfrak{t}^{18}
+ 942604 \mathfrak{t}^{19}
+ 1641943 \mathfrak{t}^{20}
+ 2817562 \mathfrak{t}^{21}
+ 4766462 \mathfrak{t}^{22}
+ 7949039 \mathfrak{t}^{23}\nonumber \\
&\quad 
+ 13075673 \mathfrak{t}^{24}
+ 21215509 \mathfrak{t}^{25}
+ 33967818 \mathfrak{t}^{26}
+ 53672334 \mathfrak{t}^{27}
+ 83721276 \mathfrak{t}^{28}
+ 128938668 \mathfrak{t}^{29} + 196113061 \mathfrak{t}^{30}\nonumber \\
&\quad
+ 294621966 \mathfrak{t}^{31}
+ 437277378 \mathfrak{t}^{32}
+ 641274985 \mathfrak{t}^{33}
+ 929417300 \mathfrak{t}^{34}
+ 1331427437 \mathfrak{t}^{35} + 1885556987 \mathfrak{t}^{36}\nonumber \\
&\quad
+ 2640188142 \mathfrak{t}^{37}
+ 3655701433 \mathfrak{t}^{38}
+ 5006129620 \mathfrak{t}^{39}
+ 6780934016 \mathfrak{t}^{40}
+ 9086244560 \mathfrak{t}^{41} + 12045972848 \mathfrak{t}^{42}\nonumber \\
&\quad
+ 15801902007 \mathfrak{t}^{43}
+ 20513360532 \mathfrak{t}^{44}
+ 26355310657 \mathfrak{t}^{45}
+ 33515711230 \mathfrak{t}^{46}
+ 42190816860 \mathfrak{t}^{47}\nonumber \\
&\quad + 52579620556 \mathfrak{t}^{48}
+ 64875914944 \mathfrak{t}^{49}
+ 79259771046 \mathfrak{t}^{50}
+ 95886714939 \mathfrak{t}^{51}
+ 114877028995 \mathfrak{t}^{52}\nonumber \\
&\quad + 136303392906 \mathfrak{t}^{53}
+ 160179911908 \mathfrak{t}^{54}
+ 186450497709 \mathfrak{t}^{55}
+ 214980384417 \mathfrak{t}^{56}
+ 245548274707 \mathfrak{t}^{57}\nonumber \\
&\quad + 277843337139 \mathfrak{t}^{58}
+ 311464045039 \mathfrak{t}^{59}
+ 345923239772 \mathfrak{t}^{60}
+ 380655501580 \mathfrak{t}^{61}
+ 415031369460 \mathfrak{t}^{62}\nonumber \\
&\quad + 448373380628 \mathfrak{t}^{63}
+ 479978300440 \mathfrak{t}^{64}
+ 509139652017 \mathfrak{t}^{65}
+ 535174708032 \mathfrak{t}^{66}
+ 557449173194 \mathfrak{t}^{67}\nonumber \\
&\quad + 575403955961 \mathfrak{t}^{68}
+ 588576717346 \mathfrak{t}^{69}
+ 596622852200 \mathfrak{t}^{70}
+ 599328799752 \mathfrak{t}^{71}
+ \cdots + \mathfrak{t}^{142}
).
\end{align}
}

\subsubsection{$U(9)$}
{\fontsize{9pt}{1pt}\selectfont
\begin{align}
&{\cal I}^{{\cal N}=8\text{ }U(9)(H)}(\mathfrak{t})\nonumber \\
&=
\frac{
1}{
(1 - \mathfrak{t})^2 (1 - \mathfrak{t}^3) (1 - \mathfrak{t}^4) (1 - \mathfrak{t}^5) (1 - \mathfrak{t}^6) (1 - \mathfrak{t}^7) (1 - \mathfrak{t}^8) (1 - \mathfrak{t}^9)
}
(1 + 2 \mathfrak{t}^{2} + \mathfrak{t}^{3} + 4 \mathfrak{t}^{4} + 3 \mathfrak{t}^{5} + 8 \mathfrak{t}^{6} + 7 \mathfrak{t}^{7} + 15 \mathfrak{t}^{8}\nonumber \\
&\quad + 15 \mathfrak{t}^{9} + 17 \mathfrak{t}^{10} + 20 \mathfrak{t}^{11} + 22 \mathfrak{t}^{12} + 22 \mathfrak{t}^{13} + 23 \mathfrak{t}^{14} + 21 \mathfrak{t}^{15} + 19 \mathfrak{t}^{16} + 15 \mathfrak{t}^{17} + 14 \mathfrak{t}^{18} + 9 \mathfrak{t}^{19} + 8 \mathfrak{t}^{20} + 5 \mathfrak{t}^{21} + 2 \mathfrak{t}^{22}\nonumber \\
&\quad + 2 \mathfrak{t}^{23} + \mathfrak{t}^{24}),\\
&{\cal I}^{U(9)\text{ ADHM-}[1](H)}(\mathfrak{t})\nonumber \\
&=
\frac{1}{
(1-\mathfrak{t})^2
(1-\mathfrak{t}^2)^2
(1-\mathfrak{t}^3)^2
(1-\mathfrak{t}^4)^2
(1-\mathfrak{t}^5)^2
(1-\mathfrak{t}^6)^2
(1-\mathfrak{t}^7)^2
(1-\mathfrak{t}^8)^2
(1-\mathfrak{t}^9)^2
}
(
1+\mathfrak{t}^{2} +2 \mathfrak{t}^{3}+4 \mathfrak{t}^{4}+6 \mathfrak{t}^{5}+12 \mathfrak{t}^{6}\nonumber \\
&\quad+18 \mathfrak{t}^{7}+33 \mathfrak{t}^{8}+52 \mathfrak{t}^{9}+79 \mathfrak{t}^{10}+120 \mathfrak{t}^{11}+188 \mathfrak{t}^{12}+270 \mathfrak{t}^{13}+398 \mathfrak{t}^{14}+562 \mathfrak{t}^{15}+788 \mathfrak{t}^{16}+1070 \mathfrak{t}^{17}+1454 \mathfrak{t}^{18}\nonumber \\
&\quad+1902 \mathfrak{t}^{19}+2484 \mathfrak{t}^{20}+3166 \mathfrak{t}^{21}+3976 \mathfrak{t}^{22}+4898 \mathfrak{t}^{23}+5974 \mathfrak{t}^{24}+7106 \mathfrak{t}^{25}+8374 \mathfrak{t}^{26}+9684 \mathfrak{t}^{27}+11040 \mathfrak{t}^{28}\nonumber \\
&\quad+12358 \mathfrak{t}^{29}+13679 \mathfrak{t}^{30}+14836 \mathfrak{t}^{31}+15914 \mathfrak{t}^{32}+16770 \mathfrak{t}^{33}+17427 \mathfrak{t}^{34}+17804 \mathfrak{t}^{35}+17980 \mathfrak{t}^{36}+\cdots+\mathfrak{t}^{72}
),\\
&{\cal I}^{U(9)\text{ ADHM-}[2](H)}(\mathfrak{t})\nonumber \\
&=
\frac{1}{
(1 - \mathfrak{t})^4 (1 - \mathfrak{t}^2)^2 (1 - \mathfrak{t}^3)^2 (1 - \mathfrak{t}^4)^3 (1 - \mathfrak{t}^5)^4 (1 - \mathfrak{t}^6)^6 (1 - \mathfrak{t}^7)^4 (1 - \mathfrak{t}^8)^4 (1 - \mathfrak{t}^9)^4 (1 - \mathfrak{t}^{10})^3
}
(1 - 2 \mathfrak{t} + 5 \mathfrak{t}^{2} + 9 \mathfrak{t}^{4}\nonumber \\
&\quad + 12 \mathfrak{t}^{5} + 45 \mathfrak{t}^{6} + 60 \mathfrak{t}^{7} + 181 \mathfrak{t}^{8} + 304 \mathfrak{t}^{9} + 702 \mathfrak{t}^{10} + 1280 \mathfrak{t}^{11} + 2690 \mathfrak{t}^{12} + 4902 \mathfrak{t}^{13} + 9703 \mathfrak{t}^{14} + 17684 \mathfrak{t}^{15}\nonumber \\
&\quad + 33246 \mathfrak{t}^{16} + 59744 \mathfrak{t}^{17} + 108583 \mathfrak{t}^{18} + 191030 \mathfrak{t}^{19} + 337079 \mathfrak{t}^{20} + 580720 \mathfrak{t}^{21} + 997077 \mathfrak{t}^{22} + 1681138 \mathfrak{t}^{23}\nonumber \\
&\quad + 2815911 \mathfrak{t}^{24} + 4644968 \mathfrak{t}^{25} + 7601434 \mathfrak{t}^{26} + 12273272 \mathfrak{t}^{27} + 19641831 \mathfrak{t}^{28} + 31055310 \mathfrak{t}^{29} + 48649506 \mathfrak{t}^{30}\nonumber \\
&\quad + 75354906 \mathfrak{t}^{31} + 115635710 \mathfrak{t}^{32} + 175563894 \mathfrak{t}^{33} + 264072778 \mathfrak{t}^{34} + 393181310 \mathfrak{t}^{35} + 580021457 \mathfrak{t}^{36}\nonumber \\
&\quad + 847313248 \mathfrak{t}^{37} + 1226552793 \mathfrak{t}^{38} + 1758809888 \mathfrak{t}^{39} + 2499538497 \mathfrak{t}^{40} + 3519778738 \mathfrak{t}^{41} + 4913038174 \mathfrak{t}^{42}\nonumber \\
&\quad + 6796841268 \mathfrak{t}^{43} + 9322154872 \mathfrak{t}^{44} + 12674892698 \mathfrak{t}^{45} + 17088163538 \mathfrak{t}^{46} + 22843002988 \mathfrak{t}^{47} + 30283276500 \mathfrak{t}^{48}\nonumber \\
&\quad + 39814280450 \mathfrak{t}^{49} + 51919771502 \mathfrak{t}^{50} + 67156174434 \mathfrak{t}^{51} + 86170630901 \mathfrak{t}^{52} + 109687969004 \mathfrak{t}^{53}\nonumber \\
&\quad + 138527956029 \mathfrak{t}^{54} + 173581619422 \mathfrak{t}^{55} + 215825927423 \mathfrak{t}^{56} + 266285676012 \mathfrak{t}^{57} + 326044199007 \mathfrak{t}^{58}\nonumber \\
&\quad + 396187935514 \mathfrak{t}^{59} + 477812574527 \mathfrak{t}^{60} + 571948958502 \mathfrak{t}^{61} + 679566373859 \mathfrak{t}^{62} + 801480390722 \mathfrak{t}^{63}\nonumber \\
&\quad + 938357696799 \mathfrak{t}^{64} + 1090609809750 \mathfrak{t}^{65} + 1258406122966 \mathfrak{t}^{66} + 1441560421328 \mathfrak{t}^{67} + 1639561580704 \mathfrak{t}^{68}\nonumber \\
&\quad + 1851461255736 \mathfrak{t}^{69} + 2075932773978 \mathfrak{t}^{70} + 2311169379340 \mathfrak{t}^{71} + 2554980063163 \mathfrak{t}^{72} + 2804706293820 \mathfrak{t}^{73}\nonumber \\
&\quad + 3057360694312 \mathfrak{t}^{74} + 3309565492672 \mathfrak{t}^{75} + 3557734759478 \mathfrak{t}^{76} + 3798032916192 \mathfrak{t}^{77} + 4026593642451 \mathfrak{t}^{78}\nonumber \\
&\quad + 4239491072816 \mathfrak{t}^{79} + 4432982685300 \mathfrak{t}^{80} + 4603479681474 \mathfrak{t}^{81} + 4747796830193 \mathfrak{t}^{82} + 4863106473860 \mathfrak{t}^{83}\nonumber \\
&\quad + 4947175498721 \mathfrak{t}^{84} + 4998288499912 \mathfrak{t}^{85} + 5015454641234 \mathfrak{t}^{86} + \cdots + \mathfrak{t}^{172}).
\end{align}
}

\subsubsection{$U(10)$}
{\fontsize{9pt}{1pt}\selectfont
\begin{align}
&{\cal I}^{{\cal N}=8\text{ }U(10)(H)}(\mathfrak{t})\nonumber \\
&=
\frac{
1}
{
(1 - \mathfrak{t}) (1 - \mathfrak{t}^2) (1 - \mathfrak{t}^3) (1 - \mathfrak{t}^4) (1 - \mathfrak{t}^5) (1 - \mathfrak{t}^6) (1 - \mathfrak{t}^7) (1 - \mathfrak{t}^8) (1 - \mathfrak{t}^9) (1 - \mathfrak{t}^{10})
}
(1 + \mathfrak{t} + 2 \mathfrak{t}^{2} + 3 \mathfrak{t}^{3} + 5 \mathfrak{t}^{4} + 7 \mathfrak{t}^{5}\nonumber \\
&\quad  + 11 \mathfrak{t}^{6} + 15 \mathfrak{t}^{7} + 22 \mathfrak{t}^{8} + 30 \mathfrak{t}^{9} + 42 \mathfrak{t}^{10} + 45 \mathfrak{t}^{11} + 56 \mathfrak{t}^{12} + 62 \mathfrak{t}^{13} + 71 \mathfrak{t}^{14} + 74 \mathfrak{t}^{15} + 80 \mathfrak{t}^{16} + 76 \mathfrak{t}^{17} + 77 \mathfrak{t}^{18} + 67 \mathfrak{t}^{19}\nonumber \\
&\quad  + 63 \mathfrak{t}^{20} + 53 \mathfrak{t}^{21} + 47 \mathfrak{t}^{22} + 34 \mathfrak{t}^{23} + 29 \mathfrak{t}^{24} + 19 \mathfrak{t}^{25} + 14 \mathfrak{t}^{26} + 8 \mathfrak{t}^{27} + 6 \mathfrak{t}^{28} + 2 \mathfrak{t}^{29} + 2 \mathfrak{t}^{30}),\\
&{\cal I}^{U(10)\text{ ADHM-}[1](H)}(\mathfrak{t})\nonumber \\
&=
\frac{1}{
(1-\mathfrak{t})^2
(1-\mathfrak{t}^2)^3
(1-\mathfrak{t}^3)^2
(1-\mathfrak{t}^4)
(1-\mathfrak{t}^5)^2
(1-\mathfrak{t}^6)^2
(1-\mathfrak{t}^7)^2
(1-\mathfrak{t}^8)^2
(1-\mathfrak{t}^9)^2
(1-\mathfrak{t}^{10})^2
}(
1+2 \mathfrak{t}^{3}+4 \mathfrak{t}^{4}+4 \mathfrak{t}^{5}\nonumber \\
&\quad +8 \mathfrak{t}^{6}+14 \mathfrak{t}^{7}+25 \mathfrak{t}^{8}+38 \mathfrak{t}^{9}+63 \mathfrak{t}^{10}+90 \mathfrak{t}^{11}+146 \mathfrak{t}^{12}+218 \mathfrak{t}^{13}+327 \mathfrak{t}^{14}+468 \mathfrak{t}^{15}+685 \mathfrak{t}^{16}+956 \mathfrak{t}^{17}+1353 \mathfrak{t}^{18}\nonumber \\
&\quad +1838 \mathfrak{t}^{19}+2510 \mathfrak{t}^{20}+3334 \mathfrak{t}^{21}+4421 \mathfrak{t}^{22}+5712 \mathfrak{t}^{23}+7373 \mathfrak{t}^{24}+9290 \mathfrak{t}^{25}+11640 \mathfrak{t}^{26}+14314 \mathfrak{t}^{27}+17494 \mathfrak{t}^{28}\nonumber \\
&\quad +20958 \mathfrak{t}^{29}+24962 \mathfrak{t}^{30}+29202 \mathfrak{t}^{31}+33904 \mathfrak{t}^{32}+38728 \mathfrak{t}^{33}+43888 \mathfrak{t}^{34}+48934 \mathfrak{t}^{35}+54138 \mathfrak{t}^{36}+58968 \mathfrak{t}^{37}\nonumber \\
&\quad +63698 \mathfrak{t}^{38}+67794 \mathfrak{t}^{39}+71538 \mathfrak{t}^{40}+74358 \mathfrak{t}^{41}+76672 \mathfrak{t}^{42}+77900 \mathfrak{t}^{43}+78460 \mathfrak{t}^{44}+\cdots+\mathfrak{t}^{88}
),\\
&{\cal I}^{U(10)\text{ ADHM-}[2](H)}(\mathfrak{t})\nonumber \\
&=
\frac{1}{
(1 - \mathfrak{t})^2 (1 - \mathfrak{t}^2)^2 (1 - \mathfrak{t}^3)^2 (1 - \mathfrak{t}^4)^4 (1 - \mathfrak{t}^5)^4 (1 - \mathfrak{t}^6)^6 (1 - \mathfrak{t}^7)^5 (1 - \mathfrak{t}^8)^4 (1 - \mathfrak{t}^9)^4 (1 - \mathfrak{t}^{10})^4 (1 - \mathfrak{t}^{11})^3
}
(
1 + 4 \mathfrak{t}^{2}\nonumber \\
&\quad + 8 \mathfrak{t}^{3} + 20 \mathfrak{t}^{4} + 46 \mathfrak{t}^{5} + 112 \mathfrak{t}^{6} + 237 \mathfrak{t}^{7} + 536 \mathfrak{t}^{8} + 1122 \mathfrak{t}^{9} + 2375 \mathfrak{t}^{10} + 4855 \mathfrak{t}^{11} + 9879 \mathfrak{t}^{12} + 19620 \mathfrak{t}^{13} + 38679 \mathfrak{t}^{14}\nonumber \\
&\quad + 74860 \mathfrak{t}^{15} + 143472 \mathfrak{t}^{16} + 271029 \mathfrak{t}^{17} + 506679 \mathfrak{t}^{18} + 935297 \mathfrak{t}^{19} + 1708843 \mathfrak{t}^{20} + 3086526 \mathfrak{t}^{21} + 5518852 \mathfrak{t}^{22}\nonumber \\
&\quad + 9763619 \mathfrak{t}^{23} + 17104095 \mathfrak{t}^{24} + 29663629 \mathfrak{t}^{25} + 50957309 \mathfrak{t}^{26} + 86699539 \mathfrak{t}^{27} + 146152876 \mathfrak{t}^{28} + 244109690 \mathfrak{t}^{29}\nonumber \\
&\quad + 404071766 \mathfrak{t}^{30} + 662906540 \mathfrak{t}^{31} + 1078079220 \mathfrak{t}^{32} + 1738146055 \mathfrak{t}^{33} + 2778609560 \mathfrak{t}^{34} + 4404648758 \mathfrak{t}^{35}\nonumber \\
&\quad + 6924612705 \mathfrak{t}^{36} + 10797379648 \mathfrak{t}^{37} + 16700556732 \mathfrak{t}^{38} + 25625442946 \mathfrak{t}^{39} + 39010872907 \mathfrak{t}^{40} + 58926587197 \mathfrak{t}^{41}\nonumber \\
&\quad + 88326377343 \mathfrak{t}^{42} + 131389131208 \mathfrak{t}^{43} + 193980258792 \mathfrak{t}^{44} + 284262632865 \mathfrak{t}^{45} + 413506977805 \mathfrak{t}^{46}\nonumber \\
&\quad + 597145860562 \mathfrak{t}^{47} + 856144451274 \mathfrak{t}^{48} + 1218750927815 \mathfrak{t}^{49} + 1722728217990 \mathfrak{t}^{50} + 2418150280550 \mathfrak{t}^{51}\nonumber \\
&\quad + 3370896108607 \mathfrak{t}^{52} + 4666941983190 \mathfrak{t}^{53} + 6417614214796 \mathfrak{t}^{54} + 8765909333158 \mathfrak{t}^{55} + 11894061685935 \mathfrak{t}^{56}\nonumber \\
&\quad + 16032448867833 \mathfrak{t}^{57} + 21470009054828 \mathfrak{t}^{58} + 28566205756582 \mathfrak{t}^{59} + 37764669482761 \mathfrak{t}^{60} + 49608442007688 \mathfrak{t}^{61}\nonumber \\
&\quad + 64756854907911 \mathfrak{t}^{62} + 84003788708149 \mathfrak{t}^{63} + 108297184277879 \mathfrak{t}^{64} + 138759297559868 \mathfrak{t}^{65}\nonumber \\
&\quad + 176707349030151 \mathfrak{t}^{66} + 223673736818777 \mathfrak{t}^{67} + 281425204939595 \mathfrak{t}^{68} + 351979775988529 \mathfrak{t}^{69}\nonumber \\
&\quad + 437620584812847 \mathfrak{t}^{70} + 540905080037550 \mathfrak{t}^{71} + 664668548396673 \mathfrak{t}^{72} + 812020181659988 \mathfrak{t}^{73}\nonumber \\
&\quad + 986330620543689 \mathfrak{t}^{74} + 1191209134203411 \mathfrak{t}^{75} + 1430469603797609 \mathfrak{t}^{76} + 1708083679494686 \mathfrak{t}^{77}\nonumber \\
&\quad + 2028120843606380 \mathfrak{t}^{78} + 2394674289765652 \mathfrak{t}^{79} + 2811773277082400 \mathfrak{t}^{80} + 3283281739154659 \mathfrak{t}^{81}\nonumber \\
&\quad + 3812785059167559 \mathfrak{t}^{82} + 4403465911672592 \mathfrak{t}^{83} + 5057972536657394 \mathfrak{t}^{84} + 5778281553589210 \mathfrak{t}^{85}\nonumber \\
&\quad + 6565560128473393 \mathfrak{t}^{86} + 7420030652750900 \mathfrak{t}^{87} + 8340843918650905 \mathfrak{t}^{88} + 9325964562681645 \mathfrak{t}^{89}\nonumber \\
&\quad + 10372075371076527 \mathfrak{t}^{90} + 11474504129686120 \mathfrak{t}^{91} + 12627179421193918 \mathfrak{t}^{92} + 13822618078083771 \mathfrak{t}^{93}\nonumber \\
&\quad + 15051949583998778 \mathfrak{t}^{94} + 16304978237027169 \mathfrak{t}^{95} + 17570286385612258 \mathfrak{t}^{96} + 18835376888365574 \mathfrak{t}^{97}\nonumber \\
&\quad + 20086855495266342 \mathfrak{t}^{98} + 21310648210451422 \mathfrak{t}^{99} + 22492251505068000 \mathfrak{t}^{100} + 23617007403650559 \mathfrak{t}^{101}\nonumber \\
&\quad + 24670398785415138 \mathfrak{t}^{102} + 25638354486582958 \mathfrak{t}^{103} + 26507557842402409 \mathfrak{t}^{104} + 27265746904898686 \mathfrak{t}^{105}\nonumber \\
&\quad + 27901999512174388 \mathfrak{t}^{106} + 28406991505001717 \mathfrak{t}^{107} + 28773222243213245 \mathfrak{t}^{108} + 28995197263795448 \mathfrak{t}^{109}\nonumber \\
&\quad + 29069564576783270 \mathfrak{t}^{110} + \cdots + \mathfrak{t}^{220}).
\end{align}
}

\subsubsection{$U(11)$}
{\fontsize{9pt}{1pt}\selectfont
\begin{align}
&{\cal I}^{{\cal N}=8\text{ }U(11)(H)}(\mathfrak{t})\nonumber \\
&=
\frac{1}{
(1 - \mathfrak{t})^2 (1 - \mathfrak{t}^3) (1 - \mathfrak{t}^4) (1 - \mathfrak{t}^5) (1 - \mathfrak{t}^6) (1 - \mathfrak{t}^7) (1 - \mathfrak{t}^8) (1 - \mathfrak{t}^9) (1 - \mathfrak{t}^{10}) (1 - \mathfrak{t}^{11})
}
(
1 + 2 \mathfrak{t}^{2} + \mathfrak{t}^{3} + 4 \mathfrak{t}^{4} + 3 \mathfrak{t}^{5}\nonumber \\
&\quad + 8 \mathfrak{t}^{6} + 7 \mathfrak{t}^{7} + 15 \mathfrak{t}^{8} + 15 \mathfrak{t}^{9} + 27 \mathfrak{t}^{10} + 29 \mathfrak{t}^{11} + 36 \mathfrak{t}^{12} + 42 \mathfrak{t}^{13} + 50 \mathfrak{t}^{14} + 55 \mathfrak{t}^{15} + 62 \mathfrak{t}^{16} + 65 \mathfrak{t}^{17} + 69 \mathfrak{t}^{18} + 68 \mathfrak{t}^{19}\nonumber \\
&\quad + 69 \mathfrak{t}^{20} + 62 \mathfrak{t}^{21} + 62 \mathfrak{t}^{22} + 54 \mathfrak{t}^{23} + 49 \mathfrak{t}^{24} + 41 \mathfrak{t}^{25} + 35 \mathfrak{t}^{26} + 27 \mathfrak{t}^{27} + 21 \mathfrak{t}^{28} + 16 \mathfrak{t}^{29} + 11 \mathfrak{t}^{30} + 8 \mathfrak{t}^{31} + 5 \mathfrak{t}^{32} + 2 \mathfrak{t}^{33}\nonumber \\
&\quad + 2 \mathfrak{t}^{34} + \mathfrak{t}^{35}),\\
&{\cal I}^{U(11)\text{ ADHM-}[1](H)}(\mathfrak{t})\nonumber \\
&=
\frac{1}{
(1-\mathfrak{t})^2
(1-\mathfrak{t}^2)^2
(1-\mathfrak{t}^3)^2
(1-\mathfrak{t}^4)^2
(1-\mathfrak{t}^5)^2
(1-\mathfrak{t}^6)^2
(1-\mathfrak{t}^7)^2
(1-\mathfrak{t}^8)^2
(1-\mathfrak{t}^9)^2
(1-\mathfrak{t}^{10})^2
(1-\mathfrak{t}^{11})^2
}
(
1+\mathfrak{t}^{2}\nonumber \\
&\quad+2 \mathfrak{t}^{3}+4 \mathfrak{t}^{4}+6 \mathfrak{t}^{5}+12 \mathfrak{t}^{6}+18 \mathfrak{t}^{7}+33 \mathfrak{t}^{8}+52 \mathfrak{t}^{9}+88 \mathfrak{t}^{10}+138 \mathfrak{t}^{11}+218 \mathfrak{t}^{12}+332 \mathfrak{t}^{13}+517 \mathfrak{t}^{14}+774 \mathfrak{t}^{15}+1160 \mathfrak{t}^{16}\nonumber \\
&\quad+1696 \mathfrak{t}^{17}+2478 \mathfrak{t}^{18}+3538 \mathfrak{t}^{19}+5037 \mathfrak{t}^{20}+7032 \mathfrak{t}^{21}+9759 \mathfrak{t}^{22}+13332 \mathfrak{t}^{23}+18073 \mathfrak{t}^{24}+24154 \mathfrak{t}^{25}+32018 \mathfrak{t}^{26}\nonumber \\
&\quad+41890 \mathfrak{t}^{27}+54317 \mathfrak{t}^{28}+69590 \mathfrak{t}^{29}+88359 \mathfrak{t}^{30}+110892 \mathfrak{t}^{31}+137930 \mathfrak{t}^{32}+169672 \mathfrak{t}^{33}+206829 \mathfrak{t}^{34}+249474 \mathfrak{t}^{35}\nonumber \\
&\quad+298185 \mathfrak{t}^{36}+352752 \mathfrak{t}^{37}+413592 \mathfrak{t}^{38}+480074 \mathfrak{t}^{39}+552285 \mathfrak{t}^{40}+629168 \mathfrak{t}^{41}+710427 \mathfrak{t}^{42}+794500 \mathfrak{t}^{43}\nonumber \\
&\quad+880751 \mathfrak{t}^{44}+967164 \mathfrak{t}^{45}+1052803 \mathfrak{t}^{46}+1135390 \mathfrak{t}^{47}+1213850 \mathfrak{t}^{48}+1285778 \mathfrak{t}^{49}+1350274 \mathfrak{t}^{50}+1405030 \mathfrak{t}^{51}\nonumber \\
&\quad+1449453 \mathfrak{t}^{52}+1481708 \mathfrak{t}^{53}+1501706 \mathfrak{t}^{54}+1508168 \mathfrak{t}^{55}+\cdots +\mathfrak{t}^{110}
),\\
&{\cal I}^{U(11)\text{ ADHM-}[2](H)}(\mathfrak{t})\nonumber \\
&=
\frac{1}{
(1 - \mathfrak{t}^2)^3 (1 - \mathfrak{t}^3)^4 (1 - \mathfrak{t}^4)^4 (1 - \mathfrak{t}^5)^5 (1 - \mathfrak{t}^6)^4 (1 - \mathfrak{t}^7)^5 (1 - \mathfrak{t}^8)^4 (1 - \mathfrak{t}^9)^4 (1 - \mathfrak{t}^{10})^4 (1 - \mathfrak{t}^{11})^4 (1 - \mathfrak{t}^{12})^3
}
(1 + 2 \mathfrak{t} + 6 \mathfrak{t}^{2}\nonumber \\
&\quad + 16 \mathfrak{t}^{3} + 42 \mathfrak{t}^{4} + 99 \mathfrak{t}^{5} + 235 \mathfrak{t}^{6} + 527 \mathfrak{t}^{7} + 1170 \mathfrak{t}^{8} + 2518 \mathfrak{t}^{9} + 5339 \mathfrak{t}^{10} + 11079 \mathfrak{t}^{11} + 22682 \mathfrak{t}^{12} + 45635 \mathfrak{t}^{13} + 90661 \mathfrak{t}^{14}\nonumber \\
&\quad + 177617 \mathfrak{t}^{15} + 343873 \mathfrak{t}^{16} + 657678 \mathfrak{t}^{17} + 1244242 \mathfrak{t}^{18} + 2328238 \mathfrak{t}^{19} + 4312594 \mathfrak{t}^{20} + 7908080 \mathfrak{t}^{21} + 14363114 \mathfrak{t}^{22}\nonumber \\
&\quad + 25842244 \mathfrak{t}^{23} + 46075765 \mathfrak{t}^{24} + 81420823 \mathfrak{t}^{25} + 142637777 \mathfrak{t}^{26} + 247759037 \mathfrak{t}^{27} + 426783456 \mathfrak{t}^{28} + 729162865 \mathfrak{t}^{29}\nonumber \\
&\quad + 1235807572 \mathfrak{t}^{30} + 2077960116 \mathfrak{t}^{31} + 3466908598 \mathfrak{t}^{32} + 5740014091 \mathfrak{t}^{33} + 9431905065 \mathfrak{t}^{34} + 15383126820 \mathfrak{t}^{35}\nonumber \\
&\quad + 24905388013 \mathfrak{t}^{36} + 40029956697 \mathfrak{t}^{37} + 63879356048 \mathfrak{t}^{38} + 101217718696 \mathfrak{t}^{39} + 159261342789 \mathfrak{t}^{40}\nonumber \\
&\quad + 248860975531 \mathfrak{t}^{41} + 386216201413 \mathfrak{t}^{42} + 595338986631 \mathfrak{t}^{43} + 911571847587 \mathfrak{t}^{44} + 1386568974296 \mathfrak{t}^{45}\nonumber \\
&\quad + 2095303360759 \mathfrak{t}^{46} + 3145842681922 \mathfrak{t}^{47} + 4692898086600 \mathfrak{t}^{48} + 6956451697745 \mathfrak{t}^{49} + 10247194999052 \mathfrak{t}^{50}\nonumber \\
&\quad + 15000996343678 \mathfrak{t}^{51} + 21825285333447 \mathfrak{t}^{52} + 31560993299330 \mathfrak{t}^{53} + 45364698777787 \mathfrak{t}^{54} + 64816737917051 \mathfrak{t}^{55}\nonumber \\
&\quad + 92062498549519 \mathfrak{t}^{56} + 129995681050081 \mathfrak{t}^{57} + 182494318472938 \mathfrak{t}^{58} + 254722434313906 \mathfrak{t}^{59}\nonumber \\
&\quad + 353512836898046 \mathfrak{t}^{60} + 487849154614141 \mathfrak{t}^{61} + 669468419217469 \mathfrak{t}^{62} + 913608495722319 \mathfrak{t}^{63}\nonumber \\
&\quad + 1239928254822698 \mathfrak{t}^{64} + 1673631423777934 \mathfrak{t}^{65} + 2246828619641834 \mathfrak{t}^{66} + 3000174533798362 \mathfrak{t}^{67}\nonumber \\
&\quad + 3984820047182998 \mathfrak{t}^{68} + 5264719973656582 \mathfrak{t}^{69} + 6919338104191495 \mathfrak{t}^{70} + 9046789263056230 \mathfrak{t}^{71}\nonumber \\
&\quad + 11767455823153968 \mathfrak{t}^{72} + 15228109653843885 \mathfrak{t}^{73} + 19606563370951145 \mathfrak{t}^{74} + 25116862024443428 \mathfrak{t}^{75}\nonumber \\
&\quad + 32015012868310531 \mathfrak{t}^{76} + 40605230333944369 \mathfrak{t}^{77} + 51246652358119298 \mathfrak{t}^{78} + 64360455051292155 \mathfrak{t}^{79}\nonumber \\
&\quad + 80437264142335401 \mathfrak{t}^{80} + 100044724215406648 \mathfrak{t}^{81} + 123835052084405979 \mathfrak{t}^{82} + 152552357179314431 \mathfrak{t}^{83}\nonumber \\
&\quad + 187039474563238026 \mathfrak{t}^{84} + 228244011102559244 \mathfrak{t}^{85} + 277223271384223712 \mathfrak{t}^{86} + 335147690002905038 \mathfrak{t}^{87}\nonumber \\
&\quad + 403302374378120616 \mathfrak{t}^{88} + 483086336113779538 \mathfrak{t}^{89} + 576008987586700824 \mathfrak{t}^{90} + 683683477728674869 \mathfrak{t}^{91}\nonumber \\
&\quad + 807816470659884681 \mathfrak{t}^{92} + 950194000396522682 \mathfrak{t}^{93} + 1112663103697175417 \mathfrak{t}^{94} + 1297109000971810935 \mathfrak{t}^{95}\nonumber \\
&\quad + 1505427707270383061 \mathfrak{t}^{96} + 1739494062327491467 \mathfrak{t}^{97} + 2001125321868805946 \mathfrak{t}^{98} + 2292040592071387932 \mathfrak{t}^{99}\nonumber \\
&\quad + 2613816573424824275 \mathfrak{t}^{100} + 2967840238037577854 \mathfrak{t}^{101} + 3355259261558707909 \mathfrak{t}^{102}\nonumber \\
&\quad + 3776931184193619400 \mathfrak{t}^{103} + 4233372459695421262 \mathfrak{t}^{104} + 4724708671551343495 \mathfrak{t}^{105}\nonumber \\
&\quad + 5250627337548513771 \mathfrak{t}^{106} + 5810334780825196163 \mathfrak{t}^{107} + 6402518615752503054 \mathfrak{t}^{108}\nonumber \\
&\quad + 7025317363267950850 \mathfrak{t}^{109} + 7676298685637647886 \mathfrak{t}^{110} + 8352447588085443561 \mathfrak{t}^{111}\nonumber \\
&\quad + 9050165803755330325 \mathfrak{t}^{112} + 9765283323309187883 \mathfrak{t}^{113} + 10493082797929962877 \mathfrak{t}^{114}\nonumber \\
&\quad + 11228337190575180450 \mathfrak{t}^{115} + 11965360739347818185 \mathfrak{t}^{116} + 12698072876271182718 \mathfrak{t}^{117}\nonumber \\
&\quad + 13420074395498132545 \mathfrak{t}^{118} + 14124734725207425555 \mathfrak{t}^{119} + 14805288822454950627 \mathfrak{t}^{120}\nonumber \\
&\quad + 15454941807623643468 \mathfrak{t}^{121} + 16066979191641944473 \mathfrak{t}^{122} + 16634880240088135549 \mathfrak{t}^{123}\nonumber \\
&\quad + 17152431878546968671 \mathfrak{t}^{124} + 17613840374788464127 \mathfrak{t}^{125} + 18013838054887265974 \mathfrak{t}^{126}\nonumber \\
&\quad + 18347782309713444850 \mathfrak{t}^{127} + 18611744346318829558 \mathfrak{t}^{128} + 18802585309084015484 \mathfrak{t}^{129}\nonumber \\
&\quad + 18918017758743384263 \mathfrak{t}^{130} + 18956650814916034454 \mathfrak{t}^{131} + \cdots + \mathfrak{t}^{262}).
\end{align}
}

\subsubsection{$U(12)$}
{\fontsize{9pt}{1pt}\selectfont
\begin{align}
&{\cal I}^{{\cal N}=8\text{ }U(12)(H)}(\mathfrak{t})\nonumber \\
&=
\frac{1}{
(1 - \mathfrak{t}) (1 - \mathfrak{t}^2) (1 - \mathfrak{t}^3) (1 - \mathfrak{t}^4) (1 - \mathfrak{t}^5) (1 - \mathfrak{t}^6) (1 - \mathfrak{t}^7) (1 - \mathfrak{t}^8) (1 - \mathfrak{t}^9) (1 - \mathfrak{t}^{10}) (1 - \mathfrak{t}^{11})(1 - \mathfrak{t}^{12})}
(
1 + \mathfrak{t} + 2 \mathfrak{t}^{2}\nonumber \\
&\quad + 3 \mathfrak{t}^{3} + 5 \mathfrak{t}^{4} + 7 \mathfrak{t}^{5} + 11 \mathfrak{t}^{6} + 15 \mathfrak{t}^{7} + 22 \mathfrak{t}^{8} + 30 \mathfrak{t}^{9} + 42 \mathfrak{t}^{10} + 56 \mathfrak{t}^{11} + 77 \mathfrak{t}^{12} + 88 \mathfrak{t}^{13} + 110 \mathfrak{t}^{14} + 129 \mathfrak{t}^{15} + 153 \mathfrak{t}^{16}\nonumber \\
&\quad + 171 \mathfrak{t}^{17} + 196 \mathfrak{t}^{18} + 209 \mathfrak{t}^{19} + 229 \mathfrak{t}^{20} + 235 \mathfrak{t}^{21} + 244 \mathfrak{t}^{22} + 238 \mathfrak{t}^{23} + 241 \mathfrak{t}^{24} + 226 \mathfrak{t}^{25} + 218 \mathfrak{t}^{26} + 198 \mathfrak{t}^{27} + 182 \mathfrak{t}^{28}\nonumber \\
&\quad + 155 \mathfrak{t}^{29} + 139 \mathfrak{t}^{30} + 111 \mathfrak{t}^{31} + 95 \mathfrak{t}^{32} + 73 \mathfrak{t}^{33} + 58 \mathfrak{t}^{34} + 41 \mathfrak{t}^{35} + 33 \mathfrak{t}^{36} + 20 \mathfrak{t}^{37} + 14 \mathfrak{t}^{38} + 8 \mathfrak{t}^{39} + 6 \mathfrak{t}^{40} + 2 \mathfrak{t}^{41}\nonumber \\
&\quad + 2 \mathfrak{t}^{42}),\\
&{\cal I}^{U(12)\text{ ADHM-}[1](H)}(\mathfrak{t})\nonumber \\
&=
\frac{1}{
(1-\mathfrak{t})^2
(1-\mathfrak{t}^2)^2
(1-\mathfrak{t}^3)^2
(1-\mathfrak{t}^4)^2
(1-\mathfrak{t}^5)^2
(1-\mathfrak{t}^6)^2
(1-\mathfrak{t}^7)^2
(1-\mathfrak{t}^8)^2
(1-\mathfrak{t}^9)^2
(1-\mathfrak{t}^{10})^2
(1-\mathfrak{t}^{11})^2
(1-\mathfrak{t}^{12})^2
}\nonumber \\
&\quad\times (
1+\mathfrak{t}^{2}+2 \mathfrak{t}^{3}+4 \mathfrak{t}^{4}+6 \mathfrak{t}^{5}+12 \mathfrak{t}^{6}+18 \mathfrak{t}^{7}+33 \mathfrak{t}^{8}+52 \mathfrak{t}^{9}+88 \mathfrak{t}^{10}+138 \mathfrak{t}^{11}+229 \mathfrak{t}^{12}+342 \mathfrak{t}^{13}+544 \mathfrak{t}^{14}+824 \mathfrak{t}^{15}\nonumber \\
&\quad+1261 \mathfrak{t}^{16}+1868 \mathfrak{t}^{17}+2798 \mathfrak{t}^{18}+4064 \mathfrak{t}^{19}+5941 \mathfrak{t}^{20}+8488 \mathfrak{t}^{21}+12119 \mathfrak{t}^{22}+16998 \mathfrak{t}^{23}+23797 \mathfrak{t}^{24}+32748 \mathfrak{t}^{25}\nonumber \\
&\quad+44926 \mathfrak{t}^{26}+60792 \mathfrak{t}^{27}+81794 \mathfrak{t}^{28}+108712 \mathfrak{t}^{29}+143664 \mathfrak{t}^{30}+187630 \mathfrak{t}^{31}+243556 \mathfrak{t}^{32}+312794 \mathfrak{t}^{33}\nonumber \\
&\quad+398998 \mathfrak{t}^{34}+503942 \mathfrak{t}^{35}+632174 \mathfrak{t}^{36}+785398 \mathfrak{t}^{37}+969096 \mathfrak{t}^{38}+1184936 \mathfrak{t}^{39}+1438637 \mathfrak{t}^{40}+1731426 \mathfrak{t}^{41}\nonumber \\
&\quad+2069296 \mathfrak{t}^{42}+2452014 \mathfrak{t}^{43}+2885325 \mathfrak{t}^{44}+3367260 \mathfrak{t}^{45}+3902193 \mathfrak{t}^{46}+4485872 \mathfrak{t}^{47}+5121164 \mathfrak{t}^{48}\nonumber \\
&\quad+5800198 \mathfrak{t}^{49}+6524105 \mathfrak{t}^{50}+7281716 \mathfrak{t}^{51}+8071384 \mathfrak{t}^{52}+8878682 \mathfrak{t}^{53}+9700200 \mathfrak{t}^{54}+10517944 \mathfrak{t}^{55}\nonumber \\
&\quad+11327328 \mathfrak{t}^{56}+12108452 \mathfrak{t}^{57}+12855732 \mathfrak{t}^{58}+13548794 \mathfrak{t}^{59}+14183157 \mathfrak{t}^{60}+14738374 \mathfrak{t}^{61}+15212608 \mathfrak{t}^{62}\nonumber \\
&\quad+15587956 \mathfrak{t}^{63}+15865271 \mathfrak{t}^{64}+16030440 \mathfrak{t}^{65}+16088968 \mathfrak{t}^{66}+\cdots +\mathfrak{t}^{132}
),\\
&{\cal I}^{U(12)\text{ ADHM-}[2](H)}(\mathfrak{t})\nonumber \\
&=
\frac{1}{
(1 + \mathfrak{t})^3 (1 - \mathfrak{t}^2) (1 - \mathfrak{t}^3)^4 (1 - \mathfrak{t}^4)^3 (1 - \mathfrak{t}^5)^5 (1 - \mathfrak{t}^6)^4 (1 - \mathfrak{t}^7)^7 (1 - \mathfrak{t}^8)^5 (1 - \mathfrak{t}^9)^4 (1 - \mathfrak{t}^{10})^4
(1 - \mathfrak{t}^{11})^4 (1 - \mathfrak{t}^{12})^4
}\nonumber \\
&\quad\times\frac{1}{(1 - \mathfrak{t}^{13})^3}
(1 + 5 \mathfrak{t} + 17 \mathfrak{t}^{2} + 51 \mathfrak{t}^{3} + 144 \mathfrak{t}^{4} + 381 \mathfrak{t}^{5} + 960 \mathfrak{t}^{6} + 2321 \mathfrak{t}^{7} + 5431 \mathfrak{t}^{8} + 12354 \mathfrak{t}^{9} + 27424 \mathfrak{t}^{10}\nonumber \\
&\quad + 59556 \mathfrak{t}^{11} + 126870 \mathfrak{t}^{12} + 265584 \mathfrak{t}^{13} + 547196 \mathfrak{t}^{14} + 1111028 \mathfrak{t}^{15} + 2225577 \mathfrak{t}^{16} + 4402361 \mathfrak{t}^{17}\nonumber \\
&\quad + 8606072 \mathfrak{t}^{18} + 16637616 \mathfrak{t}^{19} + 31827565 \mathfrak{t}^{20} + 60278563 \mathfrak{t}^{21} + 113074920 \mathfrak{t}^{22} + 210176273 \mathfrak{t}^{23} + 387228697 \mathfrak{t}^{24}\nonumber \\
&\quad + 707378062 \mathfrak{t}^{25} + 1281609918 \mathfrak{t}^{26} + 2303493614 \mathfrak{t}^{27} + 4108103311 \mathfrak{t}^{28} + 7271187179 \mathfrak{t}^{29} + 12774898730 \mathfrak{t}^{30}\nonumber \\
&\quad + 22282760092 \mathfrak{t}^{31} + 38592689290 \mathfrak{t}^{32} + 66378170756 \mathfrak{t}^{33} + 113392595542 \mathfrak{t}^{34} + 192412856560 \mathfrak{t}^{35}\nonumber \\
&\quad + 324355058694 \mathfrak{t}^{36} + 543235901068 \mathfrak{t}^{37} + 904021339784 \mathfrak{t}^{38} + 1494961731480 \mathfrak{t}^{39} + 2456853315316 \mathfrak{t}^{40}\nonumber \\
&\quad + 4012922625561 \mathfrak{t}^{41} + 6514899190700 \mathfrak{t}^{42} + 10513598701742 \mathfrak{t}^{43} + 16866382923360 \mathfrak{t}^{44} + 26899750322016 \mathfrak{t}^{45}\nonumber \\
&\quad + 42653833156139 \mathfrak{t}^{46} + 67247821658975 \mathfrak{t}^{47} + 105422825438544 \mathfrak{t}^{48} + 164343486569329 \mathfrak{t}^{49}\nonumber \\
&\quad + 254774624065590 \mathfrak{t}^{50} + 392798137838011 \mathfrak{t}^{51} + 602303503197583 \mathfrak{t}^{52} + 918579296995633 \mathfrak{t}^{53}\nonumber \\
&\quad + 1393462439121048 \mathfrak{t}^{54} + 2102678131617284 \mathfrak{t}^{55} + 3156242461945635 \mathfrak{t}^{56} + 4713121432891898 \mathfrak{t}^{57}\nonumber \\
&\quad + 7001770752414753 \mathfrak{t}^{58} + 10348752991304288 \mathfrak{t}^{59} + 15218384483971637 \mathfrak{t}^{60} + 22267355770895648 \mathfrak{t}^{61}\nonumber \\
&\quad + 32419561547088429 \mathfrak{t}^{62} + 46968048873037297 \mathfrak{t}^{63} + 67713143737557154 \mathfrak{t}^{64} + 97148563951746968 \mathfrak{t}^{65}\nonumber \\
&\quad + 138710812718890323 \mathfrak{t}^{66} + 197111539751866924 \mathfrak{t}^{67} + 278778052970678326 \mathfrak{t}^{68} + 392433991762348776 \mathfrak{t}^{69}\nonumber \\
&\quad + 549860594901452716 \mathfrak{t}^{70} + 766889307804855458 \mathfrak{t}^{71} + 1064689004961727217 \mathfrak{t}^{72}\nonumber \\
&\quad + 1471426213788717245 \mathfrak{t}^{73} + 2024394803689409942 \mathfrak{t}^{74} + 2772733053112734671 \mathfrak{t}^{75}\nonumber \\
&\quad + 3780871242269329339 \mathfrak{t}^{76} + 5132882343239835475 \mathfrak{t}^{77} + 6937942371054797732 \mathfrak{t}^{78}\nonumber \\
&\quad + 9337145839497994485 \mathfrak{t}^{79} + 12511965773672510126 \mathfrak{t}^{80} + 16694696978963385549 \mathfrak{t}^{81}\nonumber \\
&\quad + 22181275703392997971 \mathfrak{t}^{82} + 29346928183845875927 \mathfrak{t}^{83} + 38665164298936152191 \mathfrak{t}^{84}\nonumber \\
&\quad + 50730699784179417434 \mathfrak{t}^{85} + 66286959938196105031 \mathfrak{t}^{86} + 86258887733601804914 \mathfrak{t}^{87}\nonumber \\
&\quad + 111791847514317687454 \mathfrak{t}^{88} + 144297479199264795957 \mathfrak{t}^{89} + 185507413715836449312 \mathfrak{t}^{90}\nonumber \\
&\quad + 237535804191367260994 \mathfrak{t}^{91} + 302951654553054948614 \mathfrak{t}^{92} + 384861932291480002712 \mathfrak{t}^{93}\nonumber \\
&\quad + 487006429354258620341 \mathfrak{t}^{94} + 613865278063748447555 \mathfrak{t}^{95} + 770779930897495436192 \mathfrak{t}^{96}\nonumber \\
&\quad + 964088267025447550224 \mathfrak{t}^{97} + 1201274287873128315340 \mathfrak{t}^{98} + 1491132602181465541654 \mathfrak{t}^{99}\nonumber \\
&\quad + 1843947572315328353037 \mathfrak{t}^{100} + 2271686593188100100315 \mathfrak{t}^{101} + 2788206499937522492639 \mathfrak{t}^{102}\nonumber \\
&\quad + 3409471549084600566537 \mathfrak{t}^{103} + 4153780791507994520047 \mathfrak{t}^{104} + 5042001958115467285425 \mathfrak{t}^{105}\nonumber \\
&\quad + 6097808217960423920147 \mathfrak{t}^{106} + 7347913354683989547827 \mathfrak{t}^{107} + 8822300055668153281263 \mathfrak{t}^{108}\nonumber \\
&\quad + 10554435138355054480113 \mathfrak{t}^{109} + 12581464673494142292455 \mathfrak{t}^{110} + 14944381133351208379698 \mathfrak{t}^{111}\nonumber \\
&\quad + 17688153926045645924560 \mathfrak{t}^{112} + 20861814010347861710691 \mathfrak{t}^{113} + 24518482756667625905287 \mathfrak{t}^{114}\nonumber \\
&\quad + 28715334869402668854508 \mathfrak{t}^{115} + 33513485054073920753139 \mathfrak{t}^{116} + 38977788239543333262349 \mathfrak{t}^{117}\nonumber \\
&\quad + 45176543588946197824617 \mathfrak{t}^{118} + 52181093286279440017742 \mathfrak{t}^{119} + 60065308196914147535753 \mathfrak{t}^{120}\nonumber \\
&\quad + 68904953989556419309147 \mathfrak{t}^{121} + 78776933185127267609051 \mathfrak{t}^{122} + 89758400863631778305901 \mathfrak{t}^{123}\nonumber \\
&\quad + 101925754399922532369615 \mathfrak{t}^{124} + 115353500585246297585166 \mathfrak{t}^{125} + 130113006781076861779147 \mathfrak{t}^{126}\nonumber \\
&\quad + 146271146286448678553595 \mathfrak{t}^{127} + 163888851806673133848649 \mathfrak{t}^{128} + 183019594701604834327913 \mathfrak{t}^{129}\nonumber \\
&\quad + 203707811464816272056792 \mathfrak{t}^{130} + 225987302528607507369202 \mathfrak{t}^{131} + 249879631883689109248815 \mathfrak{t}^{132}\nonumber \\
&\quad + 275392559020743148033275 \mathfrak{t}^{133} + 302518537217793086763131 \mathfrak{t}^{134} + 331233314088541586205975 \mathfrak{t}^{135}\nonumber \\
&\quad + 361494671457530529014984 \mathfrak{t}^{136} + 393241341934632448127744 \mathfrak{t}^{137} + 426392138938914911361499 \mathfrak{t}^{138}\nonumber \\
&\quad + 460845335305918039112532 \mathfrak{t}^{139} + 496478322965555722810398 \mathfrak{t}^{140} + 533147582490996226361894 \mathfrak{t}^{141}\nonumber \\
&\quad + 570688986616102205736414 \mathfrak{t}^{142} + 608918456156684681045040 \mathfrak{t}^{143} + 647632980240881371121994 \mathfrak{t}^{144}\nonumber \\
&\quad + 686612005480297710255348 \mathfrak{t}^{145} + 725619190852983478257686 \mathfrak{t}^{146} + 764404516805254886593904 \mathfrak{t}^{147}\nonumber \\
&\quad + 802706728620423822769565 \mathfrak{t}^{148} + 840256085673534272448002 \mathfrak{t}^{149} + 876777380030101945375539 \mathfrak{t}^{150}\nonumber \\
&\quad + 911993180193588522660734 \mathfrak{t}^{151} + 945627248898658747994607 \mathfrak{t}^{152} + 977408077908861742660716 \mathfrak{t}^{153}\nonumber \\
&\quad + 1007072478013579479428793 \mathfrak{t}^{154} + 1034369159005379605658734 \mathfrak{t}^{155} + 1059062232497899893849798 \mathfrak{t}^{156}\nonumber \\
&\quad + 1080934570115901703861630 \mathfrak{t}^{157} + 1099790950909612362641276 \mathfrak{t}^{158} + 1115460934820765323306297 \mathfrak{t}^{159}\nonumber \\
&\quad + 1127801403615755079924864 \mathfrak{t}^{160} + 1136698716808114935531678 \mathfrak{t}^{161} + 1142070437578667659894753 \mathfrak{t}^{162}\nonumber \\
&\quad + 1143866592381786992636888 \mathfrak{t}^{163} + \cdots + \mathfrak{t}^{326}).
\end{align}
}

\subsubsection{$U(13)$}
{\fontsize{9pt}{1pt}\selectfont
\begin{align}
&{\cal I}^{{\cal N}=8\text{ }U(13)(H)}(\mathfrak{t})\nonumber \\
&=
\frac{1}{
(1 - \mathfrak{t})^2 (1 - \mathfrak{t}^3) (1 - \mathfrak{t}^4) (1 - \mathfrak{t}^5) (1 - \mathfrak{t}^6) (1 - \mathfrak{t}^7) (1 - \mathfrak{t}^8) (1 - \mathfrak{t}^9) (1 - \mathfrak{t}^{10}) (1 - \mathfrak{t}^{11}) (1 - \mathfrak{t}^{12})
(1 - \mathfrak{t}^{13})
}
(1 + 2 \mathfrak{t}^{2}\nonumber \\
&\quad + \mathfrak{t}^{3} + 4 \mathfrak{t}^{4} + 3 \mathfrak{t}^{5} + 8 \mathfrak{t}^{6} + 7 \mathfrak{t}^{7} + 15 \mathfrak{t}^{8} + 15 \mathfrak{t}^{9} + 27 \mathfrak{t}^{10} + 29 \mathfrak{t}^{11} + 48 \mathfrak{t}^{12} + 53 \mathfrak{t}^{13} + 68 \mathfrak{t}^{14} + 81 \mathfrak{t}^{15} + 99 \mathfrak{t}^{16} + 113 \mathfrak{t}^{17}\nonumber \\
&\quad + 134 \mathfrak{t}^{18} + 148 \mathfrak{t}^{19} + 168 \mathfrak{t}^{20} + 181 \mathfrak{t}^{21} + 197 \mathfrak{t}^{22} + 203 \mathfrak{t}^{23} + 216 \mathfrak{t}^{24} + 213 \mathfrak{t}^{25} + 219 \mathfrak{t}^{26} + 213 \mathfrak{t}^{27} + 210 \mathfrak{t}^{28} + 196 \mathfrak{t}^{29}\nonumber \\
&\quad + 188 \mathfrak{t}^{30} + 169 \mathfrak{t}^{31} + 155 \mathfrak{t}^{32} + 137 \mathfrak{t}^{33} + 118 \mathfrak{t}^{34} + 101 \mathfrak{t}^{35} + 86 \mathfrak{t}^{36} + 69 \mathfrak{t}^{37} + 55 \mathfrak{t}^{38} + 43 \mathfrak{t}^{39} + 33 \mathfrak{t}^{40} + 23 \mathfrak{t}^{41}\nonumber \\
&\quad + 18 \mathfrak{t}^{42} + 11 \mathfrak{t}^{43} + 8 \mathfrak{t}^{44} + 5 \mathfrak{t}^{45} + 2 \mathfrak{t}^{46} + 2 \mathfrak{t}^{47} + \mathfrak{t}^{48}),\\
&{\cal I}^{U(13)\text{ ADHM-}[1](H)}(\mathfrak{t})\nonumber \\
&=
\frac{1}{
(1-\mathfrak{t})^2
(1-\mathfrak{t}^2)^2
(1-\mathfrak{t}^3)^2
(1-\mathfrak{t}^4)^2
(1-\mathfrak{t}^5)^2
(1-\mathfrak{t}^6)^2
(1-\mathfrak{t}^7)^2
(1-\mathfrak{t}^8)^2
(1-\mathfrak{t}^9)^2
(1-\mathfrak{t}^{10})^2
(1-\mathfrak{t}^{11})^2
(1-\mathfrak{t}^{12})^2
}\nonumber \\
&\quad\times\frac{1}{
(1-\mathfrak{t}^{13})^2
}
(
1+\mathfrak{t}^{2}+2 \mathfrak{t}^{3}+4 \mathfrak{t}^{4}+6 \mathfrak{t}^{5}+12 \mathfrak{t}^{6}+18 \mathfrak{t}^{7}+33 \mathfrak{t}^{8}+52 \mathfrak{t}^{9}+88 \mathfrak{t}^{10}+138 \mathfrak{t}^{11}+229 \mathfrak{t}^{12}+354 \mathfrak{t}^{13}\nonumber \\
&\quad+555 \mathfrak{t}^{14}+854 \mathfrak{t}^{15}+1317 \mathfrak{t}^{16}+1982 \mathfrak{t}^{17}+2994 \mathfrak{t}^{18}+4432 \mathfrak{t}^{19}+6553 \mathfrak{t}^{20}+9552 \mathfrak{t}^{21}+13857 \mathfrak{t}^{22}+19858 \mathfrak{t}^{23}\nonumber \\
&\quad+28322 \mathfrak{t}^{24}+39928 \mathfrak{t}^{25}+55960 \mathfrak{t}^{26}+77674 \mathfrak{t}^{27}+107104 \mathfrak{t}^{28}+146348 \mathfrak{t}^{29}+198655 \mathfrak{t}^{30}+267340 \mathfrak{t}^{31}+357341 \mathfrak{t}^{32}\nonumber \\
&\quad+473810 \mathfrak{t}^{33}+623914 \mathfrak{t}^{34}+815244 \mathfrak{t}^{35}+1058053 \mathfrak{t}^{36}+1362882 \mathfrak{t}^{37}+1743734 \mathfrak{t}^{38}+2214954 \mathfrak{t}^{39}+2794721 \mathfrak{t}^{40}\nonumber \\
&\quad+3501486 \mathfrak{t}^{41}+4358199 \mathfrak{t}^{42}+5387314 \mathfrak{t}^{43}+6616262 \mathfrak{t}^{44}+8071214 \mathfrak{t}^{45}+9783044 \mathfrak{t}^{46}+11780242 \mathfrak{t}^{47}\nonumber \\
&\quad+14095748 \mathfrak{t}^{48}+16757780 \mathfrak{t}^{49}+19798592 \mathfrak{t}^{50}+23243448 \mathfrak{t}^{51}+27119970 \mathfrak{t}^{52}+31446264 \mathfrak{t}^{53}+36241616 \mathfrak{t}^{54}\nonumber \\
&\quad+41512180 \mathfrak{t}^{55}+47264275 \mathfrak{t}^{56}+53488164 \mathfrak{t}^{57}+60172639 \mathfrak{t}^{58}+67288316 \mathfrak{t}^{59}+74804021 \mathfrak{t}^{60}+82667844 \mathfrak{t}^{61}\nonumber \\
&\quad+90827287 \mathfrak{t}^{62}+99208604 \mathfrak{t}^{63}+107738367 \mathfrak{t}^{64}+116323224 \mathfrak{t}^{65}+124873435 \mathfrak{t}^{66}+133280566 \mathfrak{t}^{67}\nonumber \\
&\quad+141444753 \mathfrak{t}^{68}+149250810 \mathfrak{t}^{69}+156596525 \mathfrak{t}^{70}+163369272 \mathfrak{t}^{71}+169474948 \mathfrak{t}^{72}+174812546 \mathfrak{t}^{73}\nonumber \\
&\quad+179305814 \mathfrak{t}^{74}+182875734 \mathfrak{t}^{75}+185471434 \mathfrak{t}^{76}+187043244 \mathfrak{t}^{77}+187572686 \mathfrak{t}^{78}+\cdots +\mathfrak{t}^{156}
),\\
&{\cal I}^{U(13)\text{ ADHM-}[2](H)}(\mathfrak{t})\nonumber \\
&=
\frac{1}{
(1 - \mathfrak{t}) (1 - \mathfrak{t}^2)^4 (1 - \mathfrak{t}^3)^4 (1 - \mathfrak{t}^4)^2 (1 - \mathfrak{t}^5)^5 (1 - \mathfrak{t}^6)^4 (1 - \mathfrak{t}^7)^4 (1 - \mathfrak{t}^8)^5 (1 - \mathfrak{t}^9)^4(1 - \mathfrak{t}^{10})^4 (1 - \mathfrak{t}^{11})^4 (1 - \mathfrak{t}^{12})^4 }\nonumber \\
&\quad\times\frac{1}{(1 - \mathfrak{t}^{13})^4 (1 - \mathfrak{t}^{14})^3
}
(1 + \mathfrak{t} + 3 \mathfrak{t}^{2} + 9 \mathfrak{t}^{3} + 24 \mathfrak{t}^{4} + 49 \mathfrak{t}^{5} + 116 \mathfrak{t}^{6} + 254 \mathfrak{t}^{7} + 554 \mathfrak{t}^{8} + 1155 \mathfrak{t}^{9} + 2413 \mathfrak{t}^{10} + 4904 \mathfrak{t}^{11}\nonumber \\
&\quad + 9903 \mathfrak{t}^{12} + 19605 \mathfrak{t}^{13} + 38485 \mathfrak{t}^{14} + 74497 \mathfrak{t}^{15} + 142923 \mathfrak{t}^{16} + 271004 \mathfrak{t}^{17} + 509469 \mathfrak{t}^{18} + 948333 \mathfrak{t}^{19} + 1750761 \mathfrak{t}^{20}\nonumber \\
&\quad + 3203931 \mathfrak{t}^{21} + 5817807 \mathfrak{t}^{22} + 10480188 \mathfrak{t}^{23} + 18740637 \mathfrak{t}^{24} + 33264797 \mathfrak{t}^{25} + 58633887 \mathfrak{t}^{26} + 102634321 \mathfrak{t}^{27}\nonumber \\
&\quad + 178458986 \mathfrak{t}^{28} + 308258681 \mathfrak{t}^{29} + 529069080 \mathfrak{t}^{30} + 902326498 \mathfrak{t}^{31} + 1529454121 \mathfrak{t}^{32} + 2576707802 \mathfrak{t}^{33}\nonumber \\
&\quad + 4315219245 \mathfrak{t}^{34} + 7184269288 \mathfrak{t}^{35} + 11891804449 \mathfrak{t}^{36} + 19571729636 \mathfrak{t}^{37} + 32030488120 \mathfrak{t}^{38}\nonumber \\
&\quad + 52129009029 \mathfrak{t}^{39} + 84374008599 \mathfrak{t}^{40} + 135824387486 \mathfrak{t}^{41} + 217477767401 \mathfrak{t}^{42} + 346373556161 \mathfrak{t}^{43}\nonumber \\
&\quad + 548773129102 \mathfrak{t}^{44} + 864932716268 \mathfrak{t}^{45} + 1356240256066 \mathfrak{t}^{46} + 2115814270741 \mathfrak{t}^{47} + 3284179577078 \mathfrak{t}^{48}\nonumber \\
&\quad + 5072304750897 \mathfrak{t}^{49} + 7795310674660 \mathfrak{t}^{50} + 11921498311228 \mathfrak{t}^{51} + 18143333958688 \mathfrak{t}^{52}\nonumber \\
&\quad + 27479639486373 \mathfrak{t}^{53} + 41422022283512 \mathfrak{t}^{54} + 62143543662379 \mathfrak{t}^{55} + 92794682171601 \mathfrak{t}^{56}\nonumber \\
&\quad + 137920872934439 \mathfrak{t}^{57} + 204048780061033 \mathfrak{t}^{58} + 300505210149366 \mathfrak{t}^{59} + 440555568012938 \mathfrak{t}^{60}\nonumber \\
&\quad + 642978486823920 \mathfrak{t}^{61} + 934233448711306 \mathfrak{t}^{62} + 1351429798525560 \mathfrak{t}^{63} + 1946374270380633 \mathfrak{t}^{64}\nonumber \\
&\quad + 2791061655485213 \mathfrak{t}^{65} + 3985088202935256 \mathfrak{t}^{66} + 5665612468897210 \mathfrak{t}^{67} + 8020676452524912 \mathfrak{t}^{68}\nonumber \\
&\quad + 11306935116744749 \mathfrak{t}^{69} + 15873143407099951 \mathfrak{t}^{70} + 22191122602647800 \mathfrak{t}^{71} + 30896398641022115 \mathfrak{t}^{72}\nonumber \\
&\quad + 42841281956330371 \mathfrak{t}^{73} + 59163877829510339 \mathfrak{t}^{74} + 81377388185911499 \mathfrak{t}^{75} + 111485139308435490 \mathfrak{t}^{76}\nonumber \\
&\quad + 152128055992009701 \mathfrak{t}^{77} + 206772865561771414 \mathfrak{t}^{78} + 279951164647062372 \mathfrak{t}^{79} + 377561699326893800 \mathfrak{t}^{80}\nonumber \\
&\quad + 507250799504417007 \mathfrak{t}^{81} + 678888971461903731 \mathfrak{t}^{82} + 905165179999356280 \mathfrak{t}^{83}\nonumber \\
&\quad + 1202324462148739446 \mathfrak{t}^{84} + 1591079175507357204 \mathfrak{t}^{85} + 2097729530659932115 \mathfrak{t}^{86}\nonumber \\
&\quad + 2755535015409039197 \mathfrak{t}^{87} + 3606385034730323437 \mathfrak{t}^{88} + 4702824427140174797 \mathfrak{t}^{89}\nonumber \\
&\quad + 6110497625210731291 \mathfrak{t}^{90} + 7911083874298274184 \mathfrak{t}^{91} + 10205805252732465793 \mathfrak{t}^{92}\nonumber \\
&\quad + 13119598887663390753 \mathfrak{t}^{93} + 16806054860743586917 \mathfrak{t}^{94} + 21453231313859700097 \mathfrak{t}^{95}\nonumber \\
&\quad + 27290468300659880989 \mathfrak{t}^{96} + 34596331235363017502 \mathfrak{t}^{97} + 43707823445854308570 \mathfrak{t}^{98}\nonumber \\
&\quad + 55031014335134673032 \mathfrak{t}^{99} + 69053235063456634980 \mathfrak{t}^{100} + 86356996180923536106 \mathfrak{t}^{101}\nonumber \\
&\quad + 107635781362718561527 \mathfrak{t}^{102} + 133711866710605586621 \mathfrak{t}^{103} + 165556306173801787425 \mathfrak{t}^{104}\nonumber \\
&\quad + 204311208582034310473 \mathfrak{t}^{105} + 251314410961938804206 \mathfrak{t}^{106} + 308126624015314350061 \mathfrak{t}^{107}\nonumber \\
&\quad + 376561089613918295189 \mathfrak{t}^{108} + 458715744479497425107 \mathfrak{t}^{109} + 557007830160391275589 \mathfrak{t}^{110}\nonumber \\
&\quad + 674210824423429379180 \mathfrak{t}^{111} + 813493495336232079568 \mathfrak{t}^{112} + 978460794029911577036 \mathfrak{t}^{113}\nonumber \\
&\quad + 1173196208555773845465 \mathfrak{t}^{114} + 1402305096869792561551 \mathfrak{t}^{115} + 1670958406291843816858 \mathfrak{t}^{116}\nonumber \\
&\quad + 1984936067434658142322 \mathfrak{t}^{117} + 2350669228679528652711 \mathfrak{t}^{118} + 2775280370597845701264 \mathfrak{t}^{119}\nonumber \\
&\quad + 3266620216176839690605 \mathfrak{t}^{120} + 3833300230179125772998 \mathfrak{t}^{121} + 4484719389237753960736 \mathfrak{t}^{122}\nonumber \\
&\quad + 5231083801065219331983 \mathfrak{t}^{123} + 6083417668073417131857 \mathfrak{t}^{124} + 7053564024881024900716 \mathfrak{t}^{125}\nonumber \\
&\quad + 8154173643906639448333 \mathfrak{t}^{126} + 9398680494444184227221 \mathfrak{t}^{127} + 10801262172850751992041 \mathfrak{t}^{128}\nonumber \\
&\quad + 12376783788719971087539 \mathfrak{t}^{129} + 14140723909498817857006 \mathfrak{t}^{130} + 16109081325040763826129 \mathfrak{t}^{131}\nonumber \\
&\quad + 18298261609594661197575 \mathfrak{t}^{132} + 20724942718989824274623 \mathfrak{t}^{133} + 23405919180426882589502 \mathfrak{t}^{134}\nonumber \\
&\quad + 26357924793871341574503 \mathfrak{t}^{135} + 29597434181861828329316 \mathfrak{t}^{136} + 33140443975675786219920 \mathfrak{t}^{137}\nonumber \\
&\quad + 37002234923476223955828 \mathfrak{t}^{138} + 41197116721498181401229 \mathfrak{t}^{139} + 45738157914854336224263 \mathfrak{t}^{140}\nonumber \\
&\quad + 50636903755464779208890 \mathfrak{t}^{141} + 55903085452979301770573 \mathfrak{t}^{142} + 61544324770289494981251 \mathfrak{t}^{143}\nonumber \\
&\quad + 67565838411151623293146 \mathfrak{t}^{144} + 73970147079098223473764 \mathfrak{t}^{145} + 80756794469387643617571 \mathfrak{t}^{146}\nonumber \\
&\quad + 87922081741825053798888 \mathfrak{t}^{147} + 95458823231325628816079 \mathfrak{t}^{148} + 103356129236497014454878 \mathfrak{t}^{149}\nonumber \\
&\quad + 111599221710065099753046 \mathfrak{t}^{150} + 120169288510741966480081 \mathfrak{t}^{151} + 129043381598231631121781 \mathfrak{t}^{152}\nonumber \\
&\quad + 138194364115935448724392 \mathfrak{t}^{153} + 147590910753599673286016 \mathfrak{t}^{154} + 157197565072700135522068 \mathfrak{t}^{155}\nonumber \\
&\quad + 166974856665769643836831 \mathfrak{t}^{156} + 176879480069370174793691 \mathfrak{t}^{157} + 186864536325050588141351 \mathfrak{t}^{158}\nonumber \\
&\quad + 196879836949502533967375 \mathfrak{t}^{159} + 206872268910491700186904 \mathfrak{t}^{160} + 216786217977710685074913 \mathfrak{t}^{161}\nonumber \\
&\quad + 226564046612065379587093 \mathfrak{t}^{162} + 236146621342287690445875 \mathfrak{t}^{163} + 245473883444348276639710 \mathfrak{t}^{164}\nonumber \\
&\quad + 254485455653867888345499 \mathfrak{t}^{165} + 263121276697030662514697 \mathfrak{t}^{166} + 271322254582936019231043 \mathfrak{t}^{167}\nonumber \\
&\quad + 279030928951008545226592 \mathfrak{t}^{168} + 286192132264120538836973 \mathfrak{t}^{169} + 292753639367922362251981 \mathfrak{t}^{170}\nonumber \\
&\quad + 298666794841639775422685 \mathfrak{t}^{171} + 303887107724361934162452 \mathfrak{t}^{172} + 308374803543117747730749 \mathfrak{t}^{173}\nonumber \\
&\quad + 312095324164829122867994 \mathfrak{t}^{174} + 315019766760028008780606 \mathfrak{t}^{175} + 317125254162528299941240 \mathfrak{t}^{176}\nonumber \\
&\quad + 318395230040483560794845 \mathfrak{t}^{177} + 318819673612311030414584 \mathfrak{t}^{178} + \cdots + \mathfrak{t}^{356}).
\end{align}
}

\subsubsection{$U(14)$}
{\fontsize{9pt}{1pt}\selectfont
\begin{align}
&{\cal I}^{{\cal N}=8\text{ }U(14)(H)}(\mathfrak{t})\nonumber \\
&=
\frac{1}{
(1 - \mathfrak{t}) (1 - \mathfrak{t}^2) (1 - \mathfrak{t}^3) (1 - \mathfrak{t}^4) (1 - \mathfrak{t}^5) (1 - \mathfrak{t}^6) (1 - \mathfrak{t}^7) (1 - \mathfrak{t}^8) (1 - \mathfrak{t}^9) (1 - \mathfrak{t}^{10}) (1 - \mathfrak{t}^{11})
(1 - \mathfrak{t}^{12}) (1 - \mathfrak{t}^{13})}\nonumber \\
&\quad\times\frac{1}{ (1 - \mathfrak{t}^{14})
}
(1 + \mathfrak{t} + 2 \mathfrak{t}^{2} + 3 \mathfrak{t}^{3} + 5 \mathfrak{t}^{4} + 7 \mathfrak{t}^{5} + 11 \mathfrak{t}^{6} + 15 \mathfrak{t}^{7} + 22 \mathfrak{t}^{8} + 30 \mathfrak{t}^{9} + 42 \mathfrak{t}^{10} + 56 \mathfrak{t}^{11} + 77 \mathfrak{t}^{12} + 101 \mathfrak{t}^{13}\nonumber \\
&\quad + 135 \mathfrak{t}^{14} + 161 \mathfrak{t}^{15} + 202 \mathfrak{t}^{16} + 242 \mathfrak{t}^{17} + 293 \mathfrak{t}^{18} + 340 \mathfrak{t}^{19} + 400 \mathfrak{t}^{20} + 451 \mathfrak{t}^{21} + 514 \mathfrak{t}^{22} + 564 \mathfrak{t}^{23} + 623 \mathfrak{t}^{24} + 663 \mathfrak{t}^{25}\nonumber \\
&\quad + 712 \mathfrak{t}^{26} + 734 \mathfrak{t}^{27} + 767 \mathfrak{t}^{28} + 776 \mathfrak{t}^{29} + 788 \mathfrak{t}^{30} + 773 \mathfrak{t}^{31} + 766 \mathfrak{t}^{32} + 730 \mathfrak{t}^{33} + 703 \mathfrak{t}^{34} + 651 \mathfrak{t}^{35} + 609 \mathfrak{t}^{36} + 547 \mathfrak{t}^{37}\nonumber \\
&\quad + 500 \mathfrak{t}^{38} + 434 \mathfrak{t}^{39} + 384 \mathfrak{t}^{40} + 324 \mathfrak{t}^{41} + 277 \mathfrak{t}^{42} + 223 \mathfrak{t}^{43} + 187 \mathfrak{t}^{44} + 143 \mathfrak{t}^{45} + 115 \mathfrak{t}^{46} + 84 \mathfrak{t}^{47} + 65 \mathfrak{t}^{48} + 45 \mathfrak{t}^{49}\nonumber \\
&\quad + 34 \mathfrak{t}^{50} + 20 \mathfrak{t}^{51} + 14 \mathfrak{t}^{52} + 8 \mathfrak{t}^{53} + 6 \mathfrak{t}^{54} + 2 \mathfrak{t}^{55} + 2 \mathfrak{t}^{56}),\\
&{\cal I}^{U(14)\text{ ADHM-}[1](H)}(\mathfrak{t})\nonumber \\
&=
\frac{1}{
(1-\mathfrak{t})^2
(1-\mathfrak{t}^2)^3
(1-\mathfrak{t}^3)^2
(1-\mathfrak{t}^4)
(1-\mathfrak{t}^5)^2
(1-\mathfrak{t}^6)^2
(1-\mathfrak{t}^7)^2
(1-\mathfrak{t}^8)^2
(1-\mathfrak{t}^9)^2
(1-\mathfrak{t}^{10})^2
(1-\mathfrak{t}^{11})^2
(1-\mathfrak{t}^{12})^2
}\nonumber \\
&\quad\times\frac{1}{
(1-\mathfrak{t}^{13})^2
(1-\mathfrak{t}^{14})^2
}
(
1+2 \mathfrak{t}^{3}+4 \mathfrak{t}^{4}+4 \mathfrak{t}^{5}+8 \mathfrak{t}^{6}+14 \mathfrak{t}^{7}+25 \mathfrak{t}^{8}+38 \mathfrak{t}^{9}+63 \mathfrak{t}^{10}+100 \mathfrak{t}^{11}+166 \mathfrak{t}^{12}+254 \mathfrak{t}^{13}\nonumber \\
&\quad+402 \mathfrak{t}^{14}+612 \mathfrak{t}^{15}+948 \mathfrak{t}^{16}+1432 \mathfrak{t}^{17}+2173 \mathfrak{t}^{18}+3220 \mathfrak{t}^{19}+4796 \mathfrak{t}^{20}+7030 \mathfrak{t}^{21}+10285 \mathfrak{t}^{22}+14848 \mathfrak{t}^{23}\nonumber \\
&\quad+21397 \mathfrak{t}^{24}+30464 \mathfrak{t}^{25}+43222 \mathfrak{t}^{26}+60704 \mathfrak{t}^{27}+84882 \mathfrak{t}^{28}+117616 \mathfrak{t}^{29}+162172 \mathfrak{t}^{30}+221746 \mathfrak{t}^{31}+301633 \mathfrak{t}^{32}\nonumber \\
&\quad+407126 \mathfrak{t}^{33}+546497 \mathfrak{t}^{34}+728272 \mathfrak{t}^{35}+965144 \mathfrak{t}^{36}+1270160 \mathfrak{t}^{37}+1662244 \mathfrak{t}^{38}+2160946 \mathfrak{t}^{39}+2793509 \mathfrak{t}^{40}\nonumber \\
&\quad+3588160 \mathfrak{t}^{41}+4583172 \mathfrak{t}^{42}+5817804 \mathfrak{t}^{43}+7344161 \mathfrak{t}^{44}+9215188 \mathfrak{t}^{45}+11499391 \mathfrak{t}^{46}+14265678 \mathfrak{t}^{47}\nonumber \\
&\quad+17601356 \mathfrak{t}^{48}+21592540 \mathfrak{t}^{49}+26346637 \mathfrak{t}^{50}+31967288 \mathfrak{t}^{51}+38581309 \mathfrak{t}^{52}+46307944 \mathfrak{t}^{53}+55290890 \mathfrak{t}^{54}\nonumber \\
&\quad+65660178 \mathfrak{t}^{55}+77570792 \mathfrak{t}^{56}+91156002 \mathfrak{t}^{57}+106572739 \mathfrak{t}^{58}+123946604 \mathfrak{t}^{59}+143424526 \mathfrak{t}^{60}\nonumber \\
&\quad+165109712 \mathfrak{t}^{61}+189124049 \mathfrak{t}^{62}+215532234 \mathfrak{t}^{63}+244413681 \mathfrak{t}^{64}+275776914 \mathfrak{t}^{65}+309642187 \mathfrak{t}^{66}\nonumber \\
&\quad+345944560 \mathfrak{t}^{67}+384629727 \mathfrak{t}^{68}+425545912 \mathfrak{t}^{69}+468552490 \mathfrak{t}^{70}+513402518 \mathfrak{t}^{71}+559864601 \mathfrak{t}^{72}\nonumber \\
&\quad+607594956 \mathfrak{t}^{73}+656274834 \mathfrak{t}^{74}+705472294 \mathfrak{t}^{75}+754793166 \mathfrak{t}^{76}+803734928 \mathfrak{t}^{77}+851850540 \mathfrak{t}^{78}\nonumber \\
&\quad+898592918 \mathfrak{t}^{79}+943492350 \mathfrak{t}^{80}+985991614 \mathfrak{t}^{81}+1025633360 \mathfrak{t}^{82}+1061888460 \mathfrak{t}^{83}+1094350851 \mathfrak{t}^{84}\nonumber \\
&\quad+1122557198 \mathfrak{t}^{85}+1146189237 \mathfrak{t}^{86}+1164885242 \mathfrak{t}^{87}+1178444591 \mathfrak{t}^{88}+1186634326 \mathfrak{t}^{89}+1189393664 \mathfrak{t}^{90}\nonumber \\
&\quad+\cdots +\mathfrak{t}^{180}
),\\
&{\cal I}^{U(14)\text{ ADHM-}[2](H)}(\mathfrak{t})\nonumber \\
&=
\frac{1}{
(1 - \mathfrak{t})^2 (1 - \mathfrak{t}^2)^3 (1 - \mathfrak{t}^3)^2 (1 - \mathfrak{t}^4)^4 (1 - \mathfrak{t}^5)^4 (1 - \mathfrak{t}^6)^5 (1 - \mathfrak{t}^7)^4 (1 - \mathfrak{t}^8)^4 (1 - \mathfrak{t}^9)^5(1 - \mathfrak{t}^{10})^4 (1 - \mathfrak{t}^{11})^4 (1 - \mathfrak{t}^{12})^4}\nonumber \\
&\quad\times\frac{1}{ (1 - \mathfrak{t}^{13})^4 (1 - \mathfrak{t}^{14})^4 (1 - \mathfrak{t}^{15})^3}
(1 + 3 \mathfrak{t}^{2} + 8 \mathfrak{t}^{3} + 16 \mathfrak{t}^{4} + 38 \mathfrak{t}^{5} + 93 \mathfrak{t}^{6} + 192 \mathfrak{t}^{7} + 427 \mathfrak{t}^{8} + 895 \mathfrak{t}^{9} + 1863 \mathfrak{t}^{10}\nonumber \\
&\quad + 3795 \mathfrak{t}^{11} + 7678 \mathfrak{t}^{12} + 15200 \mathfrak{t}^{13} + 29912 \mathfrak{t}^{14} + 58000 \mathfrak{t}^{15} + 111427 \mathfrak{t}^{16} + 211774 \mathfrak{t}^{17} + 399080 \mathfrak{t}^{18} + 744679 \mathfrak{t}^{19}\nonumber \\
&\quad + 1378858 \mathfrak{t}^{20} + 2531539 \mathfrak{t}^{21} + 4613118 \mathfrak{t}^{22} + 8343071 \mathfrak{t}^{23} + 14983754 \mathfrak{t}^{24} + 26722472 \mathfrak{t}^{25} + 47346287 \mathfrak{t}^{26}\nonumber \\
&\quad + 83342770 \mathfrak{t}^{27} + 145796347 \mathfrak{t}^{28} + 253492752 \mathfrak{t}^{29} + 438139511 \mathfrak{t}^{30} + 752884857 \mathfrak{t}^{31} + 1286424595 \mathfrak{t}^{32}\nonumber \\
&\quad + 2185841535 \mathfrak{t}^{33} + 3693905579 \mathfrak{t}^{34} + 6209018494 \mathfrak{t}^{35} + 10381832358 \mathfrak{t}^{36} + 17269197761 \mathfrak{t}^{37} + 28579491656 \mathfrak{t}^{38}\nonumber \\
&\quad + 47059876323 \mathfrak{t}^{39} + 77106773088 \mathfrak{t}^{40} + 125720677552 \mathfrak{t}^{41} + 203995584611 \mathfrak{t}^{42} + 329426878665 \mathfrak{t}^{43}\nonumber \\
&\quad + 529475541151 \mathfrak{t}^{44} + 847039288384 \mathfrak{t}^{45} + 1348819825203 \mathfrak{t}^{46} + 2138048086828 \mathfrak{t}^{47} + 3373760016058 \mathfrak{t}^{48}\nonumber \\
&\quad + 5299843043483 \mathfrak{t}^{49} + 8288624749270 \mathfrak{t}^{50} + 12905959015157 \mathfrak{t}^{51} + 20008002927099 \mathfrak{t}^{52} + 30884425795177 \mathfrak{t}^{53}\nonumber \\
&\quad + 47469426166789 \mathfrak{t}^{54} + 72651220433424 \mathfrak{t}^{55} + 110724020575104 \mathfrak{t}^{56} + 168045113252702 \mathfrak{t}^{57}\nonumber \\
&\quad + 253986054365280 \mathfrak{t}^{58} + 382303521664179 \mathfrak{t}^{59} + 573106622661543 \mathfrak{t}^{60} + 855667886610441 \mathfrak{t}^{61}\nonumber \\
&\quad + 1272422899258457 \mathfrak{t}^{62} + 1884636834337388 \mathfrak{t}^{63} + 2780399174455224 \mathfrak{t}^{64} + 4085855613038682 \mathfrak{t}^{65}\nonumber \\
&\quad + 5980922839918619 \mathfrak{t}^{66} + 8721183944765776 \mathfrak{t}^{67} + 12668270637379547 \mathfrak{t}^{68} + 18331848654733204 \mathfrak{t}^{69}\nonumber \\
&\quad + 26427402926084308 \mathfrak{t}^{70} + 37955445384930569 \mathfrak{t}^{71} + 54309652257455782 \mathfrak{t}^{72} + 77423904394638069 \mathfrak{t}^{73}\nonumber \\
&\quad + 109971432345188725 \mathfrak{t}^{74} + 155633458763847889 \mathfrak{t}^{75} + 219460165285404642 \mathfrak{t}^{76} + 308353805978656361 \mathfrak{t}^{77}\nonumber \\
&\quad + 431712778026425114 \mathfrak{t}^{78} + 602286930146648766 \mathfrak{t}^{79} + 837308994828960578 \mathfrak{t}^{80} + 1159985506012201661 \mathfrak{t}^{81}\nonumber \\
&\quad + 1601453877951535150 \mathfrak{t}^{82} + 2203341553840306785 \mathfrak{t}^{83} + 3021099687024709563 \mathfrak{t}^{84}\nonumber \\
&\quad + 4128329245489145317 \mathfrak{t}^{85} + 5622373717261581240 \mathfrak{t}^{86} + 7631521916103173884 \mathfrak{t}^{87}\nonumber \\
&\quad + 10324249487690120475 \mathfrak{t}^{88} + 13921031566616993583 \mathfrak{t}^{89} + 18709385330592478698 \mathfrak{t}^{90}\nonumber \\
&\quad + 25062953882772294148 \mathfrak{t}^{91} + 33465626807791263858 \mathfrak{t}^{92} + 44541912964302406743 \mathfrak{t}^{93}\nonumber \\
&\quad + 59095043745319835677 \mathfrak{t}^{94} + 78154596456541055715 \mathfrak{t}^{95} + 103035795169323401307 \mathfrak{t}^{96}\nonumber \\
&\quad + 135413077904239600004 \mathfrak{t}^{97} + 177411023153109611836 \mathfrak{t}^{98} + 231716314043554098698 \mathfrak{t}^{99}\nonumber \\
&\quad + 301715094777067531269 \mathfrak{t}^{100} + 391660850291360745833 \mathfrak{t}^{101} + 506878826690176872985 \mathfrak{t}^{102}\nonumber \\
&\quad + 654014015363314971176 \mathfrak{t}^{103} + 841330857478205765148 \mathfrak{t}^{104} + 1079074094289104899816 \mathfrak{t}^{105}\nonumber \\
&\quad + 1379901599757179390291 \mathfrak{t}^{106} + 1759401587986167689275 \mathfrak{t}^{107} + 2236708291655994746608 \mathfrak{t}^{108}\nonumber \\
&\quad + 2835232054993068384455 \mathfrak{t}^{109} + 3583521771071816282926 \mathfrak{t}^{110} + 4516279704359218670156 \mathfrak{t}^{111}\nonumber \\
&\quad + 5675550959662715531975 \mathfrak{t}^{112} + 7112112160028887212273 \mathfrak{t}^{113} + 8887086248148189738257 \mathfrak{t}^{114}\nonumber \\
&\quad + 11073812682370660249330 \mathfrak{t}^{115} + 13760004610399455065750 \mathfrak{t}^{116} + 17050226803029583519678 \mathfrak{t}^{117}\nonumber \\
&\quad + 21068730145973276986871 \mathfrak{t}^{118} + 25962680226774951884663 \mathfrak{t}^{119} + 31905818921391620332527 \mathfrak{t}^{120}\nonumber \\
&\quad + 39102598761847226475870 \mathfrak{t}^{121} + 47792830134489690060159 \mathfrak{t}^{122} + 58256880873646205039985 \mathfrak{t}^{123}\nonumber \\
&\quad + 70821466440385946343645 \mathfrak{t}^{124} + 85866066446976181513665 \mathfrak{t}^{125} + 103829999655532909928653 \mathfrak{t}^{126}\nonumber \\
&\quad + 125220184571081040128288 \mathfrak{t}^{127} + 150619606217669066531612 \mathfrak{t}^{128} + 180696501463860569158984 \mathfrak{t}^{129}\nonumber \\
&\quad + 216214265224503756940577 \mathfrak{t}^{130} + 258042067868267437553362 \mathfrak{t}^{131} + 307166160123675673375371 \mathfrak{t}^{132}\nonumber \\
&\quad + 364701825611665498223915 \mathfrak{t}^{133} + 431905922832241301645113 \mathfrak{t}^{134} + 510189937989261840556716 \mathfrak{t}^{135}\nonumber \\
&\quad + 601133447547379208956965 \mathfrak{t}^{136} + 706497864982194313480658 \mathfrak{t}^{137} + 828240320039651018487903 \mathfrak{t}^{138}\nonumber \\
&\quad + 968527491204593468108636 \mathfrak{t}^{139} + 1129749183381014872709443 \mathfrak{t}^{140} + 1314531413405225460257960 \mathfrak{t}^{141}\nonumber \\
&\quad + 1525748736511234989839504 \mathfrak{t}^{142} + 1766535517804834685128778 \mathfrak{t}^{143} + 2040295824915182796320792 \mathfrak{t}^{144}\nonumber \\
&\quad + 2350711591990363770100202 \mathfrak{t}^{145} + 2701748681974190916055595 \mathfrak{t}^{146} + 3097660454471909118506093 \mathfrak{t}^{147}\nonumber \\
&\quad + 3542988431477394015053939 \mathfrak{t}^{148} + 4042559643677490949724845 \mathfrak{t}^{149} + 4601480237002589891091949 \mathfrak{t}^{150}\nonumber \\
&\quad + 5225124923413085616982791 \mathfrak{t}^{151} + 5919121872575140149097094 \mathfrak{t}^{152} + 6689332662810292622546437 \mathfrak{t}^{153}\nonumber \\
&\quad + 7541826941309960176371115 \mathfrak{t}^{154} + 8482851485559044399751676 \mathfrak{t}^{155} + 9518793410731051124223880 \mathfrak{t}^{156}\nonumber \\
&\quad + 10656137331594933624556288 \mathfrak{t}^{157} + 11901416362321428882573749 \mathfrak{t}^{158} + 13261156923098279643816677 \mathfrak{t}^{159}\nonumber \\
&\quad + 14741817418261357918958394 \mathfrak{t}^{160} + 16349720955749400378999955 \mathfrak{t}^{161} + 18090982391122424938322767 \mathfrak{t}^{162}\nonumber \\
&\quad + 19971430099554196695637462 \mathfrak{t}^{163} + 21996523004558354978105545 \mathfrak{t}^{164} + 24171263520535728518461041 \mathfrak{t}^{165}\nonumber \\
&\quad + 26500107195440112810949155 \mathfrak{t}^{166} + 28986869967227177189232379 \mathfrak{t}^{167} + 31634634070762326113589505 \mathfrak{t}^{168}\nonumber \\
&\quad + 34445653747425348892113695 \mathfrak{t}^{169} + 37421262015023480857692766 \mathfrak{t}^{170} + 40561779847571305682988636 \mathfrak{t}^{171}\nonumber \\
&\quad + 43866429190305532534445421 \mathfrak{t}^{172} + 47333251291867349950131873 \mathfrak{t}^{173} + 50959031870520667649617154 \mathfrak{t}^{174}\nonumber \\
&\quad + 54739234641912570221825080 \mathfrak{t}^{175} + 58667944720478145375259942 \mathfrak{t}^{176} + 62737823363224461585103119 \mathfrak{t}^{177}\nonumber \\
&\quad + 66940075452489205658168002 \mathfrak{t}^{178} + 71264431012457824597333877 \mathfrak{t}^{179} + 75699141923108424613946506 \mathfrak{t}^{180}\nonumber \\
&\quad + 80230994835157814026103082 \mathfrak{t}^{181} + 84845341102205494397928240 \mathfrak{t}^{182} + 89526144333255842704284928 \mathfrak{t}^{183}\nonumber \\
&\quad + 94256045933208737506590485 \mathfrak{t}^{184} + 99016448743673352073775505 \mathfrak{t}^{185} + 103787618625909078799343472 \mathfrak{t}^{186}\nonumber \\
&\quad + 108548803545874510693208641 \mathfrak{t}^{187} + 113278369433620871457147292 \mathfrak{t}^{188}\nonumber \\
&\quad + 117953951800449305001690625 \mathfrak{t}^{189} + 122552621813361614807973075 \mathfrak{t}^{190}\nonumber \\
&\quad + 127051065252601701228325709 \mathfrak{t}^{191} + 131425772520686707868549998 \mathfrak{t}^{192}\nonumber \\
&\quad + 135653237635370592009503312 \mathfrak{t}^{193} + 139710163930293864432183815 \mathfrak{t}^{194}\nonumber \\
&\quad + 143573674010102736343063578 \mathfrak{t}^{195} + 147221521366512377199054270 \mathfrak{t}^{196}\nonumber \\
&\quad + 150632300961429140991178899 \mathfrak{t}^{197} + 153785656026421341681879684 \mathfrak{t}^{198}\nonumber \\
&\quad + 156662478316236512485078472 \mathfrak{t}^{199} + 159245099089568156797614597 \mathfrak{t}^{200}\nonumber \\
&\quad + 161517468172700032888415869 \mathfrak{t}^{201} + 163465318590932862626032227 \mathfrak{t}^{202}\nonumber \\
&\quad + 165076314426660096224685038 \mathfrak{t}^{203} + 166340179779583907078776294 \mathfrak{t}^{204}\nonumber \\
&\quad + 167248806959772811936874433 \mathfrak{t}^{205} + 167796342334255101771902850 \mathfrak{t}^{206}\nonumber \\
&\quad + 167979248566902179058163124 \mathfrak{t}^{207} + \cdots + \mathfrak{t}^{414}).
\end{align}
}

\subsection{Unflavored Higgs correlators in symmetric representations}
In this subsection we list the Higgs line defect indices with one Wilson line in a symmetric representation or two Wilson lines in the same symmetric representation, in the unflavored limit.
We find that the numerators of the diagonal 2-point functions in the large $N$ limit are palindromic polynomials, which we abbreviate with ```$\cdots$''.

\subsubsection{$U(1)$}
The unflavored Wilson line correlation function for the $U(1)$ ADHM theory with $l$ flavors is given by
\begin{align}
\langle W_{n} \rangle^{U(1)\text{ ADHM-}[l](H)}(\mathfrak{t})
&=\frac{\Gamma(n+l)}{\Gamma(l)}
\frac{{}_2 \widetilde{F}_1(1-l,1+n-l;1+n;\mathfrak{t}^2)\mathfrak{t}^n(1+\mathfrak{t})^2}
{(1-\mathfrak{t}^2)^{2l}}, 
\end{align}
where $_2\widetilde{F}_1(a,b;c;z)$ $:=$ $_2F_1(a,b;c;z)/\Gamma(c)$ is the regularized hypergeometric function.

\subsubsection{$U(2)$}
\noindent \underline{1-point functions}
{\fontsize{9pt}{1pt}\selectfont
\begin{align}
&\sum_{k=0}^\infty c^k\langle W_{(k)}\rangle^{U(2)\text{ ADHM-}[1](H)}(\mathfrak{t})
=
\frac{
1+\mathfrak{t}^2-2c\mathfrak{t}^4
}{
(1-c\mathfrak{t})(1-c\mathfrak{t}^2)^2(1-\mathfrak{t})^2(1-\mathfrak{t}^2)^2
},\\
&\sum_{k=0}^\infty c^k\langle W_{(k)}\rangle^{U(2)\text{ADHM-}[2](H)}(\mathfrak{t})\nonumber \\
&=
\frac{1}{
(1 - c \mathfrak{t})^2 (1 - c \mathfrak{t}^2)^3 (1 - \mathfrak{t}) (1 - \mathfrak{t}^2)^4 (1 - \mathfrak{t}^3)^3
}
(
1 + \mathfrak{t} + 3 \mathfrak{t}^2 + 6 \mathfrak{t}^3 + 8 \mathfrak{t}^4 + 6 \mathfrak{t}^5 + 8 \mathfrak{t}^6 + 6 \mathfrak{t}^7 + 3 \mathfrak{t}^8 + \mathfrak{t}^9 + \mathfrak{t}^{10}\nonumber \\
&\quad + c \mathfrak{t}^2 (1 + \mathfrak{t} - 5 \mathfrak{t}^2 - 14 \mathfrak{t}^3 - 20 \mathfrak{t}^4 - 24 \mathfrak{t}^5 - 30 \mathfrak{t}^6 - 22 \mathfrak{t}^7 - 11 \mathfrak{t}^8 - 5 \mathfrak{t}^9 - 3 \mathfrak{t}^{10})\nonumber \\
&\quad + c^2 \mathfrak{t}^7 (2 + 9 \mathfrak{t} + 29 \mathfrak{t}^2 + 37 \mathfrak{t}^3 + 28 \mathfrak{t}^4 + 15 \mathfrak{t}^5 + 9 \mathfrak{t}^6 + 3 \mathfrak{t}^7)\nonumber \\
&\quad + c^3 \mathfrak{t}^{10} (-1 - 7 \mathfrak{t} - 9 \mathfrak{t}^2 - 10 \mathfrak{t}^3 - 9 \mathfrak{t}^4 - 7 \mathfrak{t}^5 - \mathfrak{t}^6)
),\\
&\sum_{k=0}^\infty c^k\langle W_{(k)}\rangle^{U(2)\text{ADHM-}[3](H)}(\mathfrak{t})\nonumber \\
&=
\frac{1}{
(1 - c \mathfrak{t})^3 (1 - c \mathfrak{t}^2)^4 (1 - \mathfrak{t}) (1 - \mathfrak{t}^2)^6 (1 - \mathfrak{t}^3)^5
}
(
1 + \mathfrak{t} + 6 \mathfrak{t}^{2} + 17 \mathfrak{t}^{3} + 31 \mathfrak{t}^{4} + 52 \mathfrak{t}^{5} + 92 \mathfrak{t}^{6} + 110 \mathfrak{t}^{7} + 112 \mathfrak{t}^{8} + 110 \mathfrak{t}^{9}\nonumber \\
&\quad\quad + 92 \mathfrak{t}^{10} + 52 \mathfrak{t}^{11} + 31 \mathfrak{t}^{12} + 17 \mathfrak{t}^{13} + 6 \mathfrak{t}^{14} + \mathfrak{t}^{15} + \mathfrak{t}^{16}\nonumber \\
&\quad + c \mathfrak{t}^{2} (2 + 2 \mathfrak{t} - 12 \mathfrak{t}^{2} - 44 \mathfrak{t}^{3} - 112 \mathfrak{t}^{4} - 262 \mathfrak{t}^{5} - 446 \mathfrak{t}^{6} - 554 \mathfrak{t}^{7} - 610 \mathfrak{t}^{8} - 599 \mathfrak{t}^{9} - 461 \mathfrak{t}^{10} - 277 \mathfrak{t}^{11} - 166 \mathfrak{t}^{12} - 83 \mathfrak{t}^{13}\nonumber \\
&\quad\quad - 27 \mathfrak{t}^{14} - 7 \mathfrak{t}^{15} - 4 \mathfrak{t}^{16})\nonumber \\
&\quad + c^2 \mathfrak{t}^{7} (24 + 156 \mathfrak{t} + 468 \mathfrak{t}^{2} + 804 \mathfrak{t}^{3} + 1098 \mathfrak{t}^{4} + 1317 \mathfrak{t}^{5} + 1287 \mathfrak{t}^{6} + 951 \mathfrak{t}^{7} + 612 \mathfrak{t}^{8} + 363 \mathfrak{t}^{9} + 165 \mathfrak{t}^{10} + 51 \mathfrak{t}^{11} + 18 \mathfrak{t}^{12}\nonumber \\
&\quad\quad + 6 \mathfrak{t}^{13}) \nonumber \\
&\quad + c^3 \mathfrak{t}^{9} (-21 - 135 \mathfrak{t} - 363 \mathfrak{t}^{2} - 658 \mathfrak{t}^{3} - 1051 \mathfrak{t}^{4} - 1371 \mathfrak{t}^{5} - 1337 \mathfrak{t}^{6} - 1016 \mathfrak{t}^{7} - 708 \mathfrak{t}^{8} - 410 \mathfrak{t}^{9} - 170 \mathfrak{t}^{10} - 54 \mathfrak{t}^{11}\nonumber \\
&\quad\quad - 22 \mathfrak{t}^{12} - 4 \mathfrak{t}^{13}) \nonumber \\
&\quad + c^4 \mathfrak{t}^{11} (12 + 69 \mathfrak{t} + 153 \mathfrak{t}^{2} + 289 \mathfrak{t}^{3} + 490 \mathfrak{t}^{4} + 642 \mathfrak{t}^{5} + 632 \mathfrak{t}^{6} + 557 \mathfrak{t}^{7} + 429 \mathfrak{t}^{8} + 242 \mathfrak{t}^{9} + 95 \mathfrak{t}^{10} + 36 \mathfrak{t}^{11} + 13 \mathfrak{t}^{12}\nonumber \\
&\quad\quad + \mathfrak{t}^{13}) \nonumber \\
&\quad + c^5 \mathfrak{t}^{13} (-3 - 15 \mathfrak{t} - 27 \mathfrak{t}^{2} - 54 \mathfrak{t}^{3} - 93 \mathfrak{t}^{4} - 117 \mathfrak{t}^{5} - 114 \mathfrak{t}^{6} - 117 \mathfrak{t}^{7} - 93 \mathfrak{t}^{8} - 54 \mathfrak{t}^{9} - 27 \mathfrak{t}^{10} - 15 \mathfrak{t}^{11} - 3 \mathfrak{t}^{12}) 
),\\
&\sum_{k=0}^\infty c^k\langle W_{(k)}\rangle^{U(2)\text{ADHM-}[4](H)}(\mathfrak{t})\nonumber \\
&=
\frac{1}{
(1 - c \mathfrak{t})^4 (1 - c \mathfrak{t}^2)^5 (1 - \mathfrak{t}) (1 - \mathfrak{t}^2)^8  (1 - \mathfrak{t}^3)^7
}
(
1 + \mathfrak{t} + 11 \mathfrak{t}^{2} + 34 \mathfrak{t}^{3} + 88 \mathfrak{t}^{4} + 216 \mathfrak{t}^{5} + 473 \mathfrak{t}^{6} + 797 \mathfrak{t}^{7} + 1243 \mathfrak{t}^{8}\nonumber \\
&\quad\quad + 1738 \mathfrak{t}^{9} + 2080 \mathfrak{t}^{10} + 2152 \mathfrak{t}^{11} + 2080 \mathfrak{t}^{12} + 1738 \mathfrak{t}^{13} + 1243 \mathfrak{t}^{14} + 797 \mathfrak{t}^{15} + 473 \mathfrak{t}^{16} + 216 \mathfrak{t}^{17} + 88 \mathfrak{t}^{18} + 34 \mathfrak{t}^{19}\nonumber \\
&\quad\quad + 11 \mathfrak{t}^{20} + \mathfrak{t}^{21} + \mathfrak{t}^{22}\nonumber \\
&\quad + c \mathfrak{t}^{2} (3 + 3 \mathfrak{t} - 23 \mathfrak{t}^{2} - 110 \mathfrak{t}^{3} - 452 \mathfrak{t}^{4} - 1368 \mathfrak{t}^{5} - 3025 \mathfrak{t}^{6} - 5517 \mathfrak{t}^{7} - 8983 \mathfrak{t}^{8} - 12490 \mathfrak{t}^{9} - 14824 \mathfrak{t}^{10} - 15640 \mathfrak{t}^{11}\nonumber \\
&\quad\quad - 14824 \mathfrak{t}^{12} - 12058 \mathfrak{t}^{13} - 8551 \mathfrak{t}^{14} - 5465 \mathfrak{t}^{15} - 3061 \mathfrak{t}^{16} - 1380 \mathfrak{t}^{17} - 564 \mathfrak{t}^{18} - 210 \mathfrak{t}^{19} - 59 \mathfrak{t}^{20} - 9 \mathfrak{t}^{21} - 5 \mathfrak{t}^{22})\nonumber \\
&\quad + c^2 \mathfrak{t}^{6} (6 + 186 \mathfrak{t} + 1120 \mathfrak{t}^{2} + 3586 \mathfrak{t}^{3} + 8280 \mathfrak{t}^{4} + 16306 \mathfrak{t}^{5} + 27222 \mathfrak{t}^{6} + 37872 \mathfrak{t}^{7} + 45398 \mathfrak{t}^{8} + 48494 \mathfrak{t}^{9} + 45138 \mathfrak{t}^{10}\nonumber \\
&\quad\quad + 36092 \mathfrak{t}^{11} + 25472 \mathfrak{t}^{12} + 16036 \mathfrak{t}^{13} + 8540 \mathfrak{t}^{14} + 3816 \mathfrak{t}^{15} + 1550 \mathfrak{t}^{16} + 546 \mathfrak{t}^{17} + 136 \mathfrak{t}^{18} + 30 \mathfrak{t}^{19} + 10 \mathfrak{t}^{20}) \nonumber \\
&\quad + c^3 \mathfrak{t}^{8} (-30 - 390 \mathfrak{t} - 1820 \mathfrak{t}^{2} - 5390 \mathfrak{t}^{3} - 13000 \mathfrak{t}^{4} - 26698 \mathfrak{t}^{5} - 44814 \mathfrak{t}^{6} - 62888 \mathfrak{t}^{7} - 77046 \mathfrak{t}^{8} - 82822 \mathfrak{t}^{9}\nonumber \\
&\quad\quad - 76098 \mathfrak{t}^{10} - 60508 \mathfrak{t}^{11} - 42624 \mathfrak{t}^{12} - 26180 \mathfrak{t}^{13} - 13404 \mathfrak{t}^{14} - 5960 \mathfrak{t}^{15} - 2374 \mathfrak{t}^{16} - 774 \mathfrak{t}^{17} - 180 \mathfrak{t}^{18} - 50 \mathfrak{t}^{19}\nonumber \\
&\quad\quad - 10 \mathfrak{t}^{20}) \nonumber \\
&\quad + c^4 \mathfrak{t}^{10} (60 + 504 \mathfrak{t} + 1932 \mathfrak{t}^{2} + 5388 \mathfrak{t}^{3} + 13056 \mathfrak{t}^{4} + 26356 \mathfrak{t}^{5} + 43625 \mathfrak{t}^{6} + 62061 \mathfrak{t}^{7} + 77735 \mathfrak{t}^{8} + 83614 \mathfrak{t}^{9}\nonumber \\
&\quad\quad + 76584 \mathfrak{t}^{10} + 61374 \mathfrak{t}^{11} + 43240 \mathfrak{t}^{12} + 25810 \mathfrak{t}^{13} + 12928 \mathfrak{t}^{14} + 5738 \mathfrak{t}^{15} + 2200 \mathfrak{t}^{16} + 650 \mathfrak{t}^{17} + 155 \mathfrak{t}^{18} + 45 \mathfrak{t}^{19}\nonumber \\
&\quad\quad + 5 \mathfrak{t}^{20}) \nonumber \\
&\quad + c^5 \mathfrak{t}^{12} (-60 - 380 \mathfrak{t} - 1256 \mathfrak{t}^{2} - 3416 \mathfrak{t}^{3} - 8160 \mathfrak{t}^{4} - 15836 \mathfrak{t}^{5} - 25725 \mathfrak{t}^{6} - 36881 \mathfrak{t}^{7} - 46363 \mathfrak{t}^{8} - 49538 \mathfrak{t}^{9}\nonumber \\
&\quad\quad - 45844 \mathfrak{t}^{10} - 37498 \mathfrak{t}^{11} - 26440 \mathfrak{t}^{12} - 15478 \mathfrak{t}^{13} - 7804 \mathfrak{t}^{14} - 3462 \mathfrak{t}^{15} - 1248 \mathfrak{t}^{16} - 334 \mathfrak{t}^{17} - 91 \mathfrak{t}^{18} - 21 \mathfrak{t}^{19} - \mathfrak{t}^{20}) \nonumber \\
&\quad + c^6 \mathfrak{t}^{14} (30 + 154 \mathfrak{t} + 454 \mathfrak{t}^{2} + 1240 \mathfrak{t}^{3} + 2884 \mathfrak{t}^{4} + 5340 \mathfrak{t}^{5} + 8544 \mathfrak{t}^{6} + 12288 \mathfrak{t}^{7} + 15220 \mathfrak{t}^{8} + 16084 \mathfrak{t}^{9} + 15120 \mathfrak{t}^{10}\nonumber \\
&\quad\quad + 12564 \mathfrak{t}^{11} + 8832 \mathfrak{t}^{12} + 5260 \mathfrak{t}^{13} + 2796 \mathfrak{t}^{14} + 1240 \mathfrak{t}^{15} + 414 \mathfrak{t}^{16} + 110 \mathfrak{t}^{17} + 34 \mathfrak{t}^{18} + 4 \mathfrak{t}^{19}) \nonumber \\
&\quad + c^7 \mathfrak{t}^{16} (-6 - 26 \mathfrak{t} - 70 \mathfrak{t}^{2} - 196 \mathfrak{t}^{3} - 440 \mathfrak{t}^{4} - 776 \mathfrak{t}^{5} - 1232 \mathfrak{t}^{6} - 1772 \mathfrak{t}^{7} - 2128 \mathfrak{t}^{8} - 2224 \mathfrak{t}^{9} - 2128 \mathfrak{t}^{10} - 1772 \mathfrak{t}^{11}\nonumber \\
&\quad\quad - 1232 \mathfrak{t}^{12} - 776 \mathfrak{t}^{13} - 440 \mathfrak{t}^{14} - 196 \mathfrak{t}^{15} - 70 \mathfrak{t}^{16} - 26 \mathfrak{t}^{17} - 6 \mathfrak{t}^{18})
).
\end{align}
}
For example, for $l=1$ we have
\begin{align}
&\langle W_{\ydiagram{1}} \rangle^{U(2)\text{ ADHM-}[1](H)}(\mathfrak{t})
=\frac{\mathfrak{t}}{(1-\mathfrak{t})^4},\\
&\langle W_{\ydiagram{2}} \rangle^{U(2)\text{ ADHM-}[1](H)}(\mathfrak{t})
=\frac{\mathfrak{t}^2(1+2\mathfrak{t}+4\mathfrak{t}^2-\mathfrak{t}^{4})}{(1-\mathfrak{t})^2(1-\mathfrak{t}^2)^2},\\
&\langle W_{\ydiagram{3}} \rangle^{U(2)\text{ ADHM-}[1](H)}(\mathfrak{t})
=\frac{\mathfrak{t}^3(1+3\mathfrak{t}^2-2\mathfrak{t}^3)}{(1-\mathfrak{t})^4},\\
&\langle W_{\ydiagram{4}} \rangle^{U(2)\text{ ADHM-}[1](H)}(\mathfrak{t})
=\frac{\mathfrak{t}^4(1+2\mathfrak{t}+4\mathfrak{t}^2+4\mathfrak{t}^3+4\mathfrak{t}^4-2\mathfrak{t}^5-3\mathfrak{t}^{6})}{(1-\mathfrak{t})^2(1-\mathfrak{t}^2)^2},\\
&\langle W_{\ydiagram{5}} \rangle^{U(2)\text{ ADHM-}[1](H)}(\mathfrak{t})
=\frac{\mathfrak{t}^5(1+3\mathfrak{t}^2-2\mathfrak{t}^3+5\mathfrak{t}^4-4\mathfrak{t}^5)}{(1-\mathfrak{t})^4},
\end{align}
and so on, which generalize to the symmetric representation with arbitrary dimension $k$ as
\begin{align}
\langle W_{(k)}\rangle^{U(2)\text{ ADHM-}[1](H)}(\mathfrak{t})=
\frac{
\mathfrak{t}^k (1 + \mathfrak{t} + 2 \mathfrak{t}^2 + \mathfrak{t}^{k + 1} (-2 - \mathfrak{t} - \mathfrak{t}^2 + k (-1 + \mathfrak{t}^2)))
}{
(1 - \mathfrak{t})^3 (1 - \mathfrak{t}^2)^2
}.
\end{align}

In the limit of large representation, we have

{\fontsize{9pt}{1pt}\selectfont
\begin{align}
&\langle W_{(k)}\rangle^{U(2)\text{ADHM-}[1](H)}(\mathfrak{t})
=
\frac{\mathfrak{t}^k(1 + \mathfrak{t} + 2 \mathfrak{t}^2)}{
(1 - \mathfrak{t})^3 (1 - \mathfrak{t}^2)^2}+{\cal O}(\mathfrak{t}^{2k+1}), \\
&\langle W_{(k)}\rangle^{U(2)\text{ADHM-}[2](H)}(\mathfrak{t})\nonumber \\
&=
\frac{\mathfrak{t}^k}{
(1 - \mathfrak{t})^2 (1 - \mathfrak{t}^2)^4 (1 - \mathfrak{t}^3)^3
}
(
1 - 7 \mathfrak{t}^2 - 29 \mathfrak{t}^3 - 67 \mathfrak{t}^4 - 107 \mathfrak{t}^5 - 118 \mathfrak{t}^6 - 99 \mathfrak{t}^7 - 61 \mathfrak{t}^8 - 26 \mathfrak{t}^9 - 6 \mathfrak{t}^{10} - \mathfrak{t}^{11}\nonumber \\
&\quad + k (1 + 4 \mathfrak{t} + 11 \mathfrak{t}^2 + 19 \mathfrak{t}^3 + 21 \mathfrak{t}^4 + 11 \mathfrak{t}^5 - 6 \mathfrak{t}^6 - 19 \mathfrak{t}^7 - 21 \mathfrak{t}^8 - 14 \mathfrak{t}^9 - 6 \mathfrak{t}^{10} - \mathfrak{t}^{11})
)+{\cal O}(\mathfrak{t}^{2k+2}), \\
&\langle W_{(k)}\rangle^{U(2)\text{ADHM-}[3](H)}(\mathfrak{t})\nonumber \\
&=
\frac{\mathfrak{t}^k}{
2(1 - \mathfrak{t})^2(1 - \mathfrak{t}^2)^6(1 - \mathfrak{t}^3)^5
}
(
2 + 10 \mathfrak{t}^2 + 102 \mathfrak{t}^3 + 478 \mathfrak{t}^4 + 1474 \mathfrak{t}^5 + 3320 \mathfrak{t}^6 + 5662 \mathfrak{t}^7 + 7652 \mathfrak{t}^8 + 8336 \mathfrak{t}^9 + 7318 \mathfrak{t}^{10}\nonumber \\
&\quad\quad + 5120 \mathfrak{t}^{11} + 2830 \mathfrak{t}^{12} + 1182 \mathfrak{t}^{13} + 338 \mathfrak{t}^{14} + 50 \mathfrak{t}^{15} - 2 \mathfrak{t}^{17}\nonumber \\
&\quad + k (3 + 6 \mathfrak{t} - 3 \mathfrak{t}^2 - 81 \mathfrak{t}^3 - 315 \mathfrak{t}^4 - 717 \mathfrak{t}^5 - 1110 \mathfrak{t}^6 - 1197 \mathfrak{t}^7 - 771 \mathfrak{t}^8 + 48 \mathfrak{t}^9 + 837 \mathfrak{t}^{10} + 1197 \mathfrak{t}^{11} + 1053 \mathfrak{t}^{12}\nonumber \\
&\quad\quad + 657 \mathfrak{t}^{13} + 294 \mathfrak{t}^{14} + 87 \mathfrak{t}^{15} + 12 \mathfrak{t}^{16})\nonumber \\
&\quad  + k^2 (1 + 6 \mathfrak{t} + 23 \mathfrak{t}^2 + 57 \mathfrak{t}^3 + 95 \mathfrak{t}^4 + 101 \mathfrak{t}^5 + 40 \mathfrak{t}^6 - 79 \mathfrak{t}^7 - 191 \mathfrak{t}^8 - 218 \mathfrak{t}^9 - 139 \mathfrak{t}^{10} - 11 \mathfrak{t}^{11} + 83 \mathfrak{t}^{12} + 105 \mathfrak{t}^{13}\nonumber \\
&\quad\quad + 76 \mathfrak{t}^{14} + 37 \mathfrak{t}^{15} + 12 \mathfrak{t}^{16} + 2 \mathfrak{t}^{17})
)+{\cal O}(\mathfrak{t}^{2k+3}).
\end{align}
}

\noindent \underline{Diagonal $2$-point functions}
{\fontsize{9pt}{1pt}\selectfont
\begin{align}
&\sum_{k=0}^\infty c^k\langle W_{(k)}W_{(\overline{k})}\rangle^{{\cal N}=8\text{ }U(2)(H)}(\mathfrak{t})
=
\frac{
1 + \mathfrak{t} + 2 \mathfrak{t}^2 + c \mathfrak{t}^3 (-2 - \mathfrak{t} - \mathfrak{t}^2)
}{
(1 - c) (1 - c \mathfrak{t})^2 (1 - \mathfrak{t}) (1 - \mathfrak{t}^2)
},\\
&\sum_{k=0}^\infty c^k\langle W_{(k)}W_{(\overline{k})}\rangle^{U(2)\text{ADHM-}[1](H)}(\mathfrak{t})
=
\frac{1+\mathfrak{t}^2+c\mathfrak{t}^2(1+\mathfrak{t}^2)}{(1-c)(1-c\mathfrak{t})^2(1-\mathfrak{t})^2(1-\mathfrak{t}^2)^2},\\
%
&\sum_{k=0}^\infty c^k\langle W_{(k)}W_{(\overline{k})}\rangle^{U(2)\text{ADHM-}[2](H)}(\mathfrak{t})\nonumber \\
&=\frac{1}{
(1 - c) (1 - c \mathfrak{t})^2 (1 - c \mathfrak{t}^2)^2 (1 - \mathfrak{t}) (1 - \mathfrak{t}^2)^4 (1 - \mathfrak{t}^3)^3}
(1 + \mathfrak{t} + 3 \mathfrak{t}^{2} + 6 \mathfrak{t}^{3} + 8 \mathfrak{t}^{4} + 6 \mathfrak{t}^{5} + 8 \mathfrak{t}^{6} + 6 \mathfrak{t}^{7} + 3 \mathfrak{t}^{8} + \mathfrak{t}^{9} + \mathfrak{t}^{10}\nonumber \\
&\quad + c \mathfrak{t}^2 (2 + 8 \mathfrak{t} + 8 \mathfrak{t}^2 + 4 \mathfrak{t}^3 - 2 \mathfrak{t}^4 - 12 \mathfrak{t}^5 - 22 \mathfrak{t}^6 - 16 \mathfrak{t}^7 - 8 \mathfrak{t}^8 - 4 \mathfrak{t}^9 - 2 \mathfrak{t}^{10})\nonumber \\
&\quad + c^2 \mathfrak{t}^4 (-2 - 4 \mathfrak{t} - 8 \mathfrak{t}^2 - 16 \mathfrak{t}^3 - 22 \mathfrak{t}^4 - 12 \mathfrak{t}^5 - 2 \mathfrak{t}^6 + 4 \mathfrak{t}^7 + 8 \mathfrak{t}^8 + 8 \mathfrak{t}^9 + 2 \mathfrak{t}^{10})\nonumber \\
&\quad + c^3 \mathfrak{t}^6 (1 + \mathfrak{t} + 3 \mathfrak{t}^2 + 6 \mathfrak{t}^3 + 8 \mathfrak{t}^4 + 6 \mathfrak{t}^5 + 8 \mathfrak{t}^6 + 6 \mathfrak{t}^7 + 3 \mathfrak{t}^8 + \mathfrak{t}^9 + \mathfrak{t}^{10})
),\\
&\sum_{k=0}^\infty c^k\langle W_{(k)}W_{(\overline{k})}\rangle^{U(2)\text{ADHM-}[3](H)}(\mathfrak{t})\nonumber \\
&=
\frac{1}{
(1 - c) (1 - c \mathfrak{t})^2 (1 - c \mathfrak{t}^2)^4 (1 - \mathfrak{t}) (1 - \mathfrak{t}^2)^6 (1 - \mathfrak{t}^3)^5
}
(
1 + \mathfrak{t} + 6 \mathfrak{t}^{2} + 17 \mathfrak{t}^{3} + 31 \mathfrak{t}^{4} + 52 \mathfrak{t}^{5} + 92 \mathfrak{t}^{6} + 110 \mathfrak{t}^{7} + 112 \mathfrak{t}^{8}\nonumber \\
&\quad\quad + 110 \mathfrak{t}^{9} + 92 \mathfrak{t}^{10} + 52 \mathfrak{t}^{11} + 31 \mathfrak{t}^{12} + 17 \mathfrak{t}^{13} + 6 \mathfrak{t}^{14} + \mathfrak{t}^{15} + \mathfrak{t}^{16}\nonumber \\
&\quad + c \mathfrak{t}^{2} (5 + 21 \mathfrak{t} + 29 \mathfrak{t}^{2} + 40 \mathfrak{t}^{3} + 35 \mathfrak{t}^{4} - 49 \mathfrak{t}^{5} - 218 \mathfrak{t}^{6} - 321 \mathfrak{t}^{7} - 405 \mathfrak{t}^{8} - 454 \mathfrak{t}^{9} - 385 \mathfrak{t}^{10} - 233 \mathfrak{t}^{11} - 147 \mathfrak{t}^{12}\nonumber \\
&\quad\quad - 78 \mathfrak{t}^{13} - 26 \mathfrak{t}^{14} - 6 \mathfrak{t}^{15} - 4 \mathfrak{t}^{16}) \nonumber \\
&\quad + c^2 \mathfrak{t}^{4} (-3 - 13 \mathfrak{t} - 42 \mathfrak{t}^{2} - 151 \mathfrak{t}^{3} - 290 \mathfrak{t}^{4} - 301 \mathfrak{t}^{5} - 179 \mathfrak{t}^{6} + 6 \mathfrak{t}^{7} + 347 \mathfrak{t}^{8} + 615 \mathfrak{t}^{9} + 580 \mathfrak{t}^{10} + 407 \mathfrak{t}^{11} + 280 \mathfrak{t}^{12}\nonumber \\
&\quad\quad + 143 \mathfrak{t}^{13} + 45 \mathfrak{t}^{14} + 14 \mathfrak{t}^{15} + 6 \mathfrak{t}^{16})\nonumber \\
&\quad + c^3 \mathfrak{t}^{6} (6 + 14 \mathfrak{t} + 45 \mathfrak{t}^{2} + 143 \mathfrak{t}^{3} + 280 \mathfrak{t}^{4} + 407 \mathfrak{t}^{5} + 580 \mathfrak{t}^{6} + 615 \mathfrak{t}^{7} + 347 \mathfrak{t}^{8} + 6 \mathfrak{t}^{9} - 179 \mathfrak{t}^{10} - 301 \mathfrak{t}^{11} - 290 \mathfrak{t}^{12}\nonumber \\
&\quad\quad - 151 \mathfrak{t}^{13} - 42 \mathfrak{t}^{14} - 13 \mathfrak{t}^{15} - 3 \mathfrak{t}^{16}) \nonumber \\
&\quad + c^4 \mathfrak{t}^{8} (-4 - 6 \mathfrak{t} - 26 \mathfrak{t}^{2} - 78 \mathfrak{t}^{3} - 147 \mathfrak{t}^{4} - 233 \mathfrak{t}^{5} - 385 \mathfrak{t}^{6} - 454 \mathfrak{t}^{7} - 405 \mathfrak{t}^{8} - 321 \mathfrak{t}^{9} - 218 \mathfrak{t}^{10} - 49 \mathfrak{t}^{11} + 35 \mathfrak{t}^{12}\nonumber \\
&\quad\quad + 40 \mathfrak{t}^{13} + 29 \mathfrak{t}^{14} + 21 \mathfrak{t}^{15} + 5 \mathfrak{t}^{16}) \nonumber \\
&\quad + c^5 \mathfrak{t}^{10} (1 + \mathfrak{t} + 6 \mathfrak{t}^{2} + 17 \mathfrak{t}^{3} + 31 \mathfrak{t}^{4} + 52 \mathfrak{t}^{5} + 92 \mathfrak{t}^{6} + 110 \mathfrak{t}^{7} + 112 \mathfrak{t}^{8} + 110 \mathfrak{t}^{9} + 92 \mathfrak{t}^{10} + 52 \mathfrak{t}^{11} + 31 \mathfrak{t}^{12} + 17 \mathfrak{t}^{13}\nonumber \\
&\quad\quad + 6 \mathfrak{t}^{14} + \mathfrak{t}^{15} + \mathfrak{t}^{16}) 
),\\
&\sum_{k=0}^\infty c^k\langle W_{(k)}W_{(\overline{k})}\rangle^{U(2)\text{ADHM-}[4](H)}(\mathfrak{t})\nonumber \\
&=
\frac{1}{
(1 - c) (1 - c \mathfrak{t})^2 (1 - c \mathfrak{t}^2)^6 (1 - \mathfrak{t}) (1 - \mathfrak{t}^2)^8 (1 - \mathfrak{t}^3)^7
}
(
1 + \mathfrak{t} + 11 \mathfrak{t}^{2} + 34 \mathfrak{t}^{3} + 88 \mathfrak{t}^{4} + 216 \mathfrak{t}^{5} + 473 \mathfrak{t}^{6} + 797 \mathfrak{t}^{7}\nonumber \\
&\quad\quad + 1243 \mathfrak{t}^{8} + 1738 \mathfrak{t}^{9} + 2080 \mathfrak{t}^{10} + 2152 \mathfrak{t}^{11} + 2080 \mathfrak{t}^{12} + 1738 \mathfrak{t}^{13} + 1243 \mathfrak{t}^{14} + 797 \mathfrak{t}^{15} + 473 \mathfrak{t}^{16} + 216 \mathfrak{t}^{17} + 88 \mathfrak{t}^{18}\nonumber \\
&\quad\quad + 34 \mathfrak{t}^{19} + 11 \mathfrak{t}^{20} + \mathfrak{t}^{21} + \mathfrak{t}^{22}\nonumber \\
&\quad + c \mathfrak{t}^{2} (10 + 40 \mathfrak{t} + 89 \mathfrak{t}^{2} + 205 \mathfrak{t}^{3} + 313 \mathfrak{t}^{4} + 100 \mathfrak{t}^{5} - 670 \mathfrak{t}^{6} - 2032 \mathfrak{t}^{7} - 4562 \mathfrak{t}^{8} - 7802 \mathfrak{t}^{9} - 10412 \mathfrak{t}^{10} - 11836 \mathfrak{t}^{11}\nonumber \\
&\quad\quad - 12194 \mathfrak{t}^{12} - 10572 \mathfrak{t}^{13} - 7779 \mathfrak{t}^{14} - 5143 \mathfrak{t}^{15} - 3047 \mathfrak{t}^{16} - 1404 \mathfrak{t}^{17} - 578 \mathfrak{t}^{18} - 224 \mathfrak{t}^{19} - 68 \mathfrak{t}^{20} - 8 \mathfrak{t}^{21} - 6 \mathfrak{t}^{22}) \nonumber \\
&\quad + c^2 \mathfrak{t}^{4} (3 - 9 \mathfrak{t} - 121 \mathfrak{t}^{2} - 672 \mathfrak{t}^{3} - 1710 \mathfrak{t}^{4} - 2890 \mathfrak{t}^{5} - 3867 \mathfrak{t}^{6} - 3399 \mathfrak{t}^{7} + 1052 \mathfrak{t}^{8} + 8695 \mathfrak{t}^{9} + 16627 \mathfrak{t}^{10} + 23640 \mathfrak{t}^{11}\nonumber \\
&\quad\quad + 27769 \mathfrak{t}^{12} + 25849 \mathfrak{t}^{13} + 20083 \mathfrak{t}^{14} + 13926 \mathfrak{t}^{15} + 8304 \mathfrak{t}^{16} + 3886 \mathfrak{t}^{17} + 1627 \mathfrak{t}^{18} + 631 \mathfrak{t}^{19} + 178 \mathfrak{t}^{20} + 27 \mathfrak{t}^{21}\nonumber \\
&\quad\quad + 15 \mathfrak{t}^{22}) \nonumber \\
&\quad + c^3 \mathfrak{t}^{6} (16 + 58 \mathfrak{t} + 216 \mathfrak{t}^{2} + 898 \mathfrak{t}^{3} + 2404 \mathfrak{t}^{4} + 5442 \mathfrak{t}^{5} + 10681 \mathfrak{t}^{6} + 15857 \mathfrak{t}^{7} + 16214 \mathfrak{t}^{8} + 10893 \mathfrak{t}^{9} + 373 \mathfrak{t}^{10}\nonumber \\
&\quad\quad - 14556 \mathfrak{t}^{11} - 27023 \mathfrak{t}^{12} - 30239 \mathfrak{t}^{13} - 26594 \mathfrak{t}^{14} - 20123 \mathfrak{t}^{15} - 12307 \mathfrak{t}^{16} - 5926 \mathfrak{t}^{17} - 2552 \mathfrak{t}^{18} - 986 \mathfrak{t}^{19}\nonumber \\
&\quad\quad - 256 \mathfrak{t}^{20} - 50 \mathfrak{t}^{21} - 20 \mathfrak{t}^{22}) \nonumber \\
&\quad + c^4 \mathfrak{t}^{8} (-20 - 50 \mathfrak{t} - 256 \mathfrak{t}^{2} - 986 \mathfrak{t}^{3} - 2552 \mathfrak{t}^{4} - 5926 \mathfrak{t}^{5} - 12307 \mathfrak{t}^{6} - 20123 \mathfrak{t}^{7} - 26594 \mathfrak{t}^{8} - 30239 \mathfrak{t}^{9} - 27023 \mathfrak{t}^{10}\nonumber \\
&\quad\quad - 14556 \mathfrak{t}^{11} + 373 \mathfrak{t}^{12} + 10893 \mathfrak{t}^{13} + 16214 \mathfrak{t}^{14} + 15857 \mathfrak{t}^{15} + 10681 \mathfrak{t}^{16} + 5442 \mathfrak{t}^{17} + 2404 \mathfrak{t}^{18} + 898 \mathfrak{t}^{19} + 216 \mathfrak{t}^{20}\nonumber \\
&\quad\quad + 58 \mathfrak{t}^{21} + 16 \mathfrak{t}^{22})\nonumber \\
&\quad + c^5 \mathfrak{t}^{10} (15 + 27 \mathfrak{t} + 178 \mathfrak{t}^{2} + 631 \mathfrak{t}^{3} + 1627 \mathfrak{t}^{4} + 3886 \mathfrak{t}^{5} + 8304 \mathfrak{t}^{6} + 13926 \mathfrak{t}^{7} + 20083 \mathfrak{t}^{8} + 25849 \mathfrak{t}^{9} + 27769 \mathfrak{t}^{10}\nonumber \\
&\quad\quad + 23640 \mathfrak{t}^{11} + 16627 \mathfrak{t}^{12} + 8695 \mathfrak{t}^{13} + 1052 \mathfrak{t}^{14} - 3399 \mathfrak{t}^{15} - 3867 \mathfrak{t}^{16} - 2890 \mathfrak{t}^{17} - 1710 \mathfrak{t}^{18} - 672 \mathfrak{t}^{19} - 121 \mathfrak{t}^{20}\nonumber \\
&\quad\quad - 9 \mathfrak{t}^{21} + 3 \mathfrak{t}^{22}) \nonumber \\
&\quad + c^6 \mathfrak{t}^{12} (-6 - 8 \mathfrak{t} - 68 \mathfrak{t}^{2} - 224 \mathfrak{t}^{3} - 578 \mathfrak{t}^{4} - 1404 \mathfrak{t}^{5} - 3047 \mathfrak{t}^{6} - 5143 \mathfrak{t}^{7} - 7779 \mathfrak{t}^{8} - 10572 \mathfrak{t}^{9} - 12194 \mathfrak{t}^{10} - 11836 \mathfrak{t}^{11}\nonumber \\
&\quad\quad - 10412 \mathfrak{t}^{12} - 7802 \mathfrak{t}^{13} - 4562 \mathfrak{t}^{14} - 2032 \mathfrak{t}^{15} - 670 \mathfrak{t}^{16} + 100 \mathfrak{t}^{17} + 313 \mathfrak{t}^{18} + 205 \mathfrak{t}^{19} + 89 \mathfrak{t}^{20} + 40 \mathfrak{t}^{21} + 10 \mathfrak{t}^{22}) \nonumber \\
&\quad + c^7 \mathfrak{t}^{14} (1 + \mathfrak{t} + 11 \mathfrak{t}^{2} + 34 \mathfrak{t}^{3} + 88 \mathfrak{t}^{4} + 216 \mathfrak{t}^{5} + 473 \mathfrak{t}^{6} + 797 \mathfrak{t}^{7} + 1243 \mathfrak{t}^{8} + 1738 \mathfrak{t}^{9} + 2080 \mathfrak{t}^{10} + 2152 \mathfrak{t}^{11} + 2080 \mathfrak{t}^{12}\nonumber \\
&\quad\quad + 1738 \mathfrak{t}^{13} + 1243 \mathfrak{t}^{14} + 797 \mathfrak{t}^{15} + 473 \mathfrak{t}^{16} + 216 \mathfrak{t}^{17} + 88 \mathfrak{t}^{18} + 34 \mathfrak{t}^{19} + 11 \mathfrak{t}^{20} + \mathfrak{t}^{21} + \mathfrak{t}^{22}) 
).
\end{align}
}
For example, the unflavored 2-point correlators for ${\cal N}=8$ $U(2)$ SYM theory and $U(2)$ ADHM theory with $l=1$ flavor are
\begin{align}
\label{u2N8_w1w1_2}
&\langle W_{\ydiagram{1}}W_{\overline{\ydiagram{1}}}\rangle^{\textrm{$\mathcal{N}=8$ $U(2) (H)$}}(\mathfrak{t})
=\frac{1+2\mathfrak{t}+2\mathfrak{t}^2-\mathfrak{t}^4}
{(1-\mathfrak{t})^2},\\
&\langle W_{\ydiagram{2}}W_{\overline{\ydiagram{2}}}\rangle^{\textrm{$\mathcal{N}=8$ $U(2) (H)$}}(\mathfrak{t})
=\frac{1+3\mathfrak{t}+7\mathfrak{t}^2+5\mathfrak{t}^3+\mathfrak{t}^4-3\mathfrak{t}^5-2\mathfrak{t}^6}
{(1-\mathfrak{t})(1-\mathfrak{t}^2)}, \\
&\langle W_{\ydiagram{3}}W_{\overline{\ydiagram{3}}}\rangle^{\textrm{$\mathcal{N}=8$ $U(2) (H)$}}(\mathfrak{t})
=\frac{1+2\mathfrak{t}+5\mathfrak{t}^2+4\mathfrak{t}^3+\mathfrak{t}^4-2\mathfrak{t}^5-3\mathfrak{t}^6}
{(1-\mathfrak{t})^2}, \\
&\langle W_{\ydiagram{4}}W_{\overline{\ydiagram{4}}}\rangle^{\textrm{$\mathcal{N}=8$ $U(2) (H)$}}(\mathfrak{t})
=\frac{1+3\mathfrak{t}+7\mathfrak{t}^2+9\mathfrak{t}^3+10\mathfrak{t}^4+4\mathfrak{t}^5-3\mathfrak{t}^6-7\mathfrak{t}^7-4\mathfrak{t}^8}
{(1-\mathfrak{t})(1-\mathfrak{t}^2)}, \\
&\langle W_{\ydiagram{5}}W_{\overline{\ydiagram{5}}}\rangle^{\textrm{$\mathcal{N}=8$ $U(2) (H)$}}(\mathfrak{t})
=\frac{1+2\mathfrak{t}+5\mathfrak{t}^2+4\mathfrak{t}^3+6\mathfrak{t}^4+4\mathfrak{t}^5-\mathfrak{t}^6-4\mathfrak{t}^7-5\mathfrak{t}^8}
{(1-\mathfrak{t})^2},\\
%
&\langle W_{\ydiagram{1}}W_{\overline{\ydiagram{1}}}\rangle^{\textrm{$U(2)$ ADHM-$[1] (H)$}}(\mathfrak{t})
=\frac{1+\mathfrak{t}^2}{(1-\mathfrak{t})^4},\label{u2ADHM1w1w1} \\
&\langle W_{\ydiagram{2}}W_{\overline{\ydiagram{2}}}\rangle^{\textrm{$U(2)$ ADHM-$[1] (H)$}}(\mathfrak{t})
=\frac{1+2\mathfrak{t}+5\mathfrak{t}^2+4\mathfrak{t}^3+4\mathfrak{t}^4+2\mathfrak{t}^5}
{(1-\mathfrak{t})^4(1+\mathfrak{t})^2},\label{u2ADHM1wsym2wsym2} \\
&\langle W_{\ydiagram{3}}W_{\overline{\ydiagram{3}}}\rangle^{\textrm{$U(2)$ ADHM-$[1] (H)$}}(\mathfrak{t})
=\frac{1+4\mathfrak{t}^2+3\mathfrak{t}^4}
{(1-\mathfrak{t})^4},\label{u2ADHM1wsym3wsym3} \\
&\langle W_{\ydiagram{4}}W_{\overline{\ydiagram{4}}}\rangle^{\textrm{$U(2)$ ADHM-$[1] (H)$}}(\mathfrak{t})
=\frac{1+2\mathfrak{t}+5\mathfrak{t}^2+8\mathfrak{t}^3+12\mathfrak{t}^4+10\mathfrak{t}^5+8\mathfrak{t}^6+4\mathfrak{t}^7}
{(1-\mathfrak{t})^4(1+\mathfrak{t})^2},\label{u2ADHM1wsym4wsym4} \\
&\langle W_{\ydiagram{5}}W_{\overline{\ydiagram{5}}}\rangle^{\textrm{$U(2)$ ADHM-$[1] (H)$}}(\mathfrak{t})
=\frac{1+4\mathfrak{t}^2+8\mathfrak{t}^4+5\mathfrak{t}^6}
{(1-\mathfrak{t})^4},\label{u2ADHM1wsym5wsym5}
\end{align}
and so on.

In particular, for $U(2)$ ADHM theory with flavor $l=1$ we find
\begin{align}
\langle W_{(k)}W_{(\overline{k})}\rangle^{\textrm{$U(2)$ ADHM-$[1]$}}(\mathfrak{t})
&=\frac{(1+\mathfrak{t}^2)(1+\mathfrak{t}^2-(k+2)\mathfrak{t}^{k+1}+k\mathfrak{t}^{k+3})}
{(1-\mathfrak{t})^6(1+\mathfrak{t})^2}
\end{align}
for general dimension $k$ of the symmetric representation.
Note that this formula can also be rephrased as the following decompositions of $\langle W_{(k)}W_{(\overline{k})}\rangle^{U(2)\text{ ADHM-}[1](H)}$ in terms of the index (\ref{u2ADHM1_index}) and the $2$-point function in the fundamental representation \eqref{u2ADHM1w1w1}: 
\begin{align}
&\langle W_{(2k-1)}W_{(2k-1)}\rangle^{\textrm{$U(2)$ ADHM-$[1]$}}(\mathfrak{t})
=\sum_{i=0}^{k-1}(2i+1)\langle W_{(1)}W_{(1)}\rangle^{\textrm{$U(2)$ ADHM-$[1]$}}(\mathfrak{t})
\mathfrak{t}^{2i},\\
&\langle W_{(2k)}W_{(2k)}\rangle^{\textrm{$U(2)$ ADHM-$[1]$}}(\mathfrak{t})
=
\mathcal{I}^{\textrm{$U(2)$ ADHM-$[1]$}}(\mathfrak{t})+
\sum_{i=1}^{k}(2i)\langle W_{(1)}W_{(1)}\rangle^{\textrm{$U(2)$ ADHM-$[1]$}}(\mathfrak{t})
\mathfrak{t}^{2i-1}.
\end{align}

In the limit of large representation, we have
\begin{align}
\langle W_{(\infty)}W_{(\overline{\infty})}\rangle^{\textrm{$\mathcal{N}=8$ $U(2) (H)$}}(\mathfrak{t})
&=\frac{1+2\mathfrak{t}+4\mathfrak{t}^2+2\mathfrak{t}^3+\mathfrak{t}^4}{(1-\mathfrak{t})^2(1-\mathfrak{t}^2)}. 
\end{align}

{\fontsize{9pt}{1pt}\selectfont
\begin{align}
&\langle W_{(k)}W_{(\overline{k})}\rangle^{U(2)\text{ ADHM-}[1](H)}(\mathfrak{t})
=\frac{1+2\mathfrak{t}^2+\mathfrak{t}^4}{(1-\mathfrak{t})^4(1-\mathfrak{t}^2)^2}+{\cal O}(\mathfrak{t}^{k+1}),\\
&\langle W_{(k)}W_{(\overline{k})}\rangle^{U(2)\text{ ADHM-}[2](H)}(\mathfrak{t})\nonumber \\
&=
\frac{1 + \mathfrak{t} + 7 \mathfrak{t}^2 + 16 \mathfrak{t}^3 + 27 \mathfrak{t}^4 + 37 \mathfrak{t}^5 + 46 \mathfrak{t}^6 + 37 \mathfrak{t}^7 + 27 \mathfrak{t}^8 + 16 \mathfrak{t}^9 + 7 \mathfrak{t}^{10} + \mathfrak{t}^{11} + \mathfrak{t}^{12}
}
{(1 - \mathfrak{t})^3 (1 - \mathfrak{t}^2)^4 (1 - \mathfrak{t}^3)^3}
+{\cal O}(\mathfrak{t}^{k+1}),\\
&\langle W_{(k)}W_{(\overline{k})}\rangle^{U(2)\text{ ADHM-}[3](H)}(\mathfrak{t})\nonumber \\
&=
\frac{1 + \mathfrak{t} + 15 \mathfrak{t}^2 + 42 \mathfrak{t}^3 + 111 \mathfrak{t}^4 + 241 \mathfrak{t}^5 + 449 \mathfrak{t}^6 + 640 \mathfrak{t}^7 + 834 \mathfrak{t}^8 + 906 \mathfrak{t}^9 + \cdots + \mathfrak{t}^{18}}
{
(1 - \mathfrak{t})^3 (1 - \mathfrak{t}^2)^6 (1 - \mathfrak{t}^3)^5}
+{\cal O}(\mathfrak{t}^{k+1}),\\
&\langle W_{(k)}W_{(\overline{k})}\rangle^{U(2)\text{ADHM-}[4](H)}(\mathfrak{t})\nonumber \\
&=
\frac{1}{
(1 - \mathfrak{t}) (1 - \mathfrak{t}^2)^{10} (1 - \mathfrak{t}^3)^7}
(
1 + 3 \mathfrak{t} + 30 \mathfrak{t}^2 + 135 \mathfrak{t}^3 + 514 \mathfrak{t}^4 + 1611 \mathfrak{t}^5 + 4339 \mathfrak{t}^6 + 9758 \mathfrak{t}^7 + 19215 \mathfrak{t}^8 + 33103 \mathfrak{t}^9\nonumber \\
&\quad + 50242 \mathfrak{t}^{10} + 67187 \mathfrak{t}^{11} + 80179 \mathfrak{t}^{12} + 85006 \mathfrak{t}^{13} + \cdots + \mathfrak{t}^{26}
)+{\cal O}(\mathfrak{t}^{k+1}).
\end{align}
}



\subsubsection{$U(3)$}
\noindent \underline{1-point functions}
{\fontsize{9pt}{1pt}\selectfont
\begin{align}
&\sum_{k=0}^\infty c^k\langle W_{(k)}\rangle^{U(3)\text{ ADHM-}[1](H)}(\mathfrak{t})\nonumber \\
&=
\frac{
1 + \mathfrak{t}^2 + 2 \mathfrak{t}^3 + \mathfrak{t}^4 + \mathfrak{t}^6 + c \mathfrak{t}^3 (1 - 3 \mathfrak{t}^2 - 4 \mathfrak{t}^3 - 2 \mathfrak{t}^4 - 2 \mathfrak{t}^5 - 2 \mathfrak{t}^6) + c^2 \mathfrak{t}^{10} (1 + 4 \mathfrak{t} + \mathfrak{t}^2)
}{
(1 - c \mathfrak{t}) (1 - c \mathfrak{t}^2)^2 (1 - c \mathfrak{t}^3)^2 (1 - \mathfrak{t})^2 (1 - \mathfrak{t}^2)^2 (1 - \mathfrak{t}^3)^2
},\\
&\sum_{k=0}^\infty c^k\langle W_{(k)}\rangle^{U(3)\text{ADHM-}[2](H)}(\mathfrak{t})\nonumber \\
&=
\frac{1}{
(1 - c \mathfrak{t})^2 (1 - c \mathfrak{t}^2)^4 (1 - c \mathfrak{t}^3)^3 (1 - \mathfrak{t})^2 (1 - \mathfrak{t}^2)^3 (1 - \mathfrak{t}^3)^4 (1 - \mathfrak{t}^4)^3
}
(
1 + 3 \mathfrak{t}^{2} + 6 \mathfrak{t}^{3} + 12 \mathfrak{t}^{4} + 16 \mathfrak{t}^{5} + 31 \mathfrak{t}^{6} + 36 \mathfrak{t}^{7}\nonumber \\
&\quad + 55 \mathfrak{t}^{8} + 54 \mathfrak{t}^{9} + 60 \mathfrak{t}^{10} + 54 \mathfrak{t}^{11} + 55 \mathfrak{t}^{12} + 36 \mathfrak{t}^{13} + 31 \mathfrak{t}^{14} + 16 \mathfrak{t}^{15} + 12 \mathfrak{t}^{16} + 6 \mathfrak{t}^{17} + 3 \mathfrak{t}^{18} + \mathfrak{t}^{20}\nonumber \\
&\quad + c \mathfrak{t}^{3} (3 - 5 \mathfrak{t}^{2} - 18 \mathfrak{t}^{3} - 40 \mathfrak{t}^{4} - 92 \mathfrak{t}^{5} - 159 \mathfrak{t}^{6} - 244 \mathfrak{t}^{7} - 313 \mathfrak{t}^{8} - 358 \mathfrak{t}^{9} - 384 \mathfrak{t}^{10} - 374 \mathfrak{t}^{11} - 319 \mathfrak{t}^{12} - 240 \mathfrak{t}^{13}\nonumber \\
&\quad\quad - 169 \mathfrak{t}^{14} - 104 \mathfrak{t}^{15} - 64 \mathfrak{t}^{16} - 30 \mathfrak{t}^{17} - 11 \mathfrak{t}^{18} - 4 \mathfrak{t}^{19} - 3 \mathfrak{t}^{20}) \nonumber \\
&\quad + c^2 \mathfrak{t}^{5} (-2 - 4 \mathfrak{t}^{2} - 8 \mathfrak{t}^{3} + 2 \mathfrak{t}^{4} + 49 \mathfrak{t}^{5} + 178 \mathfrak{t}^{6} + 343 \mathfrak{t}^{7} + 578 \mathfrak{t}^{8} + 772 \mathfrak{t}^{9} + 976 \mathfrak{t}^{10} + 1041 \mathfrak{t}^{11} + 1016 \mathfrak{t}^{12} + 822 \mathfrak{t}^{13}\nonumber \\
&\quad\quad + 636 \mathfrak{t}^{14} + 419 \mathfrak{t}^{15} + 266 \mathfrak{t}^{16} + 138 \mathfrak{t}^{17} + 62 \mathfrak{t}^{18} + 21 \mathfrak{t}^{19} + 12 \mathfrak{t}^{20} + 3 \mathfrak{t}^{21})\nonumber \\
&\quad + c^3 \mathfrak{t}^{8} (-2 + 8 \mathfrak{t}^{2} + 16 \mathfrak{t}^{3} + 14 \mathfrak{t}^{4} - 39 \mathfrak{t}^{5} - 142 \mathfrak{t}^{6} - 359 \mathfrak{t}^{7} - 654 \mathfrak{t}^{8} - 1036 \mathfrak{t}^{9} - 1356 \mathfrak{t}^{10} - 1495 \mathfrak{t}^{11} - 1424 \mathfrak{t}^{12}\nonumber \\
&\quad\quad - 1186 \mathfrak{t}^{13} - 896 \mathfrak{t}^{14} - 591 \mathfrak{t}^{15} - 346 \mathfrak{t}^{16} - 162 \mathfrak{t}^{17} - 70 \mathfrak{t}^{18} - 27 \mathfrak{t}^{19} - 12 \mathfrak{t}^{20} - \mathfrak{t}^{21}) \nonumber \\
&\quad + c^4 \mathfrak{t}^{14} (-2 - 4 \mathfrak{t} + 3 \mathfrak{t}^{2} + 52 \mathfrak{t}^{3} + 193 \mathfrak{t}^{4} + 430 \mathfrak{t}^{5} + 763 \mathfrak{t}^{6} + 986 \mathfrak{t}^{7} + 1154 \mathfrak{t}^{8} + 1106 \mathfrak{t}^{9} + 987 \mathfrak{t}^{10} + 730 \mathfrak{t}^{11} + 483 \mathfrak{t}^{12}\nonumber \\
&\quad\quad + 248 \mathfrak{t}^{13} + 119 \mathfrak{t}^{14} + 48 \mathfrak{t}^{15} + 20 \mathfrak{t}^{16} + 4 \mathfrak{t}^{17}) \nonumber \\
&\quad + c^5 \mathfrak{t}^{19} (-1 - 20 \mathfrak{t} - 63 \mathfrak{t}^{2} - 158 \mathfrak{t}^{3} - 259 \mathfrak{t}^{4} - 366 \mathfrak{t}^{5} - 448 \mathfrak{t}^{6} - 474 \mathfrak{t}^{7} - 427 \mathfrak{t}^{8} - 322 \mathfrak{t}^{9} - 203 \mathfrak{t}^{10} - 104 \mathfrak{t}^{11} - 57 \mathfrak{t}^{12}\nonumber \\
&\quad\quad - 20 \mathfrak{t}^{13} - 6 \mathfrak{t}^{14}) \nonumber \\
&\quad + c^6 \mathfrak{t}^{23} (4 + 10 \mathfrak{t} + 28 \mathfrak{t}^{2} + 34 \mathfrak{t}^{3} + 60 \mathfrak{t}^{4} + 68 \mathfrak{t}^{5} + 80 \mathfrak{t}^{6} + 68 \mathfrak{t}^{7} + 60 \mathfrak{t}^{8} + 34 \mathfrak{t}^{9} + 28 \mathfrak{t}^{10} + 10 \mathfrak{t}^{11} + 4 \mathfrak{t}^{12}) 
).
\end{align}
}

In the limit of large representation, we have

{\fontsize{9pt}{1pt}\selectfont
\begin{align}
&\langle W_{(k)}\rangle^{U(3)\text{ADHM-}[1](H)}(\mathfrak{t})
=
\frac{
\mathfrak{t}^k(1 + \mathfrak{t} + 4 \mathfrak{t}^2 + 6 \mathfrak{t}^3 + 7 \mathfrak{t}^4 + 5 \mathfrak{t}^5 + 5 \mathfrak{t}^6 + \mathfrak{t}^7)
}{
(1 - \mathfrak{t})^3 (1 - \mathfrak{t}^2)^3 (1 - \mathfrak{t}^3)^2
}
+{\cal O}(\mathfrak{t}^{2k+1}), \\
&\langle W_{(k)}\rangle^{U(3)\text{ADHM-}[2](H)}(\mathfrak{t})\nonumber \\
&=
\frac{\mathfrak{t}^k}{
(1 - \mathfrak{t})^3 (1 - \mathfrak{t}^2)^4 (1 - \mathfrak{t}^3)^4 (1 - \mathfrak{t}^4)^3
}
(
1 - \mathfrak{t} - 8 \mathfrak{t}^{2} - 48 \mathfrak{t}^{3} - 179 \mathfrak{t}^{4} - 513 \mathfrak{t}^{5} - 1207 \mathfrak{t}^{6} - 2399 \mathfrak{t}^{7} - 4150 \mathfrak{t}^{8} - 6258 \mathfrak{t}^{9}\nonumber \\
&\quad\quad - 8401 \mathfrak{t}^{10} - 9957 \mathfrak{t}^{11} - 10598 \mathfrak{t}^{12} - 10006 \mathfrak{t}^{13} - 8482 \mathfrak{t}^{14} - 6334 \mathfrak{t}^{15} - 4218 \mathfrak{t}^{16} - 2422 \mathfrak{t}^{17} - 1220 \mathfrak{t}^{18} - 504 \mathfrak{t}^{19}\nonumber \\
&\quad\quad - 174 \mathfrak{t}^{20} - 42 \mathfrak{t}^{21} - 8 \mathfrak{t}^{22}\nonumber \\
&\quad + k (1 + 3 \mathfrak{t} + 14 \mathfrak{t}^{2} + 38 \mathfrak{t}^{3} + 97 \mathfrak{t}^{4} + 191 \mathfrak{t}^{5} + 339 \mathfrak{t}^{6} + 491 \mathfrak{t}^{7} + 624 \mathfrak{t}^{8} + 640 \mathfrak{t}^{9} + 523 \mathfrak{t}^{10} + 255 \mathfrak{t}^{11} - 86 \mathfrak{t}^{12} - 402 \mathfrak{t}^{13}\nonumber \\
&\quad\quad - 612 \mathfrak{t}^{14} - 656 \mathfrak{t}^{15} - 580 \mathfrak{t}^{16} - 416 \mathfrak{t}^{17} - 260 \mathfrak{t}^{18} - 128 \mathfrak{t}^{19} - 56 \mathfrak{t}^{20} - 16 \mathfrak{t}^{21} - 4 \mathfrak{t}^{22})
)+{\cal O}(\mathfrak{t}^{2k+2}).
\end{align}
}

\noindent \underline{Diagonal $2$-point functions}
{\fontsize{9pt}{1pt}\selectfont
\begin{align}
&\sum_{k=0}^\infty c^k\langle W_{(k)}W_{(\overline{k})}\rangle^{{\cal N}=8\text{ }U(3)(H)}(\mathfrak{t})
=
\frac{
1 + 2 \mathfrak{t}^2 + \mathfrak{t}^3 + c \mathfrak{t}^2 (1 - 2 \mathfrak{t}^2 - 3 \mathfrak{t}^3 - 3 \mathfrak{t}^4 - 2 \mathfrak{t}^5 + \mathfrak{t}^7) + c^2 \mathfrak{t}^8 (1 + 2 \mathfrak{t} + \mathfrak{t}^3)
}{
(1 - c) (1 - c \mathfrak{t})^2 (1 - c \mathfrak{t}^2)^2 (1 - \mathfrak{t})^2 (1 - \mathfrak{t}^3)
},\\
&\sum_{k=0}^\infty c^k\langle W_{(k)}W_{(\overline{k})}\rangle^{U(3)\text{ADHM-}[1](H)}(\mathfrak{t})\nonumber \\
&
=
\frac{1}{
(1 - c) (1 - c \mathfrak{t})^2 (1 - c \mathfrak{t}^2)^3 (1 - \mathfrak{t})^2 (1 - \mathfrak{t}^2)^2 (1 - \mathfrak{t}^3)^2
}(
1 + \mathfrak{t}^2 + 2 \mathfrak{t}^3 + \mathfrak{t}^4 + \mathfrak{t}^6\nonumber \\
&\quad + c \mathfrak{t}^2 (1 + 4 \mathfrak{t} + 4 \mathfrak{t}^2 + 4 \mathfrak{t}^3 + 5 \mathfrak{t}^4 + 2 \mathfrak{t}^5 - 2 \mathfrak{t}^6)\nonumber \\
&\quad + c^2 \mathfrak{t}^6 (2 - 2 \mathfrak{t} - 5 \mathfrak{t}^2 - 4 \mathfrak{t}^3 - 4 \mathfrak{t}^4 - 4 \mathfrak{t}^5 - \mathfrak{t}^6)\nonumber \\
&\quad + c^3 \mathfrak{t}^8 (-1 - \mathfrak{t}^2 - 2 \mathfrak{t}^3 - \mathfrak{t}^4 - \mathfrak{t}^6)
),\\
&\sum_{k=0}^\infty c^k\langle W_{(k)}W_{(\overline{k})}\rangle^{U(3)\text{ADHM-}[2](H)}(\mathfrak{t})\nonumber \\
&=
\frac{1}{
(1 - c) (1 - c \mathfrak{t})^2 (1 - c \mathfrak{t}^2)^5 (1 - c \mathfrak{t}^3)^3 (1 - \mathfrak{t})^2 (1 - \mathfrak{t}^2)^3 (1 - \mathfrak{t}^3)^4 (1 - \mathfrak{t}^4)^3
}
(
1 + 3 \mathfrak{t}^{2} + 6 \mathfrak{t}^{3} + 12 \mathfrak{t}^{4} + 16 \mathfrak{t}^{5} + 31 \mathfrak{t}^{6}\nonumber \\
&\quad\quad + 36 \mathfrak{t}^{7} + 55 \mathfrak{t}^{8} + 54 \mathfrak{t}^{9} + 60 \mathfrak{t}^{10} + 54 \mathfrak{t}^{11} + 55 \mathfrak{t}^{12} + 36 \mathfrak{t}^{13} + 31 \mathfrak{t}^{14} + 16 \mathfrak{t}^{15} + 12 \mathfrak{t}^{16} + 6 \mathfrak{t}^{17} + 3 \mathfrak{t}^{18} + \mathfrak{t}^{20} \nonumber \\
&\quad + c \mathfrak{t}^{2} (2 + 13 \mathfrak{t} + 30 \mathfrak{t}^{2} + 53 \mathfrak{t}^{3} + 96 \mathfrak{t}^{4} + 142 \mathfrak{t}^{5} + 156 \mathfrak{t}^{6} + 133 \mathfrak{t}^{7} + 59 \mathfrak{t}^{8} - 39 \mathfrak{t}^{9} - 139 \mathfrak{t}^{10} - 240 \mathfrak{t}^{11} - 303 \mathfrak{t}^{12} - 291 \mathfrak{t}^{13}\nonumber \\
&\quad\quad - 238 \mathfrak{t}^{14} - 175 \mathfrak{t}^{15} - 113 \mathfrak{t}^{16} - 70 \mathfrak{t}^{17} - 33 \mathfrak{t}^{18} - 11 \mathfrak{t}^{19} - 5 \mathfrak{t}^{20} - 3 \mathfrak{t}^{21}) \nonumber \\
&\quad + c^2 \mathfrak{t}^{4} (1 - 6 \mathfrak{t} - 5 \mathfrak{t}^{2} - 34 \mathfrak{t}^{3} - 128 \mathfrak{t}^{4} - 362 \mathfrak{t}^{5} - 662 \mathfrak{t}^{6} - 998 \mathfrak{t}^{7} - 1262 \mathfrak{t}^{8} - 1317 \mathfrak{t}^{9} - 1122 \mathfrak{t}^{10} - 625 \mathfrak{t}^{11} - 62 \mathfrak{t}^{12}\nonumber \\
&\quad\quad + 403 \mathfrak{t}^{13} + 569 \mathfrak{t}^{14} + 604 \mathfrak{t}^{15} + 473 \mathfrak{t}^{16} + 331 \mathfrak{t}^{17} + 177 \mathfrak{t}^{18} + 77 \mathfrak{t}^{19} + 26 \mathfrak{t}^{20} + 15 \mathfrak{t}^{21} + 3 \mathfrak{t}^{22})\nonumber \\
&\quad + c^3 \mathfrak{t}^{7} (-3 - 33 \mathfrak{t} - 83 \mathfrak{t}^{2} - 101 \mathfrak{t}^{3} + 8 \mathfrak{t}^{4} + 318 \mathfrak{t}^{5} + 912 \mathfrak{t}^{6} + 1822 \mathfrak{t}^{7} + 2848 \mathfrak{t}^{8} + 3611 \mathfrak{t}^{9} + 3720 \mathfrak{t}^{10} + 3092 \mathfrak{t}^{11}\nonumber \\
&\quad\quad + 2002 \mathfrak{t}^{12} + 862 \mathfrak{t}^{13} - 73 \mathfrak{t}^{14} - 610 \mathfrak{t}^{15} - 709 \mathfrak{t}^{16} - 536 \mathfrak{t}^{17} - 283 \mathfrak{t}^{18} - 115 \mathfrak{t}^{19} - 42 \mathfrak{t}^{20} - 14 \mathfrak{t}^{21} - \mathfrak{t}^{22}) \nonumber \\
&\quad + c^4 \mathfrak{t}^{10} (14 + 61 \mathfrak{t} + 131 \mathfrak{t}^{2} + 253 \mathfrak{t}^{3} + 350 \mathfrak{t}^{4} + 330 \mathfrak{t}^{5} - 172 \mathfrak{t}^{6} - 1216 \mathfrak{t}^{7} - 2750 \mathfrak{t}^{8} - 4165 \mathfrak{t}^{9} - 5036 \mathfrak{t}^{10} - 5036 \mathfrak{t}^{11}\nonumber \\
&\quad\quad - 4165 \mathfrak{t}^{12} - 2750 \mathfrak{t}^{13} - 1216 \mathfrak{t}^{14} - 172 \mathfrak{t}^{15} + 330 \mathfrak{t}^{16} + 350 \mathfrak{t}^{17} + 253 \mathfrak{t}^{18} + 131 \mathfrak{t}^{19} + 61 \mathfrak{t}^{20} + 14 \mathfrak{t}^{21}) \nonumber \\
&\quad + c^5 \mathfrak{t}^{12} (-1 - 14 \mathfrak{t} - 42 \mathfrak{t}^{2} - 115 \mathfrak{t}^{3} - 283 \mathfrak{t}^{4} - 536 \mathfrak{t}^{5} - 709 \mathfrak{t}^{6} - 610 \mathfrak{t}^{7} - 73 \mathfrak{t}^{8} + 862 \mathfrak{t}^{9} + 2002 \mathfrak{t}^{10} + 3092 \mathfrak{t}^{11}\nonumber \\
&\quad\quad + 3720 \mathfrak{t}^{12} + 3611 \mathfrak{t}^{13} + 2848 \mathfrak{t}^{14} + 1822 \mathfrak{t}^{15} + 912 \mathfrak{t}^{16} + 318 \mathfrak{t}^{17} + 8 \mathfrak{t}^{18} - 101 \mathfrak{t}^{19} - 83 \mathfrak{t}^{20} - 33 \mathfrak{t}^{21} - 3 \mathfrak{t}^{22}) \nonumber \\
&\quad + c^6 \mathfrak{t}^{15} (3 + 15 \mathfrak{t} + 26 \mathfrak{t}^{2} + 77 \mathfrak{t}^{3} + 177 \mathfrak{t}^{4} + 331 \mathfrak{t}^{5} + 473 \mathfrak{t}^{6} + 604 \mathfrak{t}^{7} + 569 \mathfrak{t}^{8} + 403 \mathfrak{t}^{9} - 62 \mathfrak{t}^{10} - 625 \mathfrak{t}^{11} - 1122 \mathfrak{t}^{12}\nonumber \\
&\quad\quad - 1317 \mathfrak{t}^{13} - 1262 \mathfrak{t}^{14} - 998 \mathfrak{t}^{15} - 662 \mathfrak{t}^{16} - 362 \mathfrak{t}^{17} - 128 \mathfrak{t}^{18} - 34 \mathfrak{t}^{19} - 5 \mathfrak{t}^{20} - 6 \mathfrak{t}^{21} + \mathfrak{t}^{22}) \nonumber \\
&\quad + c^7 \mathfrak{t}^{18} (-3 - 5 \mathfrak{t} - 11 \mathfrak{t}^{2} - 33 \mathfrak{t}^{3} - 70 \mathfrak{t}^{4} - 113 \mathfrak{t}^{5} - 175 \mathfrak{t}^{6} - 238 \mathfrak{t}^{7} - 291 \mathfrak{t}^{8} - 303 \mathfrak{t}^{9} - 240 \mathfrak{t}^{10} - 139 \mathfrak{t}^{11} - 39 \mathfrak{t}^{12}\nonumber \\
&\quad\quad + 59 \mathfrak{t}^{13} + 133 \mathfrak{t}^{14} + 156 \mathfrak{t}^{15} + 142 \mathfrak{t}^{16} + 96 \mathfrak{t}^{17} + 53 \mathfrak{t}^{18} + 30 \mathfrak{t}^{19} + 13 \mathfrak{t}^{20} + 2 \mathfrak{t}^{21}) \nonumber \\
&\quad + c^8 \mathfrak{t}^{21} (1 + 3 \mathfrak{t}^{2} + 6 \mathfrak{t}^{3} + 12 \mathfrak{t}^{4} + 16 \mathfrak{t}^{5} + 31 \mathfrak{t}^{6} + 36 \mathfrak{t}^{7} + 55 \mathfrak{t}^{8} + 54 \mathfrak{t}^{9} + 60 \mathfrak{t}^{10} + 54 \mathfrak{t}^{11} + 55 \mathfrak{t}^{12} + 36 \mathfrak{t}^{13} + 31 \mathfrak{t}^{14}\nonumber \\
&\quad\quad + 16 \mathfrak{t}^{15} + 12 \mathfrak{t}^{16} + 6 \mathfrak{t}^{17} + 3 \mathfrak{t}^{18} + \mathfrak{t}^{20})
).
\end{align}
}

In the limit of large representation, we have

{\fontsize{9pt}{1pt}\selectfont
\begin{align}
&\langle W_{(k)}W_{(\overline{k})}\rangle^{U(3)\text{ADHM-}[1](H)}(\mathfrak{t})
=
\frac{
1 + 3 \mathfrak{t}^2 + 6 \mathfrak{t}^3 + 8 \mathfrak{t}^4 + 10 \mathfrak{t}^5 + 16 \mathfrak{t}^6 + 10 \mathfrak{t}^7 + 8 \mathfrak{t}^8 + 6 \mathfrak{t}^9 + 3 \mathfrak{t}^{10} + \mathfrak{t}^{12}
}{
(1 - \mathfrak{t})^4 (1 - \mathfrak{t}^2)^4 (1 - \mathfrak{t}^3)^2
}+{\cal O}(\mathfrak{t}^{k+1}),\nonumber \\
&\langle W_{(k)}W_{(\overline{k})}\rangle^{U(3)\text{ADHM-}[2](H)}(\mathfrak{t})\nonumber \\
&=
\frac{1}{
(1 - \mathfrak{t})^3 (1 - \mathfrak{t}^2)^4 (1 - \mathfrak{t}^3)^5 (1 - \mathfrak{t}^4)^4}
(
1 + \mathfrak{t} + 10 \mathfrak{t}^{2} + 31 \mathfrak{t}^{3} + 103 \mathfrak{t}^{4} + 260 \mathfrak{t}^{5} + 656 \mathfrak{t}^{6} + 1364 \mathfrak{t}^{7} + 2699 \mathfrak{t}^{8} + 4647 \mathfrak{t}^{9}\nonumber \\
&\quad + 7508 \mathfrak{t}^{10} + 10827 \mathfrak{t}^{11} + 14615 \mathfrak{t}^{12} + 17796 \mathfrak{t}^{13} + 20280 \mathfrak{t}^{14} + 20964 \mathfrak{t}^{15}
+ \cdots + \mathfrak{t}^{30}
)+{\cal O}(\mathfrak{t}^{k+1}).
\end{align}
}





\subsubsection{$U(4)$}
\noindent \underline{1-point functions}
{\fontsize{9pt}{1pt}\selectfont
\begin{align}
&\sum_{k=0}^\infty c^k\langle W_{(k)}\rangle^{U(4)\text{ ADHM-}[1](H)}(\mathfrak{t})\nonumber \\
&=
\frac{1}{
(1 - c \mathfrak{t}) (1 - c \mathfrak{t}^2)^2 (1 - c \mathfrak{t}^3)^3 (1 - c \mathfrak{t}^4)^2 (1 - \mathfrak{t})^2 (1 - \mathfrak{t}^2)^2  (1 - \mathfrak{t}^3)^2 (1 - \mathfrak{t}^4)^2
}
(
1 + \mathfrak{t}^{2} + 2 \mathfrak{t}^{3} + 4 \mathfrak{t}^{4} + 2 \mathfrak{t}^{5} + 4 \mathfrak{t}^{6} + 2 \mathfrak{t}^{7}\nonumber \\
&\quad\quad + 4 \mathfrak{t}^{8} + 2 \mathfrak{t}^{9} + \mathfrak{t}^{10} + \mathfrak{t}^{12} \nonumber \\
&\quad + c \mathfrak{t}^{4} (2 - 4 \mathfrak{t}^{2} - 8 \mathfrak{t}^{3} - 10 \mathfrak{t}^{4} - 13 \mathfrak{t}^{5} - 16 \mathfrak{t}^{6} - 17 \mathfrak{t}^{7} - 14 \mathfrak{t}^{8} - 7 \mathfrak{t}^{9} - 4 \mathfrak{t}^{10} - 3 \mathfrak{t}^{11} - 2 \mathfrak{t}^{12}) \nonumber \\
&\quad + c^{2} \mathfrak{t}^{7} (-2 - \mathfrak{t}^{3} + 4 \mathfrak{t}^{4} + 8 \mathfrak{t}^{5} + 26 \mathfrak{t}^{6} + 29 \mathfrak{t}^{7} + 32 \mathfrak{t}^{8} + 19 \mathfrak{t}^{9} + 14 \mathfrak{t}^{10} + 8 \mathfrak{t}^{11} + 6 \mathfrak{t}^{12} + \mathfrak{t}^{13})\nonumber \\
&\quad + c^{3} \mathfrak{t}^{11} (-1 + 3 \mathfrak{t}^{2} + 4 \mathfrak{t}^{3} + 2 \mathfrak{t}^{4} - 8 \mathfrak{t}^{5} - 15 \mathfrak{t}^{6} - 20 \mathfrak{t}^{7} - 18 \mathfrak{t}^{8} - 16 \mathfrak{t}^{9} - 16 \mathfrak{t}^{10} - 8 \mathfrak{t}^{11} - 3 \mathfrak{t}^{12})\nonumber \\
&\quad + c^{4} \mathfrak{t}^{22} (3 + 4 \mathfrak{t} + 10 \mathfrak{t}^{2} + 4 \mathfrak{t}^{3} + 3 \mathfrak{t}^{4}) 
).
\end{align}
}

In the limit of large representation, we have
{\fontsize{9pt}{1pt}\selectfont
\begin{align}
&\langle W_{(k)}\rangle^{U(4)\text{ADHM-}[1](H)}(\mathfrak{t})\nonumber \\
&=
\frac{\mathfrak{t}^k
(1 + \mathfrak{t} + 4 \mathfrak{t}^{2} + 9 \mathfrak{t}^{3} + 18 \mathfrak{t}^{4} + 25 \mathfrak{t}^{5} + 41 \mathfrak{t}^{6} + 46 \mathfrak{t}^{7} + 55 \mathfrak{t}^{8} + 49 \mathfrak{t}^{9} + 42 \mathfrak{t}^{10} + 27 \mathfrak{t}^{11} + 20 \mathfrak{t}^{12} + 7 \mathfrak{t}^{13} + 3 \mathfrak{t}^{14})
}{
(1 - \mathfrak{t})^3 (1 - \mathfrak{t}^2)^3 (1 - \mathfrak{t}^3)^3 (1 - \mathfrak{t}^4)^2
}\nonumber \\
&\quad +{\cal O}(\mathfrak{t}^{2k+1}).
\end{align}
}

\noindent \underline{Diagonal $2$-point functions}
{\fontsize{9pt}{1pt}\selectfont
\begin{align}
&\sum_{k=0}^\infty c^k\langle W_{(k)}W_{(\overline{k})}\rangle^{{\cal N}=8\text{ }U(4)(H)}(\mathfrak{t})\nonumber \\
&=
\frac{1}{
(1 - c) (1 - c \mathfrak{t})^2 (1 - c \mathfrak{t}^2)^3 (1 - c \mathfrak{t}^3)^2 (1 - \mathfrak{t}) (1 - \mathfrak{t}^2) (1 - \mathfrak{t}^3) (1 - \mathfrak{t}^4)
}
(
1 + \mathfrak{t} + 2 \mathfrak{t}^{2} + 3 \mathfrak{t}^{3} + 5 \mathfrak{t}^{4} + 2 \mathfrak{t}^{5} + 2 \mathfrak{t}^{6}\nonumber \\
&\quad + c \mathfrak{t}^{3} (2 + 2 \mathfrak{t} - 2 \mathfrak{t}^{2} - 9 \mathfrak{t}^{3} - 15 \mathfrak{t}^{4} - 20 \mathfrak{t}^{5} - 19 \mathfrak{t}^{6} - 11 \mathfrak{t}^{7} - 2 \mathfrak{t}^{8} + 5 \mathfrak{t}^{9} + 5 \mathfrak{t}^{10} + 2 \mathfrak{t}^{11} - \mathfrak{t}^{12} - \mathfrak{t}^{13})\nonumber \\
&\quad + c^{2} \mathfrak{t}^{5} (-2 - 2 \mathfrak{t} - 2 \mathfrak{t}^{2} - 2 \mathfrak{t}^{3} + 4 \mathfrak{t}^{4} + 12 \mathfrak{t}^{5} + 26 \mathfrak{t}^{6} + 28 \mathfrak{t}^{7} + 26 \mathfrak{t}^{8} + 12 \mathfrak{t}^{9} + 4 \mathfrak{t}^{10} - 2 \mathfrak{t}^{11} - 2 \mathfrak{t}^{12} - 2 \mathfrak{t}^{13} - 2 \mathfrak{t}^{14})\nonumber \\
&\quad + c^{3} \mathfrak{t}^{8} (-1 - \mathfrak{t} + 2 \mathfrak{t}^{2} + 5 \mathfrak{t}^{3} + 5 \mathfrak{t}^{4} - 2 \mathfrak{t}^{5} - 11 \mathfrak{t}^{6} - 19 \mathfrak{t}^{7} - 20 \mathfrak{t}^{8} - 15 \mathfrak{t}^{9} - 9 \mathfrak{t}^{10} - 2 \mathfrak{t}^{11} + 2 \mathfrak{t}^{12} + 2 \mathfrak{t}^{13})\nonumber \\
&\quad + c^{4} \mathfrak{t}^{18} (2 + 2 \mathfrak{t} + 5 \mathfrak{t}^{2} + 3 \mathfrak{t}^{3} + 2 \mathfrak{t}^{4} + \mathfrak{t}^{5} + \mathfrak{t}^{6})
),\\
&\sum_{k=0}^\infty c^k\langle W_{(k)}W_{(\overline{k})}\rangle^{U(4)\text{ADHM-}[1](H)}(\mathfrak{t})\nonumber \\
&=
\frac{1}{
(1 - c) (1 - c \mathfrak{t})^2 (1 - c \mathfrak{t}^2)^4 (1 - c \mathfrak{t}^3)^3 (1 - \mathfrak{t})^2 (1 - \mathfrak{t}^2)^2 (1 - \mathfrak{t}^3)^2 (1 - \mathfrak{t}^4)^2
}
(
1 + \mathfrak{t}^{2} + 2 \mathfrak{t}^{3} + 4 \mathfrak{t}^{4} + 2 \mathfrak{t}^{5} + 4 \mathfrak{t}^{6} + 2 \mathfrak{t}^{7}\nonumber \\
&\quad\quad + 4 \mathfrak{t}^{8} + 2 \mathfrak{t}^{9} + \mathfrak{t}^{10} + \mathfrak{t}^{12} \nonumber \\
&\quad+ c \mathfrak{t}^3 (5 + 10 \mathfrak{t} + 15 \mathfrak{t}^{2} + 21 \mathfrak{t}^{3} + 30 \mathfrak{t}^{4} + 31 \mathfrak{t}^{5} + 24 \mathfrak{t}^{6} + 12 \mathfrak{t}^{7} + 2 \mathfrak{t}^{8} + \mathfrak{t}^{9} - \mathfrak{t}^{10} - 3 \mathfrak{t}^{11} - 3 \mathfrak{t}^{12}) \nonumber \\
&\quad+ c^2 \mathfrak{t}^{5} (-2 + \mathfrak{t} + 6 \mathfrak{t}^{2} + 9 \mathfrak{t}^{3} - 3 \mathfrak{t}^{4} - 25 \mathfrak{t}^{5} - 45 \mathfrak{t}^{6} - 65 \mathfrak{t}^{7} - 74 \mathfrak{t}^{8} - 77 \mathfrak{t}^{9} - 57 \mathfrak{t}^{10} - 29 \mathfrak{t}^{11} - 3 \mathfrak{t}^{12} + \mathfrak{t}^{13} + 2 \mathfrak{t}^{14} + \mathfrak{t}^{15}) \nonumber \\
&\quad+ c^3 \mathfrak{t}^{8} (-5 - 13 \mathfrak{t} - 25 \mathfrak{t}^{2} - 31 \mathfrak{t}^{3} - 33 \mathfrak{t}^{4} - 39 \mathfrak{t}^{5} - 35 \mathfrak{t}^{6} - 15 \mathfrak{t}^{7} + 15 \mathfrak{t}^{8} + 35 \mathfrak{t}^{9} + 39 \mathfrak{t}^{10} + 33 \mathfrak{t}^{11} + 31 \mathfrak{t}^{12} + 25 \mathfrak{t}^{13}\nonumber \\
&\quad\quad + 13 \mathfrak{t}^{14} + 5 \mathfrak{t}^{15})\nonumber \\
&\quad+ c^4 \mathfrak{t}^{11} (-1 - 2 \mathfrak{t} - \mathfrak{t}^{2} + 3 \mathfrak{t}^{3} + 29 \mathfrak{t}^{4} + 57 \mathfrak{t}^{5} + 77 \mathfrak{t}^{6} + 74 \mathfrak{t}^{7} + 65 \mathfrak{t}^{8} + 45 \mathfrak{t}^{9} + 25 \mathfrak{t}^{10} + 3 \mathfrak{t}^{11} - 9 \mathfrak{t}^{12} - 6 \mathfrak{t}^{13} - \mathfrak{t}^{14} + 2 \mathfrak{t}^{15}) \nonumber \\
&\quad+ c^5 \mathfrak{t}^{16} (3 + 3 \mathfrak{t} + \mathfrak{t}^{2} - \mathfrak{t}^{3} - 2 \mathfrak{t}^{4} - 12 \mathfrak{t}^{5} - 24 \mathfrak{t}^{6} - 31 \mathfrak{t}^{7} - 30 \mathfrak{t}^{8} - 21 \mathfrak{t}^{9} - 15 \mathfrak{t}^{10} - 10 \mathfrak{t}^{11} - 5 \mathfrak{t}^{12}) \nonumber \\
&\quad+ c^6 \mathfrak{t}^{19} (-1 - \mathfrak{t}^{2} - 2 \mathfrak{t}^{3} - 4 \mathfrak{t}^{4} - 2 \mathfrak{t}^{5} - 4 \mathfrak{t}^{6} - 2 \mathfrak{t}^{7} - 4 \mathfrak{t}^{8} - 2 \mathfrak{t}^{9} - \mathfrak{t}^{10} - \mathfrak{t}^{12}) 
).
\end{align}
}

In the limit of large representation, we have

{\fontsize{9pt}{1pt}\selectfont
\begin{align}
&\langle W_{(k)}W_{(\overline{k})}\rangle^{U(4)\text{ADHM-}[1](H)}(\mathfrak{t})\nonumber \\
&=
\frac{
1 + 3 \mathfrak{t}^{2} + 8 \mathfrak{t}^{3} + 19 \mathfrak{t}^{4} + 32 \mathfrak{t}^{5} + 69 \mathfrak{t}^{6} + 108 \mathfrak{t}^{7} + 174 \mathfrak{t}^{8} + 228 \mathfrak{t}^{9} + 294 \mathfrak{t}^{10} + 328 \mathfrak{t}^{11} + 360 \mathfrak{t}^{12} + \cdots + \mathfrak{t}^{24}
}{
(1 - \mathfrak{t})^4 (1 - \mathfrak{t}^2)^4 (1 - \mathfrak{t}^3)^4 (1 - \mathfrak{t}^4)^2
}
+{\cal O}(\mathfrak{t}^{k+1}).
\end{align}
}

\subsubsection{$U(5)$}
\noindent \underline{1-point functions}
{\fontsize{9pt}{1pt}\selectfont
\begin{align}
&\sum_{k=0}^\infty c^k\langle W_{(k)}\rangle^{U(5)\text{ ADHM-}[1](H)}(\mathfrak{t})\nonumber \\
&=
\frac{1}{
(1 - c \mathfrak{t}) (1 - c \mathfrak{t}^2)^2 (1 - c \mathfrak{t}^3)^3 (1 - c \mathfrak{t}^4)^2 (1 - c \mathfrak{t}^5)^2 (1 - \mathfrak{t})^2 (1 - \mathfrak{t}^2)^2 (1 - \mathfrak{t}^3)^2
 (1 - \mathfrak{t}^4)^2 (1 - \mathfrak{t}^5)^2
}
(1 + \mathfrak{t}^{2} + 2 \mathfrak{t}^{3} + 4 \mathfrak{t}^{4}\nonumber \\
&\quad\quad + 6 \mathfrak{t}^{5} + 7 \mathfrak{t}^{6} + 8 \mathfrak{t}^{7} + 12 \mathfrak{t}^{8} + 12 \mathfrak{t}^{9} + 14 \mathfrak{t}^{10} + 12 \mathfrak{t}^{11} + 12 \mathfrak{t}^{12} + 8 \mathfrak{t}^{13} + 7 \mathfrak{t}^{14} + 6 \mathfrak{t}^{15} + 4 \mathfrak{t}^{16} + 2 \mathfrak{t}^{17} + \mathfrak{t}^{18} + \mathfrak{t}^{20}\nonumber \\
&\quad + c \mathfrak{t}^{4} (2 + 3 \mathfrak{t} + 2 \mathfrak{t}^{2} - \mathfrak{t}^{3} - 4 \mathfrak{t}^{4} - 9 \mathfrak{t}^{5} - 20 \mathfrak{t}^{6} - 34 \mathfrak{t}^{7} - 50 \mathfrak{t}^{8} - 66 \mathfrak{t}^{9} - 72 \mathfrak{t}^{10} - 75 \mathfrak{t}^{11} - 68 \mathfrak{t}^{12} - 61 \mathfrak{t}^{13} - 50 \mathfrak{t}^{14} - 39 \mathfrak{t}^{15}\nonumber \\
&\quad\quad - 26 \mathfrak{t}^{16} - 15 \mathfrak{t}^{17} - 8 \mathfrak{t}^{18} - 5 \mathfrak{t}^{19} - 2 \mathfrak{t}^{20} - 2 \mathfrak{t}^{21})\nonumber \\
&\quad + c^2 \mathfrak{t}^{11} (-8 - 15 \mathfrak{t} - 14 \mathfrak{t}^{2} - 7 \mathfrak{t}^{3} + 22 \mathfrak{t}^{4} + 58 \mathfrak{t}^{5} + 102 \mathfrak{t}^{6} + 134 \mathfrak{t}^{7} + 162 \mathfrak{t}^{8} + 169 \mathfrak{t}^{9} + 164 \mathfrak{t}^{10} + 142 \mathfrak{t}^{11} + 112 \mathfrak{t}^{12} + 76 \mathfrak{t}^{13}\nonumber \\
&\quad\quad + 48 \mathfrak{t}^{14} + 26 \mathfrak{t}^{15} + 16 \mathfrak{t}^{16} + 8 \mathfrak{t}^{17} + 4 \mathfrak{t}^{18} + \mathfrak{t}^{19}) \nonumber \\
&\quad + c^3 \mathfrak{t}^{13} (-3 - 2 \mathfrak{t} + \mathfrak{t}^{2} + 6 \mathfrak{t}^{3} + 13 \mathfrak{t}^{4} + 18 \mathfrak{t}^{5} + 19 \mathfrak{t}^{6} + 4 \mathfrak{t}^{7} - 22 \mathfrak{t}^{8} - 70 \mathfrak{t}^{9} - 123 \mathfrak{t}^{10} - 176 \mathfrak{t}^{11} - 200 \mathfrak{t}^{12} - 198 \mathfrak{t}^{13} - 167 \mathfrak{t}^{14}\nonumber \\
&\quad\quad - 122 \mathfrak{t}^{15} - 79 \mathfrak{t}^{16} - 50 \mathfrak{t}^{17} - 26 \mathfrak{t}^{18} - 16 \mathfrak{t}^{19} - 5 \mathfrak{t}^{20} - 2 \mathfrak{t}^{21}) \nonumber \\
&\quad + c^4 \mathfrak{t}^{18} (-1 + 3 \mathfrak{t}^{2} + 4 \mathfrak{t}^{3} + 2 \mathfrak{t}^{4} - 6 \mathfrak{t}^{5} - 15 \mathfrak{t}^{6} - 20 \mathfrak{t}^{7} - 9 \mathfrak{t}^{8} + 18 \mathfrak{t}^{9} + 54 \mathfrak{t}^{10} + 86 \mathfrak{t}^{11} + 107 \mathfrak{t}^{12} + 106 \mathfrak{t}^{13} + 94 \mathfrak{t}^{14} + 72 \mathfrak{t}^{15}\nonumber \\
&\quad\quad + 54 \mathfrak{t}^{16} + 30 \mathfrak{t}^{17} + 14 \mathfrak{t}^{18} + 6 \mathfrak{t}^{19} + \mathfrak{t}^{20}) \nonumber \\
&\quad + c^5 \mathfrak{t}^{33} (-3 - 8 \mathfrak{t} - 15 \mathfrak{t}^{2} - 20 \mathfrak{t}^{3} - 28 \mathfrak{t}^{4} - 20 \mathfrak{t}^{5} - 15 \mathfrak{t}^{6} - 8 \mathfrak{t}^{7} - 3 \mathfrak{t}^{8}) 
).
\end{align}
}

In the limit of large representation, we have
{\fontsize{9pt}{1pt}\selectfont
\begin{align}
&\langle W_{(k)}\rangle^{U(5)\text{ADHM-}[1](H)}(\mathfrak{t})\nonumber \\
&=
\frac{\mathfrak{t}^k}{(1 - \mathfrak{t})^3 (1 - \mathfrak{t}^2)^3 (1 - \mathfrak{t}^3)^3 (1 - \mathfrak{t}^4)^3 (1 - \mathfrak{t}^5)^2}
(
1 + \mathfrak{t} + 4 \mathfrak{t}^{2} + 9 \mathfrak{t}^{3} + 22 \mathfrak{t}^{4} + 41 \mathfrak{t}^{5} + 76 \mathfrak{t}^{6} + 125 \mathfrak{t}^{7} + 202 \mathfrak{t}^{8} + 285 \mathfrak{t}^{9}\nonumber \\
&\quad + 389 \mathfrak{t}^{10} + 480 \mathfrak{t}^{11} + 566 \mathfrak{t}^{12} + 606 \mathfrak{t}^{13} + 619 \mathfrak{t}^{14} + 575 \mathfrak{t}^{15} + 506 \mathfrak{t}^{16} + 401 \mathfrak{t}^{17} + 299 \mathfrak{t}^{18} + 198 \mathfrak{t}^{19} + 126 \mathfrak{t}^{20} + 64 \mathfrak{t}^{21}\nonumber \\
&\quad + 31 \mathfrak{t}^{22} + 11 \mathfrak{t}^{23} + 3 \mathfrak{t}^{24}
)+{\cal O}(\mathfrak{t}^{2k+1}).
\end{align}
}

\noindent \underline{Diagonal $2$-point functions}
{\fontsize{9pt}{1pt}\selectfont
\begin{align}
&\sum_{k=0}^\infty c^k\langle W_{(k)}W_{(\overline{k})}\rangle^{{\cal N}=8\text{ }U(5)(H)}(\mathfrak{t})\nonumber \\
&=
\frac{1}{
(1 - c)
(1 - c \mathfrak{t})^2
(1 - c \mathfrak{t}^2)^3
(1 - c \mathfrak{t}^3)^2
(1 - c \mathfrak{t}^4)^2
(1 - \mathfrak{t})^2
(1 - \mathfrak{t}^3)
(1 - \mathfrak{t}^4)
(1 - \mathfrak{t}^5)
}
(
1 + 2 \mathfrak{t}^{2} + \mathfrak{t}^{3} + 4 \mathfrak{t}^{4} + 3 \mathfrak{t}^{5} + 2 \mathfrak{t}^{6}\nonumber \\
&\quad\quad + 2 \mathfrak{t}^{7} + \mathfrak{t}^{8}\nonumber \\
&\quad + c \mathfrak{t}^{3} (2 + 3 \mathfrak{t} + 4 \mathfrak{t}^{2} - 3 \mathfrak{t}^{4} - 11 \mathfrak{t}^{5} - 21 \mathfrak{t}^{6} - 26 \mathfrak{t}^{7} - 25 \mathfrak{t}^{8} - 17 \mathfrak{t}^{9} - 5 \mathfrak{t}^{10} + 5 \mathfrak{t}^{11} + 10 \mathfrak{t}^{12} + 7 \mathfrak{t}^{13} + 3 \mathfrak{t}^{14} - 2 \mathfrak{t}^{15} - 3 \mathfrak{t}^{16}\nonumber \\
&\quad\quad - 2 \mathfrak{t}^{17} + \mathfrak{t}^{19})\nonumber \\
&\quad + c^{2} \mathfrak{t}^{9} (-8 - 11 \mathfrak{t} - 12 \mathfrak{t}^{2} - 3 \mathfrak{t}^{3} + 19 \mathfrak{t}^{4} + 39 \mathfrak{t}^{5} + 60 \mathfrak{t}^{6} + 56 \mathfrak{t}^{7} + 46 \mathfrak{t}^{8} + 18 \mathfrak{t}^{9} - \mathfrak{t}^{10} - 17 \mathfrak{t}^{11} - 17 \mathfrak{t}^{12} - 11 \mathfrak{t}^{13} - 5 \mathfrak{t}^{14}\nonumber \\
&\quad\quad + 2 \mathfrak{t}^{15} + 2 \mathfrak{t}^{16} + 3 \mathfrak{t}^{17})\nonumber \\
&\quad + c^{3} \mathfrak{t}^{10} (-3 - 2 \mathfrak{t} - 2 \mathfrak{t}^{2} + 5 \mathfrak{t}^{3} + 11 \mathfrak{t}^{4} + 17 \mathfrak{t}^{5} + 17 \mathfrak{t}^{6} + \mathfrak{t}^{7} - 18 \mathfrak{t}^{8} - 46 \mathfrak{t}^{9} - 56 \mathfrak{t}^{10} - 60 \mathfrak{t}^{11} - 39 \mathfrak{t}^{12} - 19 \mathfrak{t}^{13} + 3 \mathfrak{t}^{14}\nonumber \\
&\quad\quad + 12 \mathfrak{t}^{15} + 11 \mathfrak{t}^{16} + 8 \mathfrak{t}^{17})\nonumber \\
&\quad + c^{4} \mathfrak{t}^{14} (-1 + 2 \mathfrak{t}^{2} + 3 \mathfrak{t}^{3} + 2 \mathfrak{t}^{4} - 3 \mathfrak{t}^{5} - 7 \mathfrak{t}^{6} - 10 \mathfrak{t}^{7} - 5 \mathfrak{t}^{8} + 5 \mathfrak{t}^{9} + 17 \mathfrak{t}^{10} + 25 \mathfrak{t}^{11} + 26 \mathfrak{t}^{12} + 21 \mathfrak{t}^{13} + 11 \mathfrak{t}^{14} + 3 \mathfrak{t}^{15}\nonumber \\
&\quad\quad - 4 \mathfrak{t}^{17} - 3 \mathfrak{t}^{18} - 2 \mathfrak{t}^{19})\nonumber \\
&\quad + c^{5} \mathfrak{t}^{28} (-1 - 2 \mathfrak{t} - 2 \mathfrak{t}^{2} - 3 \mathfrak{t}^{3} - 4 \mathfrak{t}^{4} - \mathfrak{t}^{5} - 2 \mathfrak{t}^{6} - \mathfrak{t}^{8})
), \\
&\sum_{k=0}^\infty c^k\langle W_{(k)}W_{(\overline{k})}\rangle^{U(5)\text{ADHM-}[1](H)}(\mathfrak{t})\nonumber \\
&=
\frac{
1}{
(1 - c) (1 - c \mathfrak{t})^2 (1 - c \mathfrak{t}^2)^4 (1 - c \mathfrak{t}^3)^4 (1 - c \mathfrak{t}^4)^3 (1 - \mathfrak{t})^2 (1 - \mathfrak{t}^2)^2 (1 - \mathfrak{t}^3)^2
(1 - \mathfrak{t}^4)^2 (1 - \mathfrak{t}^5)^2
}
(
1 + \mathfrak{t}^{2} + 2 \mathfrak{t}^{3} + 4 \mathfrak{t}^{4}\nonumber \\
&\quad\quad + 6 \mathfrak{t}^{5} + 7 \mathfrak{t}^{6} + 8 \mathfrak{t}^{7} + 12 \mathfrak{t}^{8} + 12 \mathfrak{t}^{9} + 14 \mathfrak{t}^{10} + 12 \mathfrak{t}^{11} + 12 \mathfrak{t}^{12} + 8 \mathfrak{t}^{13} + 7 \mathfrak{t}^{14} + 6 \mathfrak{t}^{15} + 4 \mathfrak{t}^{16} + 2 \mathfrak{t}^{17} + \mathfrak{t}^{18} + \mathfrak{t}^{20}\nonumber \\
&\quad + c \mathfrak{t}^{3} (4 + 12 \mathfrak{t} + 24 \mathfrak{t}^{2} + 42 \mathfrak{t}^{3} + 70 \mathfrak{t}^{4} + 108 \mathfrak{t}^{5} + 148 \mathfrak{t}^{6} + 178 \mathfrak{t}^{7} + 190 \mathfrak{t}^{8} + 182 \mathfrak{t}^{9} + 160 \mathfrak{t}^{10} + 122 \mathfrak{t}^{11} + 84 \mathfrak{t}^{12} + 44 \mathfrak{t}^{13}\nonumber \\
&\quad\quad + 12 \mathfrak{t}^{14} - 10 \mathfrak{t}^{15} - 16 \mathfrak{t}^{16} - 13 \mathfrak{t}^{17} - 8 \mathfrak{t}^{18} - 6 \mathfrak{t}^{19} - 4 \mathfrak{t}^{20} - 3 \mathfrak{t}^{21}) \nonumber \\
&\quad + c^{2} \mathfrak{t}^{6} (3 + 16 \mathfrak{t} + 54 \mathfrak{t}^{2} + 100 \mathfrak{t}^{3} + 156 \mathfrak{t}^{4} + 180 \mathfrak{t}^{5} + 166 \mathfrak{t}^{6} + 76 \mathfrak{t}^{7} - 88 \mathfrak{t}^{8} - 326 \mathfrak{t}^{9} - 579 \mathfrak{t}^{10} - 782 \mathfrak{t}^{11} - 868 \mathfrak{t}^{12}\nonumber \\
&\quad\quad - 816 \mathfrak{t}^{13} - 647 \mathfrak{t}^{14} - 434 \mathfrak{t}^{15} - 238 \mathfrak{t}^{16} - 98 \mathfrak{t}^{17} - 18 \mathfrak{t}^{18} + 20 \mathfrak{t}^{19} + 23 \mathfrak{t}^{20} + 16 \mathfrak{t}^{21} + 4 \mathfrak{t}^{22}) \nonumber \\
&\quad + c^{3} \mathfrak{t}^{9} (-2 - 18 \mathfrak{t} - 46 \mathfrak{t}^{2} - 99 \mathfrak{t}^{3} - 198 \mathfrak{t}^{4} - 373 \mathfrak{t}^{5} - 628 \mathfrak{t}^{6} - 955 \mathfrak{t}^{7} - 1240 \mathfrak{t}^{8} - 1375 \mathfrak{t}^{9} - 1272 \mathfrak{t}^{10} - 906 \mathfrak{t}^{11} - 344 \mathfrak{t}^{12}\nonumber \\
&\quad\quad + 266 \mathfrak{t}^{13} + 800 \mathfrak{t}^{14} + 1127 \mathfrak{t}^{15} + 1208 \mathfrak{t}^{16} + 1052 \mathfrak{t}^{17} + 758 \mathfrak{t}^{18} + 432 \mathfrak{t}^{19} + 172 \mathfrak{t}^{20} + 32 \mathfrak{t}^{21} - 24 \mathfrak{t}^{22} - 27 \mathfrak{t}^{23} - 16 \mathfrak{t}^{24}\nonumber \\
&\quad\quad - 4 \mathfrak{t}^{25}) \nonumber \\
&\quad + c^{4} \mathfrak{t}^{13} (-8 - 47 \mathfrak{t} - 112 \mathfrak{t}^{2} - 191 \mathfrak{t}^{3} - 216 \mathfrak{t}^{4} - 94 \mathfrak{t}^{5} + 256 \mathfrak{t}^{6} + 846 \mathfrak{t}^{7} + 1622 \mathfrak{t}^{8} + 2420 \mathfrak{t}^{9} + 3058 \mathfrak{t}^{10} + 3305 \mathfrak{t}^{11}\nonumber \\
&\quad\quad + 3082 \mathfrak{t}^{12} + 2376 \mathfrak{t}^{13} + 1422 \mathfrak{t}^{14} + 450 \mathfrak{t}^{15} - 266 \mathfrak{t}^{16} - 642 \mathfrak{t}^{17} - 702 \mathfrak{t}^{18} - 581 \mathfrak{t}^{19} - 404 \mathfrak{t}^{20} - 245 \mathfrak{t}^{21} - 126 \mathfrak{t}^{22}\nonumber \\
&\quad\quad - 57 \mathfrak{t}^{23} - 22 \mathfrak{t}^{24} - 4 \mathfrak{t}^{25}) \nonumber \\
&\quad + c^{5} \mathfrak{t}^{16} (4 + 22 \mathfrak{t} + 57 \mathfrak{t}^{2} + 126 \mathfrak{t}^{3} + 245 \mathfrak{t}^{4} + 404 \mathfrak{t}^{5} + 581 \mathfrak{t}^{6} + 702 \mathfrak{t}^{7} + 642 \mathfrak{t}^{8} + 266 \mathfrak{t}^{9} - 450 \mathfrak{t}^{10} - 1422 \mathfrak{t}^{11} - 2376 \mathfrak{t}^{12}\nonumber \\
&\quad\quad - 3082 \mathfrak{t}^{13} - 3305 \mathfrak{t}^{14} - 3058 \mathfrak{t}^{15} - 2420 \mathfrak{t}^{16} - 1622 \mathfrak{t}^{17} - 846 \mathfrak{t}^{18} - 256 \mathfrak{t}^{19} + 94 \mathfrak{t}^{20} + 216 \mathfrak{t}^{21} + 191 \mathfrak{t}^{22} + 112 \mathfrak{t}^{23}\nonumber \\
&\quad\quad + 47 \mathfrak{t}^{24} + 8 \mathfrak{t}^{25})\nonumber \\
&\quad + c^{6} \mathfrak{t}^{20} (4 + 16 \mathfrak{t} + 27 \mathfrak{t}^{2} + 24 \mathfrak{t}^{3} - 32 \mathfrak{t}^{4} - 172 \mathfrak{t}^{5} - 432 \mathfrak{t}^{6} - 758 \mathfrak{t}^{7} - 1052 \mathfrak{t}^{8} - 1208 \mathfrak{t}^{9} - 1127 \mathfrak{t}^{10} - 800 \mathfrak{t}^{11} - 266 \mathfrak{t}^{12}\nonumber \\
&\quad\quad + 344 \mathfrak{t}^{13} + 906 \mathfrak{t}^{14} + 1272 \mathfrak{t}^{15} + 1375 \mathfrak{t}^{16} + 1240 \mathfrak{t}^{17} + 955 \mathfrak{t}^{18} + 628 \mathfrak{t}^{19} + 373 \mathfrak{t}^{20} + 198 \mathfrak{t}^{21} + 99 \mathfrak{t}^{22} + 46 \mathfrak{t}^{23}\nonumber \\
&\quad\quad + 18 \mathfrak{t}^{24} + 2 \mathfrak{t}^{25}) \nonumber \\
&\quad + c^{7} \mathfrak{t}^{26} (-4 - 16 \mathfrak{t} - 23 \mathfrak{t}^{2} - 20 \mathfrak{t}^{3} + 18 \mathfrak{t}^{4} + 98 \mathfrak{t}^{5} + 238 \mathfrak{t}^{6} + 434 \mathfrak{t}^{7} + 647 \mathfrak{t}^{8} + 816 \mathfrak{t}^{9} + 868 \mathfrak{t}^{10} + 782 \mathfrak{t}^{11} + 579 \mathfrak{t}^{12}\nonumber \\
&\quad\quad + 326 \mathfrak{t}^{13} + 88 \mathfrak{t}^{14} - 76 \mathfrak{t}^{15} - 166 \mathfrak{t}^{16} - 180 \mathfrak{t}^{17} - 156 \mathfrak{t}^{18} - 100 \mathfrak{t}^{19} - 54 \mathfrak{t}^{20} - 16 \mathfrak{t}^{21} - 3 \mathfrak{t}^{22}) \nonumber \\
&\quad + c^{8} \mathfrak{t}^{30} (3 + 4 \mathfrak{t} + 6 \mathfrak{t}^{2} + 8 \mathfrak{t}^{3} + 13 \mathfrak{t}^{4} + 16 \mathfrak{t}^{5} + 10 \mathfrak{t}^{6} - 12 \mathfrak{t}^{7} - 44 \mathfrak{t}^{8} - 84 \mathfrak{t}^{9} - 122 \mathfrak{t}^{10} - 160 \mathfrak{t}^{11} - 182 \mathfrak{t}^{12} - 190 \mathfrak{t}^{13}\nonumber \\
&\quad\quad - 178 \mathfrak{t}^{14} - 148 \mathfrak{t}^{15} - 108 \mathfrak{t}^{16} - 70 \mathfrak{t}^{17} - 42 \mathfrak{t}^{18} - 24 \mathfrak{t}^{19} - 12 \mathfrak{t}^{20} - 4 \mathfrak{t}^{21}) \nonumber \\
&\quad + c^{9} \mathfrak{t}^{34} (-1 - \mathfrak{t}^{2} - 2 \mathfrak{t}^{3} - 4 \mathfrak{t}^{4} - 6 \mathfrak{t}^{5} - 7 \mathfrak{t}^{6} - 8 \mathfrak{t}^{7} - 12 \mathfrak{t}^{8} - 12 \mathfrak{t}^{9} - 14 \mathfrak{t}^{10} - 12 \mathfrak{t}^{11} - 12 \mathfrak{t}^{12} - 8 \mathfrak{t}^{13} - 7 \mathfrak{t}^{14} - 6 \mathfrak{t}^{15} - 4 \mathfrak{t}^{16}\nonumber \\
&\quad\quad - 2 \mathfrak{t}^{17} - \mathfrak{t}^{18} - \mathfrak{t}^{20})).
\end{align}
}

In the limit of large representation, we have
{\fontsize{9pt}{1pt}\selectfont
\begin{align}
&\langle W_{(k)}W_{(\overline{k})}\rangle^{U(5)\text{ADHM-}[1](H)}(\mathfrak{t})\nonumber \\
&=
\frac{1}{
(1 - \mathfrak{t})^3
(1 - \mathfrak{t}^2)^4
(1 - \mathfrak{t}^3)^5
(1 - \mathfrak{t}^4)^4
(1 - \mathfrak{t}^5)^2
}
(
1 + \mathfrak{t} + 4 \mathfrak{t}^{2} + 11 \mathfrak{t}^{3} + 33 \mathfrak{t}^{4} + 78 \mathfrak{t}^{5} + 179 \mathfrak{t}^{6} + 379 \mathfrak{t}^{7} + 781 \mathfrak{t}^{8}\nonumber \\
&\quad + 1484 \mathfrak{t}^{9} + 2687 \mathfrak{t}^{10} + 4551 \mathfrak{t}^{11} + 7355 \mathfrak{t}^{12} + 11206 \mathfrak{t}^{13} + 16273 \mathfrak{t}^{14} + 22371 \mathfrak{t}^{15} + 29306 \mathfrak{t}^{16} + 36437 \mathfrak{t}^{17} + 43215 \mathfrak{t}^{18}\nonumber \\
&\quad + 48722 \mathfrak{t}^{19} + 52438 \mathfrak{t}^{20} + 53676 \mathfrak{t}^{21} + \cdots + \mathfrak{t}^{42}
)+{\cal O}(\mathfrak{t}^{k+1}).
\end{align}
}

\subsubsection{$U(6)$}
\noindent \underline{1-point functions}

{\fontsize{9pt}{1pt}\selectfont
\begin{align}
&\sum_{k=0}^\infty c^k\langle W_{(k)}\rangle^{U(6)\text{ ADHM-}[1](H)}(\mathfrak{t})\nonumber \\
&=
\frac{1}{
(1 - c \mathfrak{t}) (1 - c \mathfrak{t}^2)^2 (1 - c \mathfrak{t}^3)^3 (1 - c \mathfrak{t}^4)^4 (1 - c \mathfrak{t}^5)^2 (1 - c \mathfrak{t}^6)^2 (1 - \mathfrak{t})^2 (1 - \mathfrak{t}^2)^2
(1 - \mathfrak{t}^3)^2 (1 - \mathfrak{t}^4)^2 (1 - \mathfrak{t}^5)^2 (1 - \mathfrak{t}^6)^2}\nonumber \\
&\quad\times (
1 + \mathfrak{t}^{2} + 2 \mathfrak{t}^{3} + 4 \mathfrak{t}^{4} + 6 \mathfrak{t}^{5} + 12 \mathfrak{t}^{6} + 12 \mathfrak{t}^{7} + 21 \mathfrak{t}^{8} + 26 \mathfrak{t}^{9} + 37 \mathfrak{t}^{10} + 40 \mathfrak{t}^{11} + 55 \mathfrak{t}^{12} + 52 \mathfrak{t}^{13} + 61 \mathfrak{t}^{14} + 60 \mathfrak{t}^{15}\nonumber \\
&\quad\quad + 61 \mathfrak{t}^{16} + 52 \mathfrak{t}^{17} + 55 \mathfrak{t}^{18} + 40 \mathfrak{t}^{19} + 37 \mathfrak{t}^{20} + 26 \mathfrak{t}^{21} + 21 \mathfrak{t}^{22} + 12 \mathfrak{t}^{23} + 12 \mathfrak{t}^{24} + 6 \mathfrak{t}^{25} + 4 \mathfrak{t}^{26} + 2 \mathfrak{t}^{27} + \mathfrak{t}^{28} + \mathfrak{t}^{30}\nonumber \\
&\quad + c \mathfrak{t}^{5} (3 + 4 \mathfrak{t} + 3 \mathfrak{t}^{2} - 4 \mathfrak{t}^{4} - 14 \mathfrak{t}^{5} - 27 \mathfrak{t}^{6} - 64 \mathfrak{t}^{7} - 106 \mathfrak{t}^{8} - 172 \mathfrak{t}^{9} - 238 \mathfrak{t}^{10} - 324 \mathfrak{t}^{11} - 384 \mathfrak{t}^{12} - 460 \mathfrak{t}^{13}\nonumber \\
&\quad\quad - 498 \mathfrak{t}^{14} - 532 \mathfrak{t}^{15} - 523 \mathfrak{t}^{16} - 508 \mathfrak{t}^{17} - 445 \mathfrak{t}^{18} - 394 \mathfrak{t}^{19} - 308 \mathfrak{t}^{20} - 246 \mathfrak{t}^{21} - 178 \mathfrak{t}^{22} - 134 \mathfrak{t}^{23} - 85 \mathfrak{t}^{24}\nonumber \\
&\quad\quad - 58 \mathfrak{t}^{25} - 31 \mathfrak{t}^{26} - 18 \mathfrak{t}^{27} - 9 \mathfrak{t}^{28} - 6 \mathfrak{t}^{29} - 2 \mathfrak{t}^{30} - 2 \mathfrak{t}^{31}) \nonumber \\
&\quad + c^2 \mathfrak{t}^{9} (-4 - 3 \mathfrak{t} - 6 \mathfrak{t}^{2} - 10 \mathfrak{t}^{3} - 28 \mathfrak{t}^{4} - 46 \mathfrak{t}^{5} - 48 \mathfrak{t}^{6} - 25 \mathfrak{t}^{7} + 44 \mathfrak{t}^{8} + 164 \mathfrak{t}^{9} + 366 \mathfrak{t}^{10} + 607 \mathfrak{t}^{11} + 918 \mathfrak{t}^{12} + 1233 \mathfrak{t}^{13}\nonumber \\
&\quad\quad + 1560 \mathfrak{t}^{14} + 1805 \mathfrak{t}^{15} + 1992 \mathfrak{t}^{16} + 2023 \mathfrak{t}^{17} + 1970 \mathfrak{t}^{18} + 1783 \mathfrak{t}^{19} + 1546 \mathfrak{t}^{20} + 1257 \mathfrak{t}^{21} + 992 \mathfrak{t}^{22} + 731 \mathfrak{t}^{23}\nonumber \\
&\quad\quad + 518 \mathfrak{t}^{24} + 342 \mathfrak{t}^{25} + 214 \mathfrak{t}^{26} + 123 \mathfrak{t}^{27} + 70 \mathfrak{t}^{28} + 37 \mathfrak{t}^{29} + 20 \mathfrak{t}^{30} + 10 \mathfrak{t}^{31} + 4 \mathfrak{t}^{32} + \mathfrak{t}^{33}) \nonumber \\
&\quad + c^3 \mathfrak{t}^{13} (1 - 6 \mathfrak{t} - 6 \mathfrak{t}^{2} - 6 \mathfrak{t}^{3} + 20 \mathfrak{t}^{4} + 52 \mathfrak{t}^{5} + 94 \mathfrak{t}^{6} + 142 \mathfrak{t}^{7} + 171 \mathfrak{t}^{8} + 160 \mathfrak{t}^{9} + 34 \mathfrak{t}^{10} - 206 \mathfrak{t}^{11} - 647 \mathfrak{t}^{12} - 1236 \mathfrak{t}^{13}\nonumber \\
&\quad\quad - 2024 \mathfrak{t}^{14} - 2780 \mathfrak{t}^{15} - 3551 \mathfrak{t}^{16} - 4064 \mathfrak{t}^{17} - 4383 \mathfrak{t}^{18} - 4322 \mathfrak{t}^{19} - 4092 \mathfrak{t}^{20} - 3558 \mathfrak{t}^{21} - 2991 \mathfrak{t}^{22} - 2324 \mathfrak{t}^{23}\nonumber \\
&\quad\quad - 1741 \mathfrak{t}^{24} - 1192 \mathfrak{t}^{25} - 796 \mathfrak{t}^{26} - 476 \mathfrak{t}^{27} - 284 \mathfrak{t}^{28} - 154 \mathfrak{t}^{29} - 86 \mathfrak{t}^{30} - 40 \mathfrak{t}^{31} - 21 \mathfrak{t}^{32} - 6 \mathfrak{t}^{33} - 2 \mathfrak{t}^{34}) \nonumber \\
&\quad + c^4 \mathfrak{t}^{18} (3 + 6 \mathfrak{t} + 10 \mathfrak{t}^{2} + 6 \mathfrak{t}^{3} + 7 \mathfrak{t}^{4} - 2 \mathfrak{t}^{5} - 41 \mathfrak{t}^{6} - 114 \mathfrak{t}^{7} - 228 \mathfrak{t}^{8} - 332 \mathfrak{t}^{9} - 379 \mathfrak{t}^{10} - 264 \mathfrak{t}^{11} + 104 \mathfrak{t}^{12} + 766 \mathfrak{t}^{13}\nonumber \\
&\quad\quad + 1692 \mathfrak{t}^{14} + 2802 \mathfrak{t}^{15} + 3937 \mathfrak{t}^{16} + 4900 \mathfrak{t}^{17} + 5589 \mathfrak{t}^{18} + 5864 \mathfrak{t}^{19} + 5723 \mathfrak{t}^{20} + 5216 \mathfrak{t}^{21} + 4441 \mathfrak{t}^{22} + 3528 \mathfrak{t}^{23}\nonumber \\
&\quad\quad + 2622 \mathfrak{t}^{24} + 1816 \mathfrak{t}^{25} + 1170 \mathfrak{t}^{26} + 718 \mathfrak{t}^{27} + 411 \mathfrak{t}^{28} + 226 \mathfrak{t}^{29} + 117 \mathfrak{t}^{30} + 56 \mathfrak{t}^{31} + 21 \mathfrak{t}^{32} + 8 \mathfrak{t}^{33} + \mathfrak{t}^{34})\nonumber \\
&\quad + c^5 \mathfrak{t}^{25} (3 - 8 \mathfrak{t} - 26 \mathfrak{t}^{2} - 38 \mathfrak{t}^{3} - 38 \mathfrak{t}^{4} + 16 \mathfrak{t}^{5} + 99 \mathfrak{t}^{6} + 242 \mathfrak{t}^{7} + 362 \mathfrak{t}^{8} + 414 \mathfrak{t}^{9} + 319 \mathfrak{t}^{10} - 24 \mathfrak{t}^{11} - 597 \mathfrak{t}^{12} - 1452 \mathfrak{t}^{13}\nonumber \\
&\quad\quad - 2395 \mathfrak{t}^{14} - 3458 \mathfrak{t}^{15} - 4256 \mathfrak{t}^{16} - 4884 \mathfrak{t}^{17} - 5008 \mathfrak{t}^{18} - 4820 \mathfrak{t}^{19} - 4201 \mathfrak{t}^{20} - 3464 \mathfrak{t}^{21} - 2576 \mathfrak{t}^{22} - 1824 \mathfrak{t}^{23}\nonumber \\
&\quad\quad - 1178 \mathfrak{t}^{24} - 732 \mathfrak{t}^{25} - 406 \mathfrak{t}^{26} - 222 \mathfrak{t}^{27} - 103 \mathfrak{t}^{28} - 46 \mathfrak{t}^{29} - 15 \mathfrak{t}^{30} - 4 \mathfrak{t}^{31}) \nonumber \\
&\quad + c^6 \mathfrak{t}^{29} (-4 - 3 \mathfrak{t} + 2 \mathfrak{t}^{2} + 10 \mathfrak{t}^{3} + 18 \mathfrak{t}^{4} + 21 \mathfrak{t}^{5} + 26 \mathfrak{t}^{6} + 3 \mathfrak{t}^{7} - 34 \mathfrak{t}^{8} - 101 \mathfrak{t}^{9} - 170 \mathfrak{t}^{10} - 226 \mathfrak{t}^{11} - 208 \mathfrak{t}^{12} - 56 \mathfrak{t}^{13}\nonumber \\
&\quad\quad + 238 \mathfrak{t}^{14} + 719 \mathfrak{t}^{15} + 1282 \mathfrak{t}^{16} + 1886 \mathfrak{t}^{17} + 2340 \mathfrak{t}^{18} + 2639 \mathfrak{t}^{19} + 2650 \mathfrak{t}^{20} + 2460 \mathfrak{t}^{21} + 2072 \mathfrak{t}^{22} + 1631 \mathfrak{t}^{23}\nonumber \\
&\quad\quad + 1178 \mathfrak{t}^{24} + 803 \mathfrak{t}^{25} + 484 \mathfrak{t}^{26} + 274 \mathfrak{t}^{27} + 138 \mathfrak{t}^{28} + 62 \mathfrak{t}^{29} + 20 \mathfrak{t}^{30} + 6 \mathfrak{t}^{31}) \nonumber \\
&\quad + c^7 \mathfrak{t}^{35} (-1 + 3 \mathfrak{t}^{2} + 4 \mathfrak{t}^{3} + 2 \mathfrak{t}^{4} - 6 \mathfrak{t}^{5} - 13 \mathfrak{t}^{6} - 20 \mathfrak{t}^{7} - 15 \mathfrak{t}^{8} + 10 \mathfrak{t}^{9} + 44 \mathfrak{t}^{10} + 76 \mathfrak{t}^{11} + 81 \mathfrak{t}^{12} + 44 \mathfrak{t}^{13} - 67 \mathfrak{t}^{14}\nonumber \\
&\quad\quad - 210 \mathfrak{t}^{15} - 401 \mathfrak{t}^{16} - 564 \mathfrak{t}^{17} - 713 \mathfrak{t}^{18} - 768 \mathfrak{t}^{19} - 780 \mathfrak{t}^{20} - 696 \mathfrak{t}^{21} - 611 \mathfrak{t}^{22} - 454 \mathfrak{t}^{23} - 322 \mathfrak{t}^{24} - 196 \mathfrak{t}^{25}\nonumber \\
&\quad\quad - 115 \mathfrak{t}^{26} - 48 \mathfrak{t}^{27} - 20 \mathfrak{t}^{28} - 4 \mathfrak{t}^{29}) \nonumber \\
&\quad + c^8 \mathfrak{t}^{54} (1 + 10 \mathfrak{t} + 16 \mathfrak{t}^{2} + 40 \mathfrak{t}^{3} + 58 \mathfrak{t}^{4} + 84 \mathfrak{t}^{5} + 93 \mathfrak{t}^{6} + 116 \mathfrak{t}^{7} + 93 \mathfrak{t}^{8} + 84 \mathfrak{t}^{9} + 58 \mathfrak{t}^{10} + 40 \mathfrak{t}^{11} + 16 \mathfrak{t}^{12} + 10 \mathfrak{t}^{13} + \mathfrak{t}^{14}) 
).
\end{align}
}

In the limit of large representation, we have
{\fontsize{9pt}{1pt}\selectfont
\begin{align}
&\langle W_{(k)}\rangle^{U(6)\text{ ADHM-}[1](H)}(\mathfrak{t})\nonumber \\
&=
\frac{\mathfrak{t}^k}{
(1 - \mathfrak{t})^3 (1 - \mathfrak{t}^2)^3 (1 - \mathfrak{t}^3)^3 (1 - \mathfrak{t}^4)^3 (1 - \mathfrak{t}^5)^3 (1 - \mathfrak{t}^6)^2
}
(
1 + \mathfrak{t} + 4 \mathfrak{t}^{2} + 9 \mathfrak{t}^{3} + 22 \mathfrak{t}^{4} + 46 \mathfrak{t}^{5} + 97 \mathfrak{t}^{6} + 175 \mathfrak{t}^{7} + 327 \mathfrak{t}^{8}\nonumber \\
&\quad + 553 \mathfrak{t}^{9} + 915 \mathfrak{t}^{10} + 1405 \mathfrak{t}^{11} + 2110 \mathfrak{t}^{12} + 2960 \mathfrak{t}^{13} + 4043 \mathfrak{t}^{14} + 5216 \mathfrak{t}^{15} + 6505 \mathfrak{t}^{16} + 7699 \mathfrak{t}^{17} + 8824 \mathfrak{t}^{18}\nonumber \\
&\quad + 9580 \mathfrak{t}^{19} + 10071 \mathfrak{t}^{20} + 10053 \mathfrak{t}^{21} + 9683 \mathfrak{t}^{22} + 8842 \mathfrak{t}^{23} + 7802 \mathfrak{t}^{24} + 6489 \mathfrak{t}^{25} + 5201 \mathfrak{t}^{26} + 3915 \mathfrak{t}^{27} + 2828 \mathfrak{t}^{28}\nonumber \\
&\quad + 1894 \mathfrak{t}^{29} + 1225 \mathfrak{t}^{30} + 709 \mathfrak{t}^{31} + 395 \mathfrak{t}^{32} + 192 \mathfrak{t}^{33} + 88 \mathfrak{t}^{34} + 29 \mathfrak{t}^{35} + 11 \mathfrak{t}^{36} + \mathfrak{t}^{37}
)+{\cal O}(\mathfrak{t}^{2k+1}).
\end{align}
}

\noindent \underline{Diagonal $2$-point functions}
{\fontsize{9pt}{1pt}\selectfont
\begin{align}
&\sum_{k=0}^\infty c^k\langle W_{(k)}W_{(\overline{k})}\rangle^{{\cal N}=8\text{ }U(6)(H)}(\mathfrak{t})\nonumber \\
&=
\frac{1}{
(1 - c) (1 - c \mathfrak{t})^2 (1 - c \mathfrak{t}^2)^3 (1 - c \mathfrak{t}^3)^4 (1 - c \mathfrak{t}^4)^2 (1 - c \mathfrak{t}^5)^2 (1 - \mathfrak{t}) (1 - \mathfrak{t}^2)
(1 - \mathfrak{t}^3) (1 - \mathfrak{t}^4) (1 - \mathfrak{t}^5) (1 - \mathfrak{t}^6)
}
(
1 + \mathfrak{t}\nonumber \\
&\quad\quad + 2 \mathfrak{t}^{2} + 3 \mathfrak{t}^{3} + 5 \mathfrak{t}^{4} + 7 \mathfrak{t}^{5} + 11 \mathfrak{t}^{6} + 8 \mathfrak{t}^{7} + 9 \mathfrak{t}^{8} + 7 \mathfrak{t}^{9} + 6 \mathfrak{t}^{10} + 2 \mathfrak{t}^{11} + 2 \mathfrak{t}^{12}\nonumber \\
&\quad + c \mathfrak{t}^{4} (3 + 7 \mathfrak{t} + 10 \mathfrak{t}^{2} + 7 \mathfrak{t}^{3} - 2 \mathfrak{t}^{4} - 20 \mathfrak{t}^{5} - 41 \mathfrak{t}^{6} - 79 \mathfrak{t}^{7} - 109 \mathfrak{t}^{8} - 127 \mathfrak{t}^{9} - 117 \mathfrak{t}^{10} - 91 \mathfrak{t}^{11} - 44 \mathfrak{t}^{12} + 35 \mathfrak{t}^{14} + 44 \mathfrak{t}^{15}\nonumber \\
&\quad\quad + 35 \mathfrak{t}^{16} + 15 \mathfrak{t}^{17} - 5 \mathfrak{t}^{18} - 17 \mathfrak{t}^{19} - 16 \mathfrak{t}^{20} - 9 \mathfrak{t}^{21} - \mathfrak{t}^{22} + 5 \mathfrak{t}^{23} + 5 \mathfrak{t}^{24} + 2 \mathfrak{t}^{25} - \mathfrak{t}^{26} - \mathfrak{t}^{27})\nonumber \\
&\quad + c^{2} \mathfrak{t}^{7} (-4 - 7 \mathfrak{t} - 13 \mathfrak{t}^{2} - 18 \mathfrak{t}^{3} - 35 \mathfrak{t}^{4} - 55 \mathfrak{t}^{5} - 55 \mathfrak{t}^{6} - 13 \mathfrak{t}^{7} + 76 \mathfrak{t}^{8} + 199 \mathfrak{t}^{9} + 345 \mathfrak{t}^{10} + 453 \mathfrak{t}^{11} + 501 \mathfrak{t}^{12} + 442 \mathfrak{t}^{13}\nonumber \\
&\quad\quad + 317 \mathfrak{t}^{14} + 138 \mathfrak{t}^{15} - 22 \mathfrak{t}^{16} - 139 \mathfrak{t}^{17} - 177 \mathfrak{t}^{18} - 151 \mathfrak{t}^{19} - 89 \mathfrak{t}^{20} - 19 \mathfrak{t}^{21} + 25 \mathfrak{t}^{22} + 44 \mathfrak{t}^{23} + 36 \mathfrak{t}^{24} + 23 \mathfrak{t}^{25} + 6 \mathfrak{t}^{26}\nonumber \\
&\quad\quad - 5 \mathfrak{t}^{27} - 7 \mathfrak{t}^{28} - 4 \mathfrak{t}^{29})\nonumber \\
&\quad + c^{3} \mathfrak{t}^{10} (1 - 5 \mathfrak{t} - 11 \mathfrak{t}^{2} - 18 \mathfrak{t}^{3} + 4 \mathfrak{t}^{4} + 51 \mathfrak{t}^{5} + 119 \mathfrak{t}^{6} + 187 \mathfrak{t}^{7} + 217 \mathfrak{t}^{8} + 165 \mathfrak{t}^{9} - 23 \mathfrak{t}^{10} - 308 \mathfrak{t}^{11} - 659 \mathfrak{t}^{12} - 936 \mathfrak{t}^{13}\nonumber \\
&\quad\quad - 1103 \mathfrak{t}^{14} - 1026 \mathfrak{t}^{15} - 790 \mathfrak{t}^{16} - 414 \mathfrak{t}^{17} - 48 \mathfrak{t}^{18} + 230 \mathfrak{t}^{19} + 341 \mathfrak{t}^{20} + 316 \mathfrak{t}^{21} + 206 \mathfrak{t}^{22} + 72 \mathfrak{t}^{23} - 17 \mathfrak{t}^{24} - 60 \mathfrak{t}^{25}\nonumber \\
&\quad\quad - 48 \mathfrak{t}^{26} - 25 \mathfrak{t}^{27} - 5 \mathfrak{t}^{28} + 3 \mathfrak{t}^{29})\nonumber \\
&\quad + c^{4} \mathfrak{t}^{14} (3 + 9 \mathfrak{t} + 19 \mathfrak{t}^{2} + 22 \mathfrak{t}^{3} + 23 \mathfrak{t}^{4} + 6 \mathfrak{t}^{5} - 48 \mathfrak{t}^{6} - 141 \mathfrak{t}^{7} - 258 \mathfrak{t}^{8} - 333 \mathfrak{t}^{9} - 306 \mathfrak{t}^{10} - 114 \mathfrak{t}^{11} + 237 \mathfrak{t}^{12} + 684 \mathfrak{t}^{13}\nonumber \\
&\quad\quad + 1091 \mathfrak{t}^{14} + 1346 \mathfrak{t}^{15} + 1346 \mathfrak{t}^{16} + 1091 \mathfrak{t}^{17} + 684 \mathfrak{t}^{18} + 237 \mathfrak{t}^{19} - 114 \mathfrak{t}^{20} - 306 \mathfrak{t}^{21} - 333 \mathfrak{t}^{22} - 258 \mathfrak{t}^{23} - 141 \mathfrak{t}^{24}\nonumber \\
&\quad\quad - 48 \mathfrak{t}^{25} + 6 \mathfrak{t}^{26} + 23 \mathfrak{t}^{27} + 22 \mathfrak{t}^{28} + 19 \mathfrak{t}^{29} + 9 \mathfrak{t}^{30} + 3 \mathfrak{t}^{31})\nonumber \\
&\quad + c^{5} \mathfrak{t}^{20} (3 - 5 \mathfrak{t} - 25 \mathfrak{t}^{2} - 48 \mathfrak{t}^{3} - 60 \mathfrak{t}^{4} - 17 \mathfrak{t}^{5} + 72 \mathfrak{t}^{6} + 206 \mathfrak{t}^{7} + 316 \mathfrak{t}^{8} + 341 \mathfrak{t}^{9} + 230 \mathfrak{t}^{10} - 48 \mathfrak{t}^{11} - 414 \mathfrak{t}^{12} - 790 \mathfrak{t}^{13}\nonumber \\
&\quad\quad - 1026 \mathfrak{t}^{14} - 1103 \mathfrak{t}^{15} - 936 \mathfrak{t}^{16} - 659 \mathfrak{t}^{17} - 308 \mathfrak{t}^{18} - 23 \mathfrak{t}^{19} + 165 \mathfrak{t}^{20} + 217 \mathfrak{t}^{21} + 187 \mathfrak{t}^{22} + 119 \mathfrak{t}^{23} + 51 \mathfrak{t}^{24} + 4 \mathfrak{t}^{25}\nonumber \\
&\quad\quad - 18 \mathfrak{t}^{26} - 11 \mathfrak{t}^{27} - 5 \mathfrak{t}^{28} + \mathfrak{t}^{29})\nonumber \\
&\quad + c^{6} \mathfrak{t}^{23} (-4 - 7 \mathfrak{t} - 5 \mathfrak{t}^{2} + 6 \mathfrak{t}^{3} + 23 \mathfrak{t}^{4} + 36 \mathfrak{t}^{5} + 44 \mathfrak{t}^{6} + 25 \mathfrak{t}^{7} - 19 \mathfrak{t}^{8} - 89 \mathfrak{t}^{9} - 151 \mathfrak{t}^{10} - 177 \mathfrak{t}^{11} - 139 \mathfrak{t}^{12} - 22 \mathfrak{t}^{13}\nonumber \\
&\quad\quad + 138 \mathfrak{t}^{14} + 317 \mathfrak{t}^{15} + 442 \mathfrak{t}^{16} + 501 \mathfrak{t}^{17} + 453 \mathfrak{t}^{18} + 345 \mathfrak{t}^{19} + 199 \mathfrak{t}^{20} + 76 \mathfrak{t}^{21} - 13 \mathfrak{t}^{22} - 55 \mathfrak{t}^{23} - 55 \mathfrak{t}^{24} - 35 \mathfrak{t}^{25}\nonumber \\
&\quad\quad - 18 \mathfrak{t}^{26} - 13 \mathfrak{t}^{27} - 7 \mathfrak{t}^{28} - 4 \mathfrak{t}^{29})\nonumber \\
&\quad + c^{7} \mathfrak{t}^{28} (-1 - \mathfrak{t} + 2 \mathfrak{t}^{2} + 5 \mathfrak{t}^{3} + 5 \mathfrak{t}^{4} - \mathfrak{t}^{5} - 9 \mathfrak{t}^{6} - 16 \mathfrak{t}^{7} - 17 \mathfrak{t}^{8} - 5 \mathfrak{t}^{9} + 15 \mathfrak{t}^{10} + 35 \mathfrak{t}^{11} + 44 \mathfrak{t}^{12} + 35 \mathfrak{t}^{13} - 44 \mathfrak{t}^{15} - 91 \mathfrak{t}^{16}\nonumber \\
&\quad\quad - 117 \mathfrak{t}^{17} - 127 \mathfrak{t}^{18} - 109 \mathfrak{t}^{19} - 79 \mathfrak{t}^{20} - 41 \mathfrak{t}^{21} - 20 \mathfrak{t}^{22} - 2 \mathfrak{t}^{23} + 7 \mathfrak{t}^{24} + 10 \mathfrak{t}^{25} + 7 \mathfrak{t}^{26} + 3 \mathfrak{t}^{27})\nonumber \\
&\quad + c^{8} \mathfrak{t}^{47} (2 + 2 \mathfrak{t} + 6 \mathfrak{t}^{2} + 7 \mathfrak{t}^{3} + 9 \mathfrak{t}^{4} + 8 \mathfrak{t}^{5} + 11 \mathfrak{t}^{6} + 7 \mathfrak{t}^{7} + 5 \mathfrak{t}^{8} + 3 \mathfrak{t}^{9} + 2 \mathfrak{t}^{10} + \mathfrak{t}^{11} + \mathfrak{t}^{12})
).
\end{align}
}

\subsection{Unflavored Higgs correlators in antisymmetric representations}
In this subsection we list the Higgs line defect indices with one Wilson line in a antisymmetric representation or two Wilson lines in the same antisymmetric representation, in the unflavored limit.
Here we display only the results which are independent of the correlators of symmetric Wilson lines after taking into account the symmetry \eqref{Higgssymmetrymanifestinintegral}.
We find that the numerators of the 1-point functions for any flavors $l$ and the diagonal 2-point functions for $l\ge 2$ are palindromic polynomials, which we abbreviate with ```$\cdots$''.

\subsubsection{$U(2)$}
\noindent \underline{1-point functions}
{\fontsize{9pt}{1pt}\selectfont
\begin{align}
&\langle W_{\ydiagram{1,1}}\rangle^{U(2)\text{ADHM-}[1](H)}(\mathfrak{t})
=
\frac{2\mathfrak{t}^3}{(1-\mathfrak{t})^2(1-\mathfrak{t}^2)^2},\\
&\langle W_{\ydiagram{1,1}}\rangle^{U(2)\text{ADHM-}[2](H)}(\mathfrak{t})
=
\frac{
\mathfrak{t}^2 (1 + 7 \mathfrak{t} + 9 \mathfrak{t}^2 + 10 \mathfrak{t}^3 + 9 \mathfrak{t}^4 + 7 \mathfrak{t}^5 + \mathfrak{t}^6)
}{
(1 - \mathfrak{t}) (1 - \mathfrak{t}^2)^4 (1 - \mathfrak{t}^3)^3
},\\
&\langle W_{\ydiagram{1,1}}\rangle^{U(2)\text{ADHM-}[3](H)}(\mathfrak{t})
=
\frac{
3 \mathfrak{t}^2 
(1 + 5 \mathfrak{t} + 9 \mathfrak{t}^2 + 18 \mathfrak{t}^3 + 31 \mathfrak{t}^4 + 39 \mathfrak{t}^5 + 38 \mathfrak{t}^6 + 39 \mathfrak{t}^7 + 31 \mathfrak{t}^8 + 18 \mathfrak{t}^9 + 9 \mathfrak{t}^{10} + 5 \mathfrak{t}^{11} + \mathfrak{t}^{12})
}{
(1 - \mathfrak{t}) (1 - \mathfrak{t}^2)^6 (1 - \mathfrak{t}^3)^5},\\
&\langle W_{\ydiagram{1,1}}\rangle^{U(2)\text{ADHM-}[4](H)}(\mathfrak{t})\nonumber \\
&=
\frac{
2 \mathfrak{t}^{2}
(3 + 13 \mathfrak{t} + 35 \mathfrak{t}^{2} + 98 \mathfrak{t}^{3} + 220 \mathfrak{t}^{4} + 388 \mathfrak{t}^{5} + 616 \mathfrak{t}^{6} + 886 \mathfrak{t}^{7} + 1064 \mathfrak{t}^{8} + 1112 \mathfrak{t}^{9} + \cdots + 3 \mathfrak{t}^{18})
}{
(1 - \mathfrak{t}) (1 - \mathfrak{t}^2)^8 (1 - \mathfrak{t}^3)^7
}.
\end{align}
}

\subsubsection{$U(3)$}
\noindent \underline{1-point functions}
{\fontsize{9pt}{1pt}\selectfont
\begin{align}
&\langle W_{\ydiagram{1,1}}\rangle^{U(3)\text{ADHM-}[1](H)}(\mathfrak{t})
=
\frac{2\mathfrak{t}^3}{(1-\mathfrak{t})^4(1-\mathfrak{t}^2)^2},\\
&\langle W_{\ydiagram{1,1,1}}\rangle^{U(3)\text{ADHM-}[1](H)}(\mathfrak{t})
=
\frac{\mathfrak{t}^5(1+4\mathfrak{t}+\mathfrak{t}^2)}{(1-\mathfrak{t})^2(1-\mathfrak{t}^2)^2(1-\mathfrak{t}^3)^2},\\
&\langle W_{\ydiagram{1,1}}\rangle^{U(3)\text{ADHM-}[2](H)}(\mathfrak{t})
=
\frac{
\mathfrak{t}^2 
(1 + 7 \mathfrak{t} + 10 \mathfrak{t}^{2} + 25 \mathfrak{t}^{3} + 38 \mathfrak{t}^{4} + 61 \mathfrak{t}^{5} + 63 \mathfrak{t}^{6} + 78 \mathfrak{t}^{7} + \cdots + \mathfrak{t}^{14})
}
{
(1 - \mathfrak{t})^3 (1 - \mathfrak{t}^2)^3 (1 - \mathfrak{t}^3)^3 (1 - \mathfrak{t}^4)^3
}
,\\
&\langle W_{\ydiagram{1,1,1}}\rangle^{U(3)\text{ADHM-}[2](H)}(\mathfrak{t})
=
\frac{
2 \mathfrak{t}^4 
(2 + 5 \mathfrak{t} + 14 \mathfrak{t}^{2} + 17 \mathfrak{t}^{3} + 30 \mathfrak{t}^{4} + 34 \mathfrak{t}^{5} + 40 \mathfrak{t}^{6} + \cdots + 2 \mathfrak{t}^{12})
}{
(1 - \mathfrak{t})^2 (1 - \mathfrak{t}^2)^3 (1 - \mathfrak{t}^3)^4 (1 - \mathfrak{t}^4)^3
},\\
&\langle W_{\ydiagram{1,1}}\rangle^{U(3)\text{ADHM-}[3](H)}(\mathfrak{t})\nonumber \\
&=
\frac{
3 \mathfrak{t}^{2}
}{
(1 - \mathfrak{t})^2 (1 - \mathfrak{t}^2)^5 (1 - \mathfrak{t}^3)^6 (1 - \mathfrak{t}^4)^5
}
(1 + 6 \mathfrak{t} + 17 \mathfrak{t}^{2} + 56 \mathfrak{t}^{3} + 139 \mathfrak{t}^{4} + 322 \mathfrak{t}^{5} + 632 \mathfrak{t}^{6} + 1174 \mathfrak{t}^{7} + 1922 \mathfrak{t}^{8} + 2974 \mathfrak{t}^{9}\nonumber \\
&\quad + 4137 \mathfrak{t}^{10} + 5412 \mathfrak{t}^{11} + 6464 \mathfrak{t}^{12} + 7280 \mathfrak{t}^{13} + 7496 \mathfrak{t}^{14} + \cdots + \mathfrak{t}^{28}),\\
%
&\langle W_{\ydiagram{1,1,1}}\rangle^{U(3)\text{ADHM-}[3](H)}(\mathfrak{t})\nonumber \\
&=
\frac{
\mathfrak{t}^{3}
}{
(1 - \mathfrak{t})^2 (1 - \mathfrak{t}^2)^5 (1 - \mathfrak{t}^3)^6 (1 - \mathfrak{t}^4)^5
}
(1 + 16 \mathfrak{t} + 40 \mathfrak{t}^{2} + 132 \mathfrak{t}^{3} + 274 \mathfrak{t}^{4} + 626 \mathfrak{t}^{5} + 1109 \mathfrak{t}^{6} + 1946 \mathfrak{t}^{7} + 2882 \mathfrak{t}^{8}\nonumber \\
&\quad + 4210 \mathfrak{t}^{9} + 5355 \mathfrak{t}^{10} + 6628 \mathfrak{t}^{11} + 7235 \mathfrak{t}^{12} + 7660 \mathfrak{t}^{13} + \cdots + \mathfrak{t}^{26}).
\end{align}
}

\subsubsection{$U(4)$}
\noindent \underline{1-point functions}
{\fontsize{9pt}{1pt}\selectfont
\begin{align}
&\langle W_{\ydiagram{1,1}}\rangle^{U(4)\text{ADHM-}[1](H)}(\mathfrak{t})
=
\frac{
2 \mathfrak{t}^3 (1 + \mathfrak{t}^2)}
{(1 - \mathfrak{t})^4 (1 - \mathfrak{t}^2)^4
},\\
&\langle W_{\ydiagram{1,1,1}}\rangle^{U(4)\text{ADHM-}[1](H)}(\mathfrak{t})
=
\frac{
 \mathfrak{t}^5 (1 + 4 \mathfrak{t} + \mathfrak{t}^2)}
{(1 - \mathfrak{t})^4 (1 - \mathfrak{t}^2)^2 (1 - \mathfrak{t}^3)^2
},\\
&\langle W_{\ydiagram{1,1,1,1}}\rangle^{U(4)\text{ADHM-}[1](H)}(\mathfrak{t})
=
\frac{
 \mathfrak{t}^8 (3 + 4 \mathfrak{t} + 10 \mathfrak{t}^2 + 4 \mathfrak{t}^3 + 3 \mathfrak{t}^4)}
{(1 - \mathfrak{t})^2 (1 - \mathfrak{t}^2)^2 (1 - \mathfrak{t}^3)^2 (1 - \mathfrak{t}^4)^2
},\\
&\langle W_{\ydiagram{1,1}}\rangle^{U(4)\text{ADHM-}[2](H)}(\mathfrak{t})\nonumber \\
&=
\frac{
\mathfrak{t}^{2} 
}{
(1 - \mathfrak{t})^2 (1 - \mathfrak{t}^2)^4 (1 - \mathfrak{t}^3)^3 (1 - \mathfrak{t}^4)^4 (1 - \mathfrak{t}^5)^3
}
(1 + 8 \mathfrak{t} + 20 \mathfrak{t}^{2} + 59 \mathfrak{t}^{3} + 134 \mathfrak{t}^{4} + 293 \mathfrak{t}^{5} + 559 \mathfrak{t}^{6} + 992 \mathfrak{t}^{7} + 1600 \mathfrak{t}^{8}\nonumber \\
&\quad + 2441 \mathfrak{t}^{9} + 3419 \mathfrak{t}^{10} + 4536 \mathfrak{t}^{11} + 5597 \mathfrak{t}^{12} + 6532 \mathfrak{t}^{13} + 7122 \mathfrak{t}^{14} + 7362 \mathfrak{t}^{15} + \cdots + \mathfrak{t}^{30}),\\
%
&\langle W_{\ydiagram{1,1,1}}\rangle^{U(4)\text{ADHM-}[2](H)}(\mathfrak{t})\nonumber \\
&=
\frac{
2 \mathfrak{t}^{4} 
}{
(1 - \mathfrak{t})^3 (1 - \mathfrak{t}^2)^2 (1 - \mathfrak{t}^3)^4 (1 - \mathfrak{t}^4)^4 (1 - \mathfrak{t}^5)^3
}
(2 + 7 \mathfrak{t} + 23 \mathfrak{t}^{2} + 51 \mathfrak{t}^{3} + 113 \mathfrak{t}^{4} + 211 \mathfrak{t}^{5} + 366 \mathfrak{t}^{6} + 576 \mathfrak{t}^{7} + 840 \mathfrak{t}^{8}\nonumber \\
&\quad + 1133 \mathfrak{t}^{9} + 1431 \mathfrak{t}^{10} + 1680 \mathfrak{t}^{11} + 1858 \mathfrak{t}^{12} + 1915 \mathfrak{t}^{13} + \cdots + 2 \mathfrak{t}^{26}),\\
%
&\langle W_{\ydiagram{1,1,1,1}}\rangle^{U(4)\text{ADHM-}[2](H)}(\mathfrak{t})\nonumber \\
&=
\frac{
\mathfrak{t}^{6} 
}{
(1 - \mathfrak{t}) (1 - \mathfrak{t}^2)^4 (1 - \mathfrak{t}^3)^4 (1 - \mathfrak{t}^4)^4 (1 - \mathfrak{t}^5)^3
}
(6 + 26 \mathfrak{t} + 82 \mathfrak{t}^{2} + 186 \mathfrak{t}^{3} + 385 \mathfrak{t}^{4} + 679 \mathfrak{t}^{5} + 1107 \mathfrak{t}^{6} + 1647 \mathfrak{t}^{7}\nonumber \\
&\quad + 2279 \mathfrak{t}^{8} + 2892 \mathfrak{t}^{9} + 3444 \mathfrak{t}^{10} + 3796 \mathfrak{t}^{11} + 3936 \mathfrak{t}^{12} + \cdots + 6 \mathfrak{t}^{24}).
\end{align}
}

\noindent \underline{Diagonal $2$-point functions}
{\fontsize{9pt}{1pt}\selectfont
\begin{align}
&\langle
W_{\ydiagram{1,1}}
W_{\overline{\ydiagram{1,1}}}
\rangle^{U(4)\text{ADHM-}[1](H)}(\mathfrak{t})
=
\frac{
(1+\mathfrak{t}^2)(1+2\mathfrak{t}^2+4\mathfrak{t}^3+5\mathfrak{t}^4)}{(1-\mathfrak{t})^4(1-\mathfrak{t}^2)^4},\\
&\langle W_{\ydiagram{1,1}}W_{\overline{\ydiagram{1,1}}}\rangle^{U(4)\text{ ADHM-}[2](H)}(\mathfrak{t})\nonumber \\
&=
\frac{1}{
(1 - \mathfrak{t})^2 (1 - \mathfrak{t}^2)^4 (1 - \mathfrak{t}^3)^3 (1 - \mathfrak{t}^4)^4 (1 - \mathfrak{t}^5)^3}
(
1 + 2 \mathfrak{t} + 12 \mathfrak{t}^{2} + 41 \mathfrak{t}^{3} + 125 \mathfrak{t}^{4} + 321 \mathfrak{t}^{5} + 766 \mathfrak{t}^{6} + 1621 \mathfrak{t}^{7} + 3176 \mathfrak{t}^{8}\nonumber \\
&\quad + 5639 \mathfrak{t}^{9} + 9332 \mathfrak{t}^{10} + 14263 \mathfrak{t}^{11} + 20370 \mathfrak{t}^{12} + 27093 \mathfrak{t}^{13} + 33843 \mathfrak{t}^{14} + 39527 \mathfrak{t}^{15} + 43447 \mathfrak{t}^{16} + 44770 \mathfrak{t}^{17}\nonumber \\
&\quad+\cdots+\mathfrak{t}^{34}).
\end{align}
}

\subsubsection{$U(5)$}
\noindent \underline{1-point functions}
{\fontsize{9pt}{1pt}\selectfont
\begin{align}
&\langle W_{\ydiagram{1,1}}\rangle^{U(5)\text{ADHM-}[1](H)}(\mathfrak{t})
=
\frac{2 \mathfrak{t}^3 (1 + \mathfrak{t}^2 + 2 \mathfrak{t}^3 + \mathfrak{t}^4 + \mathfrak{t}^6)}
{(1 - \mathfrak{t})^4 (1 - \mathfrak{t}^2)^4 (1 - \mathfrak{t}^3)^2},\\
&\langle W_{\ydiagram{1,1,1}}\rangle^{U(5)\text{ADHM-}[1](H)}(\mathfrak{t})
=
\frac{\mathfrak{t}^5 (1 + \mathfrak{t}^2) (1 + 4 \mathfrak{t} + \mathfrak{t}^2)}
{(1 - \mathfrak{t})^4 (1 - \mathfrak{t}^2)^4 (1 - \mathfrak{t}^3)^2},\\
&\langle W_{\ydiagram{1,1,1,1}}\rangle^{U(5)\text{ADHM-}[1](H)}(\mathfrak{t})
=
\frac{\mathfrak{t}^8 (3 + 4 \mathfrak{t} + 10 \mathfrak{t}^2 + 4 \mathfrak{t}^3 + 3 \mathfrak{t}^4)}
{(1 - \mathfrak{t})^4 (1 - \mathfrak{t}^2)^2 (1 - \mathfrak{t}^3)^2 (1 - \mathfrak{t}^4)^2},\\
&\langle W_{\ydiagram{1,1,1,1,1}}\rangle^{U(5)\text{ADHM-}[1](H)}(\mathfrak{t})
=
\frac{\mathfrak{t}^{11} (3 + 8 \mathfrak{t} + 15 \mathfrak{t}^2 + 20 \mathfrak{t}^3 + 28 \mathfrak{t}^4 + 20 \mathfrak{t}^5 + 15 \mathfrak{t}^6 + 8 \mathfrak{t}^7 + 3 \mathfrak{t}^8)}
{(1 - \mathfrak{t})^2 (1 - \mathfrak{t}^2)^2 (1 - \mathfrak{t}^3)^2 (1 - \mathfrak{t}^4)^2 (1 - \mathfrak{t}^5)^2},\\
&\langle W_{\ydiagram{1,1}}\rangle^{U(5)\text{ADHM-}[2](H)}(\mathfrak{t})\nonumber \\
&=
\frac{
\mathfrak{t}^{2} (1 + \mathfrak{t})^3
}{
(1 - \mathfrak{t}^2)^6
(1 - \mathfrak{t}^3)^4
(1 - \mathfrak{t}^4)^4
(1 - \mathfrak{t}^5)^3
(1 - \mathfrak{t}^6)^3
}
(1 + 7 \mathfrak{t} + 13 \mathfrak{t}^{2} + 49 \mathfrak{t}^{3} + 110 \mathfrak{t}^{4} + 273 \mathfrak{t}^{5} + 573 \mathfrak{t}^{6} + 1195 \mathfrak{t}^{7}\nonumber \\
&\quad + 2219 \mathfrak{t}^{8} + 4090 \mathfrak{t}^{9} + 6911 \mathfrak{t}^{10} + 11401 \mathfrak{t}^{11} + 17614 \mathfrak{t}^{12} + 26454 \mathfrak{t}^{13} + 37531 \mathfrak{t}^{14} + 51762 \mathfrak{t}^{15} + 67866 \mathfrak{t}^{16} + 86359 \mathfrak{t}^{17}\nonumber \\
&\quad + 105015 \mathfrak{t}^{18} + 123922 \mathfrak{t}^{19} + 139989 \mathfrak{t}^{20} + 153577 \mathfrak{t}^{21} + 161502 \mathfrak{t}^{22} + 164886 \mathfrak{t}^{23} + \cdots + \mathfrak{t}^{46}),\\
&\langle W_{\ydiagram{1,1,1}}\rangle^{U(5)\text{ADHM-}[2](H)}(\mathfrak{t})\nonumber \\
&=
\frac{
2 \mathfrak{t}^{4} (1 + \mathfrak{t})^3
}{
(1 - \mathfrak{t}^2)^6
(1 - \mathfrak{t}^3)^4
(1 - \mathfrak{t}^4)^4
(1 - \mathfrak{t}^5)^3
(1 - \mathfrak{t}^6)^3
}
(2 + 7 \mathfrak{t} + 27 \mathfrak{t}^{2} + 65 \mathfrak{t}^{3} + 175 \mathfrak{t}^{4} + 375 \mathfrak{t}^{5} + 804 \mathfrak{t}^{6} + 1523 \mathfrak{t}^{7} + 2809 \mathfrak{t}^{8}\nonumber \\
&\quad + 4740 \mathfrak{t}^{9} + 7776 \mathfrak{t}^{10} + 11889 \mathfrak{t}^{11} + 17588 \mathfrak{t}^{12} + 24569 \mathfrak{t}^{13} + 33196 \mathfrak{t}^{14} + 42586 \mathfrak{t}^{15} + 52905 \mathfrak{t}^{16} + 62657 \mathfrak{t}^{17}\nonumber \\
&\quad + 71844 \mathfrak{t}^{18} + 78776 \mathfrak{t}^{19} + 83640 \mathfrak{t}^{20} + 84970 \mathfrak{t}^{21} + \cdots + 2 \mathfrak{t}^{42}),\\
&\langle W_{\ydiagram{1,1,1,1}}\rangle^{U(5)\text{ADHM-}[2](H)}(\mathfrak{t})\nonumber \\
&=
\frac{
\mathfrak{t}^{6} (1 + \mathfrak{t})^3
}{
(1 - \mathfrak{t}^2)^6
(1 - \mathfrak{t}^3)^4
(1 - \mathfrak{t}^4)^4
(1 - \mathfrak{t}^5)^3
(1 - \mathfrak{t}^6)^3
}
(6 + 26 \mathfrak{t} + 88 \mathfrak{t}^{2} + 220 \mathfrak{t}^{3} + 540 \mathfrak{t}^{4} + 1104 \mathfrak{t}^{5} + 2196 \mathfrak{t}^{6} + 3936 \mathfrak{t}^{7}\nonumber \\
&\quad + 6737 \mathfrak{t}^{8} + 10693 \mathfrak{t}^{9} + 16339 \mathfrak{t}^{10} + 23375 \mathfrak{t}^{11} + 32253 \mathfrak{t}^{12} + 42112 \mathfrak{t}^{13} + 53011 \mathfrak{t}^{14} + 63465 \mathfrak{t}^{15} + 73411 \mathfrak{t}^{16}\nonumber \\
&\quad + 80895 \mathfrak{t}^{17} + 86185 \mathfrak{t}^{18} + 87692 \mathfrak{t}^{19} + \cdots + 6 \mathfrak{t}^{38}),\\
&\langle W_{\ydiagram{1,1,1,1,1}}\rangle^{U(5)\text{ADHM-}[2](H)}(\mathfrak{t})\nonumber \\
&=
\frac{
2 \mathfrak{t}^8 (1 + \mathfrak{t})^2 
}{
(1 - \mathfrak{t}^2)^5
(1 - \mathfrak{t}^3)^4
(1 - \mathfrak{t}^4)^4
(1 - \mathfrak{t}^5)^4
(1 - \mathfrak{t}^6)^3
}
(2 + 15 \mathfrak{t} + 42 \mathfrak{t}^{2} + 127 \mathfrak{t}^{3} + 286 \mathfrak{t}^{4} + 614 \mathfrak{t}^{5} + 1164 \mathfrak{t}^{6} + 2124 \mathfrak{t}^{7}\nonumber \\
&\quad + 3512 \mathfrak{t}^{8} + 5618 \mathfrak{t}^{9} + 8376 \mathfrak{t}^{10} + 12020 \mathfrak{t}^{11} + 16286 \mathfrak{t}^{12} + 21296 \mathfrak{t}^{13} + 26432 \mathfrak{t}^{14} + 31716 \mathfrak{t}^{15} + 36298 \mathfrak{t}^{16} + 40143 \mathfrak{t}^{17}\nonumber \\
&\quad + 42438 \mathfrak{t}^{18} + 43420 \mathfrak{t}^{19} + \cdots + 2 \mathfrak{t}^{38}).
\end{align}
}

\noindent \underline{Diagonal $2$-point functions}
{\fontsize{9pt}{1pt}\selectfont
\begin{align}
&\langle
W_{\ydiagram{1,1}}
W_{\overline{\ydiagram{1,1}}}
\rangle^{U(5)\text{ADHM-}[1](H)}(\mathfrak{t})
=
\frac{
1 + 3 \mathfrak{t}^2 + 6 \mathfrak{t}^3 + 11 \mathfrak{t}^4 + 16 \mathfrak{t}^5 + 22 \mathfrak{t}^6 + 22 \mathfrak{t}^7 + 22 \mathfrak{t}^8 + 12 \mathfrak{t}^9 + 5 \mathfrak{t}^{10}
}{
(1 - \mathfrak{t})^4 (1 - \mathfrak{t}^2)^4 (1 - \mathfrak{t}^3)^2
}.
\end{align}
}

\subsubsection{$U(6)$}
\noindent \underline{1-point functions}
{\fontsize{9pt}{1pt}\selectfont
\begin{align}
&\langle W_{\ydiagram{1,1}}\rangle^{U(6)\text{ADHM-}[1](H)}(\mathfrak{t})
=
\frac{
2 \mathfrak{t}^{3} 
(1 + \mathfrak{t}^{2} + 2 \mathfrak{t}^{3} + 4 \mathfrak{t}^{4} + 2 \mathfrak{t}^{5} + 4 \mathfrak{t}^{6} + 2 \mathfrak{t}^{7} + 4 \mathfrak{t}^{8} + 2 \mathfrak{t}^{9} + \mathfrak{t}^{10} + \mathfrak{t}^{12})
}{
(1 - \mathfrak{t})^4 (1 - \mathfrak{t}^2)^4 (1 - \mathfrak{t}^3)^2 (1 - \mathfrak{t}^4)^2
},\\
&\langle W_{\ydiagram{1,1,1}}\rangle^{U(6)\text{ADHM-}[1](H)}(\mathfrak{t})
=
\frac{
\mathfrak{t}^{5} (1 + 4 \mathfrak{t} + \mathfrak{t}^{2}) (1 + \mathfrak{t}^{2} + 2 \mathfrak{t}^{3} + \mathfrak{t}^{4} + \mathfrak{t}^{6})
}{
(1 - \mathfrak{t})^4 (1 - \mathfrak{t}^2)^4 (1 - \mathfrak{t}^3)^4
},\\
&\langle W_{\ydiagram{1,1,1,1}}\rangle^{U(6)\text{ADHM-}[1](H)}(\mathfrak{t})
=
\frac{
\mathfrak{t}^{8} (3 + 4 \mathfrak{t} + 10 \mathfrak{t}^{2} + 4 \mathfrak{t}^{3} + 3 \mathfrak{t}^{4})
}{
(1 - \mathfrak{t})^4 (1 - \mathfrak{t}^2)^5 (1 - \mathfrak{t}^3)^2 (1 - \mathfrak{t}^4)
},\\
&\langle W_{\ydiagram{1,1,1,1,1}}\rangle^{U(6)\text{ADHM-}[1](H)}(\mathfrak{t})
=
\frac{
\mathfrak{t}^{11} 
(3 + 8 \mathfrak{t} + 15 \mathfrak{t}^{2} + 20 \mathfrak{t}^{3} + 28 \mathfrak{t}^{4} + 20 \mathfrak{t}^{5} + 15 \mathfrak{t}^{6} + 8 \mathfrak{t}^{7} + 3 \mathfrak{t}^{8})
}{
(1 - \mathfrak{t})^4 (1 - \mathfrak{t}^2)^2 (1 - \mathfrak{t}^3)^2 (1 - \mathfrak{t}^4)^2 (1 - \mathfrak{t}^5)^2
},\\
&\langle W_{\ydiagram{1,1,1,1,1,1}}\rangle^{U(6)\text{ADHM-}[1](H)}(\mathfrak{t})
=
\frac{
\mathfrak{t}^{14}
(1 + 10 \mathfrak{t} + 16 \mathfrak{t}^{2} + 40 \mathfrak{t}^{3} + 58 \mathfrak{t}^{4} + 84 \mathfrak{t}^{5} + 93 \mathfrak{t}^{6} + 116 \mathfrak{t}^{7} + \cdots + \mathfrak{t}^{14})
}{
(1 - \mathfrak{t})^2 (1 - \mathfrak{t}^2)^2 (1 - \mathfrak{t}^3)^2 (1 - \mathfrak{t}^4)^2 (1 - \mathfrak{t}^5)^2 (1 - \mathfrak{t}^6)^2
}.
\end{align}
}

\noindent \underline{Diagonal $2$-point functions}
{\fontsize{9pt}{1pt}\selectfont
\begin{align}
&\langle
W_{\ydiagram{1,1}}
W_{\overline{\ydiagram{1,1}}}
\rangle^{U(6)\text{ADHM-}[1](H)}(\mathfrak{t})\nonumber \\
&=
\frac{
(1 + 2 \mathfrak{t}^2 + 6 \mathfrak{t}^3 + 12 \mathfrak{t}^4 + 16 \mathfrak{t}^5 + 31 \mathfrak{t}^6 + 40 \mathfrak{t}^7 + 56 \mathfrak{t}^8 + 56 \mathfrak{t}^9 + 56 \mathfrak{t}^{10} + 38 \mathfrak{t}^{11} + 29 \mathfrak{t}^{12} + 12 \mathfrak{t}^{13} + 5 \mathfrak{t}^{14})
}{
(1 - \mathfrak{t})^4 (1 - \mathfrak{t}^2)^5 (1 - \mathfrak{t}^3)^2 (1 - \mathfrak{t}^4)
}
,\\
&\langle
W_{\ydiagram{1,1,1}}
W_{\overline{\ydiagram{1,1,1}}}
\rangle^{U(6)\text{ADHM-}[1](H)}(\mathfrak{t})\nonumber \\
&=
\frac{
(1 + 3 \mathfrak{t}^2 + 8 \mathfrak{t}^3 + 15 \mathfrak{t}^4 + 28 \mathfrak{t}^5 + 52 \mathfrak{t}^6 + 72 \mathfrak{t}^7 + 101 \mathfrak{t}^8 + 120 \mathfrak{t}^9 + 119 \mathfrak{t}^{10} + 96 \mathfrak{t}^{11} + 71 \mathfrak{t}^{12} + 28 \mathfrak{t}^{13} + 6 \mathfrak{t}^{14})
}{
(1 - \mathfrak{t})^4 (1 - \mathfrak{t}^2)^4 (1 - \mathfrak{t}^3)^4
}.
\end{align}
}

\subsubsection{$U(7)$}
\noindent \underline{Diagonal $2$-point functions}
{\fontsize{9pt}{1pt}\selectfont
\begin{align}
&\langle
W_{\ydiagram{1,1}}
W_{\overline{\ydiagram{1,1}}}
\rangle^{U(7)\text{ADHM-}[1](H)}(\mathfrak{t})\nonumber \\
&=
\frac{1}{
(1 - \mathfrak{t})^4 (1 - \mathfrak{t}^2)^4 (1 - \mathfrak{t}^3)^2 (1 - \mathfrak{t}^4)^2 (1 - \mathfrak{t}^5)^2
}
(
1 + 3 \mathfrak{t}^{2} + 6 \mathfrak{t}^{3} + 14 \mathfrak{t}^{4} + 26 \mathfrak{t}^{5} + 51 \mathfrak{t}^{6} + 86 \mathfrak{t}^{7} + 143 \mathfrak{t}^{8} + 212 \mathfrak{t}^{9}\nonumber \\
&\quad + 306 \mathfrak{t}^{10} + 394 \mathfrak{t}^{11} + 488 \mathfrak{t}^{12} + 540 \mathfrak{t}^{13} + 573 \mathfrak{t}^{14} + 552 \mathfrak{t}^{15} + 495 \mathfrak{t}^{16} + 400 \mathfrak{t}^{17} + 305 \mathfrak{t}^{18} + 202 \mathfrak{t}^{19} + 126 \mathfrak{t}^{20} + 66 \mathfrak{t}^{21}\nonumber \\
&\quad + 34 \mathfrak{t}^{22} + 12 \mathfrak{t}^{23} + 5 \mathfrak{t}^{24}
),\\
&\langle
W_{\ydiagram{1,1,1}}
W_{\overline{\ydiagram{1,1,1}}}
\rangle^{U(7)\text{ADHM-}[1](H)}(\mathfrak{t})\nonumber \\
&=
\frac{1}{
(1 - \mathfrak{t})^4 (1 - \mathfrak{t}^2)^4 (1 - \mathfrak{t}^3)^4 (1 - \mathfrak{t}^4)^2
}
(1 + 3 \mathfrak{t}^{2} + 8 \mathfrak{t}^{3} + 18 \mathfrak{t}^{4} + 34 \mathfrak{t}^{5} + 73 \mathfrak{t}^{6} + 124 \mathfrak{t}^{7} + 210 \mathfrak{t}^{8} + 316 \mathfrak{t}^{9} + 449 \mathfrak{t}^{10}\nonumber \\
&\quad + 564 \mathfrak{t}^{11} + 666 \mathfrak{t}^{12} + 682 \mathfrak{t}^{13} + 645 \mathfrak{t}^{14} + 526 \mathfrak{t}^{15} + 371 \mathfrak{t}^{16} + 212 \mathfrak{t}^{17} + 102 \mathfrak{t}^{18} + 30 \mathfrak{t}^{19} + 6 \mathfrak{t}^{20}).
\end{align}
}

\subsubsection{$U(8)$}
\noindent \underline{Diagonal $2$-point functions}
{\fontsize{9pt}{1pt}\selectfont
\begin{align}
&\langle
W_{\ydiagram{1,1}}
W_{\overline{\ydiagram{1,1}}}
\rangle^{U(8)\text{ADHM-}[1](H)}(\mathfrak{t})
=
\frac{
(1 + \mathfrak{t})^2
}{
(1 - \mathfrak{t})^2 (1 - \mathfrak{t}^2)^7 (1 - \mathfrak{t}^3)^2 (1 - \mathfrak{t}^4) (1 - \mathfrak{t}^5)^2 (1 - \mathfrak{t}^6)^2
}
(1 + 2 \mathfrak{t}^{2} + 6 \mathfrak{t}^{3} + 12 \mathfrak{t}^{4} + 20 \mathfrak{t}^{5}\nonumber \\
&\quad + 44 \mathfrak{t}^{6} + 76 \mathfrak{t}^{7} + 138 \mathfrak{t}^{8} + 214 \mathfrak{t}^{9} + 342 \mathfrak{t}^{10} + 496 \mathfrak{t}^{11} + 716 \mathfrak{t}^{12} + 928 \mathfrak{t}^{13} + 1195 \mathfrak{t}^{14} + 1424 \mathfrak{t}^{15} + 1655 \mathfrak{t}^{16} + 1776 \mathfrak{t}^{17}\nonumber \\
&\quad + 1865 \mathfrak{t}^{18} + 1804 \mathfrak{t}^{19} + 1707 \mathfrak{t}^{20} + 1486 \mathfrak{t}^{21} + 1262 \mathfrak{t}^{22} + 980 \mathfrak{t}^{23} + 745 \mathfrak{t}^{24} + 506 \mathfrak{t}^{25} + 341 \mathfrak{t}^{26} + 202 \mathfrak{t}^{27} + 117 \mathfrak{t}^{28}\nonumber \\
&\quad + 54 \mathfrak{t}^{29} + 29 \mathfrak{t}^{30} + 12 \mathfrak{t}^{31} + 5 \mathfrak{t}^{32}),\\
&\langle
W_{\ydiagram{1,1,1}}
W_{\overline{\ydiagram{1,1,1}}}
\rangle^{U(8)\text{ADHM-}[1](H)}(\mathfrak{t})
=
\frac{1}{
(1 - \mathfrak{t})^4 (1 - \mathfrak{t}^2)^4 (1 - \mathfrak{t}^3)^4 (1 - \mathfrak{t}^4)^2 (1 - \mathfrak{t}^5)^2
}
(1 + 3 \mathfrak{t}^{2} + 8 \mathfrak{t}^{3} + 18 \mathfrak{t}^{4} + 38 \mathfrak{t}^{5} + 81 \mathfrak{t}^{6}\nonumber \\
&\quad + 154 \mathfrak{t}^{7} + 289 \mathfrak{t}^{8} + 492 \mathfrak{t}^{9} + 810 \mathfrak{t}^{10} + 1242 \mathfrak{t}^{11} + 1807 \mathfrak{t}^{12} + 2448 \mathfrak{t}^{13} + 3147 \mathfrak{t}^{14} + 3762 \mathfrak{t}^{15} + 4229 \mathfrak{t}^{16} + 4410 \mathfrak{t}^{17}\nonumber \\
&\quad + 4299 \mathfrak{t}^{18} + 3846 \mathfrak{t}^{19} + 3194 \mathfrak{t}^{20} + 2406 \mathfrak{t}^{21} + 1659 \mathfrak{t}^{22} + 1018 \mathfrak{t}^{23} + 560 \mathfrak{t}^{24} + 258 \mathfrak{t}^{25} + 105 \mathfrak{t}^{26} + 30 \mathfrak{t}^{27} + 6 \mathfrak{t}^{28}), \\
&\langle
W_{\ydiagram{1,1,1,1}}
W_{\overline{\ydiagram{1,1,1,1}}}
\rangle^{U(8)\text{ADHM-}[1](H)}(\mathfrak{t})
=
\frac{1}{
(1 - \mathfrak{t})^4 (1 - \mathfrak{t}^2)^4 (1 - \mathfrak{t}^3)^4 (1 - \mathfrak{t}^4)^4
}
(1 + 3 \mathfrak{t}^{2} + 8 \mathfrak{t}^{3} + 21 \mathfrak{t}^{4} + 40 \mathfrak{t}^{5} + 94 \mathfrak{t}^{6} + 176 \mathfrak{t}^{7}\nonumber \\
&\quad + 344 \mathfrak{t}^{8} + 580 \mathfrak{t}^{9} + 982 \mathfrak{t}^{10} + 1488 \mathfrak{t}^{11} + 2204 \mathfrak{t}^{12} + 2928 \mathfrak{t}^{13} + 3762 \mathfrak{t}^{14} + 4364 \mathfrak{t}^{15} + 4809 \mathfrak{t}^{16} + 4740 \mathfrak{t}^{17} + 4385 \mathfrak{t}^{18}\nonumber \\
&\quad + 3552 \mathfrak{t}^{19} + 2667 \mathfrak{t}^{20} + 1664 \mathfrak{t}^{21} + 928 \mathfrak{t}^{22} + 396 \mathfrak{t}^{23} + 146 \mathfrak{t}^{24} + 32 \mathfrak{t}^{25} + 6 \mathfrak{t}^{26}).
\end{align}
}

\subsubsection{$U(9)$}
\noindent \underline{Diagonal $2$-point functions}
{\fontsize{9pt}{1pt}\selectfont
\begin{align}
&\langle
W_{\ydiagram{1,1}}
W_{\overline{\ydiagram{1,1}}}
\rangle^{U(9)\text{ADHM-}[1](H)}(\mathfrak{t})
=
\frac{
(1 + t)^2
}{
(1 - \mathfrak{t})^2 (1 - \mathfrak{t}^2)^6 (1 - \mathfrak{t}^3)^2 (1 - \mathfrak{t}^4)^2 (1 - \mathfrak{t}^5)^2 (1 - \mathfrak{t}^6)^2 (1 - \mathfrak{t}^7)^2
}
(1 + 3 \mathfrak{t}^{2} + 6 \mathfrak{t}^{3} + 14 \mathfrak{t}^{4}\nonumber \\
&\quad + 26 \mathfrak{t}^{5} + 56 \mathfrak{t}^{6} + 102 \mathfrak{t}^{7} + 194 \mathfrak{t}^{8} + 338 \mathfrak{t}^{9} + 580 \mathfrak{t}^{10} + 942 \mathfrak{t}^{11} + 1505 \mathfrak{t}^{12} + 2280 \mathfrak{t}^{13} + 3384 \mathfrak{t}^{14} + 4800 \mathfrak{t}^{15} + 6625 \mathfrak{t}^{16}\nonumber \\
&\quad  + 8786 \mathfrak{t}^{17} + 11335 \mathfrak{t}^{18} + 14078 \mathfrak{t}^{19} + 16988 \mathfrak{t}^{20} + 19770 \mathfrak{t}^{21} + 22331 \mathfrak{t}^{22} + 24362 \mathfrak{t}^{23} + 25803 \mathfrak{t}^{24} + 26374 \mathfrak{t}^{25} + 26174 \mathfrak{t}^{26}\nonumber \\
&\quad  + 25068 \mathfrak{t}^{27} + 23290 \mathfrak{t}^{28} + 20864 \mathfrak{t}^{29} + 18119 \mathfrak{t}^{30} + 15142 \mathfrak{t}^{31} + 12262 \mathfrak{t}^{32} + 9530 \mathfrak{t}^{33} + 7166 \mathfrak{t}^{34} + 5152 \mathfrak{t}^{35} + 3583 \mathfrak{t}^{36}\nonumber \\
&\quad  + 2364 \mathfrak{t}^{37} + 1509 \mathfrak{t}^{38} + 906 \mathfrak{t}^{39} + 525 \mathfrak{t}^{40} + 280 \mathfrak{t}^{41} + 146 \mathfrak{t}^{42} + 66 \mathfrak{t}^{43} + 34 \mathfrak{t}^{44} + 12 \mathfrak{t}^{45} + 5 \mathfrak{t}^{46}), \\
&\langle
W_{\ydiagram{1,1,1}}
W_{\overline{\ydiagram{1,1,1}}}
\rangle^{U(9)\text{ADHM-}[1](H)}(\mathfrak{t})
=
\frac{
(1 + \mathfrak{t})^2
}{
(1 - \mathfrak{t})^2 (1 - \mathfrak{t}^2)^6 (1 - \mathfrak{t}^3)^4 (1 - \mathfrak{t}^4)^2 (1 - \mathfrak{t}^5)^2 (1 - \mathfrak{t}^6)^2
}
(1 + 3 \mathfrak{t}^{2} + 8 \mathfrak{t}^{3} + 18 \mathfrak{t}^{4} + 38 \mathfrak{t}^{5}\nonumber \\
&\quad + 86 \mathfrak{t}^{6} + 164 \mathfrak{t}^{7} + 328 \mathfrak{t}^{8} + 598 \mathfrak{t}^{9} + 1060 \mathfrak{t}^{10} + 1780 \mathfrak{t}^{11} + 2910 \mathfrak{t}^{12} + 4472 \mathfrak{t}^{13} + 6676 \mathfrak{t}^{14} + 9474 \mathfrak{t}^{15} + 12914 \mathfrak{t}^{16}\nonumber \\
&\quad + 16788 \mathfrak{t}^{17} + 20986 \mathfrak{t}^{18} + 24932 \mathfrak{t}^{19} + 28438 \mathfrak{t}^{20} + 30890 \mathfrak{t}^{21} + 32054 \mathfrak{t}^{22} + 31646 \mathfrak{t}^{23} + 29858 \mathfrak{t}^{24} + 26654 \mathfrak{t}^{25}\nonumber \\
&\quad + 22666 \mathfrak{t}^{26} + 18222 \mathfrak{t}^{27} + 13851 \mathfrak{t}^{28} + 9874 \mathfrak{t}^{29} + 6650 \mathfrak{t}^{30} + 4136 \mathfrak{t}^{31} + 2401 \mathfrak{t}^{32} + 1280 \mathfrak{t}^{33} + 621 \mathfrak{t}^{34} + 262 \mathfrak{t}^{35} + 105 \mathfrak{t}^{36}\nonumber \\
&\quad + 30 \mathfrak{t}^{37} + 6 \mathfrak{t}^{38}),\\
&\langle
W_{\ydiagram{1,1,1,1}}
W_{\overline{\ydiagram{1,1,1,1}}}
\rangle^{U(9)\text{ADHM-}[1](H)}(\mathfrak{t})
=
\frac{1}{
(1 - \mathfrak{t})^4 (1 - \mathfrak{t}^2)^4 (1 - \mathfrak{t}^3)^4 (1 - \mathfrak{t}^4)^4 (1 - \mathfrak{t}^5)^2
}
(
1 + 3 \mathfrak{t}^{2} + 8 \mathfrak{t}^{3} + 21 \mathfrak{t}^{4} + 44 \mathfrak{t}^{5} + 102 \mathfrak{t}^{6}\nonumber \\
&\quad + 206 \mathfrak{t}^{7} + 423 \mathfrak{t}^{8} + 788 \mathfrak{t}^{9} + 1435 \mathfrak{t}^{10} + 2460 \mathfrak{t}^{11} + 4073 \mathfrak{t}^{12} + 6352 \mathfrak{t}^{13} + 9522 \mathfrak{t}^{14} + 13506 \mathfrak{t}^{15} + 18308 \mathfrak{t}^{16} + 23470 \mathfrak{t}^{17}\nonumber \\
&\quad + 28630 \mathfrak{t}^{18} + 32936 \mathfrak{t}^{19} + 35879 \mathfrak{t}^{20} + 36694 \mathfrak{t}^{21} + 35299 \mathfrak{t}^{22} + 31670 \mathfrak{t}^{23} + 26511 \mathfrak{t}^{24} + 20446 \mathfrak{t}^{25} + 14542 \mathfrak{t}^{26}\nonumber \\
&\quad  + 9372 \mathfrak{t}^{27}+ 5467 \mathfrak{t}^{28} + 2798 \mathfrak{t}^{29} + 1261 \mathfrak{t}^{30} + 466 \mathfrak{t}^{31} + 149 \mathfrak{t}^{32} + 32 \mathfrak{t}^{33} + 6 \mathfrak{t}^{34}).
\end{align}
}

\subsection{Unflavored Higgs correlators of charged Wilson lines}
\subsubsection{$U(2)$}
\begin{align}
%
&\langle W_{2} \rangle^{\textrm{$U(2)$ ADHM-$[2] (H)$}}(\mathfrak{t})\nonumber \\
&=\frac{\mathfrak{t}^2(2+4\mathfrak{t}+17\mathfrak{t}^2+25\mathfrak{t}^3+26\mathfrak{t}^4
+15\mathfrak{t}^5+8\mathfrak{t}^6-\mathfrak{t}^7-4\mathfrak{t}^8-3\mathfrak{t}^9-\mathfrak{t}^{10})}
{(1-\mathfrak{t})(1-\mathfrak{t}^2)^4(1-\mathfrak{t}^3)^3}, \\
&\langle W_{3} \rangle^{\textrm{$U(2)$ ADHM-$[2] (H)$}}(\mathfrak{t})
=\frac{2\mathfrak{t}^3(1-\mathfrak{t}+7\mathfrak{t}^2+6\mathfrak{t}^3+4\mathfrak{t}^4
-2\mathfrak{t}^5-2\mathfrak{t}^6-2\mathfrak{t}^7)}
{(1-\mathfrak{t})^3(1-\mathfrak{t}^2)^2(1-\mathfrak{t}^3)^3}.
\end{align}

\bibliographystyle{utphys}
\bibliography{ref}

\end{document}